\documentclass[twocolumn, dvipdfmx]{aastex631_rev}

\usepackage{amsmath}
\usepackage{color}
\usepackage{comment}
\usepackage{url}

\usepackage{graphicx}

\received{May 10, 2022}
\revised{July 25, 2022}
\accepted{August 7, 2022}
\submitjournal{ApJ}

\graphicspath{{./}{figures/}}

\begin{document}

%

%\title{Template \aastex Article with Examples: v6.31\footnote{Released on March, 1st, 2021}}

%\title{The Molecular Composition of the Protosolar Disk Midplanes with Shadow Structures beyond the Water Snowline}
\title{The Molecular Composition of Shadowed Protosolar Disk Midplanes beyond the Water Snowline}

\correspondingauthor{Shota Notsu}
\email{shota.notsu@riken.jp}

\author[0000-0003-2493-912X]{Shota Notsu}
%\author[]{Shota Notsu}[0000-0003-2493-912X]
%%
%\author{Shota Notsu}
\altaffiliation{RIKEN Special Postdoctoral Researcher (SPDR, Fellow)}
\affiliation{Star and Planet Formation Laboratory, RIKEN Cluster for Pioneering Research, 2-1 Hirosawa, Wako, Saitama 351-0198, Japan}

%%\email{shota.notsu@riken.jp}

\author[0000-0003-3290-6758]{Kazumasa Ohno}
%\author{Kazumasa Ohno}
\affiliation{Department of Astronomy and Astrophysics, University of California, Santa Cruz, 1156 High St, Santa Cruz, CA 95064, USA}

\author[0000-0003-4902-222X]{Takahiro Ueda}
%\author{Takahiro Ueda}
\affiliation{Max Planck Institute for Astronomy, K{\"o}nigstuhl 17, D-69117 Heidelberg, Germany}
\affiliation{National Astronomical Observatory of Japan, 2-21-1 Osawa, Mitaka, Tokyo 181-8588, Japan}

\author[0000-0001-6078-786X]{Catherine Walsh}
%\author{Catherine Walsh}
\affiliation{School of Physics and Astronomy, University of Leeds, Leeds, LS2 9JT, UK}

\author[0000-0002-8743-1318]{Christian Eistrup}
%\author{Christian Eistrup}
\affiliation{Max Planck Institute for Astronomy, K{\"o}nigstuhl 17, D-69117 Heidelberg, Germany}

\author[0000-0002-7058-7682]{Hideko Nomura}
%\author{Hideko Nomura}
\affiliation{National Astronomical Observatory of Japan, 2-21-1 Osawa, Mitaka, Tokyo 181-8588, Japan}

\begin{abstract}
The disk midplane temperature is potentially affected by the dust traps/rings.
%%.
The dust depletion beyond the water snowline will cast a shadow.
%%, 
%%%
%%%
%%%
%%%
In this study, we adopt a detailed gas-grain chemical reaction network, and
investigate the radial gas and ice abundance distributions of dominant carbon-, oxygen-, and nitrogen-bearing molecules in disks with shadow structures beyond the water snowline around a protosolar-like star.
In shadowed disks, the dust grains at $r\sim3-8$ au are predicted to have more than $\sim5-10$ times amounts of
ices of organic molecules such as H$_{2}$CO, CH$_{3}$OH, and NH$_{2}$CHO, saturated hydrocarbon ices such as CH$_{4}$ and C$_{2}$H$_{6}$, in addition to H$_{2}$O, CO, CO$_{2}$, NH$_{3}$, N$_{2}$, and HCN ices, compared with those in non-shadowed disks.
In the shadowed regions, we find that hydrogenation (especially of CO ice) is the dominant formation mechanism of complex organic molecules.
%%, 
%
The gas-phase N/O ratios show much larger spatial variations than the gas-phase C/O ratios, thus the N/O ratio is predicted to be a useful tracer of the shadowed region. 
N$_{2}$H$^{+}$ line emission is a potential tracer of the shadowed region.
We conclude that a shadowed region allows the recondensation of key volatiles onto dust grains, provides a region of chemical enrichment of ices that is much closer to the star than within a non-shadowed disk, and may explain to some degree the trapping of O$_{2}$ ice in dust grains that formed comet 67P/Churyumov-Gerasimenko.
We discuss that, if formed in a shadowed disk, Jupiter does not need to have migrated vast distances.

\end{abstract}

%%\keywords{Classical Novae (251) --- Ultraviolet astronomy(1736) --- History of astronomy(1868) --- Interdisciplinary astronomy(804)}


%%%20220223 20:35
%%
\section{Introduction}\label{sec:1}
%% \label{sec:intro}
%%%
Protoplanetary disks are composed of bare and ice-coated refractory grains, and gas, which are the ingredients of planetesimals and planets (e.g., \citealt{Williams2011, Oberg2021}).
Disks are chemically active environments which create simple and complex molecules, including organic material.
The physical and chemical conditions of protoplanetary disks determine the properties of forming planets, including mass and chemical composition  (e.g., \citealt{Oberg2011, Oberg2016, Oberg2021, Madhusudhan2014, Pontoppidan2014, Eistrup2016, Eistrup2018, Cridland2017, Booth2019, Notsu2020, Schneider2021, Turrini2021, Molliere2022}).
In addition, molecular abundances in comets and other primitive small bodies in our solar system are determined by the combination of 
chemical evolution in the protosolar disk and inheritance from molecular clouds (e.g., \citealt{Mumma2011, Caselli2012, Walsh2014, Eistrup2016, Eistrup2018, Altwegg2019, Drozdovskaya2019, Oberg2021}).
%%%
%%%
\begin{comment}
\end{comment}
%%%
%%%
\\ \\
The disk thermal structure plays a predominant role in controlling the disk chemical structure.
The midplane temperature $T(r)$ in protoplanetary disks is determined by heating due to
viscous dissipation (only in the innermost region) and the stellar irradiation grazing the disk surface (e.g., \citealt{Oka2011, Mori2021}).
%%%
Since the amount of heating per unit volume from both of these sources decreases with increasing disk radius $r$, 
in a smooth (non-shadowed) disk the temperature monotonically decreases as the disk radius increases \citep{Kusaka1970, Kenyon1987, Chiang1997, Oka2011}.
Inside the water snowline H$_{2}$O ice evaporates from the dust grain surfaces into the gas phase, whereas outside it is frozen out onto the dust grain surfaces \citep{Hayashi1981}. 
In addition, the cold outer disk is needed for the formation of various complex organic molecules, since the sequential hydrogenation of CO on the cold ($T(r)\sim10-30$ K) dust grain surfaces leads to the formation of H$_{2}$CO and CH$_{3}$OH, which are key feedstock molecules that produce more complex organic molecules (e.g., \citealt{Tielens1982, Watanabe2002, Cuppen2009, Fuchs2009, Drozdovskaya2014, Furuya2014, Walsh2014, Walsh2016, Chuang2016, Bosman2018, Aikawa2020}).
%%%
%%
\\ \\
Under the assumption of classical monotonically decreasing disk temperature profiles, previous studies have discussed the disk radial locations of formation of planets.
Here we introduce the examples of Jupiter in the Solar System.
Recent extensive observations towards Jupiter's atmosphere (by the Galileo probe, Cassini, and Juno spacecraft) have revealed uniform enrichment patterns of the elemental abundances
from protosolar abundances by a factor of two to four (see e.g., \citealt{Atreya2020, Li2020}).
The heavy element enrichment has been proposed to originate from planetesimals and/or pebbles dissolved in the atmospheres (e.g., \citealt{Pollack1986, Iaroslavitz2007, Hori2011, Venturini2016, Shibata2022}) and/or core erosion (e.g., \citealt{Moll2017}).
It has been noted that the abundances of highly volatile elements, such as N and noble gas elements, are comparable to the other elemental abundances.
\citet{Owen1999} suggested that this uniform enrichment originates from planetesimals formed in extremely cold ($T(r)<30$ K) environments, where nitrogen and noble gases can freeze.
\citet{Oberg2019} and \citet{Bosman2019} suggested that such uniform enrichment could be explained if the Jovian core had formed at $r>30$ au, where the disk temperature is extremely cold ($T(r)<30$ K, outside the N$_{2}$ snowline) under the classical monotonically decreasing disk temperature profile.
However, according to theoretical studies on core formation and migration \citep{Bitsch2019},
the occurrence rate of the migration of a core from $r>30$ au to 5 au is extremely low, and the core migration time scale is of the order of a few Myr 
even if such a migration occurs. 
\citet{Kruijer2017, Kruijer2020} discussed that Jupiter's core may be formed within 1 Myr to demarcate the inner and outer Solar System, on the basis of the isotope analyses of meteorites.
%%
%%%
\\ \\
However, the disk temperature profile may not have a monotonically decreasing distribution if a disk has substructures, as found
in recent observations of radial gas and dust distributions on many disks (e.g., \citealt{Isella2018, Andrews2020, Oberg2021MAPS}).
Several previous studies have suggested that a shadowed region where the direct stellar light does not reach is potentially generated, depending on the inner disk density structure.
A puffed inner disk rim can block the radiation from the central star and cast a shadow, so-called self-shadowing (e.g., \citealt{Dullemond2001, Dullemond2004, Dullemond2010, Flock2016, Ueda2017}).
\citet{Ueda2019} showed that a dust pileup at the inner edge of the MRI (magnetorotational instability) dead-zone casts a shadow behind it, producing cold regions of $T(r)\sim$50 K at $r\sim2-7$ au around Herbig Ae/Be stars.
\citet{Jang-Condell2012}, \citet{Turner2012}, \citet{IsellaTurner2018}, and \citet{Okuzumi2022} described that the outer wall of the gap produced by a giant planet receives extra starlight heating and puffs up, throwing a shadow across the disk beyond.
\citet{IsellaTurner2018} showed that the surface brightness contrast between the outer wall and shadow for the 1000 $M_{\Earth}$ ($\sim3.1$ $M_{\mathrm{Jupiter}}$) planet is an order of magnitude greater than a model neglecting the temperature disturbances. 
%%%
In addition, \citet{Ueda2021} and \citet{Okuzumi2022} discussed that a small puffed-up rim and outer shadowed region will result in the formation of ring and gap structures in disks by thermal wave instabilities (TWI).
Recently, \citet{Ohashi2022} found a steep temperature decrease outside the dust clumps at $r\sim20$ au in the disk around Class 0/I protostar L1527 IRS. They suggest that the dust clumps create a shadowed region outside, resulting in the sudden drop in temperature.
\\ \\
A dust pileup at the water snowline may also cast a shadow behind it.
The dust surface density just inside the water snowline can be enhanced by orders of magnitude. This is because efficient fragmentation slows the radial drift of silicate grains, leading to enhanced production of small grains within the water snowline (e.g., \citealt{Birnstiel2010, Banzatti2015, Cieza2016, Pinilla2017, Muller2021}).
%%%
%%.
Such fragmented dust, with higher surface density and scale height, will cast a shadow behind the water snowline and provide cold environments where volatile materials can freeze \citep{Ueda2019, Ohno2021}. 
%
%%%
%%.
\\ \\
These disk shadow structures may have a significant effect on the atmospheric composition of planets forming within.
\citet{Ohno2021} computed the temperature structure of a proto-solar (T Tauri) disk which has a shadowed region beyond the water snowline, and investigated the radial volatile distributions. %%.
They found that the vicinity of the current orbit of Jupiter ($r\sim5$ au) could be $T(r)<$30 K if the small-dust surface density decreases by a factor of $\gtrsim30$ across the water snowline.
They discussed that the shadow can cause the condensation of most volatile substances, namely N$_{2}$ and noble gases, and that the dissolution of shadowed solids can explain the elemental abundance patterns of the Jovian atmosphere even if Jupiter formed near the current orbit.
%%.
However, \citet{Ohno2021} included limited carbon, nitrogen, oxygen-bearing molecules (H$_{2}$O, CO, CO$_{2}$, C$_{2}$H$_{6}$, N$_{2}$, and NH$_{3}$ only, see also \citealt{Oberg2019}).
%%%
In addition, they fixed the total (gas$+$ice) abundances of each volatile and calculated the balance between thermal desorption and freeze-out onto dust grains within each molecular species only.
%%%
In order to investigate the effects of the shadow beyond the water snowline on disk chemical evolution including H$_{2}$CO, CH$_{3}$OH, and more complex organic molecules (such as e.g., NH$_{2}$CHO and HCOOCH$_{3}$), more detailed gas-grain chemical modeling
is needed (e.g., \citealt{Walsh2014, Walsh2015, Eistrup2016, Eistrup2018, Bosman2018, Notsu2021}).
%%
%%%
\begin{comment}
%%%
\end{comment}
\\ \\
In this study, we calculate the chemical structure of a shadowed disk midplane around a T Tauri star (a protosolar-like star), using a detailed gas-grain chemical reaction network.
We investigate the radial abundance distributions of dominant carbon-, oxygen-, and nitrogen-bearing molecules and the radial distributions of elemental abundance ratios (C/O and N/O ratios) in the gas and ice of disks with shadow structures.
We also investigate the dependance of the disk chemical structures on ionisation rates and initial abundances.
We discuss the effects of disk shadowing on chemical evolution of complex organic molecules and forming planetary atmospheres. 
We also discuss the implications of our results for the chemical composition of comets and asteroids in the Solar System.
The outline of our model calculations are explained in Section \ref{sec:2}.
The results and discussion of our calculations are described in Sections \ref{sec:3} and \ref{sec:4}, respectively. 
The conclusions are presented in Section \ref{sec:5c}.
%%%
%%%20220228 15:24
%%%
\section{Methods}\label{sec:2}
%%%
\subsection{The physical model of the protoplanetary disk midplane}\label{sec:2-1}
%%%
%%% 
%%\begin{figure}[ht!]
\begin{figure*}[hbtp]
\begin{center}
\vspace{1cm}
%\plotone{cost.pdf}
%
\includegraphics[scale=0.55]{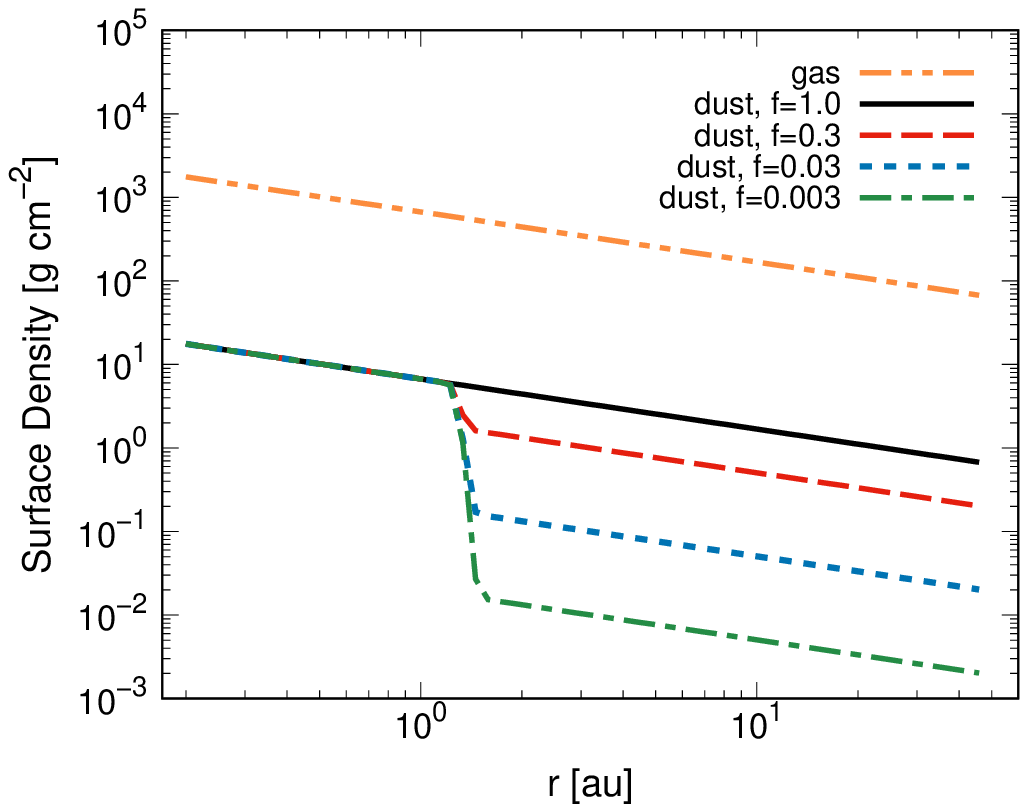}
\includegraphics[scale=0.55]{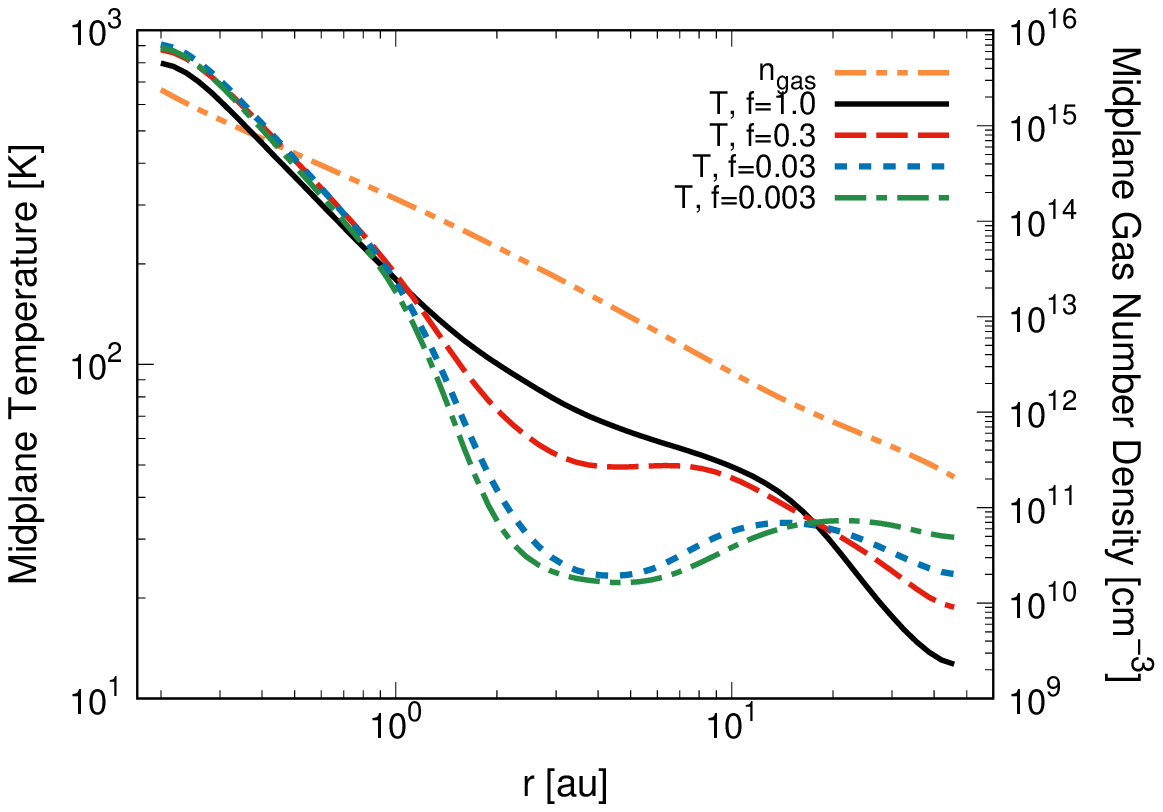}
\end{center}
\vspace{-0.2cm}
\caption{
Physical structures of the adopted disk models with shadow structures beyond the water snowline. 
[Left panel]: The radial profiles of the gas surface density $\Sigma_{\mathrm{gas}}(r)$ [g cm$^{-2}$] (orange dashed double-dotted line) and the dust surface density $\Sigma_{\mathrm{dust}}(r)$ [g cm$^{-2}$].
[Right panel]: The radial profiles of the gas number density $n_{\mathrm{gas}}(r)$ [cm$^{-3}$] (orange dashed double-dotted line) and the temperature $T(r)$ [K] in the disk midplane.
In the profiles of $\Sigma_{\mathrm{dust}}(r)$ and $T(r)$, the black solid lines, the red dashed lines, the blue dotted lines, and the green dashed dotted lines show the profiles for different values of the parameter $f$ (=1.0, 0.3, 0.03, and 0.003), respectively.
%%.
%%vspace{0.2cm}
\\ \\
}\label{Figure1}
\end{figure*}
\vspace{0.5cm}
%%%
%%%
%% 
We adopt the physical model of a steady, axisymmetric Keplerian disk around a T Tauri star (a protosolar-like star) with mass $M_{\mathrm{*}}$=1.0$M_{\bigodot}$, radius $R_{\mathrm{*}}$=2.6$R_{\bigodot}$, 
and effective temperature $T_{\mathrm{*}}$=4300K.
%
%%%
For the disk density structure, we adopted the same parameterized disk model as \citet{Ohno2021} in which the dust surface density steeply varies around the water snowline.
The radial surface density profiles
are described by the following equations, 
\begin{equation}\label{equation1}
\Sigma_{\mathrm{gas}}(r)=670\left(\frac{r}{1\ \mathrm{au}}\right)^{-3/5}\ \mathrm{g}\ \mathrm{cm}^{-2}, 
\end{equation}
\begin{equation}\label{equation2}
\Sigma_{\mathrm{dust}}(r)=
\begin{cases}
0.01\Sigma_{\mathrm{gas}}(r) & (r<R_{\mathrm{SL}}(\mathrm{H}_{2}\mathrm{O}))\\
0.01f\Sigma_{\mathrm{gas}}(r) & (r\geq R_{\mathrm{SL}}(\mathrm{H}_{2}\mathrm{O})),
\end{cases}
\end{equation}
where $r$ is the disk radius from the central star, $\Sigma_{\mathrm{gas}}$ is the gas surface density, $\Sigma_{\mathrm{dust}}$ is the surface density of the small dust ($0.1-100$ $\mu$m) which contribute to the dust opacity, $R_{\mathrm{SL}}$(H$_{2}$O)$=1.3$ au is the assumed radial position of the water snowline.
In protoplanetary disks, the maximum grain size is expected to be larger than that of the typical ISM dust due to significant dust growth. However, large grains ($\gtrsim100$ $\mu$m) have little impact on the disk temperature structure because they have small opacity and are depleted at the disk surface where stellar irradiation is absorbed.
The power index of the radial gas surface density profile ($=-3/5$) is calculated under the assumption of a disk with steady accretion and viscous heating (e.g., \citealt{Nakamoto1994, Oka2011}).
%%
%%%
We adopted various values of the parameter $f$ (=1.0 to 0.3, 0.03, and 0.003) to investigate the effects of the magnitude of shadowing on the disk thermal and chemical structures.
The actual values of the parameter $f$ depend on the efficiency of fragmentation of silicate and icy grains, which is controlled by the turbulence viscosity strength and stickiness of the dust grains (e.g., \citealt{Birnstiel2010, Banzatti2015, Pinilla2017}).
\citet{Banzatti2015} reported dust surface density variations of $f\sim$0.1, 0.001, and 0.01 around the water snowline for turbulence viscosity strengths of $\alpha_{\mathrm{t}}$\footnote{$\alpha_{\mathrm{t}}$ is a dimensionless parameter to quantify the disk viscosity \citep{Shakura1973}.}$=$$10^{-2}$, $10^{-3}$, and $10^{-4}$, respectively.
 %%
%%
%%%
\\ \\
The disk midplane temperature profiles were extracted from the results of the 2D thermal model calculation in \citet{Ohno2021} (see also Section 2 in \citealt{Ueda2019}).
In this calculation, the above radial density profiles (see Equations \ref{equation1} and \ref{equation2}) are adopted and the Monte Carlo radiative transfer code RADMC-3D \citep{Dullemond2012} is used.
Scattering is assumed to be isotropic because full-scattering treatment is computationally expensive. We expect that the temperature structure is not significantly modified even if we consider the full scattering because the temperature at the shadowed region seems to be insensitive to the presence of scattering \citep{Ueda2019}.
The vertical dust density distribution and the temperature structure are mutually dependent, thus the radiative transfer calculations were iteratively performed to obtain a self-consistent disk structure (for more details, see \citealt{Ueda2019}).
The calculations also include the internal radiative flux produced by viscous accretion, $q_{\mathrm{acc}}$, determined by the following equation \citep{Lynden-Bell1974, Pringle1981, Nomura2005},
\begin{equation}
q_{\mathrm{acc}}=\frac{9}{4}\alpha_{\mathrm{t}}\rho_{\mathrm{gas}}c_{\mathrm{s}}^{2}\Omega_{\mathrm{K}},
\end{equation}
where 
$\alpha_{\mathrm{t}}$ is set to be $3\times10^{-4}$, 
$\rho_{\mathrm{gas}}$ is the gas mass density, $c_{\mathrm{s}}$ is the sound speed, and $\Omega_{\mathrm{K}}$ is the Keplerian frequency.
The surface density coefficient ($\Sigma_{0}$=670 g cm$^{-2}$)
and $\alpha_{\mathrm{t}}$
in Equation \ref{equation1} are adjusted so that the water snowline position is outside the current Earth orbit (=1.3 au) in the calculation of the temperature profiles. 
We note that we have adopted the same dust surface density within the water snowline in all models so that the position of the water snowline is fixed. This is because in this study we focus on the effects of shadowing on disk chemical structures beyond the water snowline.
%%
%%%
\citet{Eistrup2016} adopted the lower surface density model ($\Sigma_{\mathrm{gas}}$(5.2 au)=16 g cm$^{-2}$) than our adopted model ($\Sigma_{\mathrm{gas}}$(5.2 au)=249 g cm$^{-2}$), and the water snowline position in their model is around 0.7 au.  
%% 
%%%
We assume the same dust and gas temperatures ($T_{\mathrm{dust}}=T_{\mathrm{gas}}=T$), which is a valid assumption in the dense molecular regions of the disk.
The radial profile of the midplane gas mass density $\rho_{\mathrm{0, gas}}(r)$ is given by the following equation,
\begin{equation}
\rho_{\mathrm{0, gas}}(r)=\frac{\Sigma_{\mathrm{gas}}(r)}{\sqrt{2\pi}h_{\mathrm{gas}}},
\end{equation}
where $h_{\mathrm{gas}}=c_{\mathrm{s}}/\Omega_{\mathrm{K}}$ is the gas scale height.
We adopted the DSHARP dust opacity model (\citealt{Birnstiel2018}, see also e.g., \citealt{Henning1996, Draine2003, Warren2008}), and assumed a dust-size distribution that follows a power law with an index of -3.5, similar to the MRN distribution \citep{Mathis1977}, with minimum and maximum dust radii of 0.1 and 100 $\mu$m, respectively.
We assume that larger grains have been removed from the disk atmosphere by settling and radial drift.
%%%
The bulk densities of dust grains $\rho_{\mathrm{internal}}$ are set to be 3.0 g cm$^{-3}$ for the region where the icy component is evaporated ($T>$160 K) and 1.4 g cm$^{-3}$ elsewhere.
%%%
%%%
We note that icy grains can efficiently coagulate into larger ($\gg$1 mm) dust particles and cm-size pebbles outside the water snowline (e.g., \citealt{Ros2013, Sato2016, Drazkowska2017, Pinilla2017}), but such large dust particles and pebbles have negligible contribution to the opacity compared with that of smaller dust particles \citep{Miyake1993}, and they are confined to the disk midplane.
Thus, $\Sigma_{\mathrm{dust}}$ is not necessarily the same as the total surface density of all solid components (dust grains and pebbles), for example outside the water snowline \citep{Banzatti2015}.
We note that in our disk chemical modeling (see Section \ref{sec:2-2}), we use the dust density profile in the disk midplane calculated from this small dust (0.1-100  $\mu$m) surface density profile $\Sigma_{\mathrm{dust}}$.
%%%
\\ \\
The left panel of Figure \ref{Figure1} shows $\Sigma_{\mathrm{gas}}(r)$ and $\Sigma_{\mathrm{dust}}(r)$ for different values of the parameter $f$ (=1.0, 0.3, 0.03, and 0.003).
The right panel of Figure \ref{Figure1} shows the radial profiles of the gas number density $n_{\mathrm{gas}}(r)$ and temperature $T(r)$ in the disk midplane for different values of the parameter $f$.
$n_{\mathrm{gas}}(r)$ is calculated by dividing $\rho_{\mathrm{0, gas}}(r)$ by the mean particle mass (=2.33 amu).
The values of $n_{\mathrm{gas}}$ are largest in the innermost region ($\gtrsim10^{14}$ cm$^{-3}$ at $r\lesssim1.4$ au), and decrease as the value of $r$ increases ($\lesssim10^{12}$ cm$^{-3}$ at $r\gtrsim17$ au).
The value of $T$ within the water snowline is independent of $f$, and is primally controlled by viscous heating, yielding the radial dependance of $\propto$ $r^{-9/10}$ \citep{Oka2011}.
\\ \\
As demonstrated in \citet{Ohno2021}, the depletion of small dust grains outside the water snowline casts a shadow, changing the disk midplane temperature significantly.
In the cases of smaller values of $f$, the decrease of dust surface density outside the water snowline results in reduced dust opacity, which reduces the amount of stellar radiation received per unit volume and cools the outer region compared with the case of the non-shadowed disk ($f=1.0$).
The values of $T$ at $r\sim1.3-17$ au decrease with decreasing $f$, as the shadow extends to farther-out.
Even in the case of a small variation in the dust density profile with $f=0.3$, $T$ is around 20 K lower than that in the non-shadowed disk with $f=1.0$.
For $f\geq0.03$, $T$ is $<30$ K at $r\sim3-8$ au, and $<25$ K at around the current orbit of Jupiter ($\sim5$ au).
In contrast, the values of $T$ at $r>20$ au increase with decreasing $f$, although the degree of increase is small and $T$ is $<35$ K even for $f=0.003$.
We interpret that reduced dust surface density beyond the water snowline allows the reprocessed stellar radiation to enter the outer disk midplane from the upper disk surface \citep{Cleeves2016}.
At the outer disk midplane ($r>20$ au), the heating effect by such radiation overcomes the cooling effect which reduces as $r$ increases.
%%%
The disk temperature is expected to be higher in the outer disk (beyond $r>20$ au) if we assume lower surface densities than we assumed.
The values of disk masses are determined by the density profile in the outer region of the disk.
We note that we have adopted the surface densities following a single power-law profile for the sake of simplicity (see Equations \ref{equation1} and \ref{equation2}).
The disk gas mass within 20 au and 40 au is $\sim2.2\times10^{-2}$$M_{\bigodot}$ and $\sim5.8\times10^{-2}$$M_{\bigodot}$, respectively.
\citet{Eistrup2016} adopted a disk surface density profile with an outer exponential cutoff radius of $r_{\mathrm{c}}=20$ au. 
If we include such outer exponential cutoff with $r_{\mathrm{c}}\sim20$ au in the disk surface density profile, the total disk mass within 40 au will be around half of the value without cutoff.
Recently, \citet{Trapman2022} estimated from HD and C$^{18}$O line observations that the disk masses for some typical Class II T Tauri disks (such as TW Hya) are around a few $\times10^{-2}$$M_{\bigodot}$.
\\ \\
We assumed that the water snowline position as 1.3 au, but it might be outside the current orbit of Mars (1.52 au). 
\citet{Hansen2009} discussed that Mars forms from a protoplanet that has moved by scattering from near the current orbit of Earth. 
\\ \\
In our calculations, the FUV and X-ray radiation from both the central T Tauri star and interstellar radiation fields are neglected, since 
the disk has a high surface density ($\Sigma_{\mathrm{gas}}(r)$$\sim10^{2}-10^{3}$ g cm$^{-3}$) and the disk midplane is effectively shielded (e.g., \citealt{Nomura2005, Nomura2007, Walsh2015}).
\\ \\
\citet{Cleeves2014}, \citet{Eistrup2016, Eistrup2018}, and \citet{Schwarz2018, Schwarz2019} discussed that disk chemical evolution (such as chemical processing of CO) are affected by the degree of (cosmic-ray) ionization.
In order to investigate the effects of ionization on chemistry in the shadowed disk, we adopt two values of radially constant ionization rates $\xi_{\mathrm{CR}}(r)=$$1.0\times10^{-17}$ [s$^{-1}$] and $1.0\times10^{-18}$ [s$^{-1}$] (e.g., \citealt{Cleeves2014, Eistrup2016, Eistrup2018}).
The former high ionization value includes contributions from the decay products of short-lived radionuclides (SLRs) \citep{Cleeves2013a, Cleeves2013b} and from cosmic-rays, originating externally to the disk (e.g., \citealt{Umebayashi2009, Padovani2018}),
whereas the latter low ionisation value considers contribution from the decay products of SLRs only.
We note that, in reality, the ionization rate has radial dependance, since the contribution from SLRs decreases as $r$ increases \citep{Cleeves2013b} and that from cosmic-rays increases as $r$ increases (e.g., \citealt{Umebayashi2009, Padovani2018, Aikawa2021, Seifert2021, Fujii2022}).
Using the Equation (30) of \citet{Cleeves2013b} and our assumed $\Sigma_{\mathrm{gas}}(r)$ profile, the ionization rates for the contribution from SLRs are estimated to be $\sim5\times10^{-19}$ [s$^{-1}$] at $r\sim5$ au and $\sim7\times10^{-19}$ [s$^{-1}$] at $r\sim1$ au. 
Nonetheless, we assume a constant value of the ionization rate at all radii to clarify its impacts on the disk chemical evolution.
%% 
%%%
%%%
%%%
%%%
%%
%\\ \\
%
\subsection{Calculation of the disk chemical structure}\label{sec:2-2}
We calculated the chemical evolution of the shadowed disk midplane around a T Tauri star (a protosolar-like star) using a detailed gas-grain chemical reaction network.
%%%
The chemical reaction network adopted in this work is basically similar to that in \citet{Notsu2021}.
%%%
%%%
The detailed background theories and procedures are also discussed in our previous works (e.g., \citealt{Walsh2010, Walsh2012, Walsh2014, Walsh2015}, \citealt{Heinzeller2011}, \citealt{Eistrup2016, Eistrup2018, Eistrup2019}, \citealt{Notsu2016, Notsu2017, Notsu2018, Notsu2019}), although there are some differences between the models in these studies and our adopted model which we explain in this Section.
%%%
%% 
Our model calculates the time-dependent chemical evolution at the disk midplane of each radial distance
and does not include physical mass transport in the radial direction by viscous accretion and in the vertical direction by diffusive turbulent mixing and disk winds \citep{Heinzeller2011}. 
%%
%%%
\subsubsection{Gas-phase reactions}\label{sec:2-2-1}
Our gas-phase chemistry includes the complete network from the release of the UMIST Database for Astrochemistry (UDfA), called RATE12, and is publicly available\footnote{\url{http://udfa.ajmarkwick.net}} \citep{McElroy2013}.
RATE12 includes gas-phase two-body reactions, photoionization and photodissociation, direct cosmic-ray ionization, and cosmic-ray-induced photoionization and photodissociation.
Since the FUV radiation fields from the central T Tauri star and interstellar FUV fields are neglected in our calculations (see also Section \ref{sec:2-1}), the photodissociation and photoionization by FUV radiation are effectively zero.
In the cosmic-ray-induced photoreactions, UV photons are generated internally via the interaction of secondary electrons produced by cosmic-rays with H$_{2}$ molecules \citep{Gredel1987, Gredel1989}. 
%%. 
\\ \\
As in \citet{Walsh2015}, we also added a set of three-body reactions and ``hot'' H$_{2}$ chemistry, although they are not expected to be important around the water snowline ($\sim100-200$ K).
Moreover, the gas phase chemical network is supplemented with reactions for important species, for example the CH$_{3}$O radical, which are not included in RATE12. 
The gas-phase formation and destruction reactions for these species are from the Ohio State University (OSU) network \citep{Garrod2008}.
%%%
%%%
%%%
\subsubsection{Gas-grain interactions}\label{sec:2-2-2}
%%%
\begin{comment}
\end{comment}
%%%%
\begin{table}
\caption{The binding energies for the major molecules $E_{\mathrm{des}}$($j$) and their estimated snowline positions $R_{\mathrm{SL}}$($j$) in the non-shadowed disk ($f=1.0$)}              %%% title of Table
\label{Table:1}      %%% is used to refer this table in the text
\centering                                      %%% used for centering table
\begin{tabular}{c c c}          %%% centered columns (4 columns)
\hline\hline                        %%% inserts double horizontal lines
Species $j$ & $E_{\mathrm{des}}$($j$) [K] & $R_{\mathrm{SL}}$($j$) [au]\\    %%% table heading
\hline                                   %%% inserts single horizontal line
%    H &$3.807\times10^{-5}$&$4.458\times10^{-17}$&650\\%\tablefootmark{a}\\
%    H$_{2}$ &$4.997\times10^{-1}$&$4.140\times10^{-5}$&430\\%\tablefootmark{b}\\
    NH$_{2}$CHO&5560&$\sim1.2$\\
    H$_{2}$O &4880&1.3\\%\tablefootmark{c}\\      %%% inserting body of the table
    CH$_{3}$OH &3820&$\sim1.5$\\%\tablefootmark{i}\\ 
    HCN &3610&$\sim1.5$\\%\tablefootmark{f}\\
    H$_{2}$CO &3260&$\sim1.8$\\%\tablefootmark{e}\\   
    NH$_{3}$ &2715&$\sim2.3$\\%\tablefootmark{i}\\
    C$_{2}$H$_{6}$&2320&$\sim3.2$\\
    CO$_{2}$ &2267&$\sim3.3$\\%\tablefootmark{e}\\
     CH$_{4}$ &1252&$\sim15$\\%\tablefootmark{h}\\
    O$_{2}$ &898&$\sim20$\\%\tablefootmark{e}\\    
    CO &855&$\sim22$\\%\tablefootmark{g}\\
%    O &1660&\\%\tablefootmark{d}\\
%%5
%%%
%    OH & $5.164\times10^{-8}$&$6.019\times10^{-14}$&3210\\%\tablefootmark{d}\\   
%   C &$2.571\times10^{-8}$&$1.310\times10^{-16}$&715\\%\tablefootmark{f}\\
%    HCO$^{+}$ &$3.553\times10^{-9}$& --- &---\\   
%%%
    N$_{2}$ &790&$\sim24$\\%\tablefootmark{g}\\
%%%
   H &650&$\sim34$\\%\tablefootmark{g}\\
  H$_{2}$ &430&$\sim35$\\%\tablefootmark{g}\\
%%%
%%%5
%    C$_{2}$H &$1.776\times10^{-10}$&$5.537\times10^{-17}$&1330\\%\tablefootmark{f}\\   
%%    C$_{2}$H$_{2}$ &2090&\\%\tablefootmark{i}\\   
%    N &$2.105\times10^{-5}$&$5.531\times10^{-14}$&715\\%\tablefootmark{f}\\   
%%
%    CN &1355&\\%\tablefootmark{f}\\
%%
\hline                                             %%%inserts single line
%%%
%%
\end{tabular}
%\tablebib{$^{a}$\citet{Al-Halabi2007}; $^{b}$\citet{Acharyya2014}; $^{c}$\citet{Dulieu2013}; $^{d}$\citet{He2014}, \citet{He&Vidali2014}; 
%$^{e}$\citet{Noble2012}; $^{f}$Average between \citet{Hasegawa1993} and \citet{Aikawa1996} values; $^{g}$\citet{Oberg2005}; $^{h}$\citet{Smith2016}; 
%$^{i}$Estimated from \citet{Colling2004}
%}
%
\end{table}
%%%%
As in \citet{Notsu2021}, we include the freeze-out of gas-phase molecules on dust grains, and the thermal and non-thermal desorption of molecules from dust grains \citep{Hasegawa1992, Walsh2010, Walsh2012, Notsu2016}.
%%\.
As non-thermal desorption mechanisms,
cosmic-ray-induced photodesorption, reactive desorption (see Section \ref{sec:2-2-3}), and direct cosmic-ray-induced (thermal) desorption \citep{Leger1985, Hasegawa1993, Hollenbach2009} are adopted.
We note that the direct cosmic-ray-induced (thermal) desorption has no significant impact on chemistry, since its reaction timescale is typically much longer ($\gg10^{7}$ years) than the age of Class II disk \citep{Hollenbach2009}. 
We adopt the value for the integrated cosmic-ray-induced UV photon flux as $10^{4}$ photons cm$^{-2}$ s$^{-1}$ \citep{Prasad1983, Walsh2014}.
We scale the internal UV photon flux by the cosmic-ray ionization rate.
%%%
On the basis of the adopted dust-size distribution (see Section \ref{sec:2-1}), we assume compact spherical grains with an average radius $a$ of 0.1 $\mu$m.
We note that the total dust surface area is mainly dominated by the smallest dust grains, under the MRN distribution.
%%
%%, 
The values of photodesorption yields adopted in this work, $Y_{\mathrm{des}}$($j$), are the same as those adopted in \citet{Notsu2021}.
We use experimentally determined photodesorption yields, where available (e.g., \citealt{Oberg2007, Oberg2009a, Oberg2009b, Fillion2014, Bertin2016, Cuppen2017}, see Table 1 of \citealt{Notsu2021}).
For all species without experimentally determined photodesorption yields, a value of $10^{-3}$ molecules photon$^{-1}$ is used.
As in \citet{Notsu2021}, we include the fragmentation pathways for photodesorption of water and methanol molecules (e.g., \citealt{Oberg2009b, Arasa2010, Arasa2015, Bertin2016, Cruz-Diaz2016}).
\\ \\
The sticking coefficient is assumed to be 1 for all species, except for H, which leads to H$_{2}$ formation (for more details, see Appendix B.2 of \citealt{Bosman2018}).
We adopt the same values of the binding (desorption) energies $E_{\mathrm{des}}$($j$) for all molecules as used in \citet{Notsu2021}.
In Table \ref{Table:1}, $E_{\mathrm{des}}$($j$) for several dominant molecules and their estimated snowline positions $R_{\mathrm{SL}}$($j$) in the non-shadowed disk ($f=1.0$) are listed (see also Section \ref{sec:3}).
We defined the snowline positions as the radii where the gas and ice abundances of each molecule are the same because of the balance between thermal desorption and freeze-out.
We note that the binding energies of molecules depend on the chemical compositions and physical structures (e.g., crystal or amorphous) of the ice mantles on the dust grains (e.g., \citealt{Cuppen2017, Penteado2017, Kouchi2021}, see also Section \ref{sec:4-2}).
For CO and N$_{2}$, we assume the values of pure CO and N$_{2}$ ices, respectively \citep{Oberg2005}.
%%
%%%
\subsubsection{Grain-surface reactions}\label{sec:2-2-3}
For the grain-surface reactions, we adopt the same reaction network as \citet{Notsu2021} used. 
Those reactions are mostly based on the Ohio State University (OSU) network \citep{Garrod2008}, with including some extended network for some molecules (such as CH$_{3}$OH and NH$_{2}$CHO, see Section 2.2.3 of \citealt{Notsu2021} and e.g., \citealt{Noble2015, Chuang2016}).
In addition to grain-surface two-body reactions \citep{Hasegawa1992} and reactive desorption, grain-surface cosmic-ray-induced photodissociation is also included in our calculations \citep{Garrod2008, Walsh2014, Walsh2015}. 
%%%%
%
Only the top two monolayers of the ice mantle are chemically active.
We assume that the size of the barrier to surface diffusion is $0.3\times$$E_{\mathrm{des}}$($j$) \citep{Walsh2015}.
For the lightest reactants, H and H$_{2}$, we adopt either the classical diffusion rate or the quantum tunneling rate depending on which is fastest \citep{Hasegawa1992,Bosman2018}.
For the quantum tunneling rates, we adopt a rectangular barrier of width 1.0 \AA \citep{Hasegawa1992, Bosman2018}.
%%%%
%%% 20211216 20:49
%%%
\subsubsection{Initial abundances}\label{sec:2-2-4}
%%%%
\begin{table}
\caption{Initial gas and ice molecular abundances with respect to total H nuclei assumed in the inheritance scenario.}              %%% title of Table
\label{Table:2}      %%% is used to refer this table in the text
\centering                                      %%% used for centering table
\begin{tabular}{c c c}          %%% centered columns (4 columns)
\hline\hline                        %%% inserts double horizontal lines
Species $j$ & $n_{j, \mathrm{gas}}$/$n_{\mathrm{H}}$ &  $n_{j, \mathrm{ice}}$/$n_{\mathrm{H}}$ \\%& $E_{\mathrm{des}}$($j$) [K]\\    %%% table heading
\hline                                   %%% inserts single horizontal line
    H &$3.807\times10^{-5}$&$4.458\times10^{-17}$\\%&650\tablefootmark{a}\\
    H$_{2}$ &$4.997\times10^{-1}$&$4.140\times10^{-5}$\\%&430\tablefootmark{b}\\
    He &$9.750\times10^{-2}$&$7.823\times10^{-20}$\\%&650\tablefootmark{a}\\
    H$_{2}$O &$7.080\times10^{-7}$&$1.984\times10^{-4}$\\%&4880\tablefootmark{c}\\      %%% inserting body of the table
    O &0.0& $2.073\times10^{-13}$\\%&1660\tablefootmark{d}\\
    O$_{2}$ &0.0&$4.035\times10^{-12}$\\%&898\tablefootmark{e}\\
    OH & $5.164\times10^{-8}$&$6.019\times10^{-14}$\\%&3210\tablefootmark{d}\\   
    C &$2.571\times10^{-8}$&$1.310\times10^{-16}$\\%&715\tablefootmark{f}\\
    CO &$7.532\times10^{-5}$&$2.946\times10^{-5}$\\%&855\tablefootmark{g}\\
    CO$_{2}$ &$7.487\times10^{-7}$&$2.856\times10^{-7}$\\%&2267\tablefootmark{e}\\
%    HCO$^{+}$ &$3.553\times10^{-9}$& --- \\%&---\\   
    CH$_{4}$ &$1.120\times10^{-6}$&$7.384\times10^{-6}$\\%&1252\tablefootmark{h}\\
    C$_{2}$H$_{6}$ &$1.152\times10^{-7}$&$2.417\times10^{-6}$\\
    H$_{2}$CO &$1.108\times10^{-7}$&$8.437\times10^{-6}$\\%&3260\tablefootmark{e}\\               
    CH$_{3}$OH &$3.558\times10^{-9}$&$6.027\times10^{-7}$\\%&3820\tablefootmark{i}\\
    N &$2.105\times10^{-5}$&$5.531\times10^{-14}$\\%&715\tablefootmark{f}\\   
    N$_{2}$ &$9.765\times10^{-6}$&$5.411\times10^{-6}$\\%&790\tablefootmark{g}\\
    NH$_{3}$ &$2.933\times10^{-7}$&$1.327\times10^{-5}$\\%&2715\tablefootmark{i}\\
    HCN &$7.718\times10^{-8}$&$2.772\times10^{-6}$\\%&3610\tablefootmark{f}\\
    NH$_{2}$CHO &$2.811\times10^{-10}$&$4.155\times10^{-7}$\\    
    CN &$3.016\times10^{-9}$&$1.406\times10^{-15}$\\%&1355\tablefootmark{f}\\
    NO &$3.453\times10^{-8}$&$7.510\times10^{-17}$\\
    OCN &$5.379\times10^{-10}$&$8.189\times10^{-19}$\\
    C$_{2}$H &$1.776\times10^{-10}$&$5.537\times10^{-17}$\\%&1330\tablefootmark{f}\\   
    C$_{2}$H$_{2}$ &$7.440\times10^{-8}$&$3.291\times10^{-10}$\\%&2090\tablefootmark{i}\\   
    C$_{2}$H$_{4}$ &$1.516\times10^{-8}$&$2.009\times10^{-10}$\\%&2090\tablefootmark{i}\\   
    C$_{3}$H$_{2}$ &$5.416\times10^{-8}$&$8.463\times10^{-10}$\\
%%%
    CH$_{3}$CCH &$1.514\times10^{-8}$&$5.170\times10^{-7}$\\ 
    CH$_{2}$CCH$_{2}$ &$1.698\times10^{-8}$&$5.171\times10^{-7}$\\    
    CH$_{3}$CHCH$_{2}$ &$3.296\times10^{-8}$&$1.691\times10^{-9}$\\  
%%%
    %%%
    CH$_{3}$NH$_{2}$ &5.982$\times10^{-10}$&$4.645\times10^{-7}$\\ 
    CH$_{2}$NH &4.834$\times10^{-10}$&$1.185\times10^{-10}$\\ 
    CH$_{3}$CN &$1.029\times10^{-9}$&$1.033\times10^{-8}$\\
    HC$_{3}$N &$6.041\times10^{-9}$&$8.024\times10^{-9}$\\%&3610\tablefootmark{f}\\ 
    NH$_{2}$OH & $2.862\times10^{-9}$ &$4.322\times10^{-6}$ \\
    HNCO &$8.720\times10^{-10}$ & $1.097\times10^{-9}$\\
    %%%
    CH$_{3}$CHO & $3.539\times10^{-9}$&$2.133\times10^{-9}$\\
    CH$_{3}$OCH$_{3}$ & $1.710\times10^{-13}$&$3.966\times10^{-14}$\\
    HCOOH & $3.622\times10^{-10}$&$1.061\times10^{-10}$\\
    HCOOCH$_{3}$ & $2.869\times10^{-13}$&$3.789\times10^{-14}$\\
    CH$_{3}$COOH & $8.920\times10^{-24}$&$6.340\times10^{-21}$\\
    C$_{2}$H$_{5}$OH & $2.455\times10^{-12}$&$1.251\times10^{-13}$\\
     %%
%    S &$5.835\times10^{-8}$&$3.574\times10^{-19}$\\
%    SO &$1.022\times10^{-8}$&$9.073\times10^{-11}$\\
%    SO$_{2}$ &$1.020\times10^{-9}$&$1.384\times10^{-18}$\\
%    H$_{2}$S &$2.159\times10^{-9}$&$1.569\times10^{-10}$\\ 
%    H$_{2}$CS &$5.399\times10^{-10}$&$3.516\times10^{-9}$\\ 
    %%
    %%
\hline                                             %%%inserts single line
%%%
%%
\end{tabular}
\tablecomments{These initial abundances are the same as \citet{Notsu2021} used in their chemical modeling.}
%\tablebib{aa}
%%\tablebib{$^{a}$\citet{Al-Halabi2007}; $^{b}$\citet{Acharyya2014}; $^{c}$\citet{Dulieu2013}; $^{d}$\citet{He2014}, \citet{He&Vidali2014}; 
%$^{e}$\citet{Noble2012}; $^{f}$Average between \citet{Hasegawa1993} and \citet{Aikawa1996} values; $^{g}$\citet{Oberg2005}; $^{h}$\citet{Smith2016}; 
%$^{i}$Estimated from \citet{Colling2004}
%}
%
\end{table}
%%%%
As \citet{Eistrup2016, Eistrup2018} conducted, the disk chemical evolution is calculated with two different sets of initial abundances: molecular species and atomic species.
All abundances in our calculations of this paper are with respect to the total number density of H nuclei.
%%%
For both sets of initial abundances (molecular and atomic) the elemental ratios are consistent.
The choice of these initial abundances is motivated by the following two extreme scenarios about the history of the disk midplane material \citep{Oberg2021, vanDishoeck2021}. 
%%%20210828
\\ \\
The use of molecular initial abundances assumes that the disk material is wholly inherited from the molecular cloud and the pre-stellar core from which the central star formed, implying a more quiescent mode of disk formation.
The enhanced deuterium fractionation of water and organic molecules in disks is considered to be a probe of pre-stellar inheritance, 
since efficient deuteration of water and organic molecular ices is only possible in cold and embedded pre-stellar cloud cores where UV radiation is almost completely attenuated and CO is frozen-out onto dust grains
 (e.g., \citealt{Cleeves2014Sci, Furuya2016, Furuya2017, Drozdovskaya2019, Jensen2021}).
\citet{Drozdovskaya2019} described that the relative abundances of volatile organic molecules (with respect to e.g., methanol) correlate, with some scatter, between ALMA data towards the protostar IRAS 16293-2422 B and Rosetta's in-situ monitoring data of comet 67P/Churyumov-Gerasimenko.
This result implies that the volatile composition of solar system comets may be partially inherited from the pre-stellar and protostellar phases.
%%
%%%
 %%% 
In contrast, the use of atomic initial abundances implies that full chemical reset has occurred during disk formation.
Throughout disk formation, infalling material is subject to radiation from the central star and accretion shocks, which may alter the chemical composition of material (e.g., \citealt{Visser2009, Drozdovskaya2016, Miura2017, Notsu2021}).
\citet{Oberg2021} and \citet{vanDishoeck2021} discussed that neither a complete inheritance nor a complete reset scenario can explain the results of recent disk observations (by e.g., ALMA) and the full solar system record.
They described that both must have played an important role during disk formation.
%%
%%
%%%%%%
\\ \\
As \citet{Walsh2015} and \citet{Notsu2021} adopted, 
the values of volatile elemental abundances for He, O, C, N, and S are respectively $9.75\times10^{-2}$, $3.20\times10^{-4}$, $1.40\times10^{-4}$, $7.50\times10^{-5}$, and $8.00\times10^{-8}$ relative to total hydrogen nuclei density.
%%%%%
For other elements, we use the low-metallicity
%%%%%%
elemental abundances from \citet{Graedel1982}.
In the reset scenario, we use the above elemental abundances as initial abundances of our disk chemical calculations, 
except hydrogen.
We assume that the abundances for H$_{2}$ and H are respectively $5.0\times10^{-1}$ and $5.0\times10^{-5}$.
\\ \\
Table \ref{Table:2} shows the gas and ice fractional molecular abundances with respect to total hydrogen nuclei density 
which are used as molecular initial abundances in the inheritance scenario.
In this Table \ref{Table:2}, we include dominant species and molecules which are important in disk chemical evolution.
We adopt the same molecular abundances as \citet{Notsu2021} used for the initial abundances in their chemical modeling. 
Previous chemical calculations (e.g., \citealt{Walsh2015, Eistrup2016, Eistrup2018, Drozdovskaya2016}) adopted a similar water-rich molecular abundances as initial conditions.
%%%
%%%
%%%
\section{Results}\label{sec:3}
%%%%
In Sections \ref{sec:3-1rev}-\ref{sec:3-2rev}, we investigate the results for our standard model (assuming molecular initial abundances and $\xi_{\mathrm{CR}}(r)=$$1.0\times10^{-17}$ [s$^{-1}$]), and describe the effects of a disk shadow on the radial molecular abundance distributions.
In Appendix \ref{Asec:A}, we also study the disk chemical structures for different disk ionization rates and for atomic initial abundances (see also Section \ref{sec:3-3rev}).
In addition, in Sections \ref{sec:3-1rev}-\ref{sec:3-2rev} we show the chemical abundance distributions at $t=10^{6}$ years, which is the typical age of Class II disks and consistent with previous studies (e.g., \citealt{Eistrup2016}). 
In Appendix \ref{Bsec:B}, we also investigate the time evolution of radial molecular abundance distributions (see also Section \ref{sec:3-4rev}).
\subsection{Dominant carbon-, oxygen-, and nitrogen-carriers around the current orbit of Jupiter}\label{sec:3-1rev}
%%%
\begin{figure*}[hbtp]
\begin{center}
\vspace{1cm}
%%\vspace{-0.2cm}
%\plotone{cost.pdf}
%
%%%
\includegraphics[scale=0.55]{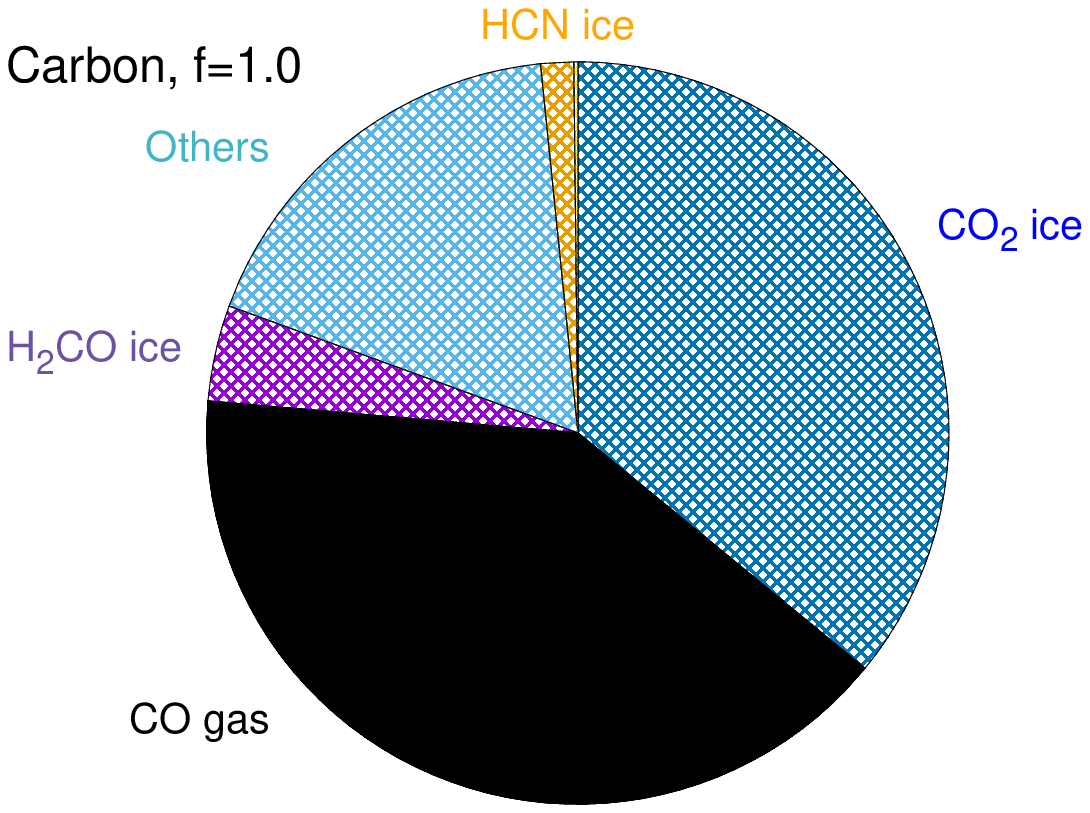}
\includegraphics[scale=0.55]{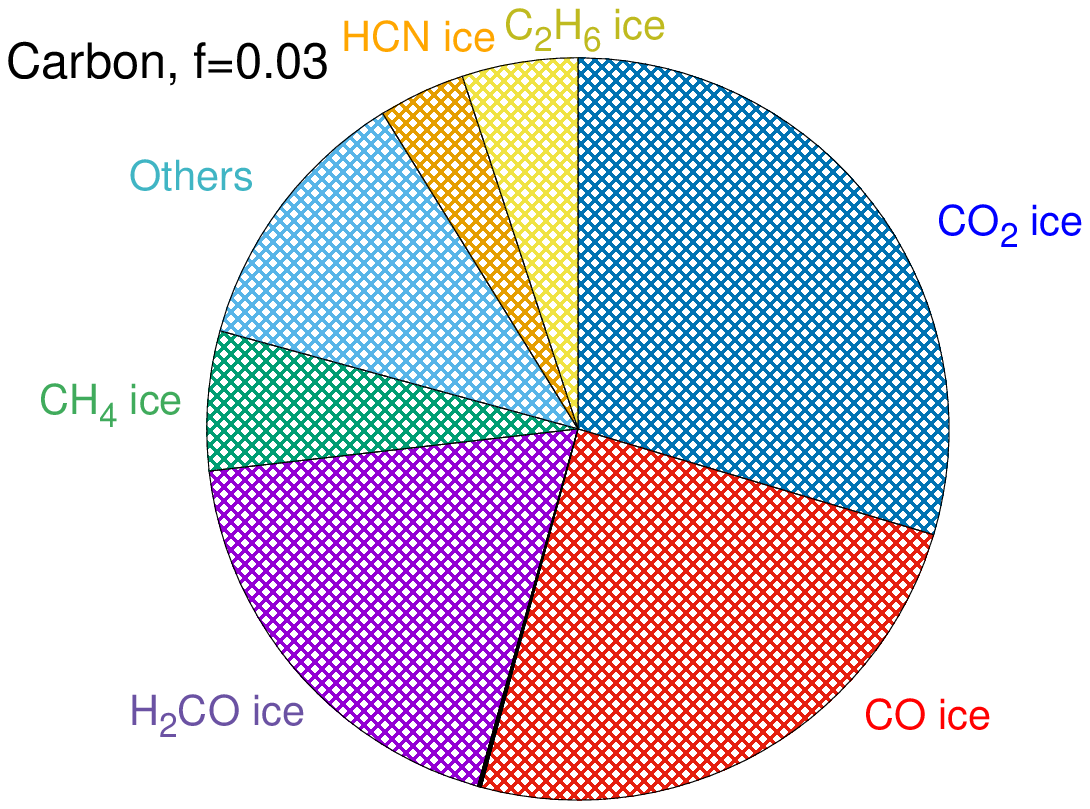}
\includegraphics[scale=0.55]{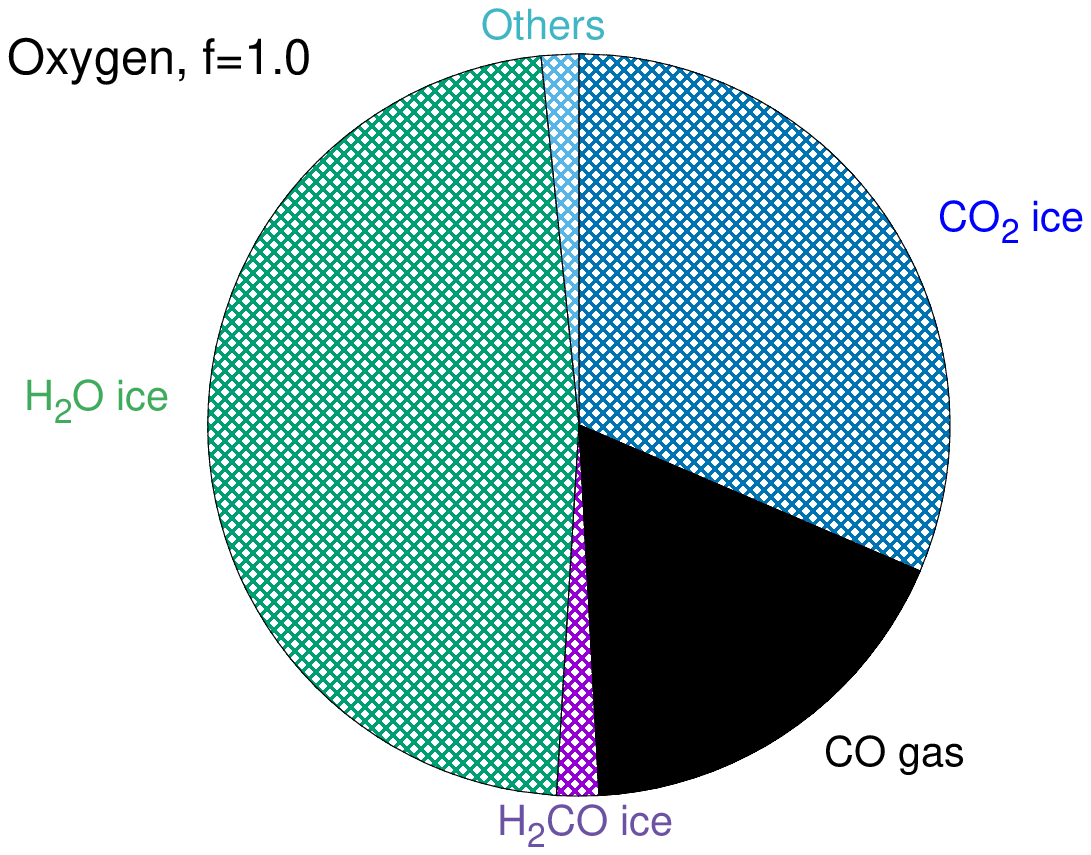}
\includegraphics[scale=0.55]{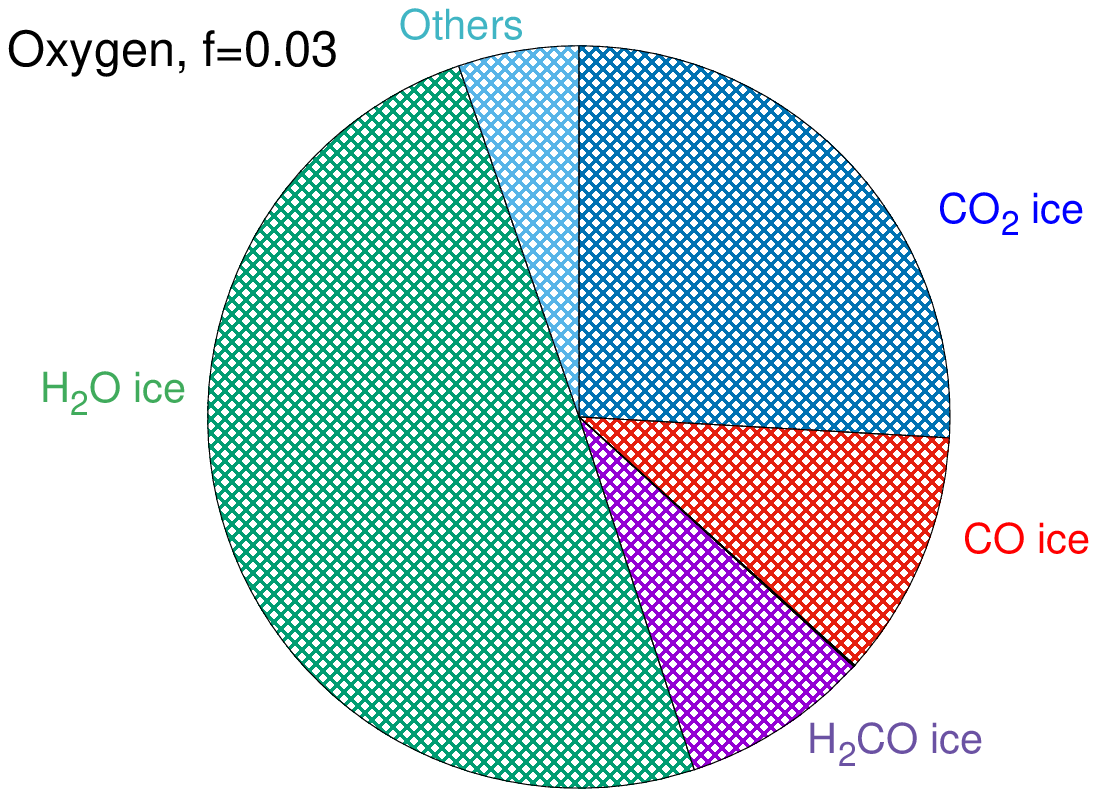}
\includegraphics[scale=0.55]{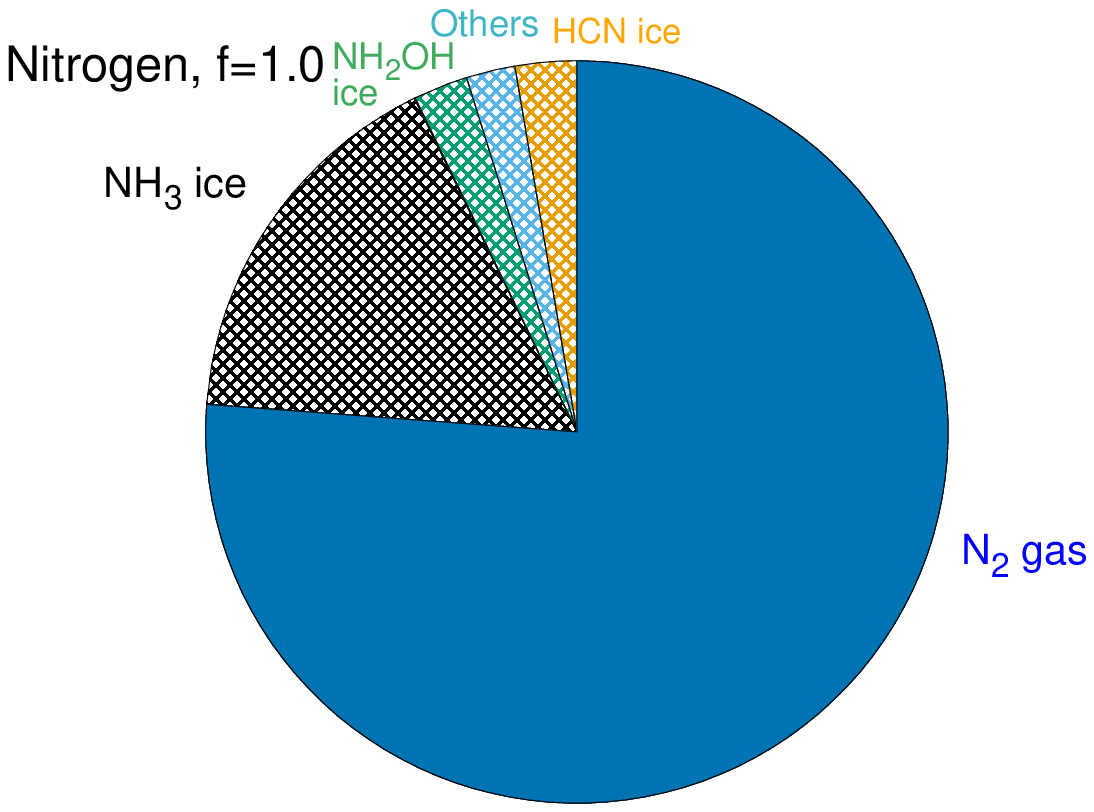}
\includegraphics[scale=0.55]{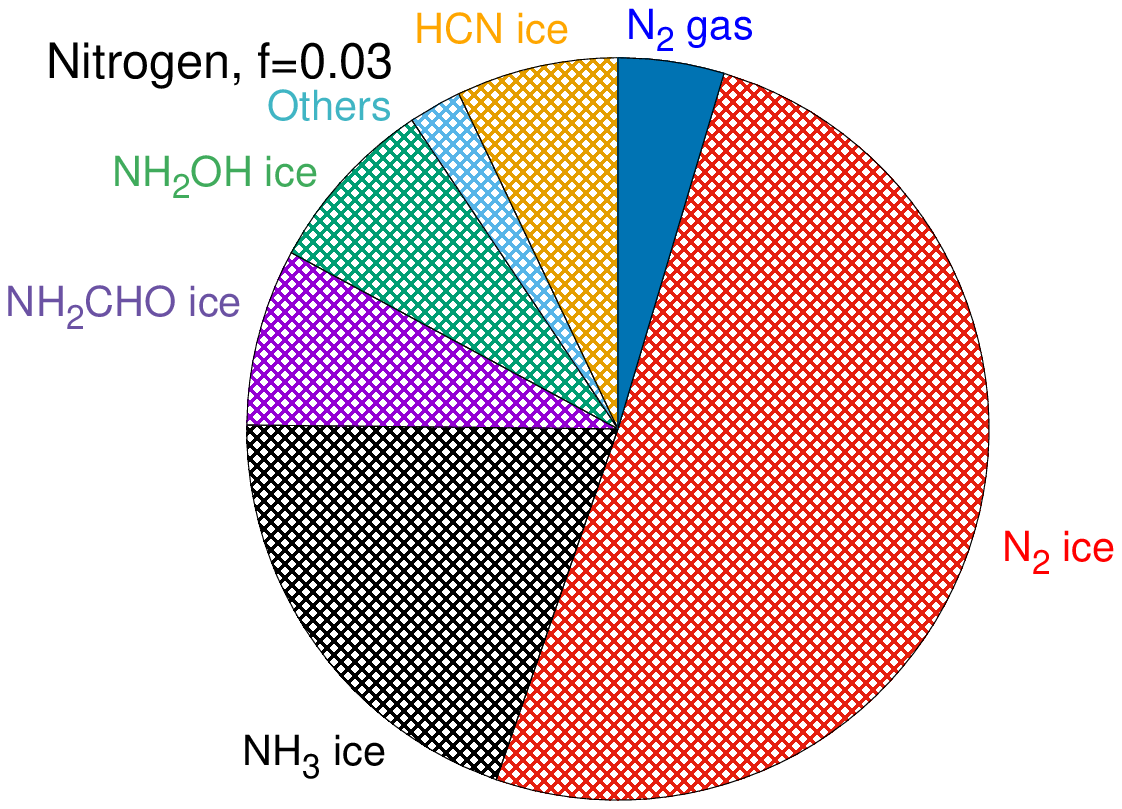}
%%
%%%
%%%
%%%
%%%
\end{center}
\vspace{-0.2cm}
\caption{
Pie charts of the percentage contributions of the dominant carbon-bearing molecules to the total elemental carbon abundance ($=1.4\times10^{-4}$, top panels), 
the dominant oxygen-bearing molecules to the total elemental oxygen abundance ($=3.2\times10^{-4}$, middle panels), and 
the dominant nitrogen-bearing molecules to the total elemental nitrogen abundance ($=7.5\times10^{-5}$, bottom panels), at $r=$5.3 au (around the current orbit of Jupiter) and t=$10^{6}$ years.
The left panels show the contributions for the disk midplane with the monotonically decreasing density and temperature profile ($f=1.0$), whereas the right panel shows the contributions for the shadowed disk midplane ($f=0.03$).
These panels show the results for molecular initial abundances (the ``inheritance'' scenario) and $\xi_{\mathrm{CR}}(r)=$$1.0\times10^{-17}$ [s$^{-1}$].
The filled and hatched slices are respectively the contributions of gaseous and icy molecules.
%%. 
[Top panels]: The dark blue, red, black, purple, green, orange, yellow, and light blue slices are respectively the contributions of CO$_{2}$ ice, CO ice, CO gas, H$_{2}$CO ice, CH$_{4}$ ice, HCN ice, C$_{2}$H$_{6}$ ice, and other molecules (such as NH$_{2}$CHO ice and CH$_{3}$OH ice for $f=0.03$, and CH$_{3}$CCH ice, CH$_{2}$CCH$_{2}$ ice, and CH$_{3}$CHCH$_{2}$ ice for $f=1.0$).
[Middle panels]: The dark blue, red, black, purple, green, and light blue slices are respectively the contributions of CO$_{2}$ ice, CO ice, CO gas, H$_{2}$CO ice, H$_{2}$O ice, and other molecules (such as NH$_{2}$CHO ice and CH$_{3}$OH ice).
%%%
[Bottom panels]: The dark blue, red, black, purple, green, orange, and light blue slices are respectively the contributions of N$_{2}$ gas, N$_{2}$ ice, NH$_{3}$ ice, NH$_{2}$CHO ice, NH$_{2}$OH ice, HCN ice, and other molecules (such as HC$_{3}$N ice).
}\label{Figure2rev_pie_CNO}
\end{figure*}
%%%
\begin{comment}
%%%
\end{comment}
%%%
%
%%\\ \\
%%%
In this Section, we show the composition of the dominant carbon-, oxygen-, and nitrogen-carriers around the current orbit of Jupiter.
%% .
Figure \ref{Figure2rev_pie_CNO} shows pie charts of the percentage contributions of the dominant carbon-, oxygen-, and nitrogen-bearing molecules to the total elemental abundances ($=1.4\times10^{-4}$ for carbon, $=3.2\times10^{-4}$ for oxygen, $=7.5\times10^{-5}$ for nitrogen) at $r=$5.3 au (around the current orbit of Jupiter) and t=$10^{6}$ years, in the non-shadowed disk ($f=1.0$) and a shadowed disk ($f=0.03$).
These results are from the model in which we assume molecular initial abundances and a high ionisation rate (see Figures \ref{Figure3_rev_radial}-\ref{Figure5_rev_radial} in Section \ref{sec:3-2rev}).
\\ \\
In the non-shadowed disk, the dominant carbon carriers at $r=$5.3 au are CO$_{2}$ ice ($\sim$36\%) and CO gas ($\sim$40\%).
In the shadowed disk, CO freezes out onto the dust grain surface, and the amounts of CO$_{2}$ ice ($\sim$30\%), CO ice ($\sim$25\%), and ices of unsaturated hydrocarbon molecules (such as C$_{3}$H$_{4}$) decrease, 
whereas those of H$_{2}$CO ice ($\sim$19\%), CH$_{4}$ ice ($\sim$6\%), C$_{2}$H$_{6}$ ice ($\sim$5\%), HCN ice ($\sim4$\%), NH$_{2}$CHO ice ($\sim4$\%), and CH$_{3}$OH ice ($\sim3$\%) increase.
%% .
The enhanced abundances of these organic molecules in the shadowed disks are owing to the reaction pathways in which the sequential hydrogenation of CO on the dust grain surfaces and cosmic-ray-induced photodissociation of CH$_{3}$OH are the key reactions, as we explain in Sections \ref{sec:3-2-2rev} and \ref{sec:3-2-3rev}.
%%.
\\ \\
In the non-shadowed disk, the dominant oxygen carriers at $r=$5.3 au are H$_{2}$O ice ($\sim47$\%), CO$_{2}$ ice ($\sim$31\%) and CO gas ($\sim$18\%).
In the shadowed disk, the amounts of CO$_{2}$ ice ($\sim$26\%) and CO ice ($\sim$11\%) decrease, 
whereas that of H$_{2}$O ice is similar ($\sim$50\%), and that of H$_{2}$CO ice ($\sim$8\%) increases.
\\ \\
In the non-shadowed disk, the dominant nitrogen carriers at $r=$5.3 au are N$_{2}$ gas ($\sim76$\%) and NH$_{3}$ ice ($\sim17$\%).
In the shadowed disk, N$_{2}$ mostly freezes-out onto the dust grain surface, and the amount of N$_{2}$ gas ($\sim5$\%) decreases,
whereas that of NH$_{3}$ ice is almost similar ($\sim20$\%), and those of N$_{2}$ ice ($\sim$51\%) and other nitrogen-bearing molecules such as NH$_{2}$OH, NH$_{2}$CHO, and HCN ices ($\sim7-8$\% each) increase. 
The enhanced abundances of NH$_{2}$OH, NH$_{2}$CHO, and HCN in the shadowed disks are due to 
hydrogenation reactions on the dust grain surfaces (see Section \ref{sec:3-2-3rev}). 
%%.
\\ \\
According to these results, in shadowed disks the dust grains at around the current orbit of Jupiter ($r\sim$5.3 au) are expected to have significant amounts of saturated hydrocarbon ices such as CH$_{4}$ and C$_{2}$H$_{6}$, ices of organic molecules such as H$_{2}$CO, NH$_{2}$CHO, and CH$_{3}$OH, in addition to  H$_{2}$O, CO, CO$_{2}$, NH$_{3}$, N$_{2}$, HCN, and NH$_{2}$OH ices, compared with those in the non-shadowed disks (mostly ices of H$_{2}$O, CO$_{2}$, NH$_{3}$, and unsaturated hydrocarbon molecules).
These abundant saturated hydrocarbons and various organic molecules such as H$_{2}$CO, NH$_{2}$CHO, and CH$_{3}$OH were not reported by \citet{Ohno2021}, since they approximated various organic molecules by considering C$_{2}$H$_{6}$ alone.
%.
In addition, our results provide intriguing implications for the compositions of small bodies (such as comets and asteroids) formed at around the current orbit of Jupiter, as we discuss in Section \ref{sec:4-2}.
%%
%% .
%%%
\subsection{Radial molecular abundance distributions for our standard model}\label{sec:3-2rev}
In this Section, we describe the effects of disk shadowing on the radial molecular abundance distributions.
%%.
%%\\ \\
The radial profiles of fractional abundances with respect to total hydrogen nuclei densities $n_{\mathrm{X}}$/$n_{\mathrm{H}}$ at $t=10^{6}$ years for dominant carbon-, oxygen-, nitrogen-bearing molecules are shown in Figures \ref{Figure3_rev_radial}-\ref{Figure5_rev_radial} in Sections \ref{sec:3-2-1rev}-\ref{sec:3-2-3rev}, with separate panels for each molecule, for different values of the parameter $f$ (=1.0, 0.3, 0.03, and 0.003), respectively.
In Figures \ref{Figure6_rev_radial}-\ref{Figure8_rev_radial} in Section \ref{sec:3-2-4rev}, the radial fractional abundance distributions at $t=10^{6}$ years of other major carbon-, nitrogen-, and oxygen-bearing molecules, mainly complex organic molecules, are shown. 
%%.
In Figure \ref{Figure9_COMs_rev_radial} in Section \ref{sec:3-2-5rev}, the radial profiles of fractional abundances with respect to total hydrogen nuclei densities at t=$10^{6}$ years for sums of larger (complex) organic molecules are shown. We show this to demonstrate the efficiency of conversion from more simple to more complex molecules for different dust depletion factors.
The panels in these Figures \ref{Figure3_rev_radial}-\ref{Figure9_COMs_rev_radial} show the results for the radially constant cosmic-ray ionization rate of $\xi_{\mathrm{CR}}(r)=$$1.0\times10^{-17}$ [s$^{-1}$] with assuming molecular initial abundances (the ``inheritance'' scenario).
%%. 
%%
%%
%%
%.
\\ \\
Since previous studies (e.g., \citealt{Eistrup2016, Eistrup2018}) have discussed the chemical structure of the non-shadowed disk ($f=1.0$) in detail, in this paper we mainly focus on the effects of shadowing on disk chemical structures.
We note that the results of our calculations for the non-shadowed disk are largely consistent with those in \citet{Eistrup2016, Eistrup2018} with some exceptions, such as HCN (see Section \ref{sec:3-2-3rev} of this paper).
%%\\ \\
%%%%.
%%%
%%\subsection{Molecular abundance distributions for our standard model}\label{sec:3-2-1rev}
\subsubsection{H$_{2}$O, CO, CO$_{2}$, and O$_{2}$}\label{sec:3-2-1rev}
%%%
\begin{figure*}[hbtp]
\begin{center}
\vspace{1cm}
%%\vspace{-0.4cm}
%\plotone{cost.pdf}
%\
\includegraphics[scale=0.63]{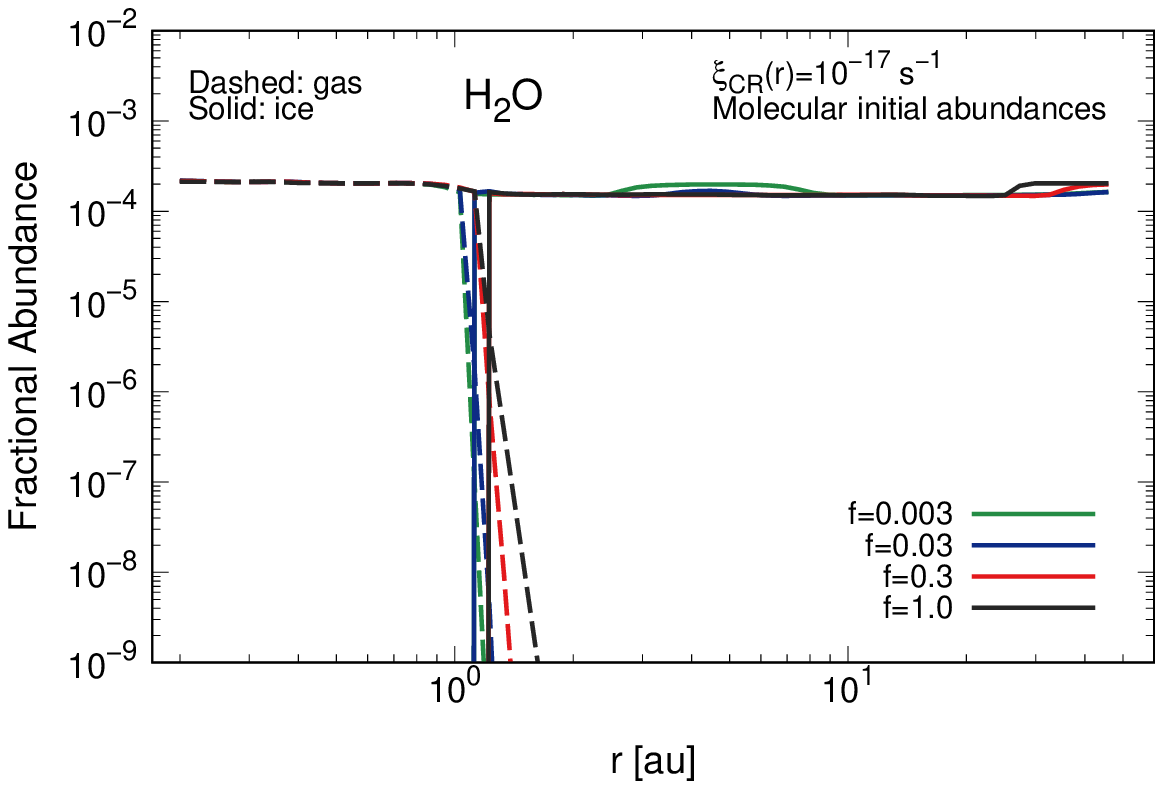}
\includegraphics[scale=0.63]{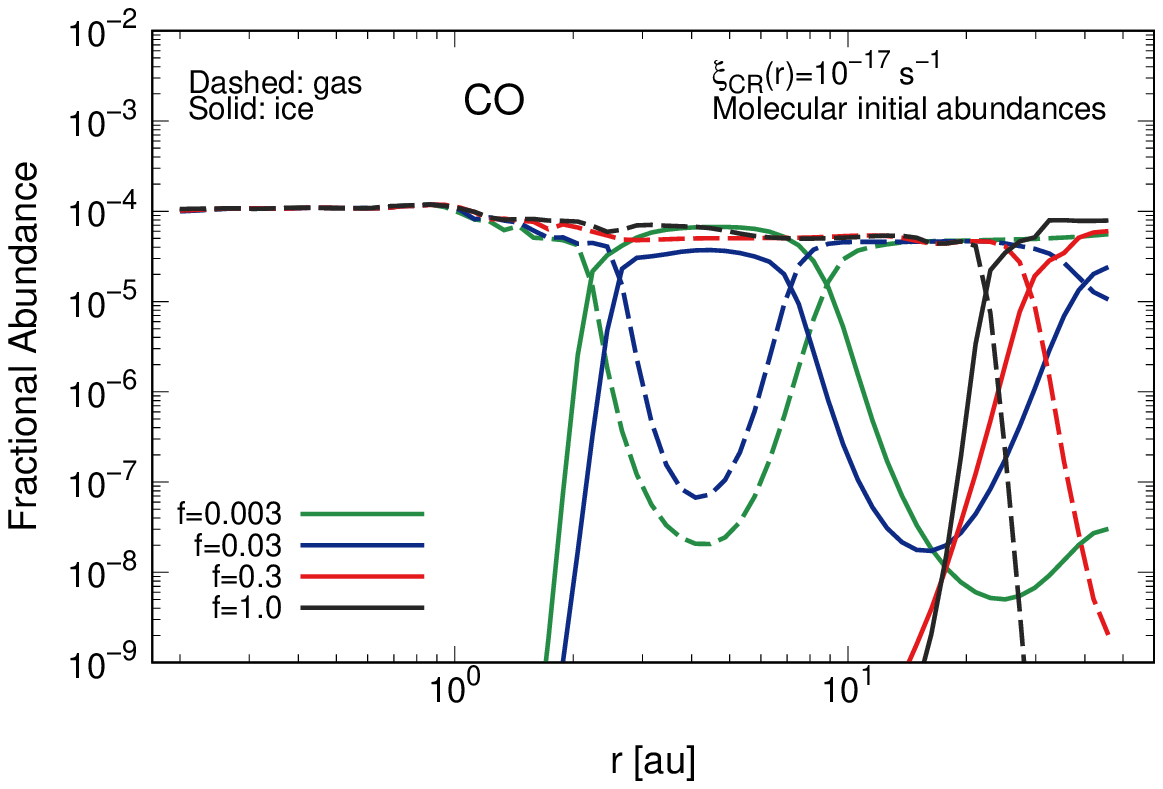}
\includegraphics[scale=0.63]{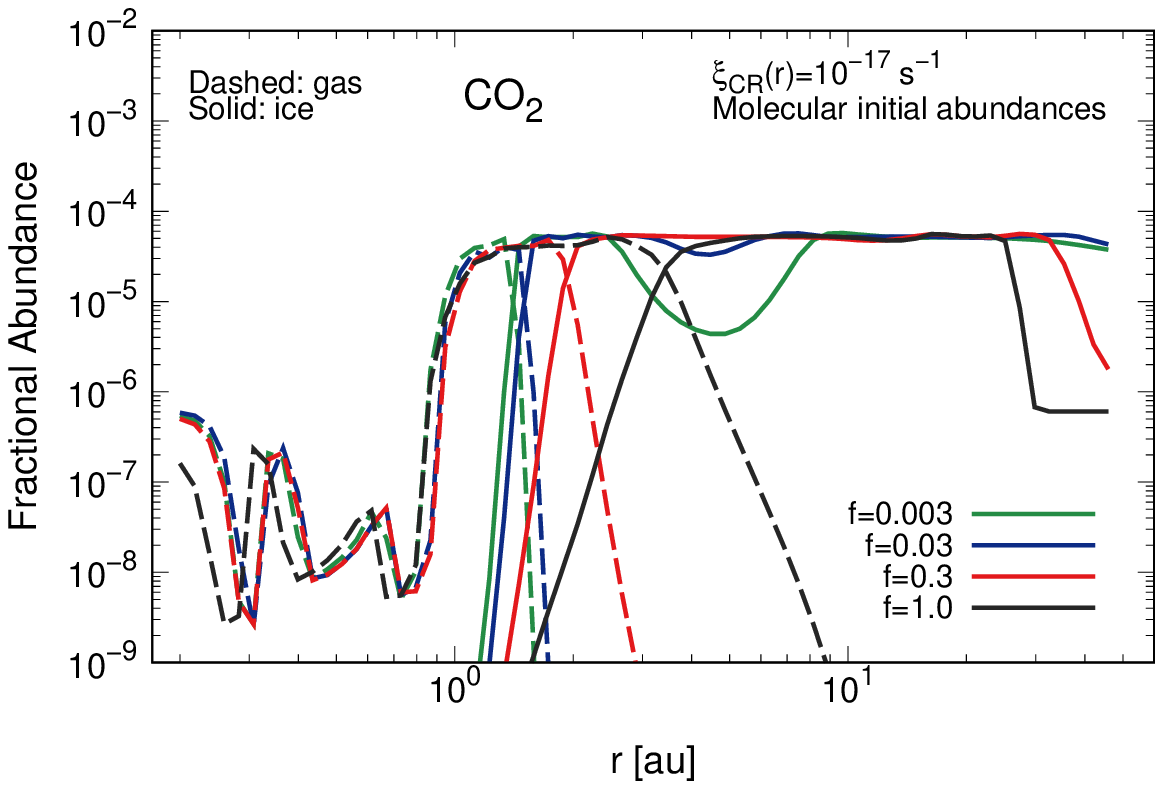}
%\%includegraphics[scale=0.58]{20210902_t3_ds_0.1um_atomic-ini_17_Xray-rev_enhanced-water_nH-rev_1.0e6yr_CO2gi.eps}
\includegraphics[scale=0.63]{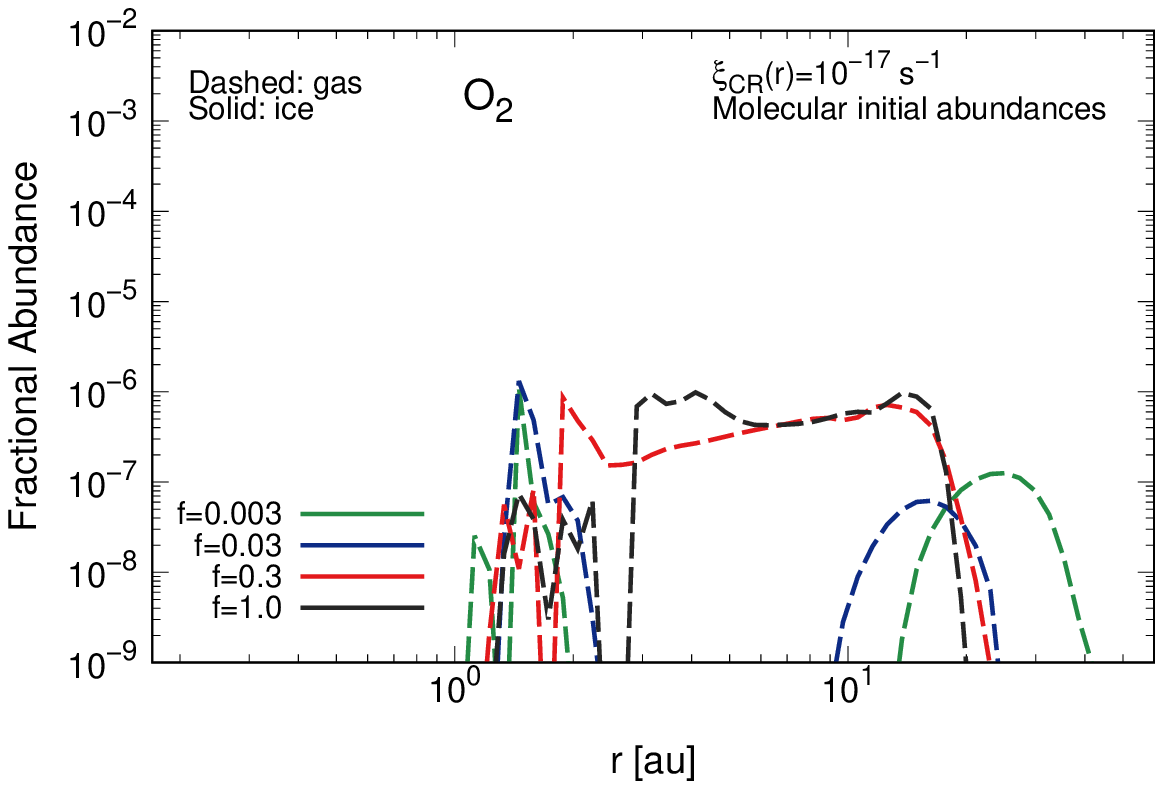}
%%\includegraphics[scale=0.58]{20210902_t3_ds_0.1um_atomic-ini_17_Xray-rev_enhanced-water_nH-rev_1.0e6yr_O2gi.eps}
%%20220428 16:04
\end{center}
\vspace{-0.2cm}
\caption{
The radial profiles of fractional abundances with respect to total hydrogen nuclei densities at t=$10^{6}$ years for H$_{2}$O ($n_{\mathrm{H}_{2}\mathrm{O}}$/$n_{\mathrm{H}}$, top left panel), CO ($n_{\mathrm{CO}}$/$n_{\mathrm{H}}$, top right panel), CO$_{2}$ ($n_{\mathrm{CO}_{2}}$/$n_{\mathrm{H}}$, bottom left panel), and O$_{2}$ ($n_{\mathrm{O}_{2}}$/$n_{\mathrm{H}}$, bottom right panel).
These panels show the results for the radially constant cosmic-ray ionization rate $\xi_{\mathrm{CR}}(r)=$$1.0\times10^{-17}$ [s$^{-1}$] and molecular initial abundances (the ``inheritance'' scenario).
The dashed and solid lines show the profiles for gaseous and icy molecules, respectively.
The black, red, blue, and green lines show the profiles for different values of the parameter $f$ (=1.0, 0.3, 0.03, and 0.003), respectively.
\\ \\
%.
%%\vspace{0.2cm}
}\label{Figure3_rev_radial}
\end{figure*}
\vspace{0.2cm}
Figure \ref{Figure3_rev_radial} shows the radial profiles of fractional abundances for dominant oxygen-bearing molecules; H$_{2}$O (water), CO (carbon monoxide), CO$_{2}$ (carbon dioxide), and O$_{2}$ (molecular oxygen).
The water snowline position ($\sim$1.3 au, $T(r)\sim140$ K) and the H$_{2}$O gas abundances ($\sim2\times10^{-4}$) within the water snowline do not change for various values of $f$, since the disk midplane temperature is significantly changed only beyond the water snowline.
We note that the gas abundances of other molecules are also unchanged within the water snowline.
%%%
\\ \\
%%%
%%%
%%,
Outside the water snowline, H atoms are supplied onto the dust grain surface from the gas or produced in situ within the grain mantles.
We note that in the gas-phase, H atoms are mainly produced by cosmic-ray-induced photodissociation of H$_{2}$.
Since the fractional abundances of H$_{2}$ with respect to total H nuclei ($\sim5\times10^{-1}$) are much larger than those of other molecules, the gas-phase production is dominant.
Thus, if we conduct disk chemical modeling with atomic initial abundances (see Appendix \ref{Asec:A}), H$_{2}$O ice is efficiently formed outside the water snowline via the following reaction \citep{Eistrup2016},
\begin{equation}\label{Rec5}
\mathrm{OH}_{\mathrm{ice}} + \mathrm{H}_{\mathrm{ice}} \rightarrow \mathrm{H}_{2}\mathrm{O}_{\mathrm{ice}}.
\end{equation}
However, because we have adopted abundant initial water ice abundances ($=1.984\times10^{-4}$, see Table \ref{Table:1}), further water ice formation in the outer disk midplane does not proceed efficiently. 
Thus, the H$_{2}$O ice abundances beyond the water snowline are relatively constant ($\sim(1-2)\times10^{-4}$) for various values of $f$.
We note that the H$_{2}$O ice abundances increase (from $\sim1.5\times10^{-4}$ to $\sim2\times10^{-4}$) in the coldest regions with $T(r)\lesssim20$ K 
(at $r\sim3-8$ au for $f=0.003$ and at $r\gtrsim25$ au for $f=1.0$), where the formation of CO$_{2}$ ice does not proceed efficiently (see below).
%%, 
\begin{comment}
\end{comment}
\\ \\
In the non-shadowed disk ($f=1.0$), the CO snowline position is $r\sim22$ au ($T(r)\sim25$ K),
and 
CO gas abundances inside the water snowline are $10^{-4}$.
In addition, in the non-shadowed disk CO gas abundances between the H$_{2}$O and CO snowlines are $\sim(4-8)\times10^{-5}$.
%%
%%%
\\ \\
%.
As introduced later, the chemical conversion of CO plays vital roles in producing other carbon-bearing molecules.
%% .
%%
CO molecules accreting onto the dust grain surfaces can react with OH radicals produced on the ice by cosmic-ray-induced photodissociation of H$_{2}$O ice (e.g., \citealt{Drozdovskaya2016, Eistrup2016, Eistrup2018, Schwarz2018, Schwarz2019}).
This produces CO$_{2}$ ice via the following grain-surface reaction,
\begin{equation}\label{Rec6}
\mathrm{CO}_{\mathrm{ice}} + \mathrm{OH}_{\mathrm{ice}} \rightarrow \mathrm{CO}_{2 \ \mathrm{ice}} +  \mathrm{H}_{\mathrm{ice}}.
\end{equation}
In addition, for the ISM level ionization rate ($\xi_{\mathrm{CR}}(r)=$$10^{-17}$ [s$^{-1}$]), the following destruction pathway of gas-phase CO by He$^{+}$ is also efficient, 
\begin{equation}\label{Rec7}
\mathrm{CO} + \mathrm{He}^{+} \rightarrow \mathrm{C}^{+} + \mathrm{O} + \mathrm{He}, 
\end{equation}
which leads to the formation of CH$_{4}$, C$_{2}$H$_{6}$, and other hydrocarbons (e.g., \citealt{Aikawa1999, Furuya2014, Eistrup2016, Eistrup2018, Yu2016, Bosman2018}, see also Sections \ref{sec:3-2-2rev} and \ref{sec:3-2-4rev}).
We note that He$^{+}$ is produced by the direct cosmic-ray ionisation of He.
Moreover, the sequential hydrogenation of CO on the dust grain surfaces leads to the formation of H$_{2}$CO and CH$_{3}$OH ices (e.g., \citealt{Watanabe2002, Drozdovskaya2014, Bosman2018}, see also Section \ref{sec:3-2-2rev} and Appendix \ref{Asec:A-2}).
Several studies have suggested that the above chemical processes (partly) explain the depletion of CO in the Class II disks reported by recent observations with 
ALMA (see e.g., \citealt{Nomura2016, Nomura2021, Schwarz2016, Krijt2018, Krijt2020, Bergner2020, Zhang2020, Zhang2021}). 
%\\ \\
\begin{comment}
\end{comment}
%%
%%
\\ \\
%%
%% 
%%%
In the shadowed disk ($f\leq0.03$), CO freezes-out onto dust grains at around the current orbit of Jupiter ($r\sim3-8$ au, $T(r)\lesssim25$ K), and CO ice abundances are $\sim(3-6)\times10^{-5}$.
%.
In addition, CO returns to gas phase at $r>8$ au.
At $r>20$ au, CO (re-)freezes out onto dust grains, and the adsorption front moves outward with decreasing $f$, since the values of $T(r)$ at $r>20$ au increase ($\lesssim35$ K at the maximum) with decreasing $f$ (see also Section 2.1).
This CO abundance profile is consistent with the results reported in \citet{Ohno2021}. 
We suggest that reducing the dust surface density beyond the water snowline allows the reprocessed stellar radiation to enter the outer disk midplane from the disk upper layers, and that the penetration of such radiation causes the temperature rise in the outer disk ($r>20$ au).
\\ \\
In the non-shadowed disk ($f=1.0$), the CO$_{2}$ snowline position is $r\sim3.3$ au ($T(r)\sim75$ K).
CO$_{2}$ ice abundances are $\sim5\times10^{-5}$ between the CO$_{2}$ and the CO snowlines, and decrease ($<10^{-6}$) beyond the CO snowline.
The CO$_{2}$ snowline position moves inward with decreasing $f$.
In the shadowed disk ($f\leq0.03$), CO$_{2}$ ice abundances decrease at $r\sim3-8$ au (from $\sim5\times10^{-5}$ to $<10^{-5}$) with decreasing $f$,
whereas H$_{2}$O ice abundances at such radii slightly increase (from $\sim1.5\times10^{-4}$ to $\sim2\times10^{-4}$).
\citet{Eistrup2016} discussed that at coldest conditions ($T(r)\lesssim20$ K), the formation of H$_{2}$O ice (Reaction \ref{Rec5}) is faster than that of CO$_{2}$ ice (Reaction \ref{Rec6}).
This is because at the coldest conditions ($T(r)\lesssim20$ K) the mobility of H is higher than those of CO and OH, and because the adsorption rate of atomic hydrogen onto dust grains at $T(r)\lesssim20$ K becomes larger than that at $T(r)>20$ K ($E_{\mathrm{des}}$(H)=650 K).
%%%
%%
%%%
\\ \\
In our standard disk model, O$_{2}$ abundances are much smaller ($\lesssim10^{-6}$ for gas and $\lesssim10^{-9}$ for ice) than those of H$_{2}$O, CO, and CO$_{2}$.
%%.
O$_{2}$ is formed in the gas-phase via the following reaction \citep{Walsh2015, Eistrup2016, Notsu2021},
\begin{equation}\label{Rec8}
\mathrm{O} + \mathrm{OH} \rightarrow \mathrm{O}_{2} +  \mathrm{H}.
\end{equation}
Since O$_{2}$ is very volatile (the binding energy $E_{\mathrm{des}}(\mathrm{O}_{2})=898$ K, \citealt{Noble2012}), it remains in the gas-phase at $r\lesssim20$ au in the non-shadowed disk ($f=1.0$).
The O$_{2}$ gas abundances between the water and O$_{2}$ snowlines are 
$\sim10^{-7}-10^{-6}$.
%% 
%%%
\\ \\
In the shadowed disk ($f\leq0.03$), O$_{2}$ freezes-out onto dust grains at $r\sim2-10$ au, and returns to the gas phase at $r>10$ au.
%%%
Outside the O$_{2}$ snowline, O$_{2}$ ice is formed via the following grain-surface reaction \citep{Taquet2016, Eistrup2019},
\begin{equation}\label{Rec8.5}
\mathrm{O}_{\mathrm{ice}} + \mathrm{O}_{\mathrm{ice}} \rightarrow \mathrm{O}_{2 \ \mathrm{ice}}.
\end{equation}
In our standard disk model, the initial atomic oxygen is zero (see Table \ref{Table:2}).
In addition, gas-phase production route of atomic oxygen from CO by Reaction \ref{Rec7} is not efficient outside the O$_{2}$ snowline since the binding energies of CO and O$_{2}$ are almost similar (see Table \ref{Table:1}).
Thus, in our standard disk model, O$_{2}$ ice abundances are low ($<<10^{-9}$) both in the shadowed and non-shadowed disks.
%\\ \\
\\ \\
We note that our calculations also yield higher O$_{2}$ ice abundances ($\sim10^{-5}-10^{-4}$) for atomic initial abundances and low ionisation rates.
Such larger abundances are consistent with the measured cometary abundances (\citealt{Bieler2015, Rubin2015}, see also Section \ref{sec:4-2} of this paper).
%%. 
Moreover, we find that in the shadowed disk such an O$_{2}$ abundant region is located in the inner region ($r\sim2-10$ au) compared with that in the non-shadowed disk ($r\gtrsim20$ au), as detailed in Section \ref{sec:4-2} and Appendix \ref{Asec:A-1}.
%
%%%
%%%
%%%
%%%%%
%
\subsubsection{Other dominant carbon-bearing molecules}\label{sec:3-2-2rev}
%%%%
\begin{figure*}[hbtp]
\begin{center}
\vspace{1cm}
%\plotone{cost.pdf}
%\plotone[scale=0.63]{{20210624_surface-density.eps}
%\plotone[scale=0.63]{{20210624_ngas-T.eps}
%\includegraphics[scale=0.58]{20210902_t1_ds_0.1um_mol-ini_17_Xray-rev_enhanced-water_nH-rev_1.0e6yr_O2gi.eps}
%\includegraphics[scale=0.58]{20210902_t3_ds_0.1um_atomic-ini_17_Xray-rev_enhanced-water_nH-rev_1.0e6yr_O2gi.eps}
\includegraphics[scale=0.63]{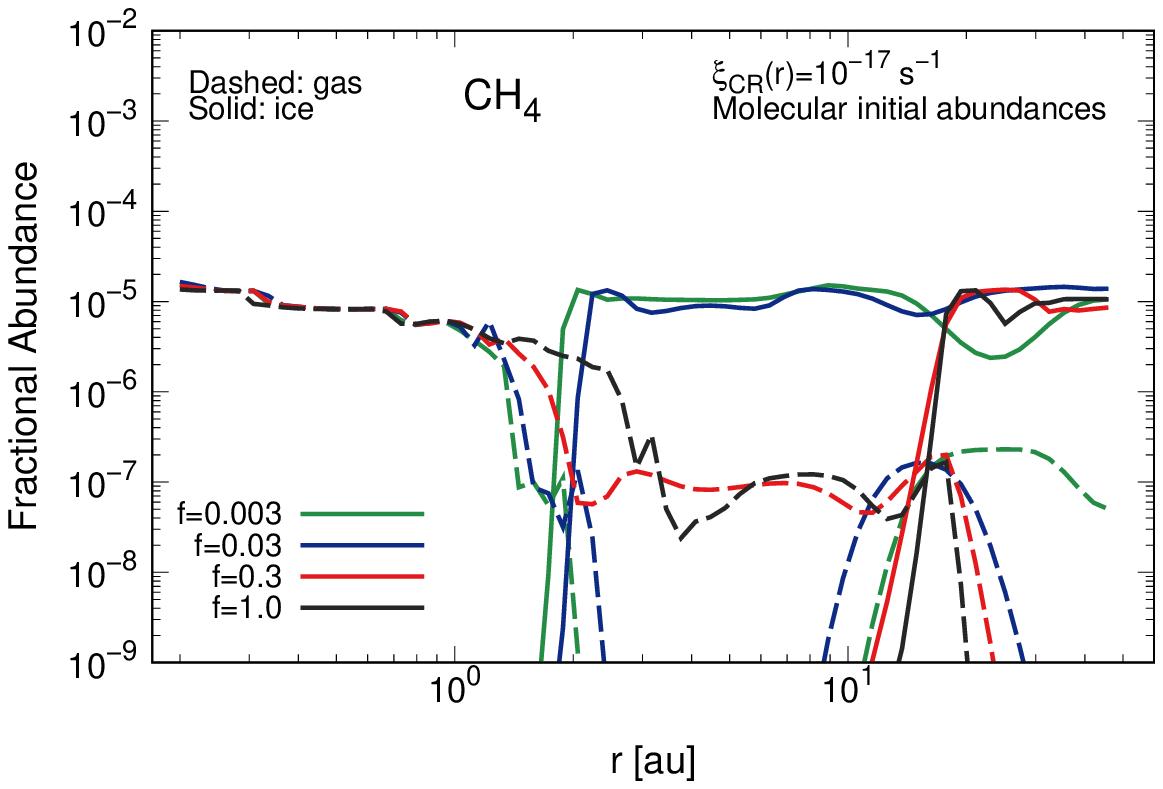}
\includegraphics[scale=0.63]{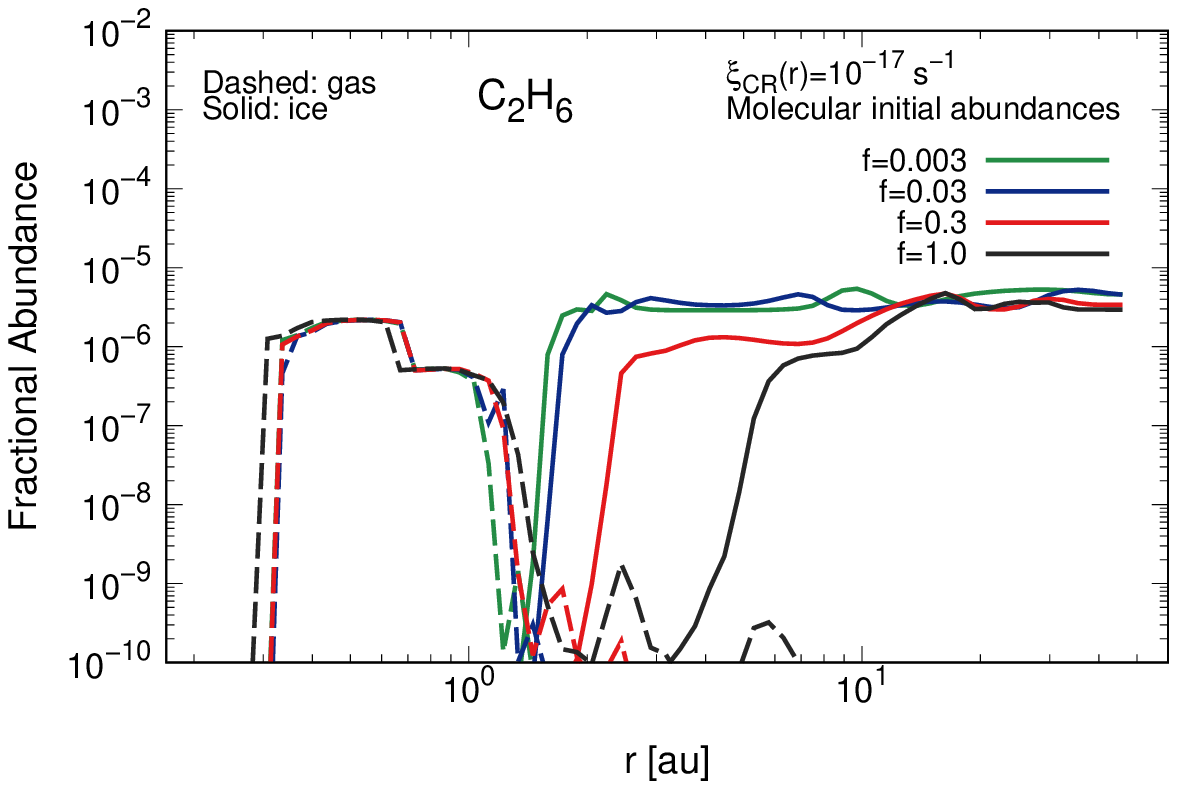}
\includegraphics[scale=0.63]{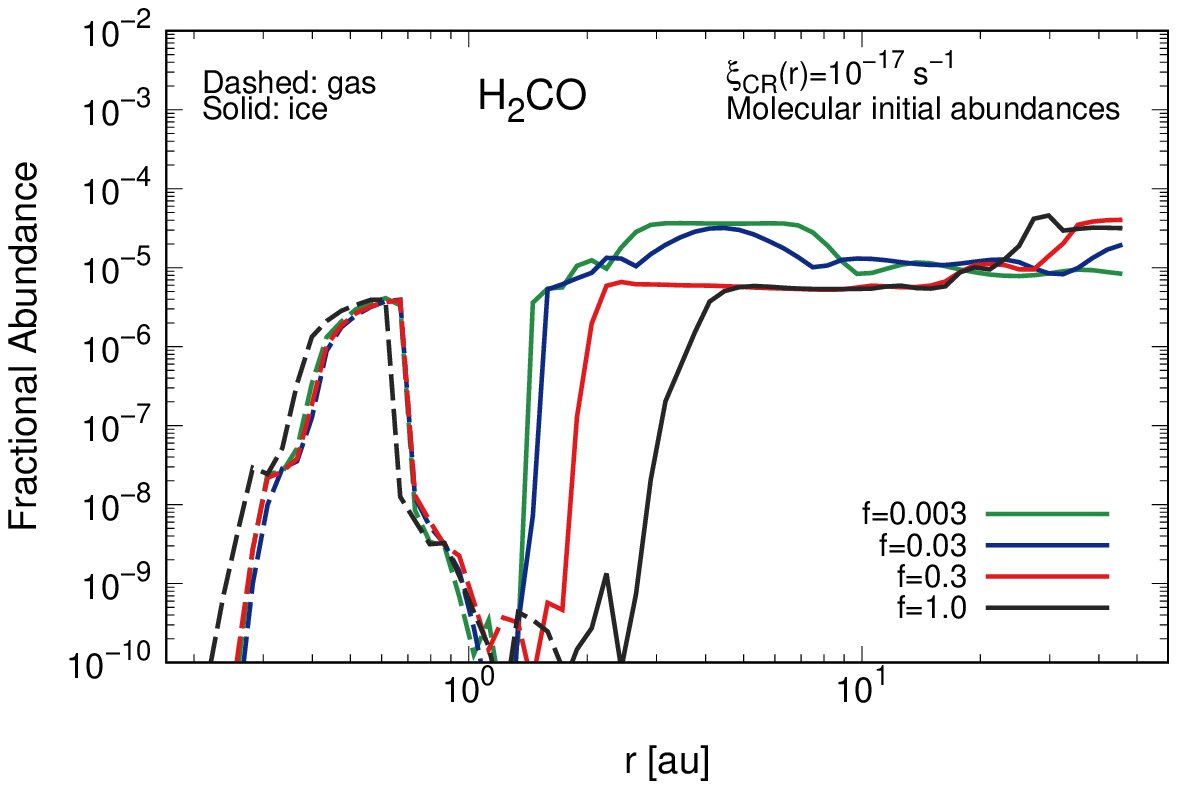}
\includegraphics[scale=0.63]{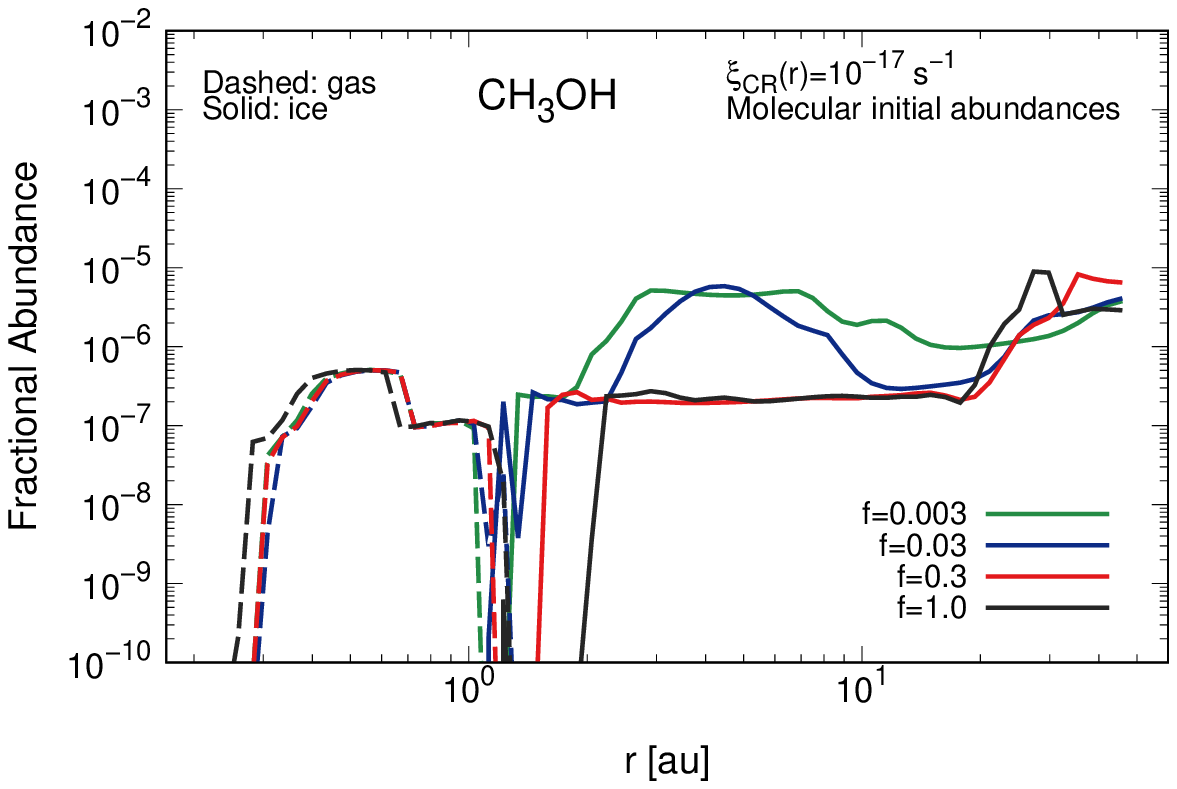}
\end{center}
\vspace{-0.2cm}
\caption{
%The radial profiles of fractional abundances with respect to total hydrogen nuclei densities at t=$10^{6}$ years for 
Same as Figure \ref{Figure3_rev_radial}, but for 
CH$_{4}$ ($n_{\mathrm{CH}_{4}}$/$n_{\mathrm{H}}$, top left panel), 
C$_{2}$H$_{6}$ ($n_{\mathrm{C}_{2}\mathrm{H}_{6}}$/$n_{\mathrm{H}}$, top right panel),
H$_{2}$CO ($n_{\mathrm{H}_{2}\mathrm{CO}}$/$n_{\mathrm{H}}$, bottom left panel), and
CH$_{3}$OH ($n_{\mathrm{CH}_{3}\mathrm{OH}}$/$n_{\mathrm{H}}$, bottom right panel).
\\ \\
%%.
}\label{Figure4_rev_radial}
\end{figure*}
\vspace{0.2cm}
%%%%
%%%20211224 19:57
%%%
Figure \ref{Figure4_rev_radial} shows the radial profiles of fractional abundances for other dominant carbon-bearing molecules (CH$_{4}$, C$_{2}$H$_{6}$, H$_{2}$CO, and CH$_{3}$OH).
We note that CH$_{4}$ (methane) and C$_{2}$H$_{6}$ (ethane) are dominant acyclic saturated hydrocarbon molecules (alkanes).
In the non-shadowed disk ($f=1.0$), the CH$_{4}$ snowline position is $r\sim15$ au ($T(r)\sim38$ K), 
%%and for molecular initial abundances, 
and the CH$_{4}$ gas abundances between the CO$_{2}$ and CH$_{4}$
snowlines are $\sim10^{-7}$.
These values are much smaller than those in the disk with a lower ionisation rate ($=$$10^{-18}$ [s$^{-1}$], see Appendix \ref{Asec:A}) and that for the initial CH$_{4}$ gas abundance ($=1.120\times10^{-6}$).
Outside the CH$_{4}$ snowline, CH$_{4}$ ice abundances are slightly enhanced ($\sim10^{-5}$) compared with the initial CH$_{4}$ ice abundances ($=7.384\times10^{-6}$).
\\ \\
Here we describe the dominant formation and destruction routes of CH$_{4}$ both in the gas-phase and icy-phase.
The destruction of gas-phase CO by He$^{+}$ (see Section \ref{sec:3-2-1rev}) produces C$^{+}$, and a sequence of reactions with e.g., H$_{2}$ transform C$^{+}$ to e.g., CH$_{2}$$^{+}$/CH$_{3}$$^{+}$/CH$_{5}$$^{+}$, which lead to CH$_{4}$ gas and ice (e.g., \citealt{Aikawa1999, Furuya2014, Bosman2018}, see Section \ref{sec:3-2-1rev}).
In addition, \citet{Bosman2018} and \citet{Eistrup2018} discussed that CH$_{4}$ ice is also produced both inside/outside the CO snowline through the hydrogenation of CH$_{3}$ ice, which is formed by cosmic-ray-induced photodissociation of CH$_{3}$OH (see also below).
Inside the CH$_{4}$ snowline, CH$_{4}$ gas is destroyed by cosmic-ray-induced photodissociation and ion-molecule reactions (such as CH$_{4}$ $+$ C$^{+}$), and the carbon is thus converted from CH$_{4}$ gas to CO$_{2}$, H$_{2}$CO, and hydrocarbons such as e.g., C$_{2}$H$_{2}$, C$_{2}$H$_{4}$, and C$_{3}$H$_{4}$ (\citealt{Aikawa1999, Eistrup2016, Eistrup2018, Yu2016}, see also Section \ref{sec:3-2-4rev}).
%%%
%%%
%%
%%%
%%
\\ \\
The CH$_{4}$ snowline positions moves inward with decreasing $f$.
In the shadowed disk ($f\leq0.03$), CH$_{4}$ snowline positions are $r\lesssim2$ au, and CH$_{4}$ ice abundances at $r\sim2-15$ au are around $\sim10^{-5}$.
%%.
This result indicates that the CH$_{4}$ ice abundances are comparable in both CO sublimation and frozen regions.
Thus, we interpret that the above reaction pathway starting from cosmic-ray-induced photodissociation of CH$_{3}$OH is efficient for CH$_{4}$ ice formation in this region.
\\ \\
In the non-shadowed disk ($f=1.0$), the C$_{2}$H$_{6}$ snowline position is $r\sim3.2$ au, which is similar to that for CO$_{2}$ ($E_{\mathrm{des}}(\mathrm{CO}_{2})=2267$ K, $E_{\mathrm{des}}(\mathrm{C}_{2}\mathrm{H}_{6})=2320$ K). 
%%, 
%%%
The C$_{2}$H$_{6}$ ice abundances outside its snowline increase with increasing $r$. The ice abundances are $\sim(3-6)\times10^{-6}$ outside the CH$_{4}$ snowline, which is larger than the initial ice abundance ($=2.417\times10^{-6}$). 
\\ \\
%%% 
Previous modeling studies suggested that unsaturated hydrocarbon molecules (such as C$_{2}$H$_{2}$, C$_{3}$H$_{2}$, C$_{3}$H$_{4}$) are efficiently produced with 
the chemical reaction pathways starting from CH$_{4}$ $+$ C$^{+}$, where C$^{+}$ is formed by CO $+$ He$^{+}$ (e.g., \citealt{Aikawa1999, Furuya2014, Eistrup2016, Eistrup2018, Yu2016, Bosman2018}, see Section \ref{sec:3-2-1rev}). 
%%%
In addition, such unsaturated hydrocarbon molecules are also supplied by the cosmic-ray-induced photodissociation of CH$_{4}$ and subsequent gas-phase combination reactions of CH$_{x}$ radicals (e.g., \citealt{Aikawa1999, Eistrup2016, Yu2016}).
These reaction pathways are efficient in the inner warm disk, at least within the CH$_{4}$ snowline (see also Section \ref{sec:3-2-4rev}).
The saturated hydrocarbons including C$_{2}$H$_{6}$ are formed on the grain-surfaces by the hydrogenation of such unsaturated hydrocarbon ices (e.g., \citealt{Aikawa1999, Bosman2018}).
In addition, \citet{Bosman2018} and \citet{Eistrup2018} discussed that C$_{2}$H$_{6}$ ice is also produced both inside/outside the CO snowline by CH$_{3}$$_{\mathrm{ice}}$ $+$ CH$_{3}$$_{\mathrm{ice}}$, where CH$_{3}$ ice is formed by cosmic-ray-induced photodissociation of CH$_{3}$OH ice. 
Thus, we conclude that grain-surface reactions are needed to form icy-phase saturated hydrocarbon molecules, including C$_{2}$H$_{6}$.
%%
%
%%%
\\ \\
In the shadowed disk ($f\leq0.03$), C$_{2}$H$_{6}$ snowline positions are $r<2$ au, and C$_{2}$H$_{6}$ ice abundances at $r\sim2-10$ au significantly increase ($\sim5\times10^{-6}$)
compared with the values in the non-shadowed disk. We interpret that the reaction pathway starting from CH$_{3}$OH ice destruction becomes more important at these radii of the shadowed disk, since the abundances of unsaturated hydrocarbon ices (such as C$_{2}$H$_{2}$, C$_{3}$H$_{2}$, C$_{3}$H$_{4}$) decrease in the shadowed disk (see Section \ref{sec:3-2-4rev} for details).
\\ \\
%%%
%%%
In the non-shadowed disk ($f=1.0$), the H$_{2}$CO (formaldehyde) and CH$_{3}$OH (methanol) snowline positions are $r\sim1.8$ au ($T(r)\sim100$ K) and $\sim1.5$ au ($T(r)\sim120$ K), respectively, and they are just outside the water snowline ($r=1.3$ au).
The H$_{2}$CO and CH$_{3}$OH ice abundances outside their snowlines increase with increasing $r$.
%%For molecular initial abundances, t
The H$_{2}$CO ice abundances between the H$_{2}$CO and CO snowlines are $\sim8\times10^{-6}$ at most, which is similar to the initial H$_{2}$CO ice abundance ($=8.437\times10^{-6}$). 
The CH$_{3}$OH ice abundances between the CH$_{3}$OH and CO snowlines are $\sim2\times10^{-7}$, which is smaller than the initial CH$_{3}$OH ice abundance ($=6.027\times10^{-7}$).
We suggest that the differences in the CH$_{3}$OH and H$_{2}$CO abundances with respect to the initial abundances inside the CO snowline are also related to whether or not gas-phase formation pathways are present (e.g., \citealt{Fockenberg2002, Atkinson2006, Loomis2015, Walsh2016, Pegues2020}).
Both H$_{2}$CO and CH$_{3}$OH ice abundances increase outside the CO snowline ($\sim(2-5)\times10^{-5}$ for H$_{2}$CO ice and $\sim(2-9)\times10^{-6}$ for CH$_{3}$OH ice).
\\ \\
%%%,
%%%%
The freeze-out of CO onto dust grains plays a central role in producing organic molecules, such as H$_{2}$CO and CH$_{3}$OH.
According to previous studies (e.g., \citealt{Tielens1982, Watanabe2002, Cuppen2009, Fuchs2009, Drozdovskaya2014, Furuya2014, Walsh2014, Walsh2016, Chuang2016, Bosman2018, Aikawa2020}, see also Appendix \ref{Asec:A-2}), the following sequential hydrogenation of CO on the dust grain surfaces leads to the formation of H$_{2}$CO and CH$_{3}$OH ices,
\begin{equation}\label{Rec9}
\mathrm{CO}_{\mathrm{ice}} \xrightarrow{+\mathrm{H}} \mathrm{HCO}_{\mathrm{ice}} \xrightarrow{+\mathrm{H}} \mathrm{H}_{2}\mathrm{CO}_{\mathrm{ice}}, 
\end{equation}
and
\begin{equation}\label{Rec10}
\mathrm{H}_{2}\mathrm{CO}_{\mathrm{ice}}
\xrightarrow{+\mathrm{H}} \mathrm{CH}_{2}\mathrm{OH}_{\mathrm{ice}} / \mathrm{CH}_{3}\mathrm{O}_{\mathrm{ice}} \xrightarrow{+\mathrm{H}} \mathrm{CH}_{3}\mathrm{OH}_{\mathrm{ice}}.
\end{equation}
This grain-surface reaction pathway produces a large amount of H$_{2}$CO and CH$_{3}$OH ices in the cold regions where CO freezes-out onto dust grains.
%%.
The cosmic-ray-induced photodissociation of CH$_{3}$OH ice produces many radicals such as CH$_{3}$O, CH$_{3}$, CH$_{2}$OH ices, and radical-radical reactions on the warmer grains also create more complex organic molecules, in addition to atom addition reactions on the colder grains \citep{Walsh2014}.
We note that our chemical reaction network includes both hydrogenation and abstraction pathways along the methanol formation route \citep{Chuang2016}, and CH$_{2}$OH ice is the dominant methanol H-atom abstraction product.
%%
\begin{comment}
\end{comment}
\\ \\
In the shadowed disk ($f\leq0.03$), the H$_{2}$CO and CH$_{3}$OH ice abundances at $r\sim2-10$ au significantly increase ($\sim(1-5)\times10^{-5}$ for H$_{2}$CO ice and $\sim(2-9)\times10^{-6}$ for CH$_{3}$OH ice) compared with the values in the non-shadowed disk, since the temperature is low ($T(r)\lesssim30$ K) and the sequential hydrogenation of CO ice (Reactions \ref{Rec9} and \ref{Rec10}) proceeds efficiently. 
In contrast, their ice abundances at $r\gtrsim20$ au become a bit smaller ($\sim10^{-5}$ for H$_{2}$CO ice and $\sim(1-3)\times10^{-6}$ for CH$_{3}$OH ice), since the CO adsorption front moves outward with decreasing $f$.
%\\ \\
%
%%
\subsubsection{Dominant nitrogen-bearing molecules}\label{sec:3-2-3rev}
%%\subsubsection{N$_{2}$, NH$_{3}$, HCN, and NH$_{2}$CHO}\label{sec:3-2-3rev}
%%%%
\begin{figure*}[hbtp]
\begin{center}
\vspace{1cm}
%\hspace{1cm}
%\plotone{cost.pdf}
%\plotone[scale=0.63]{{20210624_surface-density.eps}
%\plotone[scale=0.63]{{20210624_ngas-T.eps}
\includegraphics[scale=0.63]{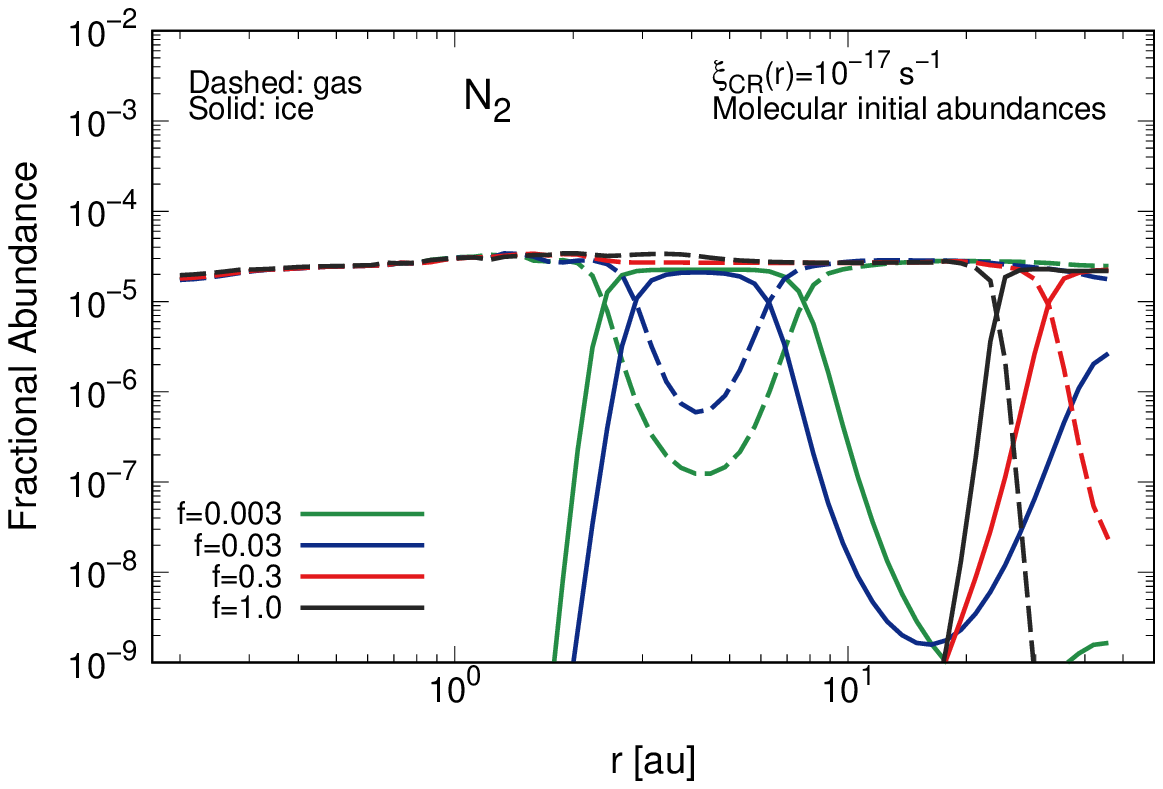}
\includegraphics[scale=0.63]{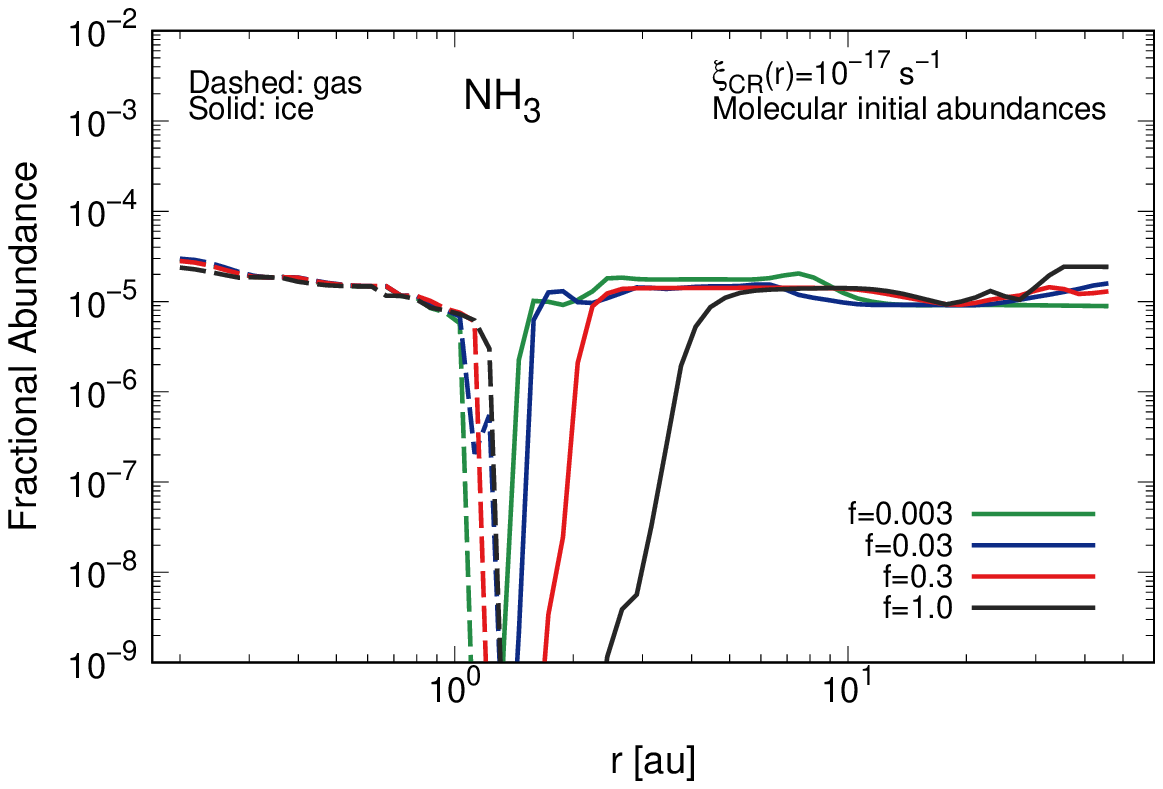}
\includegraphics[scale=0.63]{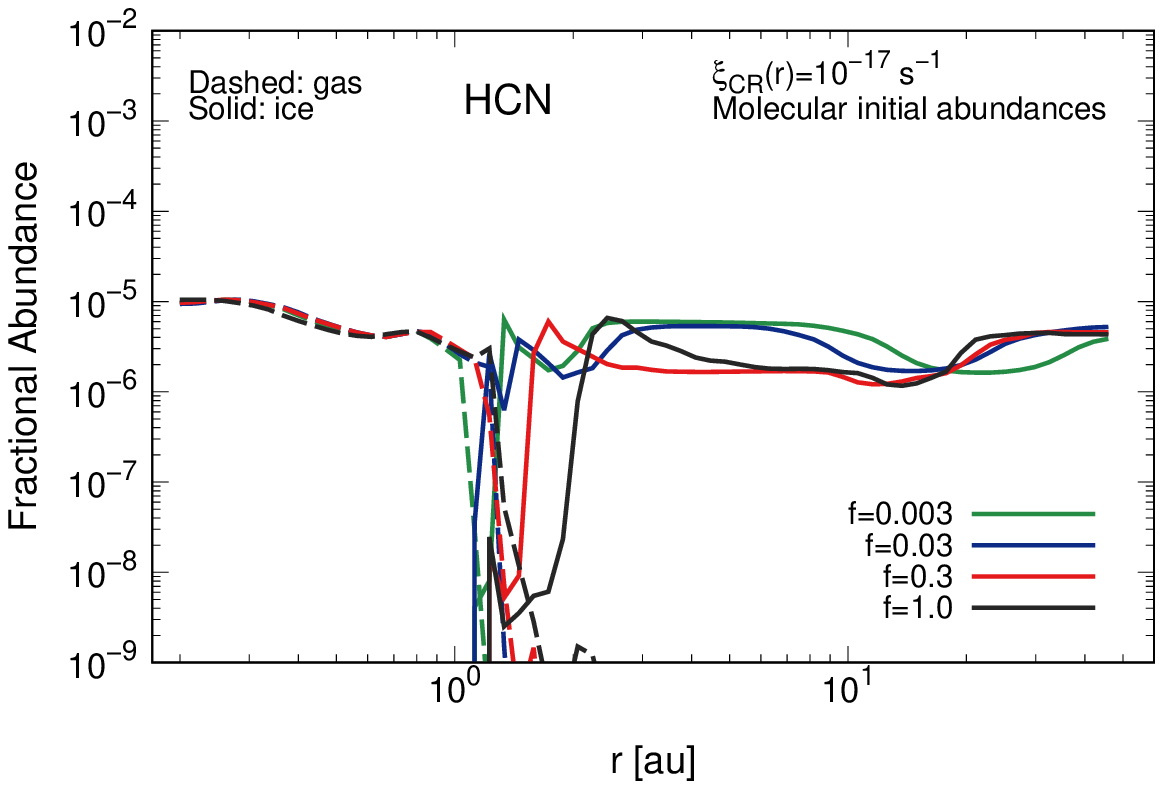}
\includegraphics[scale=0.63]{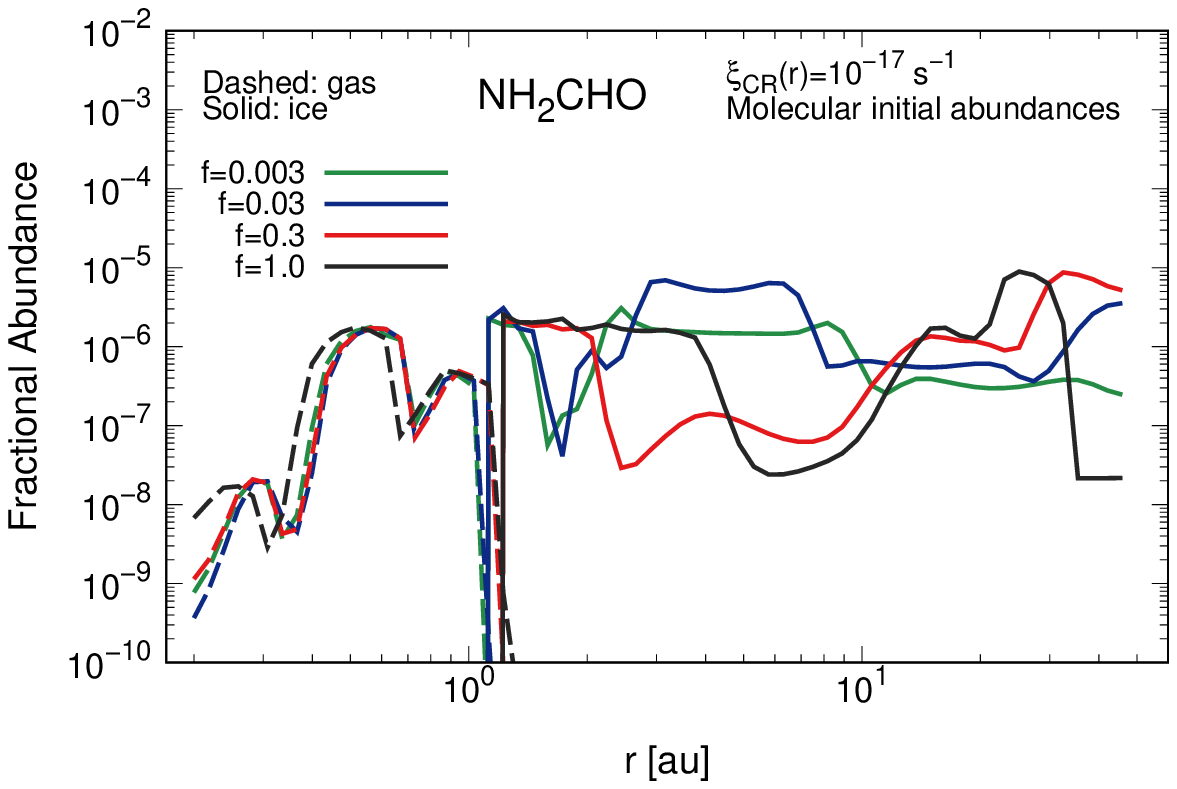}
%\includegraphics[scale=0.58]{20210902_t3_ds_0.1um_atomic-ini_17_Xray-rev_enhanced-water_nH-rev_1.0e6yr_NH2CHOgi.eps}
%%\includegraphics[scale=0.58]{20210902_t1_ds_0.1um_mol-ini_17_Xray-rev_enhanced-water_nH-rev_1.0e6yr_NH2OHgi.eps}
%%\includegraphics[scale=0.58]{20210902_t3_ds_0.1um_atomic-ini_17_Xray-rev_enhanced-water_nH-rev_1.0e6yr_NH2OHgi.eps}
%%\includegraphics[scale=0.58]{20210902_t1_ds_0.1um_mol-ini_17_Xray-rev_enhanced-water_nH-rev_1.0e6yr_HC3Ngi.eps}
%%\includegraphics[scale=0.58]{20210902_t3_ds_0.1um_atomic-ini_17_Xray-rev_enhanced-water_nH-rev_1.0e6yr_HC3Ngi.eps}
%%%
\end{center}
\vspace{-0.2cm}
\caption{
%The radial profiles of fractional abundances with respect to total hydrogen nuclei densities at t=$10^{6}$ years for 
Same as Figure \ref{Figure3_rev_radial}, but for 
N$_{2}$ ($n_{\mathrm{N}_{2}}$/$n_{\mathrm{H}}$, top left panel), 
NH$_{3}$ ($n_{\mathrm{NH}_{3}}$/$n_{\mathrm{H}}$, top right panel),
HCN ($n_{\mathrm{HCN}}$/$n_{\mathrm{H}}$, bottom left panel), and
NH$_{2}$CHO ($n_{\mathrm{NH}_{2}\mathrm{CHO}}$/$n_{\mathrm{H}}$, bottom right panel).
%%The dashed and solid lines show the profiles for gaseous and icy molecules, respectively.
%\vspace{0.2cm}
\\ \\
}\label{Figure5_rev_radial}
\end{figure*}
\vspace{0.2cm}
%%
%%%20211001 22:47
%%%
Figure \ref{Figure5_rev_radial} shows the radial profiles of fractional abundances for dominant nitrogen-bearing molecules (N$_{2}$, NH$_{3}$, HCN, and NH$_{2}$CHO).
In the non-shadowed disk ($f=1.0$), the N$_{2}$ (molecular nitrogen) snowline position is $r\sim24$ au ($T(r)\sim23$ K), which is slightly outside the CO snowline.
%%%%%
For the atomic initial abundances, atomic N quickly forms N$_{2}$ gas and it freezes out onto dust grains outside its snowline \citep{Schwarz2014, Eistrup2016}.
In addition, for the molecular initial abundances, the initial atomic N abundance ($=2.1\times10^{-5}$, see Table \ref{Table:1}) is similar to the initial N$_{2}$ gas $+$ ice abundance ($=1.5\times10^{-5}$, see Table \ref{Table:1}), and this atomic N is also quickly converted to N$_{2}$ in the disk.
%%%%
Thus, the N$_{2}$ gas $+$ ice abundances are relatively constant ($\sim(2-3)\times10^{-5}$) throughout the disk for various values of $f$ and various initial conditions (see also Appendix \ref{Asec:A-3}).
\\ \\
In the shadowed disk ($f\leq0.03$), like CO, N$_{2}$ freezes-out onto dust grains at around the current orbit of Jupiter ($r\sim3-8$ au), and N$_{2}$ ice abundances are $\sim2\times10^{-5}$. 
In addition, N$_{2}$ returns to the gas phase at $r>8$ au.
At $r>20$ au, N$_{2}$ (re-)freezes out onto dust grains, and the adsorption front moves outward with decreasing $f$, since $T(r)$ at $r>20$ au increases ($\lesssim35$ K at the maximum) with decreasing $f$ (see also Section 2.1).
This N$_{2}$ abundance profile is consistent with the results reported in \citet{Ohno2021}.
\\ \\
%%%
In the non-shadowed disk ($f=1.0$), the NH$_{3}$ (ammonia) snowline position is $r\sim2.3$ au, which is slightly outside the water snowline ($r=1.3$ au).
NH$_{3}$ ice abundances just outside the NH$_{3}$ snowline ($r\sim2-4$ au) are $\ll10^{-6}$, and those at $r>5$ au are $\sim1\times10^{-5}$, which is similar to the initial NH$_{3}$ ice abundance ($=1.327\times10^{-5}$).
We suggest that on the warm grain surfaces at $r\sim2-4$ au, NH$_{3}$ ice is converted to NH$_{2}$CHO ice via the following grain-surface reaction route \citep{Jones2011, Walsh2014, Lopez-Sepulcre2015},
\begin{equation}\label{Rec11.5}
%%%
 \mathrm{NH}_{2 \ \mathrm{ice}} + \mathrm{HCO}_{\mathrm{ice}} \rightarrow \mathrm{NH}_{2}\mathrm{CHO}_{\mathrm{ice}},
\end{equation}
where NH$_{2}$ ice is formed by the cosmic-ray-induced photodissociation of NH$_{3}$ ice.
%.
\\ \\
In the shadowed disk ($f\leq0.03$), NH$_{3}$ snowline positions are $r\lesssim2$ au.
NH$_{3}$ ice abundances at $r\sim2-5$ au are significantly enhanced, since the disk temperature decreases and the above  destruction route of NH$_{3}$ ice (see Reaction \ref{Rec11.5}) becomes inefficient.
%%%.
\\ \\
%%% 
Formamide (NH$_{2}$CHO) has been proposed as a key precursor of various (pre)metabolic and (pre)genetic molecules such as pyruvic acid and adenine \citep{Saladino2012}.
Gas-phase NH$_{2}$CHO has been detected in pre-stellar cores and protostar envelopes (e.g., \citealt{Kahane2013, Lopez-Sepulcre2015, Okoda2021}) and comets (e.g., \citealt{Bockelee-Morvan2000}), but has not yet been detected in Class II disks.
The NH$_{2}$CHO snowline position ($r\sim1.2$ au, $T(r)\sim150$ K) is slightly inside the water snowline ($E_{\mathrm{des}}$(NH$_{2}$CHO)$=5560$ K, $E_{\mathrm{des}}$(H$_{2}$O)$=4880$ K), thus the NH$_{2}$CHO snowline position does not move in our chemical modeling.
\\ \\
In the non-shadowed disk ($f=1.0$), the NH$_{2}$CHO ice abundances are $\gtrsim10^{-6}$ in the outer cold disk (around the CO/N$_{2}$ snowlines, $r\sim14-33$ au), and $<10^{-7}$ at $r\lesssim10$ au.
In addition, NH$_{2}$CHO is efficiently formed ($\gtrsim10^{-6}$) just outside its snowline ($r\sim1.2-4$ au).
We interpret that just outside its snowline, the above radical-radical formation route on the warm dust grains is efficient (see Reaction \ref{Rec11.5}).
%%%
\\ \\
In the shadowed disk ($f\leq0.03$), NH$_{2}$CHO ice abundances at around the current orbit of Jupiter ($r\sim3-8$ au) are enhanced ($\gtrsim10^{-6}$). 
%%. 
We propose that the following hydrogenation routes become efficient in this region.
\citet{Garrod2008}, \citet{Walsh2014}, and \citet{Lopez-Sepulcre2015} discussed that grain-surface NH$_{2}$CHO can form via the atom addition reactions in the cold region, such as the following reaction routes, 
\begin{equation}\label{Rec12}
\mathrm{OCN}_{\mathrm{ice}} \xrightarrow{+\mathrm{H}} \mathrm{HNCO}_{\mathrm{ice}} \xrightarrow{+\mathrm{H}} 
\mathrm{NH}_{2}\mathrm{CO}_{\mathrm{ice}}/\mathrm{NHCHO}_{\mathrm{ice}} \\
\end{equation}
and 
\begin{equation}\label{Rec12-1}
\mathrm{NH}_{2}\mathrm{CO}_{\mathrm{ice}}/\mathrm{NHCHO}_{\mathrm{ice}} \xrightarrow{+\mathrm{H}}  \mathrm{NH}_{2}\mathrm{CHO}_{\mathrm{ice}},
\end{equation}
where OCN$_{\mathrm{ice}}$ is formed via the grain-surface reactions (CN$_{\mathrm{ice}}$+O$_{\mathrm{ice}}$ and ON$_{\mathrm{ice}}$+C$_{\mathrm{ice}}$) or gas-phase radical-radical reactions.
We note that our chemical reaction network includes a competing hydrogenation abstraction pathway during hydrogenation from HNCO to NH$_{2}$CHO, on the basis of the experimental results in \citet{Noble2015}.
%%%
%%
\\ \\
%%%
HCN (hydrogen cyanide) and CN (cyanide) are the simplest molecules containing both C and N atoms, and reactions including HCN lead to the formation of more complex cyanides (such as CH$_{3}$CN and aminoacetonitrile H$_{2}$NCH$_{2}$CN) \citep{Oberg2011, Noble2013}.
In the non-shadowed disk ($f=1.0$), the HCN snowline position is $r\sim1.5$ au ($T(r)\sim120$ K), which is just outside the water snowline ($r=1.3$ au).
HCN ice abundances are $\sim(1-3)\times10^{-6}$ at $r\sim3-20$ au and $\sim4\times10^{-6}$ at $r\gtrsim20$ au, which 
are slightly smaller/larger than the initial HCN ice abundance ($=2.772\times10^{-6}$).
%%.
\citet{Eistrup2016} showed much lower HCN ice abundances ($\ll10^{-7}$ outside the HCN snowline) for their model calculations with molecular initial abundances, since they set the initial HCN ice abundance to zero.
\\ \\
In the shadowed disk ($f\leq0.03$), the HCN ice abundances at around the current orbit of Jupiter ($r\sim3-8$ au) are slightly enhanced ($\sim5\times10^{-6}$).
%%%. 
According to \citet{Aikawa1999} and \citet{Eistrup2016}, HCN is formed through the gas-phase reaction of HCO with N atom, with subsequent freeze-out onto dust grains.
HCO gas is efficiently formed by hydrogenation of CO on the cold dust-grain surface and cosmic-ray-induced desorption of HCO ice (see also Sections \ref{sec:3-2-1rev} and \ref{sec:3-2-2rev}).
\citet{Schwarz2014} and \citet{Eistrup2018} showed that HCN ice becomes abundant in the outer cold region of the disk where CO freezes-out onto dust grains, although it is efficiently produced by a few Myr.
Our calculations for both non-shadowed and shadowed disks also show that HCN ice is efficiently formed at the cold regions where CO freezes out onto dust grains. 
Thus we confirm that the above HCN formation route which starts from hydrogenation of CO is dominant in the disks.
%%.
\subsubsection{Other organic molecules}\label{sec:3-2-4rev}
%%%
%%%
\begin{figure*}[hbtp]
\begin{center}
\vspace{1cm}
%\hspace{1cm}
%\plotone{cost.pdf}
%\plotone[scale=0.63]{{20210624_surface-density.eps}
%\plotone[scale=0.63]{{20210624_ngas-T.eps}
%%\includegraphics[scale=0.58]{20210902_t1_ds_0.1um_mol-ini_17_Xray-rev_enhanced-water_nH-rev_1.0e6yr_O2gi.eps}
%%\includegraphics[scale=0.58]{20210902_t3_ds_0.1um_atomic-ini_17_Xray-rev_enhanced-water_nH-rev_1.0e6yr_O2gi.eps}
%%\includegraphics[scale=0.58]{20210902_t1_ds_0.1um_mol-ini_17_Xray-rev_enhanced-water_nH-rev_1.0e6yr_CH4gi.eps}
%%\includegraphics[scale=0.58]{20210902_t3_ds_0.1um_atomic-ini_17_Xray-rev_enhanced-water_nH-rev_1.0e6yr_CH4gi.eps}
\includegraphics[scale=0.63]{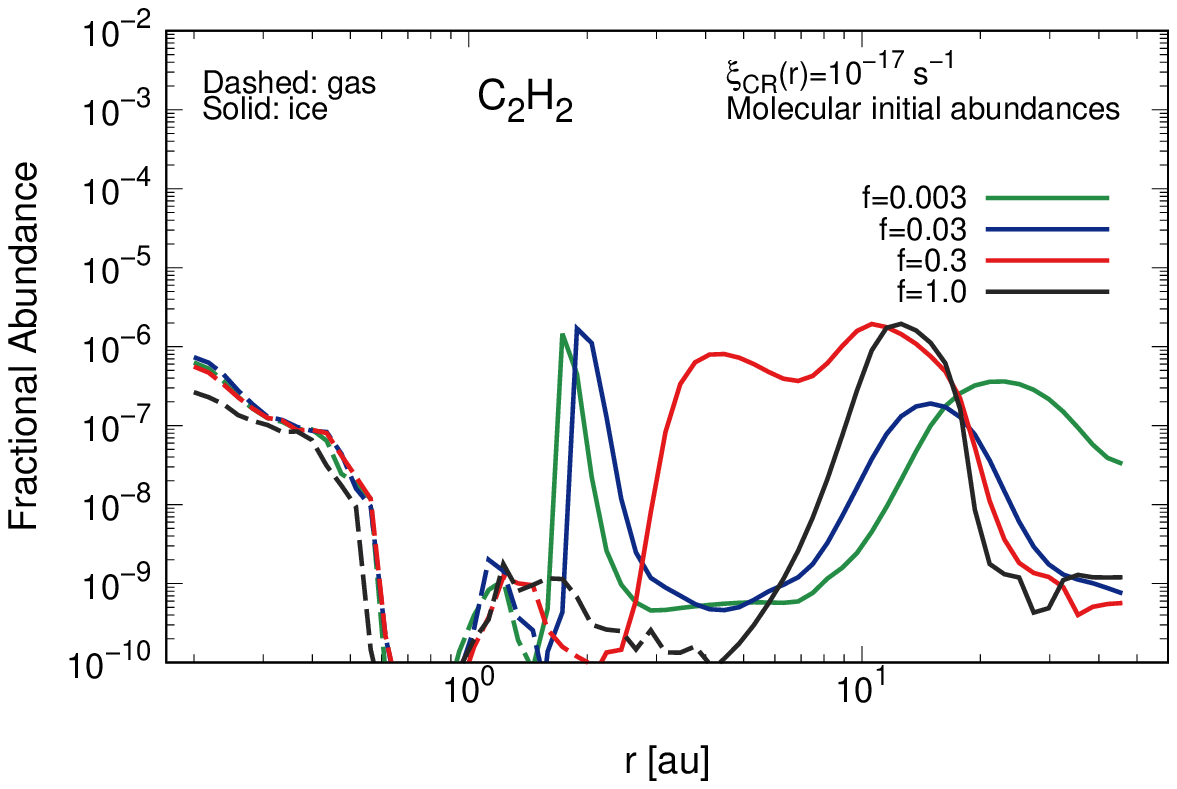}
\includegraphics[scale=0.63]{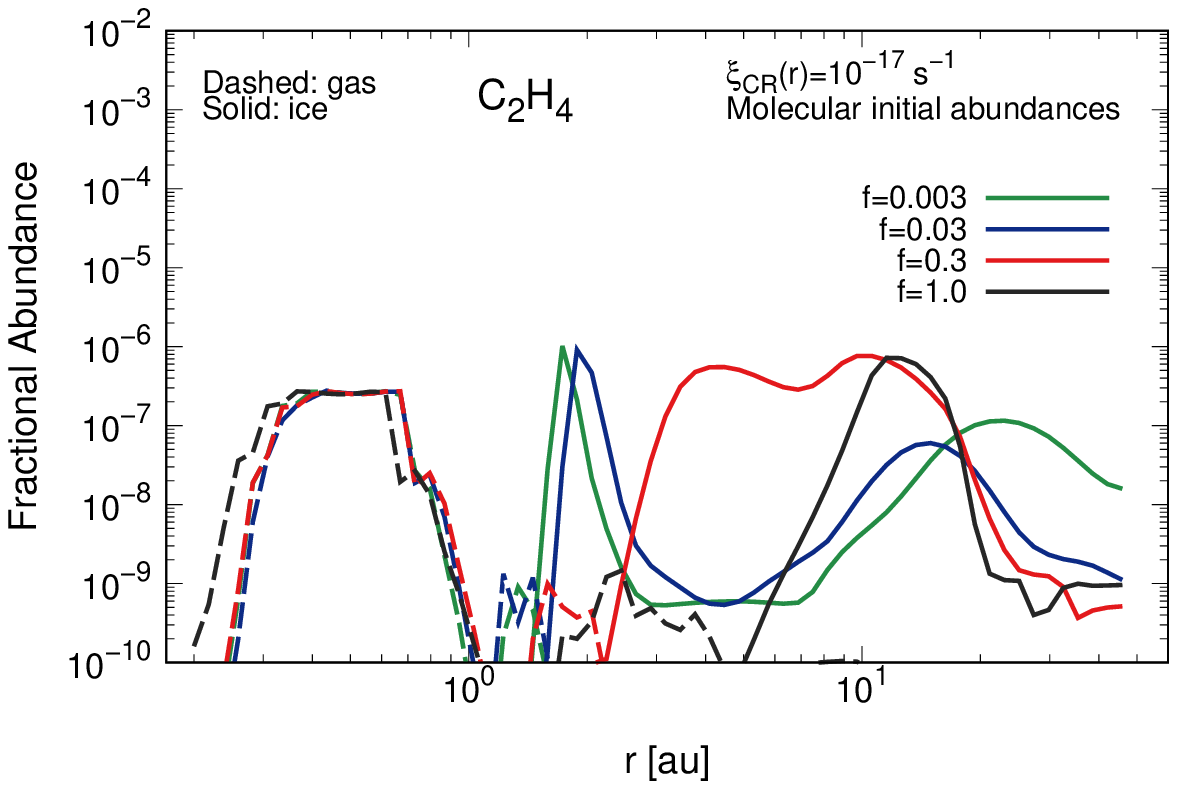}
\includegraphics[scale=0.63]{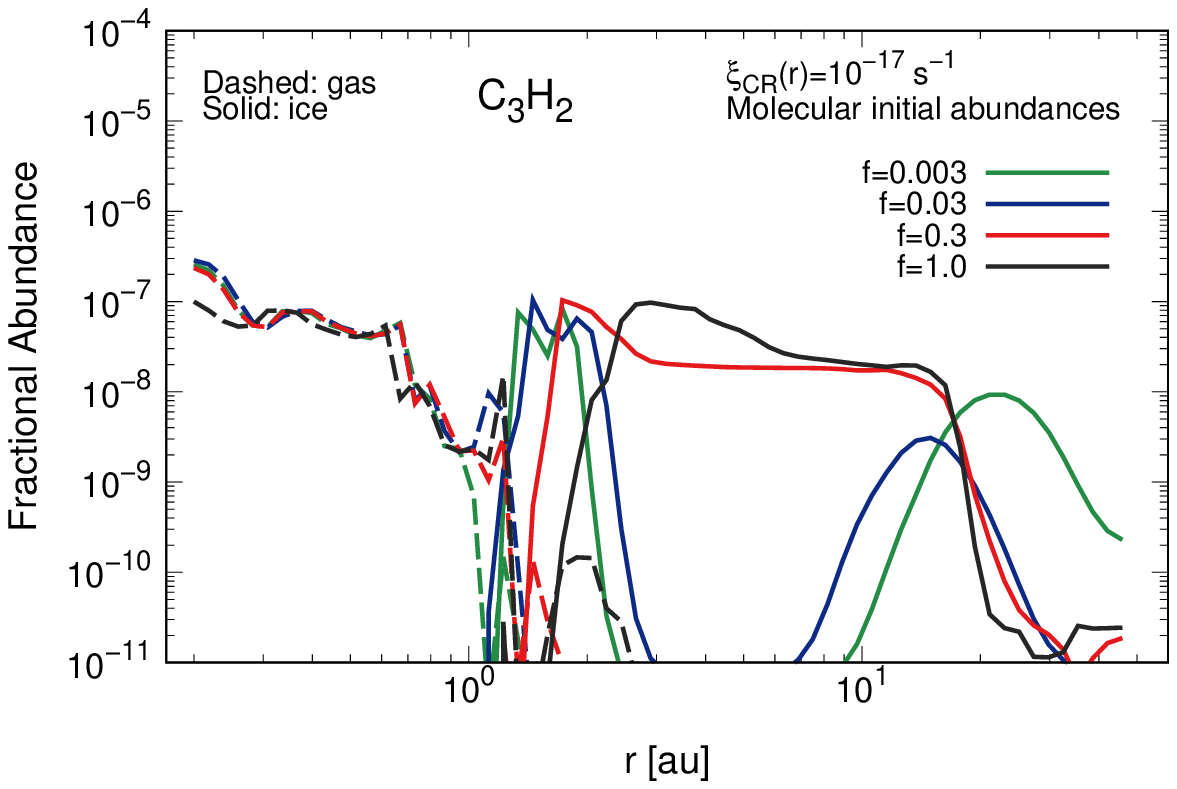}
\includegraphics[scale=0.63]{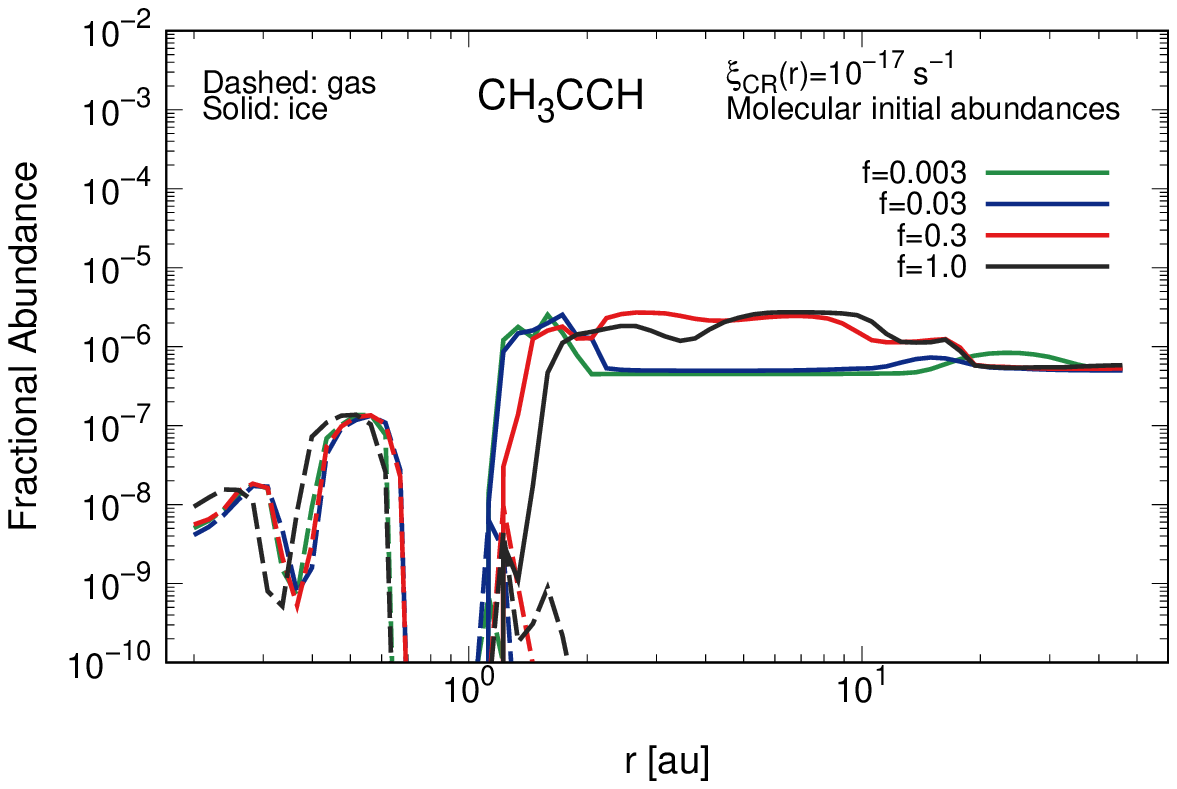}
\includegraphics[scale=0.63]{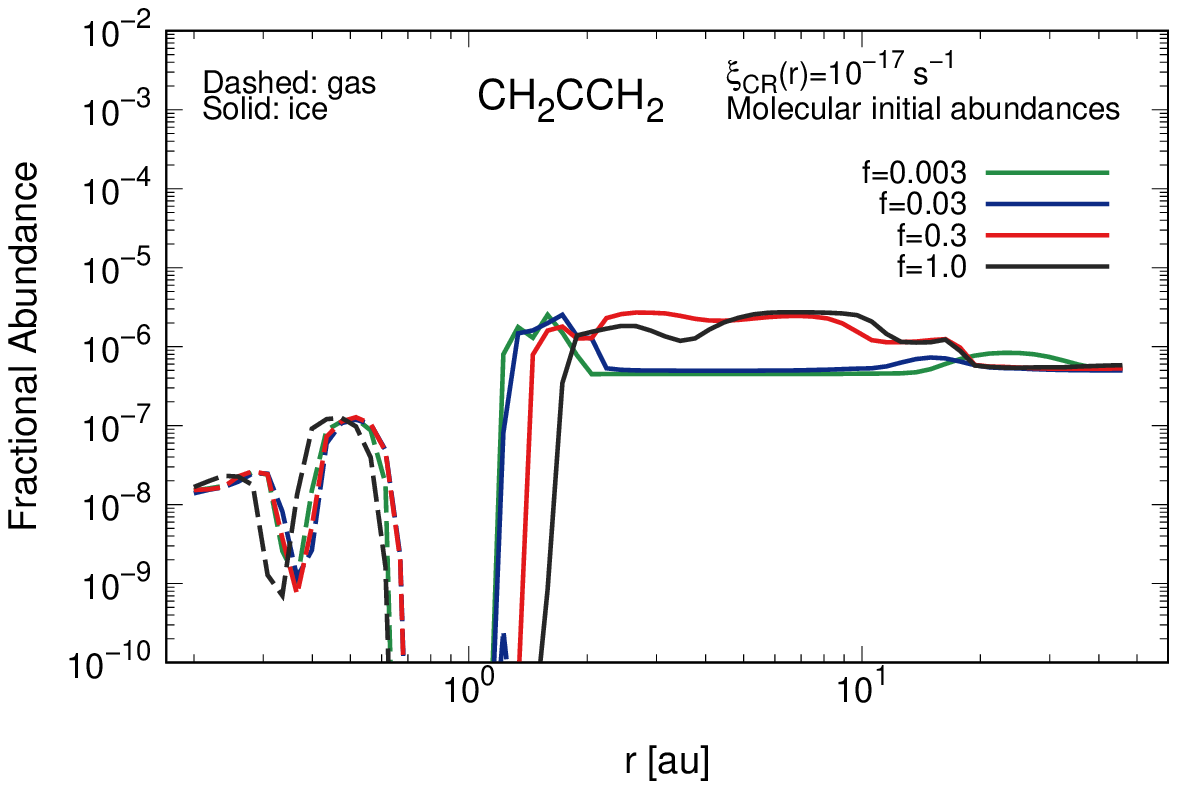}
\includegraphics[scale=0.63]{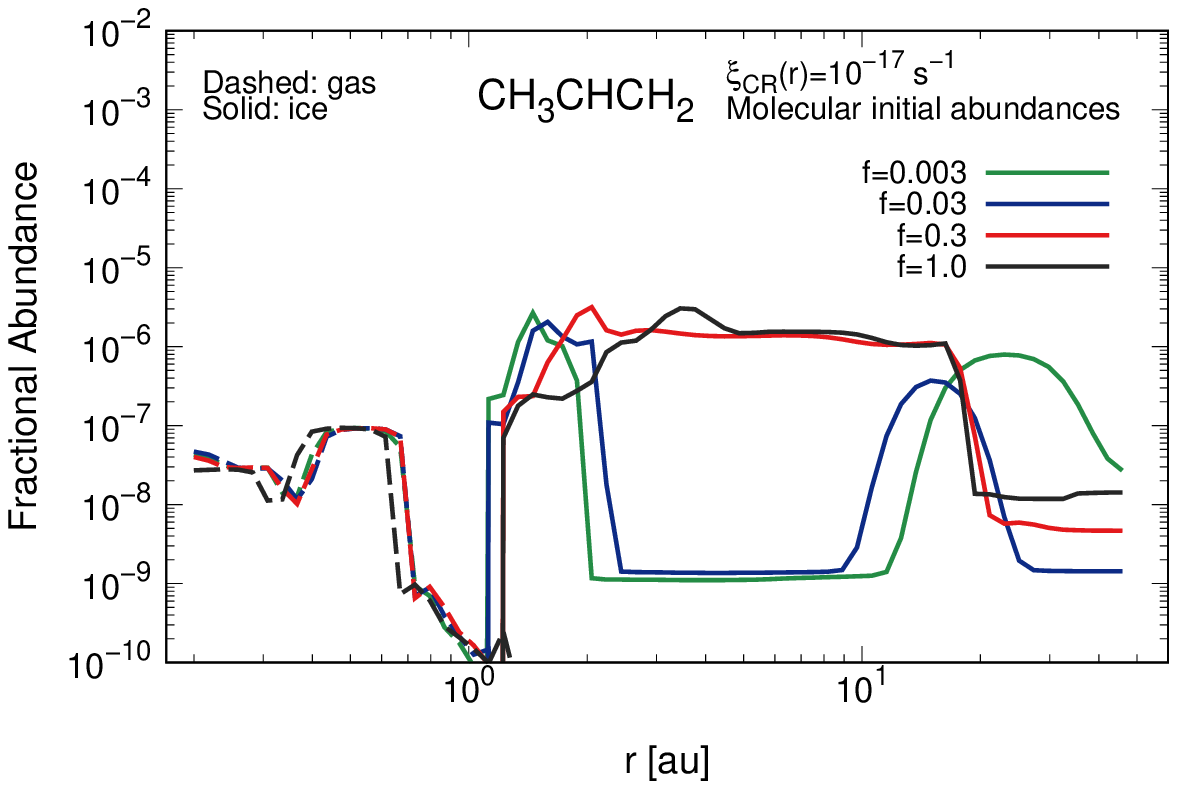}
\end{center}
\vspace{-0.2cm}
\caption{
%%The radial profiles of fractional abundances with respect to total hydrogen nuclei densities at t=$10^{6}$ years for 
Same as Figure \ref{Figure3_rev_radial}, but for 
%%CH$_{4}$ ($n_{\mathrm{CH}_{4}}$/$n_{\mathrm{H}}$, top panels), 
C$_{2}$H$_{2}$ ($n_{\mathrm{C}_{2}\mathrm{H}_{2}}$/$n_{\mathrm{H}}$, top left panel),
C$_{2}$H$_{4}$ ($n_{\mathrm{C}_{2}\mathrm{H}_{4}}$/$n_{\mathrm{H}}$, top right panel),
C$_{3}$H$_{2}$ ($n_{\mathrm{C}_{3}\mathrm{H}_{2}}$/$n_{\mathrm{H}}$, middle left panel),
CH$_{3}$CCH ($n_{\mathrm{CH}_{3}\mathrm{CCH}}$/$n_{\mathrm{H}}$, middle right panel),
CH$_{2}$CCH$_{2}$ ($n_{\mathrm{CH}_{2}\mathrm{CCH}_{2}}$/$n_{\mathrm{H}}$, bottom left panel), and
CH$_{3}$CHCH$_{2}$ ($n_{\mathrm{CH}_{3}\mathrm{CHCH}_{2}}$/$n_{\mathrm{H}}$, bottom right panel).
%%In the profiles in Figures \ref{Figure6_rev_radial}-\ref{Figure8_rev_radial}, only the profiles for molecular initial abundances are shown.
%%
\\ \\
%%\vspace{0.2cm}
}\label{Figure6_rev_radial}
\end{figure*}
%%%
%%%
%%\\ \\
\begin{comment}
\end{comment}
In this Section \ref{sec:3-2-4rev}, we show the radial abundance distributions of other major carbon-, nitrogen-, and oxygen-bearing molecules, mainly larger organic molecules \footnote{We note that NH$_{2}$OH is technically an inorganic molecule since it does not include carbon.}.
\\ \\
Figure \ref{Figure6_rev_radial} shows the radial profiles of fractional abundances for C$_{2}$H$_{2}$ (acetylene), C$_{2}$H$_{4}$ (ethylene), C$_{3}$H$_{2}$ (cyclopropyne), 
C$_{3}$H$_{4}$ (propyne CH$_{3}$CCH and allene CH$_{2}$CCH$_{2}$), and C$_{3}$H$_{6}$ (propylene, CH$_{3}$CHCH$_{2}$). 
These are dominant unsaturated hydrocarbon molecules.
%%.
In the non-shadowed disk (f = 1.0), C$_{2}$H$_{2}$ and C$_{2}$H$_{4}$ ice abundances are largest ($\sim(1-2)\times10^{-6}$ and $\sim(3-8)\times10^{7}$, respectively) at $r\sim10-15$ au, which is just inside the CH$_{4}$ snowline.
In the shadowed disk ($f\leq0.03$), these ice abundances significantly decrease ($\ll10^{-7}$) at $r\sim3-15$ au, and increase ($\sim10^{-6}$) at $r\sim2$ au.
Moreover, in the non-shadowed disk the ice abundances of other molecules (C$_{3}$H$_{2}$, C$_{3}$H$_{4}$, and C$_{3}$H$_{6}$) are enhanced at $r\sim2-15$ au, whereas in the shadowed disk they decrease at $r\sim2-15$ au and increase at $r\sim1-2$ au.
\\ \\
As we describe in Section \ref{sec:3-2-2rev} (see also e.g., \citealt{Aikawa1999, Furuya2014, Eistrup2016, Eistrup2018, Yu2016, Bosman2018}), the formation pathways of these unsaturated hydrocarbon molecules are efficient within the CH$_{4}$ snowline, since the gas-phase chemical reaction pathways starting from CH$_{4}$ $+$ C$^{+}$, and the cosmic-ray-induced photodissociation of CH$_{4}$ and subsequent gas-phase combination reactions of CH$_{x}$ radicals are active.
%%with CH$_{4}$ and CH$_{x}$ radicals are important.
In the shadowed disks ($f\leq0.03$), the CH$_{4}$ snowline positions are $r\lesssim2$ au and such gas-phase formation reactions do not proceed efficiently at $r\gtrsim2$ au. Thus, the unsaturated hydrocarbon ices are deficient at $r\sim3-15$ au (including the current orbit of Jupiter), in contrast to saturated hydrocarbons such as CH$_{4}$ and C$_{2}$H$_{6}$.
%%%
%%%
%
%%%
%%%
\begin{figure*}[hbtp]
\begin{center}
\vspace{1cm}
%\hspace{1cm}
%\plotone{cost.pdf}
%\plotone[scale=0.63]{{20210624_surface-density.eps}
%\plotone[scale=0.63]{{20210624_ngas-T.eps}
%\includegraphics[scale=0.58]{20210902_t1_ds_0.1um_mol-ini_17_Xray-rev_enhanced-water_nH-rev_1.0e6yr_H2COgi.eps}
%\includegraphics[scale=0.58]{20210902_t3_ds_0.1um_atomic-ini_17_Xray-rev_enhanced-water_nH-rev_1.0e6yr_H2COgi.eps}
%\includegraphics[scale=0.58]{20210902_t1_ds_0.1um_mol-ini_17_Xray-rev_enhanced-water_nH-rev_1.0e6yr_CH3OHgi.eps}
%\includegraphics[scale=0.58]{20210902_t3_ds_0.1um_atomic-ini_17_Xray-rev_enhanced-water_nH-rev_1.0e6yr_CH3OHgi.eps}
%%%
\includegraphics[scale=0.63]{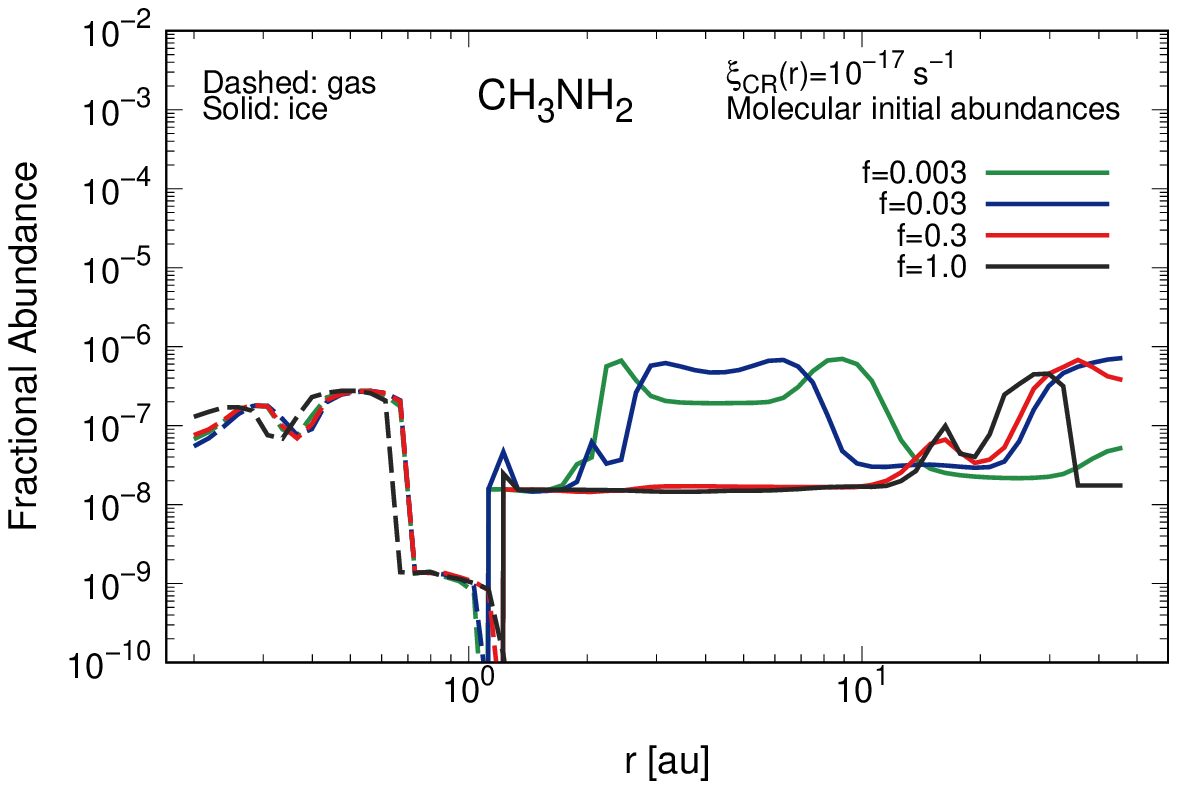}
\includegraphics[scale=0.63]{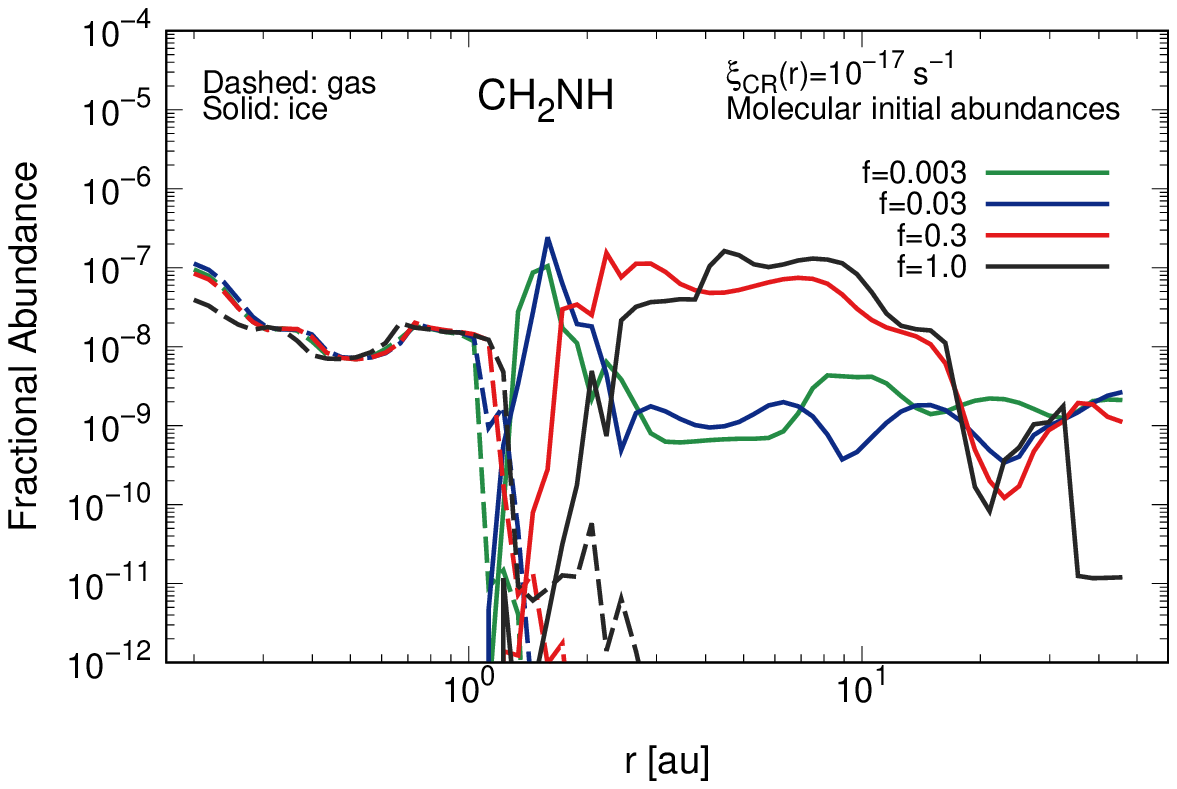}
\includegraphics[scale=0.63]{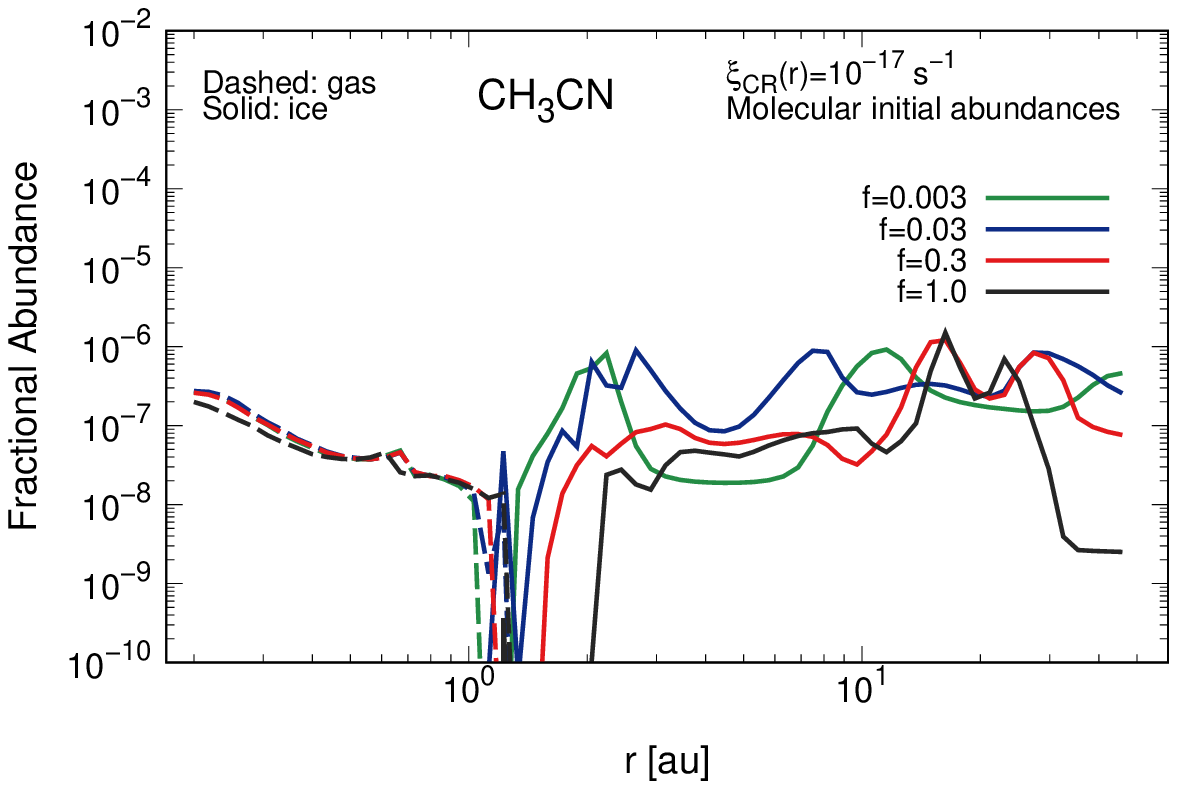}
\includegraphics[scale=0.63]{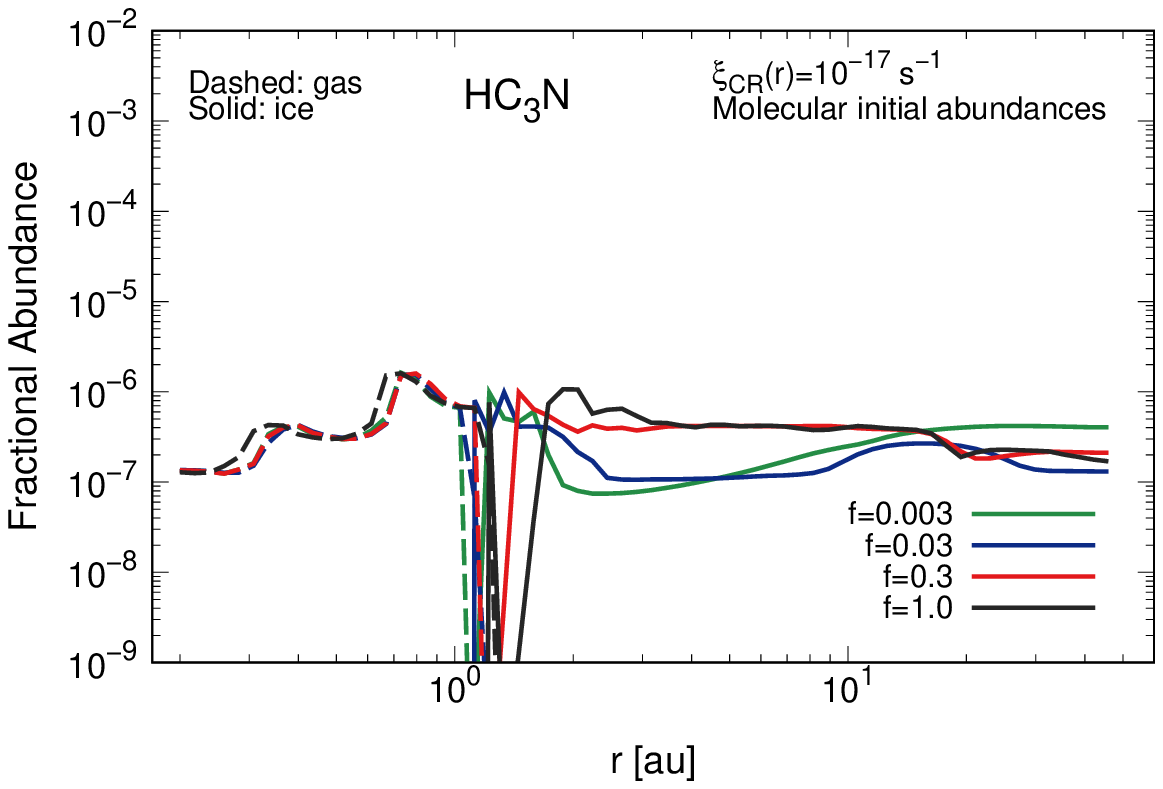}
\includegraphics[scale=0.63]{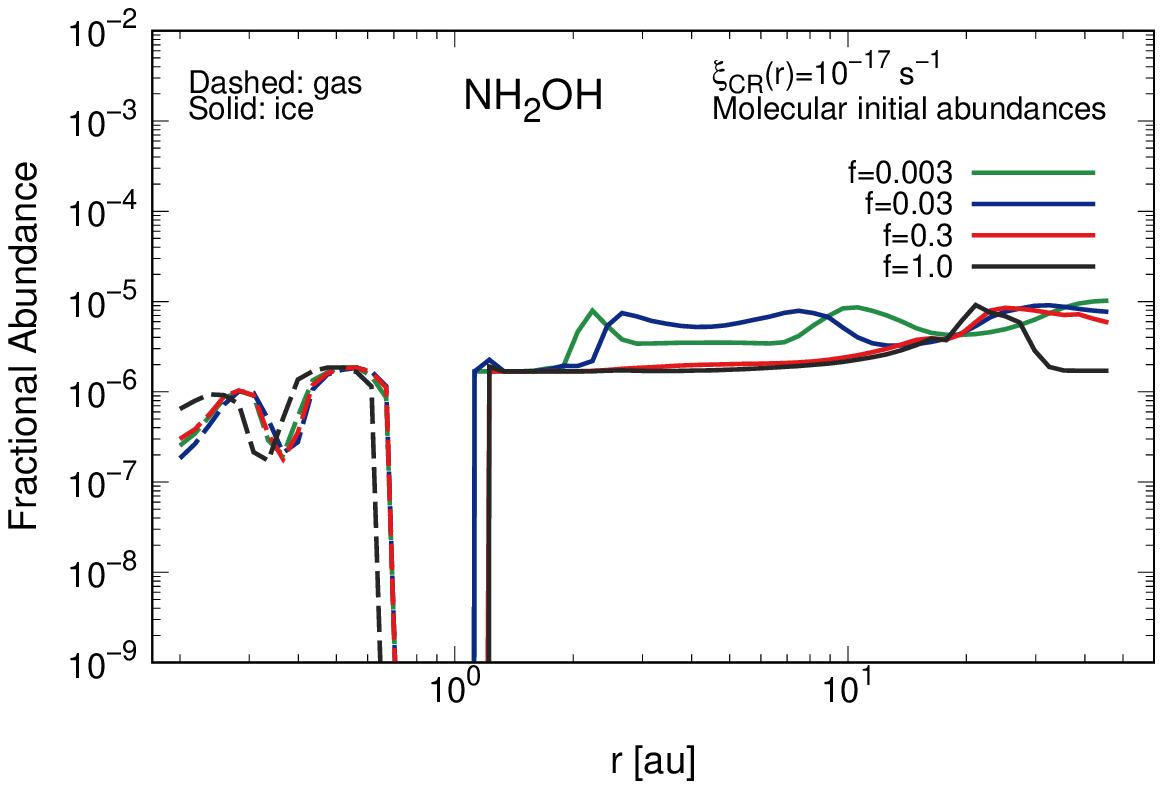}
%%%
\includegraphics[scale=0.63]{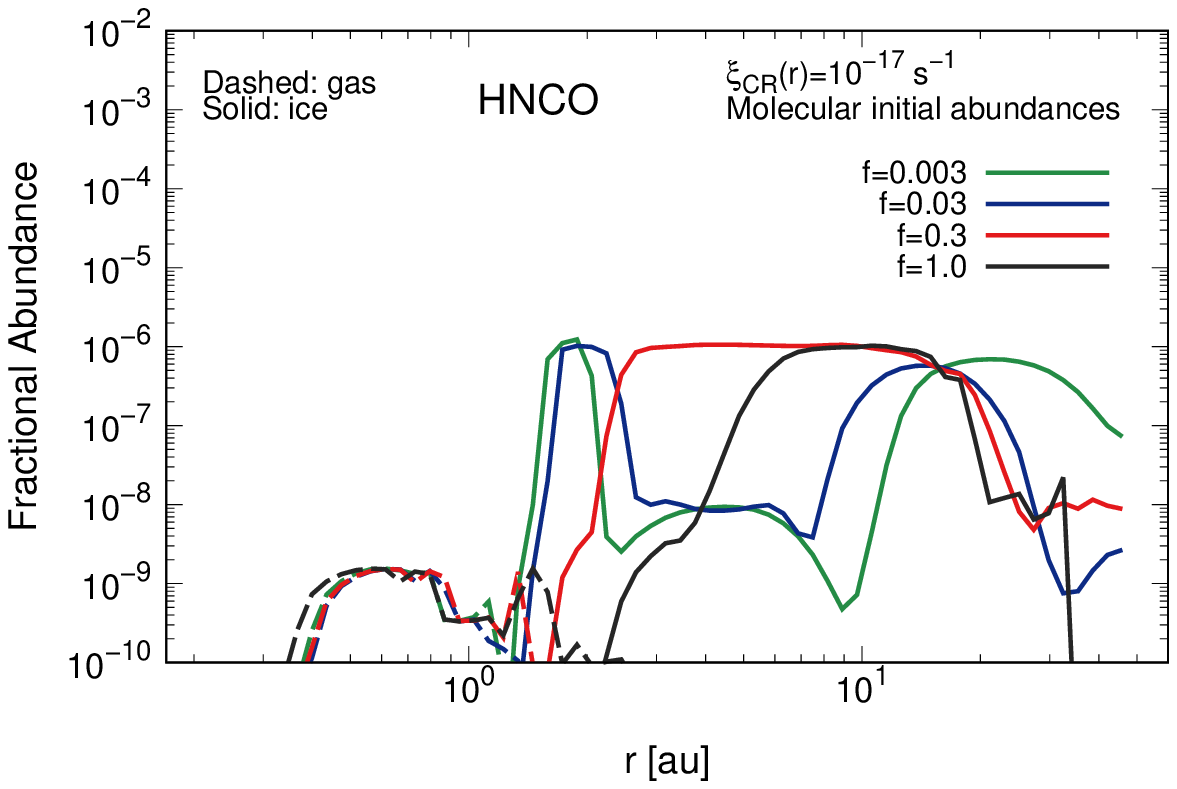}
%%%
\end{center}
\vspace{-0.2cm}
\caption{
%The radial profiles of fractional abundances with respect to total hydrogen nuclei densities at t=$10^{6}$ years for 
Same as Figure \ref{Figure3_rev_radial}, but for 
CH$_{3}$NH$_{2}$ ($n_{\mathrm{CH}_{3}\mathrm{NH}_{2}}$/$n_{\mathrm{H}}$, top left panel),
CH$_{2}$NH ($n_{\mathrm{CH}_{2}\mathrm{NH}}$/$n_{\mathrm{H}}$, top right panel),
CH$_{3}$CN ($n_{\mathrm{CH}_{3}\mathrm{CN}}$/$n_{\mathrm{H}}$, middle left panel),
HC$_{3}$N ($n_{\mathrm{HC}_{3}\mathrm{N}}$/$n_{\mathrm{H}}$, middle right panel),
NH$_{2}$OH ($n_{\mathrm{NH}_{2}\mathrm{OH}}$/$n_{\mathrm{H}}$, bottom left panel), and
HNCO ($n_{\mathrm{HNCO}}$/$n_{\mathrm{H}}$, bottom right panel).
\\ \\
%%%%
}\label{Figure7_rev_radial}
\end{figure*}
%%%
%%%
%%%
\begin{figure*}[hbtp]
\begin{center}
\vspace{1cm}
%\hspace{1cm}
%\plotone{cost.pdf}
%\plotone[scale=0.63]{{20210624_surface-density.eps}
%\plotone[scale=0.63]{{20210624_ngas-T.eps}
\includegraphics[scale=0.63]{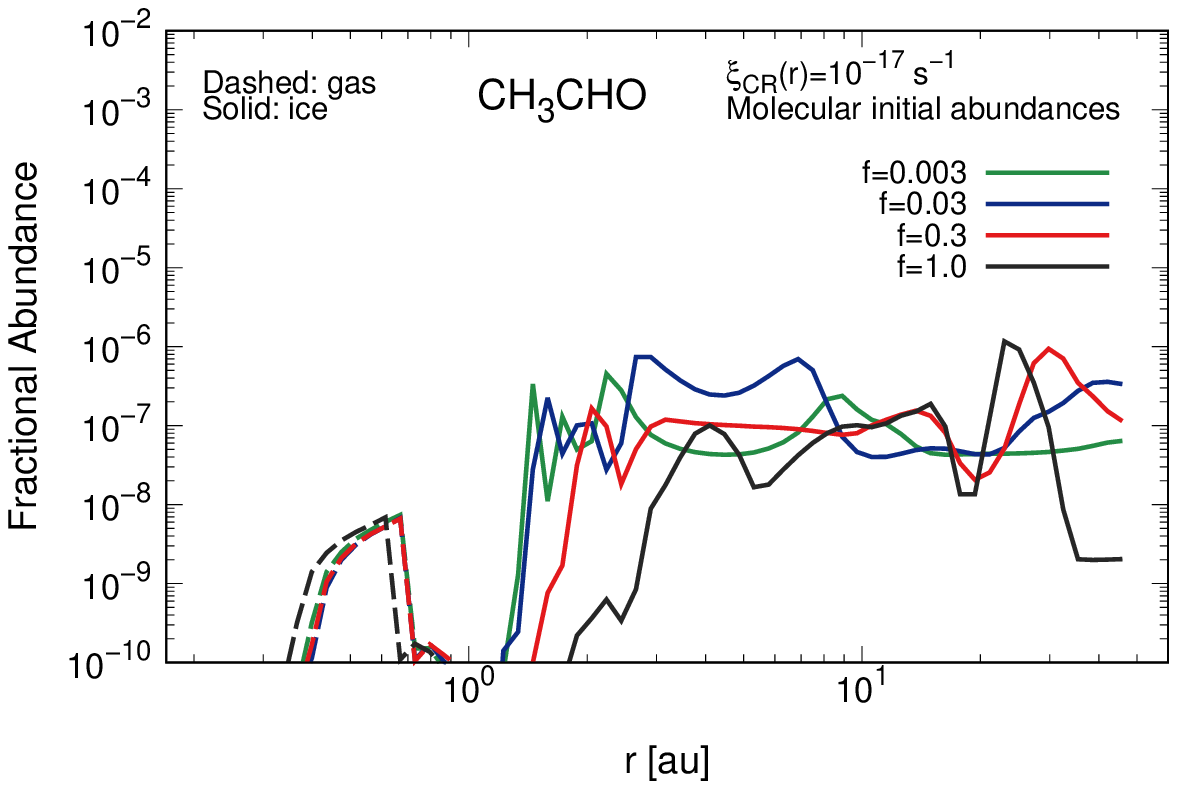}
\includegraphics[scale=0.63]{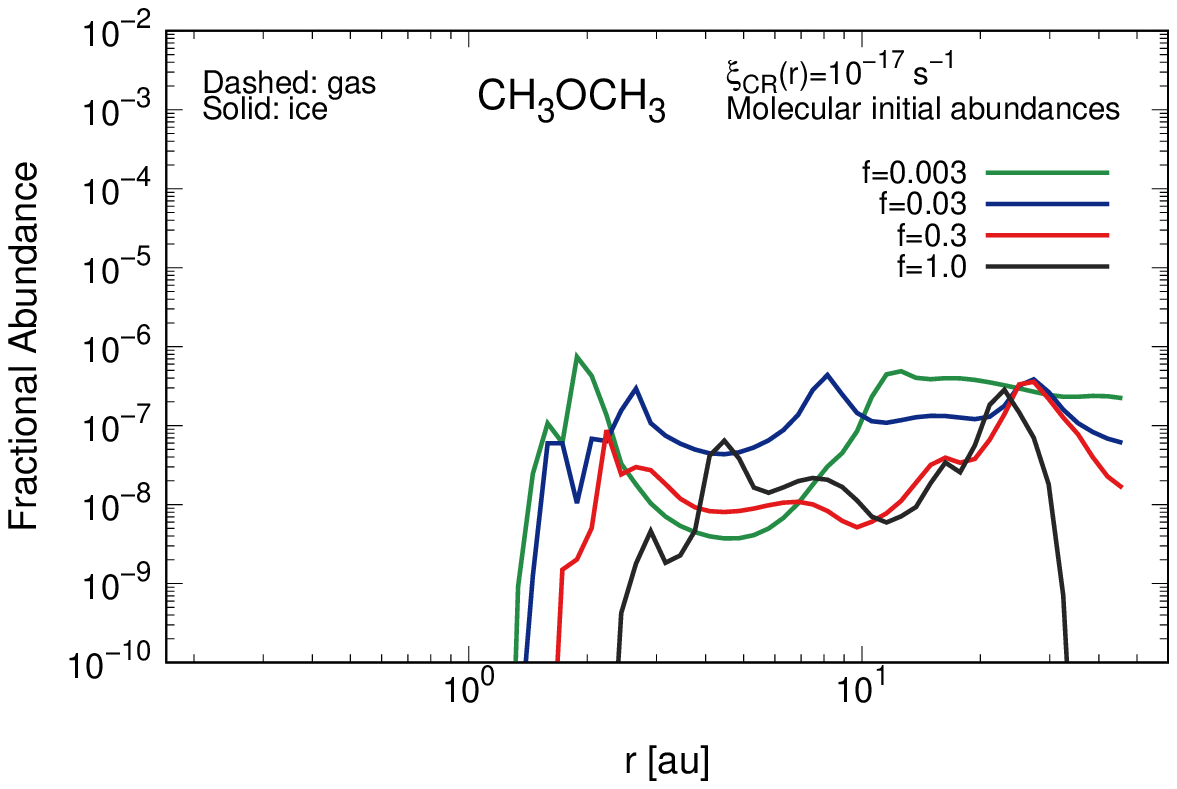}
\includegraphics[scale=0.63]{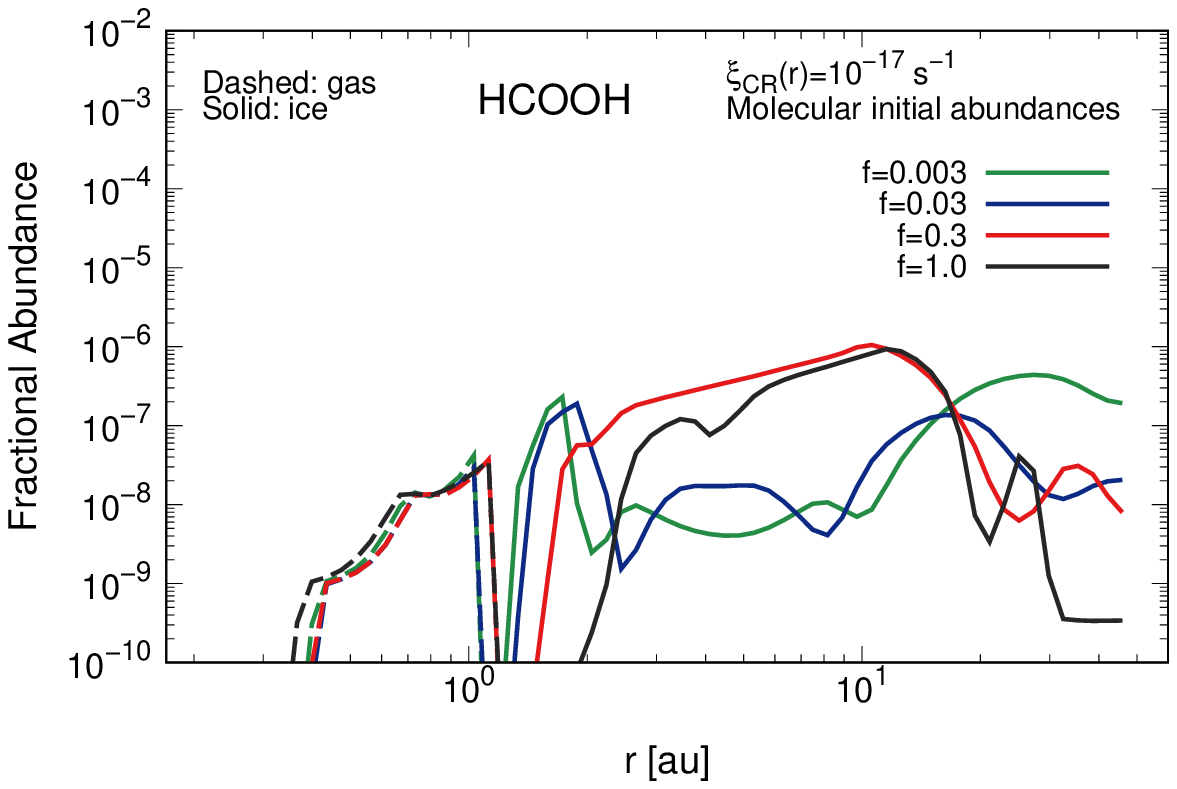}
\includegraphics[scale=0.63]{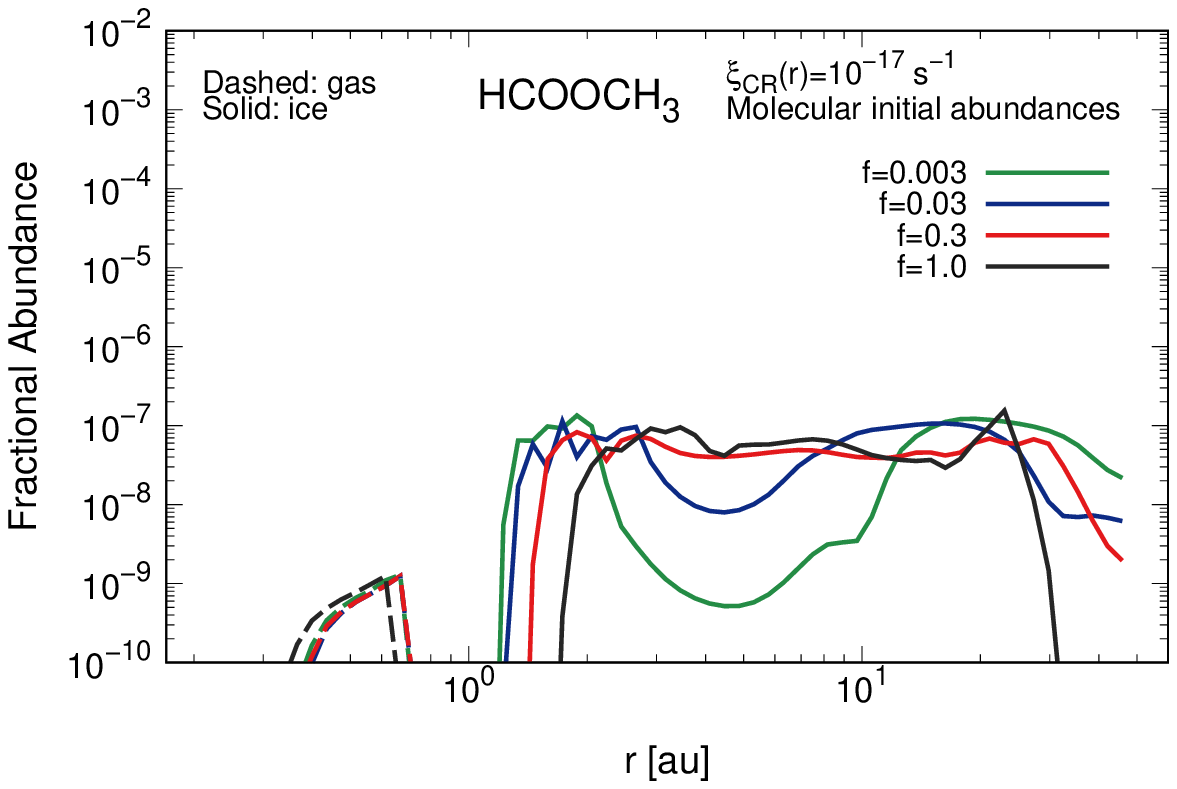}
\includegraphics[scale=0.63]{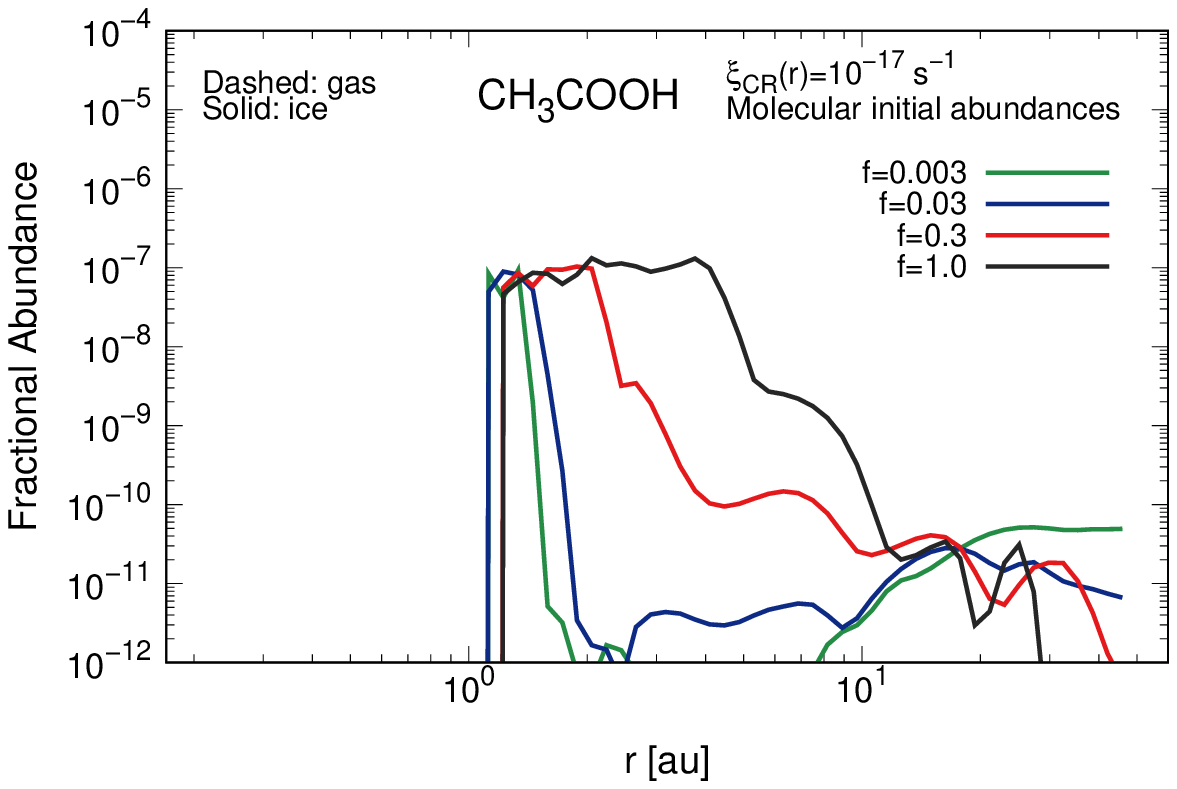}
\includegraphics[scale=0.63]{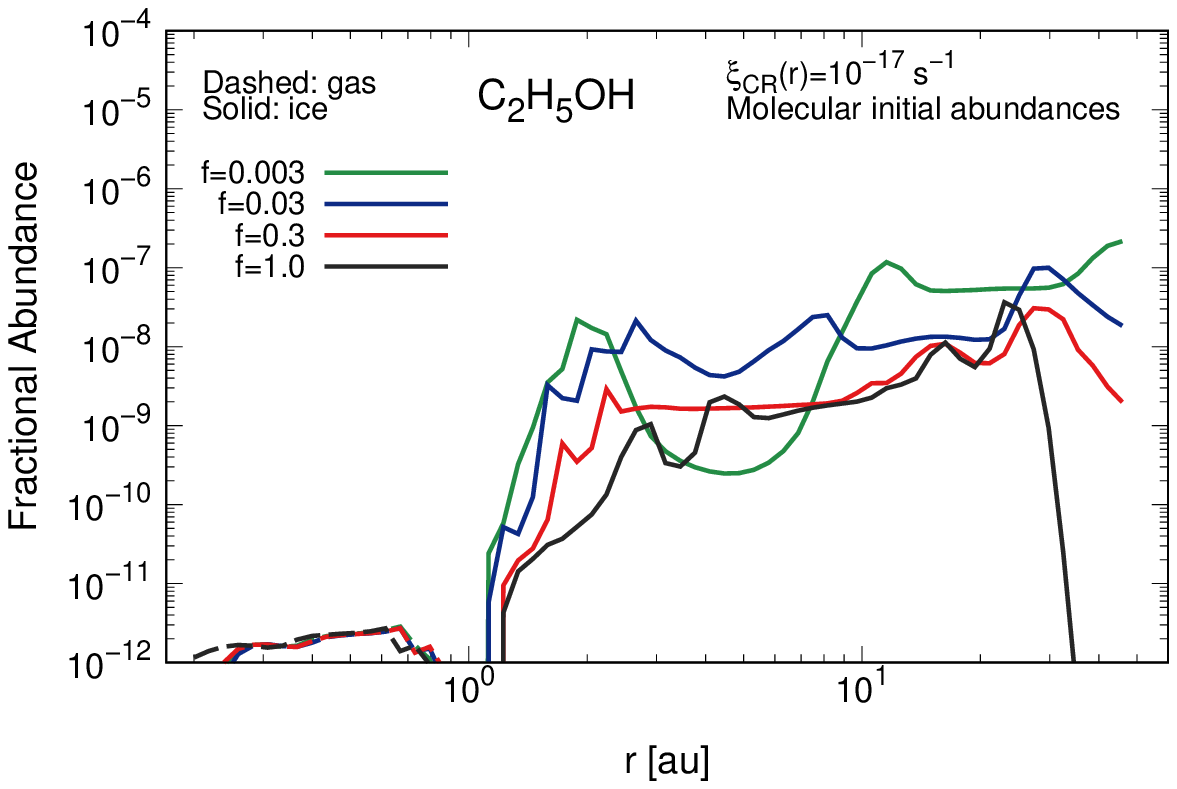}
%%%
\end{center}
\vspace{-0.2cm}
\caption{
%The radial profiles of fractional abundances with respect to total hydrogen nuclei densities at t=$10^{6}$ years for 
Same as Figure \ref{Figure3_rev_radial}, but for 
CH$_{3}$CHO ($n_{\mathrm{CH}_{3}\mathrm{CHO}}$/$n_{\mathrm{H}}$, top left panel),
CH$_{3}$OCH$_{3}$ ($n_{\mathrm{CH}_{3}\mathrm{OCH}_{3}}$/$n_{\mathrm{H}}$, top right panel),
HCOOH ($n_{\mathrm{HCOOH}}$/$n_{\mathrm{H}}$, middle left panel),
HCOOCH$_{3}$ ($n_{\mathrm{HCOOCH}_{3}}$/$n_{\mathrm{H}}$, middle right panel),
CH$_{3}$COOH ($n_{\mathrm{CH}_{3}\mathrm{COOH}}$/$n_{\mathrm{H}}$, bottom left panel), and
C$_{2}$H$_{5}$OH ($n_{\mathrm{C}_{2}\mathrm{H}_{5}OH}$/$n_{\mathrm{H}}$, bottom right panel).
%H$_{2}$CO ($n_{\mathrm{H}_{2}\mathrm{CO}}$/$n_{\mathrm{H}}$, top panels), 
%CH$_{3}$OH ($n_{\mathrm{CH}_{3}\mathrm{OH}}$/$n_{\mathrm{H}}$, second row panels),
%CH$_{3}$CN ($n_{\mathrm{CH}_{3}\mathrm{CN}}$/$n_{\mathrm{H}}$, third row panels), and
%NH$_{2}$CHO ($n_{\mathrm{NH}_{2}\mathrm{CHO}}$/$n_{\mathrm{H}}$, bottom panels). 
%%
\\ \\
}\label{Figure8_rev_radial}
\end{figure*}
%%%%
\\ \\
Figure \ref{Figure7_rev_radial} shows the radial profiles of fractional abundances for 
CH$_{3}$NH$_{2}$ (methylamine), CH$_{2}$NH (methylene imine), CH$_{3}$CN (acetonitrile), HC$_{3}$N (cyanoacetylene), NH$_{2}$OH (hydroxylamine), and HNCO (isocyanic acid).
%%They are nitrogen-bearing complex organic molecules, except NH$_{2}$OH.
Figure \ref{Figure8_rev_radial} shows the radial profiles of fractional abundances for CH$_{3}$CHO (acetaldehyde), CH$_{3}$OCH$_{3}$ (dimethyl ether), HCOOH (formic acid), HCOOCH$_{3}$ (methyl formate), CH$_{3}$COOH (acetic acid), and C$_{2}$H$_{5}$OH (ethanol).
The ice abundances of CH$_{3}$NH$_{2}$ and NH$_{2}$OH at $r\sim3-8$ au (around the Jupiter orbit) are smaller in the non-shadowed disk ($\sim10^{-8}$ and $\sim2\times10^{-6}$ for $f=1.0$, respectively) than those in the shadowed disk ($\sim(2-7)\times10^{-6}$ and $\sim(3-8)\times10^{-6}$ for $f\leq0.03$, respectively). These are saturated molecules.
%%\\ \\
In contrast, the ice abundances of CH$_{2}$NH, HC$_{3}$N, HCOOH, HCOOCH$_{3}$, CH$_{3}$COOH, and HNCO at $r\sim3-8$ au are larger in the non-shadowed disk (e.g., $\sim10^{-7}-10^{-6}$ for HCOOH and $f=1.0$) than those in the shadowed disk (e.g., $\lesssim10^{-8}$ for HCOOH and $f\leq0.03$).
Moreover, the ice abundances of CH$_{3}$CN, CH$_{3}$CHO, CH$_{3}$OCH$_{3}$, and C$_{2}$H$_{5}$OH at $r\sim3-8$ au are larger for $f=0.03$ (e.g., $\sim10^{-7}-10^{-6}$ for CH$_{3}$CN) than those for $f=$1.0, 0.3, and 0.003 (e.g., $<10^{-7}$ for CH$_{3}$CN).
\\ \\
\citet{Walsh2014} discussed that CH$_{3}$NH$_{2}$ is formed via the sequential hydrogenations of CH$_{2}$NH ice, where CH$_{2}$NH ice is originated from atom addition to small hydrocarbon radicals (CH$_{3}$, CH$_{2}$) on the dust grain surfaces.
In addition, \citet{Garrod2008} and \citet{Walsh2014} described that on the warm dust grain surfaces the association of the methyl and amine radicals (CH$_{3}$$_\mathrm{ice}$ + NH$_{2}$$_\mathrm{ice}$) is also dominant.
%%%
Such small hydrocarbon radicals are efficiently formed both in the warm and cold grain surfaces by cosmic-ray-induced photodissociation of CH$_{3}$OH ice, which is efficiently produced by Reactions \ref{Rec9} and \ref{Rec10} (see Section \ref{sec:3-2-2rev} and \citealt{Eistrup2018}).
Thus, we find that the former atom addition reaction route is efficient in the shadowed region ($T(r)<30$ K).
%%%
\\ \\
\citet{Garrod2008} explained that NH$_{2}$OH is formed initially by NH + OH addition on grains, followed by hydrogenation.
In addition, the addition reaction of OH+NH$_{2}$ becomes dominant on the warm dust grains (see also e.g., \citealt{Molyarova2018}).
We propose that the former reaction route is efficient in the shadowed region.
%%%
\\ \\
We find that the dependance of CH$_{3}$CN and HC$_{3}$N ice abundances on the values of $f$ are determined by the combination of the following reaction routes.
\citet{Walsh2014} described that similar to H$_{2}$CO, CH$_{3}$CN and HC$_{3}$N can form in the gas-phase through multiple pathways.
On the cold dust grain surfaces ($T(r)\sim10-30$ K), CH$_{3}$CN ice is formed via the sequential hydrogenations of C$_{2}$N ice, and C$_{2}$N can form via C$_{\mathrm{ice}}$ + CN$_{\mathrm{ice}}$ and N$_{\mathrm{ice}}$ + C$_{2}$$_{\mathrm{ice}}$, or freeze out from the gas phase \citep{Walsh2014}.
In warmer regions, CH$_{3}$CN ice can also form via the following radical-radical reaction with no reaction barrier, CN$_{\mathrm{ice}}$ + CH$_{3}$$_{\mathrm{ice}}$.
%%.
In addition, on the cold dust grain surface ($T(r)\sim10-30$ K), HC$_{3}$N ice is formed via the hydrogenation of C$_{3}$N ice.
%%\\ \\
%%%% 
\\ \\
\citet{Garrod2008}, \citet{Walsh2014}, and \citet{Lopez-Sepulcre2015} described that HNCO is formed on the cold dust grain surfaces by hydrogenation of OCN, but is further hydrogenated to NH$_{2}$CHO (see Reactions \ref{Rec12} and \ref{Rec12-1} in Section \ref{sec:3-2-3rev}).
\\ \\
For other organic molecules, grain-surface association of large radical-radical reactions in the warm regions ($T(r)\sim50$ K) are needed for their formation (e.g., \citealt{Garrod2006, Garrod2008, Herbst2009, Vasyunin2013, Walsh2014}).
In addition, the dependance of ice abundances on the values of $f$ are determined by the combination of following radical-radical reactions and radical formation reactions from CO and/or CH$_{3}$OH on the dust grain surfaces.
CH$_{3}$CHO, CH$_{3}$OCH$_{3}$, HCOOCH$_{3}$, and C$_{2}$H$_{5}$OH ices can form via the following radical-radical reactions with no reaction barriers on the dust grain surfaces, respectively \citep{Garrod2008, Walsh2014};
%%%
\begin{equation}\label{Rec14}
%\mathrm{CH}
\mathrm{CH}_{3\ \mathrm{ice}} + \mathrm{HCO}_{\mathrm{ice}} \rightarrow \mathrm{CH}_{3}\mathrm{CHO}_{\mathrm{ice}}, 
\end{equation}
\begin{equation}\label{Rec15}
\mathrm{CH}_{3\ \mathrm{ice}} + \mathrm{CH}_{3}\mathrm{O}_{\mathrm{ice}} \rightarrow \mathrm{CH}_{3}\mathrm{OCH}_{3\ \mathrm{ice}},
\end{equation}
\begin{equation}\label{Rec16}
\mathrm{HCO}_{\mathrm{ice}} + \mathrm{CH}_{3}\mathrm{O}_{\mathrm{ice}} \rightarrow \mathrm{HCOOCH}_{3\ \mathrm{ice}},
\end{equation}
and
\begin{equation}\label{Rec17}
\mathrm{CH}_{3\ \mathrm{ice}} + \mathrm{CH}_{2}\mathrm{OH}_{\mathrm{ice}} \rightarrow \mathrm{C}_{2}\mathrm{H}_{5}\mathrm{OH}_{\mathrm{ice}}.
\end{equation}
%%% 
%%such as 
The radicals of HCO, CH$_{3}$, CH$_{3}$O, and CH$_{2}$OH are produced during the sequential hydrogenation of CO to form CH$_{3}$OH and/or the cosmic-ray-induced photodissociation of CH$_{3}$OH ice.
The abundance of CH$_{3}$OH ice at $r\sim3-8$ au is larger in the shadowed disk than that in the non-shadowed disk (see Section \ref{sec:3-2-2rev}), whereas the 
shadowed disks with lower values of $f$ ($=0.003$) exhibit lower ice abundances for CH$_{3}$CHO, CH$_{3}$OCH$_{3}$, HCOOCH$_{3}$, and C$_{2}$H$_{5}$OH. This trend is owing to the temperature dependence of the radical-radical reaction efficiency:
%%radical-radical reactions are efficient on the warm dust grain surfaces ($T(r)\sim50$ K). 
the mobility of such radicals is lower than that of hydrogen and light atoms, especially on the cold dust grain surfaces ($T(r)\sim10-30$ K).
%%%%
\\ \\
HCOOH and CH$_{3}$COOH ices can also form via the radial-radical reaction routes on the warm dust grain surfaces. 
However, CH$_{3}$CO, the precursor of CH$_{3}$COOH and CH$_{3}$CHO (the hydrogenation route), cannot form on the cold dust grain surfaces, 
since the reaction barrier of CH$_{3}$$_{\mathrm{ice}}$ + CO$_{\mathrm{ice}}$ is large ($=3460$ K, \citealt{Walsh2014}).
Moreover, COOH, the precursor of CH$_{3}$COOH and HCOOH, cannot form on the cold dust grain surfaces,
since the reaction barrier of OH$_{\mathrm{ice}}$ + CO$_{\mathrm{ice}}$ is large 
($=3000$ K, \citealt{Walsh2014}).
We suggest that HCOOH ice can be formed via OH$_{\mathrm{ice}}$$+$HCO$_{\mathrm{ice}}$, although the mobility of both ices is low on the cold dust grain surfaces \citep{Garrod2008}.
%%
%%%
\begin{comment}
%%%
\end{comment}
%%%
%%%
%%%
%%%
\subsubsection{The total complex organic reservoir}\label{sec:3-2-5rev}
%%%
\begin{figure*}[hbtp]
\begin{center}
\vspace{1cm}
%\hspace{1cm}
%\plotone{cost.pdf}
%
\includegraphics[scale=0.63]{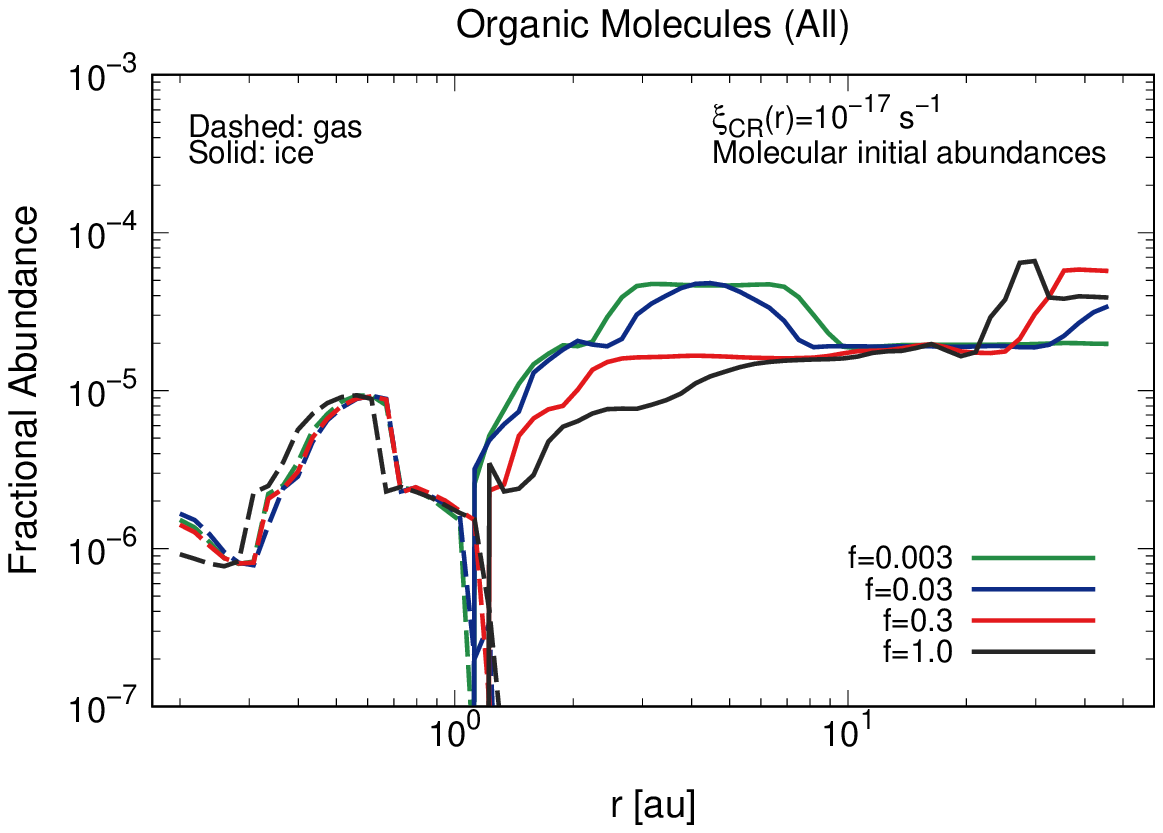}
\includegraphics[scale=0.63]{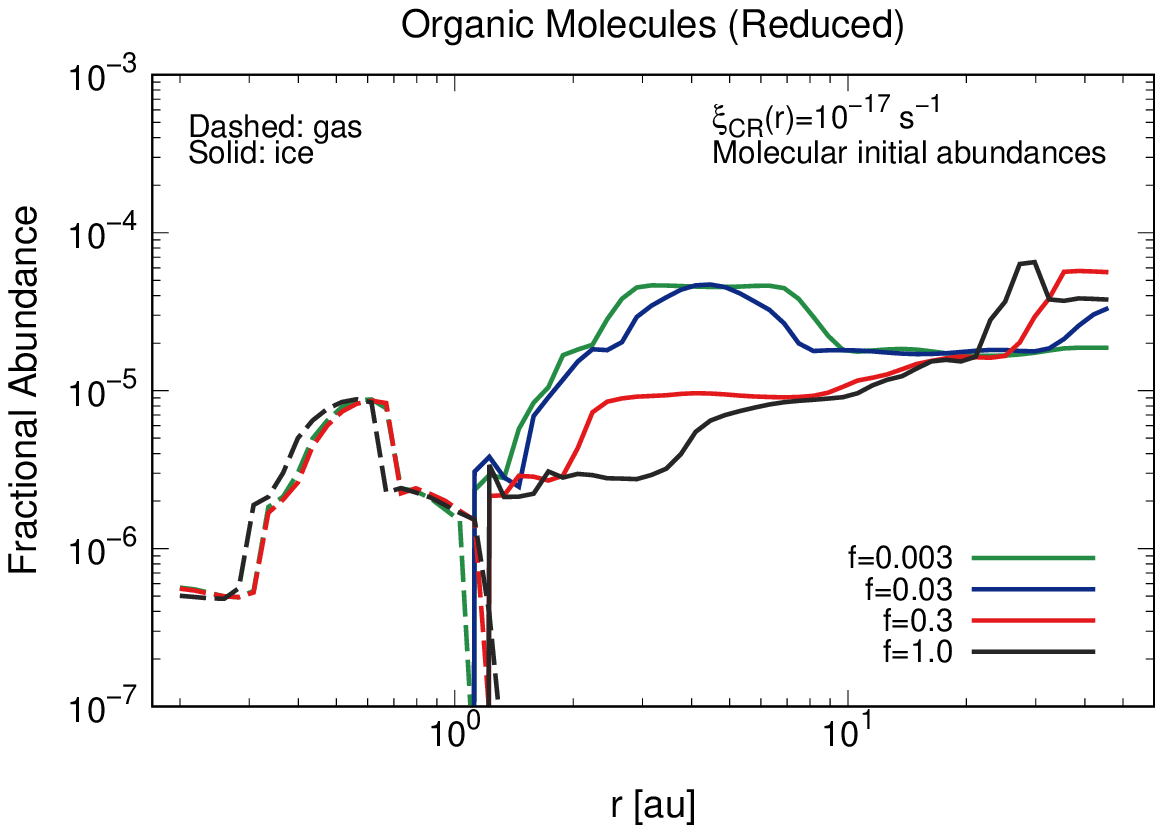}
%%%
\end{center}
\vspace{-0.2cm}
\caption{
%%%%
Same as Figure \ref{Figure3_rev_radial}, but for the total reservoir of larger (complex) organic molecules.
The left panel (``All'') shows the sums of all of larger organic molecules for which we show the results in Section \ref{sec:3-2rev} (Figures \ref{Figure4_rev_radial}-\ref{Figure8_rev_radial}).
The right panel (``Reduced'') shows the sums of all of larger organic molecules after we subtract dominant unsaturated hydrocarbon molecules for which we show the results in Figure \ref{Figure6_rev_radial}.
%
%\vspace{0.2cm}
\\ \\
}\label{Figure9_COMs_rev_radial}
\end{figure*}
%%%
Figure \ref{Figure9_COMs_rev_radial} shows the radial profiles of sums of fractional abundances with respect to total hydrogen nuclei densities at t=$10^{6}$ years for larger (complex) organic molecules.
In the left (``All'') panel of Figure \ref{Figure9_COMs_rev_radial}, we show the sums of all larger (complex) organic molecules introduced in Section \ref{sec:3-2rev} (Figures \ref{Figure4_rev_radial}-\ref{Figure8_rev_radial}); C$_{2}$H$_{6}$, H$_{2}$CO, CH$_{3}$OH, NH$_{2}$CHO, C$_{2}$H$_{2}$, C$_{2}$H$_{4}$, C$_{3}$H$_{2}$, 
CH$_{3}$CCH, CH$_{2}$CCH$_{2}$, CH$_{3}$CHCH$_{2}$, HC$_{3}$N, CH$_{3}$CN, CH$_{3}$CHO, CH$_{3}$OCH$_{3}$, CH$_{3}$NH$_{2}$, HCOOH, HCOOCH$_{3}$, CH$_{3}$COOH, CH$_{2}$NH, HNCO, C$_{2}$H$_{5}$OH, 
We note that CO, H$_{2}$O, O$_{2}$, CH$_{4}$, N$_{2}$, NH$_{3}$, HCN, and NH$_{2}$OH are not included in this Figure, since they are smaller molecules and/or inorganic molecules.
\\ \\
In the non-shadowed disks, the total icy fractional abundances of the listed organic molecules is $\sim3\times10^{-6}$ just outside the water snowline ($r\sim1.5$ au), and gradually increases with increasing $r$. It is $(1-2)\times10^{-5}$ at $r\sim3-8$ au, around the current orbit of Jupiter, and is $(4-6)\times10^{-5}$ outside the CO snowline.
In the shadowed disk, it is $(3-5)\times10^{-5}$ at $r\sim3-8$ au, and $2\times10^{-5}$ at $r\gtrsim10$ au.
Thus in the shadowed disk, our model predicts that the dust grains at $r\sim3-8$ au contain $\sim2-5$ times the amount of larger (saturated$+$unsaturated) organic molecules in total compared with those in the non-shadowed disks.
\\ \\
In the right (``Reduced'') panel of Figure \ref{Figure9_COMs_rev_radial}, we show the sums of all larger (complex) organic molecules but exclude the dominant unsaturated hydrocarbon molecules shown in Figure \ref{Figure6_rev_radial}; C$_{2}$H$_{2}$, C$_{2}$H$_{4}$, C$_{3}$H$_{2}$, CH$_{3}$CCH, CH$_{2}$CCH$_{2}$, CH$_{3}$CHCH$_{2}$.
In the non-shadowed disks, at $r\sim3-8$ au, the icy abundances of sums of these ``Reduced'' organic molecules are smaller ($\sim(3-9)\times10^{-6}$) than those of ``All'' organic molecules ($(1-2)\times10^{-5}$).
In contrast, in the shadowed disks, the icy abundances of sums of organic molecules are similar ($(3-5)\times10^{-5}$) between these two cases.
This is because such dominant unsaturated hydrocarbon molecules are frozen onto the dust grains at $r\sim3-8$ au, and their ice abundances significantly decrease with decreasing $f$ (see Figure \ref{Figure6_rev_radial}).
%%%
This result demonstrates that the shadowed region promotes the synthesis of saturated organic molecules rather than unsaturated hydrocarbons.
%%%
Thus in the shadowed disk, the dust grains at $r\sim3-8$ au are predicted to have around $5-10$ times the amount of larger saturated organic molecules in total compared with those in the non-shadowed disks.
%%%
\\ \\
%%%
%\\ \\
%.
\subsection{The dependance of the disk chemical structure on ionization rate and initial abundances}\label{sec:3-3rev}
%%%
In this subsection, we briefly summarize key results of disk chemical evolution for atomic initial compositions and different ionization rates, which are detailed in Appendix \ref{Asec:A}.
%%%
Figures \ref{Figure14rev3_appendix}-\ref{Figure16rev3_appendix} in Appendix \ref{Asec:A} show the radial profiles of fractional abundances at t=$10^{6}$ years for dominant oxygen-, carbon-, nitrogen-bearing molecules (H$_{2}$O, CO, CO$_{2}$, O$_{2}$, CH$_{4}$, C$_{2}$H$_{6}$, H$_{2}$CO, CH$_{3}$OH, N$_{2}$, NH$_{3}$, HCN, and NH$_{2}$CHO)
in the shadowed and non-shadowed disk midplane ($f=1.0$ and $f=0.03$, respectively).
%%%.
In these Figures, we assume either molecular or atomic initial abundances and either low or high ionization rates ($\xi_{\mathrm{CR}}(r)=$$10^{-18}$, $10^{-17}$ [s$^{-1}$]).
We note that Figures \ref{Figure3_rev_radial}-\ref{Figure5_rev_radial} in Section \ref{sec:3-2rev} show the radial abundance profiles for the same molecules with molecular initial abundances and $\xi_{\mathrm{CR}}(r)=10^{-17}$ [s$^{-1}$].
\\ \\
In Appendix \ref{Asec:A}, we show these results and explain the dependance of the disk chemical evolution on disk ionization rates and initial abundances in detail. 
%%%Here we summarize some key results.
According to our calculations, CO/CO$_{2}$ abundances become smaller/larger with increasing ionization rates, respectively.
CH$_{4}$ and C$_{2}$H$_{6}$ gas abundances within their snowline become smaller as the ionization rates become larger.
In addition, abundances of H$_{2}$O and organic molecules are larger for molecular initial abundances than those for atomic initial abundances. 
It is worth noting that O$_{2}$ ice abundances are $\sim10^{-5}-10^{-4}$ (consistent with the measured cometary abundances, see also Section \ref{sec:4-2}) only for atomic initial abundances and the low ionization rates.
%%%
\subsection{Time evolution of molecular abundances}\label{sec:3-4rev}
In this subsection, we briefly summarize some key results for the time evolution of molecular abundances, which are detailed in Appendix \ref{Bsec:B}.
%%%
Figures \ref{Figure17rev3_time_appendix}-\ref{Figure19rev3_time_appendix} in Appendix \ref{Bsec:B} show the time evolution of the radial profiles of fractional abundances for dominant oxygen-, carbon-, nitrogen-bearing molecules (H$_{2}$O, CO, CO$_{2}$, O$_{2}$, CH$_{4}$, C$_{2}$H$_{6}$, H$_{2}$CO, CH$_{3}$OH, N$_{2}$, NH$_{3}$, HCN, and NH$_{2}$CHO)
in the shadowed and non-shadowed disk midplane ($f=1.0$ and 0.03), when assuming molecular initial abundances and high ionization rates ($\xi_{\mathrm{CR}}(r)=$$10^{-17}$ [s$^{-1}$]).
These initial conditions are similar to those in Sections \ref{sec:3-1rev} and \ref{sec:3-2rev}.
\\ \\
In Appendix \ref{Bsec:B}, we describe these results and explain the time evolution of molecular abundances in detail.
%%.
According to our calculations, in the shadowed region ($r\sim3-8$ au) the icy abundances of CO$_{2}$ and organic molecules such as H$_{2}$CO, CH$_{3}$OH, and NH$_{2}$CHO become larger with time. 
Thus, we find that
if the shadowed region is maintained for a relatively long time ($t\sim10^{6}$ years), chemical evolution 
may produce dust grains and solid objects with large amounts of CO$_{2}$ and organic molecular ices (see also Section \ref{sec:4-2}).
In addition, the ice abundances of these molecules can be a clue in constraining the formation age of solid bodies in the shadowed region.
%%
%%%20220201 21:23
%%%20220405 19:12 ok.
%%%
\section{Discussion}\label{sec:4}
\subsection{The C/O and N/O ratios and implication for planetary atmospheres}\label{sec:4-1}
%%%
\begin{figure*}[hbtp]
\begin{center}
\vspace{1cm}
%\hspace{1cm}
%\plotone{cost.pdf}
%
\includegraphics[scale=0.63]{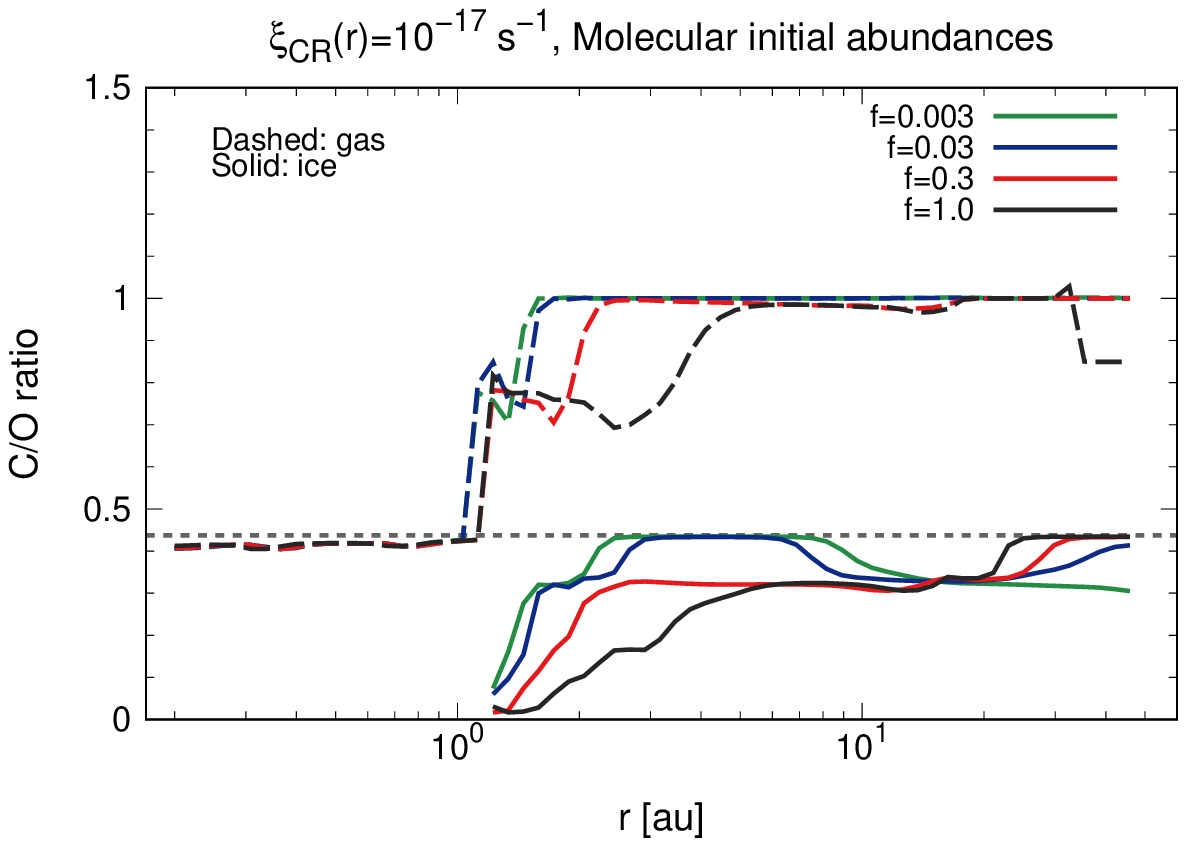}
\includegraphics[scale=0.63]{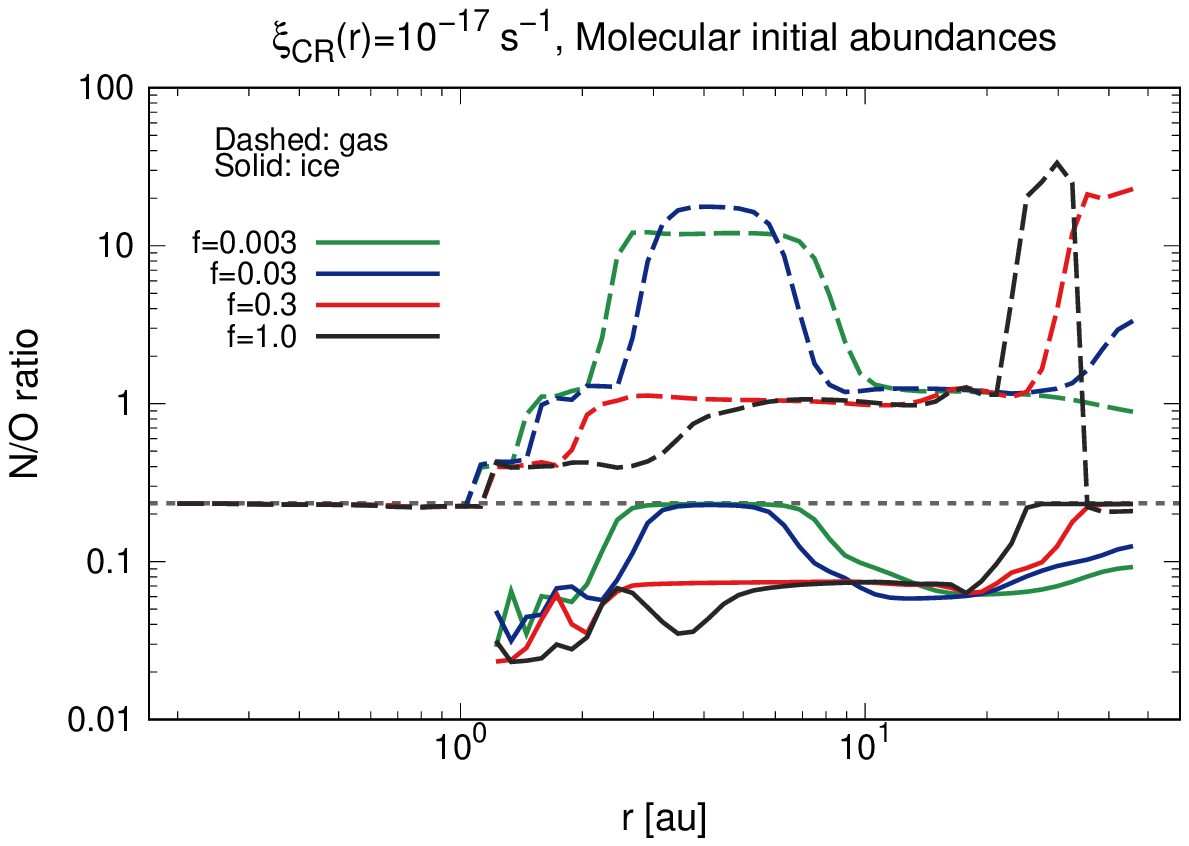}
\includegraphics[scale=0.63]{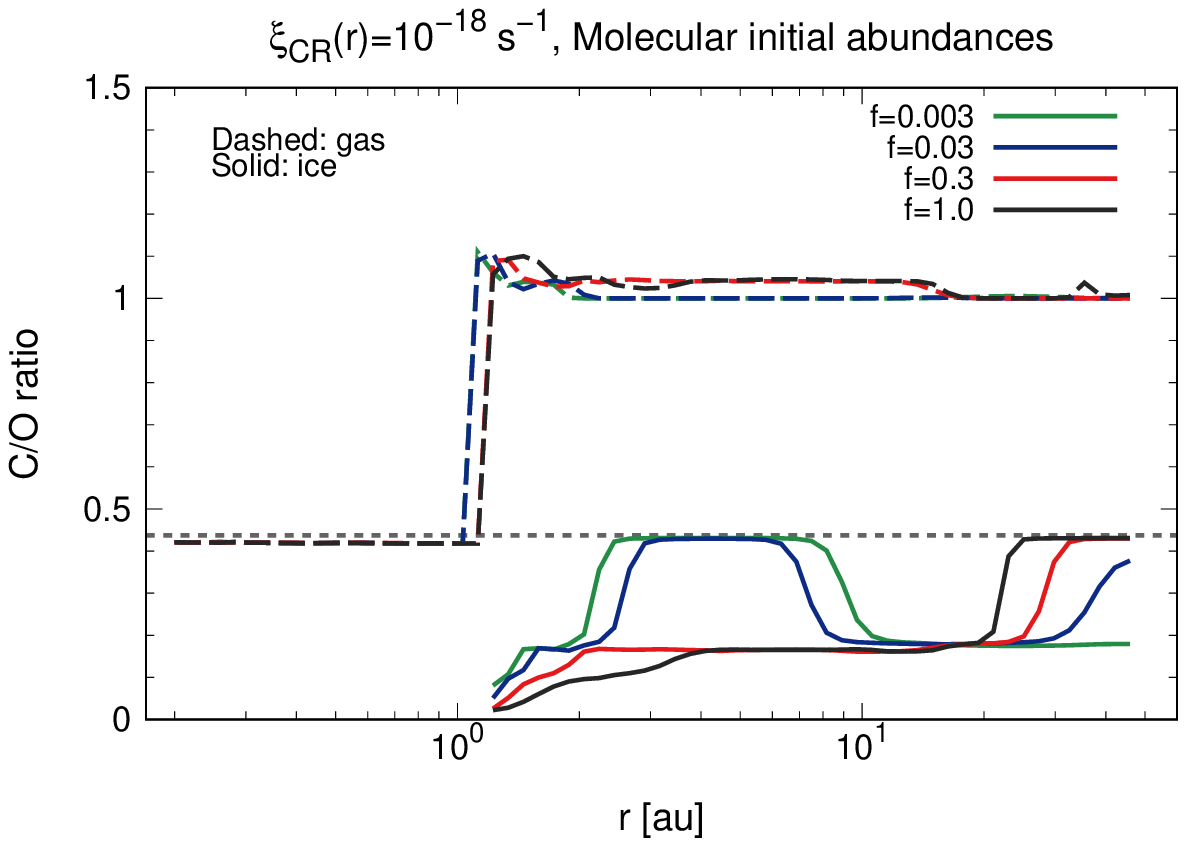}
\includegraphics[scale=0.63]{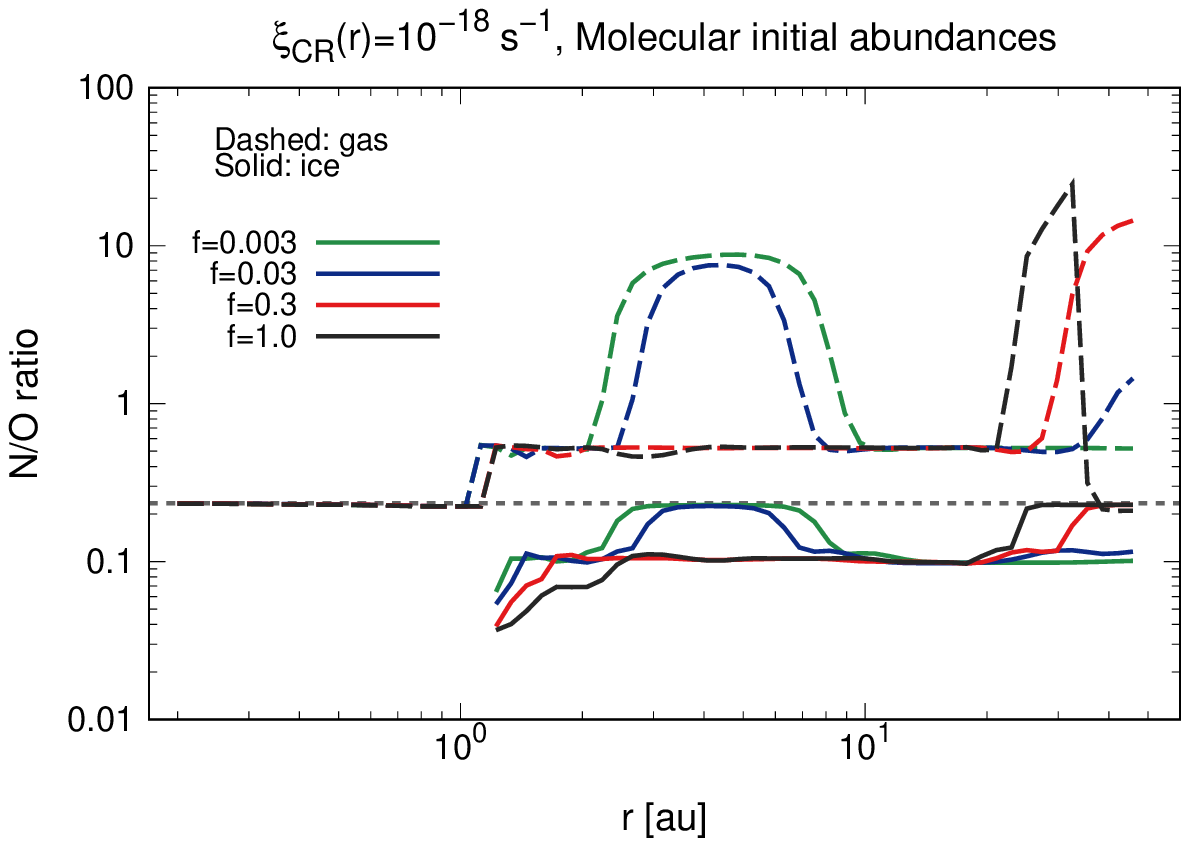}
%\includegraphics[scale=0.62]{20211115r2_CtoOratio_t3_ds_0.1um_atomic-ini_Xray-rev_enhanced-water_nH-rev_1.0e6yr.eps}
%\includegraphics[scale=0.62]{20211115r2_NtoOratio_t3_ds_0.1um_atomic-ini_Xray-rev_enhanced-water_nH-rev_1.0e6yr.eps}
%\includegraphics[scale=0.62]{20211115r2_CtoOratio_t4_ds_0.1um_atomic-ini_Xray-rev_enhanced-water_nH-rev_1.0e6yr.eps}
%\includegraphics[scale=0.62]{20211115r2_NtoOratio_t4_ds_0.1um_atomic-ini_Xray-rev_enhanced-water_nH-rev_1.0e6yr.eps}
%%%
%%%
%%%
\end{center}
\vspace{-0.2cm}
\caption{
The radial profiles of C/O ratios (left panels) and N/O ratios (right panels) at t=$10^{6}$ years and for molecular initial abundances (the ``inheritance'' scenario).
The dashed and solid lines show the profiles for gaseous and icy molecules, respectively.
The black, red, blue, and green lines show the profiles for different values of the parameter $f$ (=1.0, 0.3, 0.03, and 0.003), respectively.
%%%%%.
Top panels show the results for $\xi_{\mathrm{CR}}(r)=$$1.0\times10^{-17}$ [s$^{-1}$], whereas bottom panels show the results for $\xi_{\mathrm{CR}}(r)=$$1.0\times10^{-18}$ [s$^{-1}$]. The horizontal dotted lines show the values of the initial elemental abundance ratios (C/O ratio $=0.44$, N/O ratio$=0.23$).
%%%%.
%%%
\\ \\
%\vspace{0.2cm}
}\label{Figure10rev_CtoO-NtoO_inheritance}
\end{figure*}
\begin{figure*}[hbtp]
\begin{center}
\vspace{1cm}
%\hspace{1cm}
%\plotone{cost.pdf}
%\plotone[scale=0.63]{{20210624_surface-density.eps}
%\plotone[scale=0.63]{{20210624_ngas-T.eps}
%\includegraphics[scale=0.62]{20210902r_CtoOratio_t1_ds_0.1um_mol-ini_Xray-rev_enhanced-water_nH-rev_1.0e6yr.eps}
%\includegraphics[scale=0.62]{20210902r_NtoOratio_t1_ds_0.1um_mol-ini_Xray-rev_enhanced-water_nH-rev_1.0e6yr.eps}
%\includegraphics[scale=0.62]{20210902r_CtoOratio_t2_ds_0.1um_mol-ini_Xray-rev_enhanced-water_nH-rev_1.0e6yr.eps}
%\includegraphics[scale=0.62]{20210902r_NtoOratio_t2_ds_0.1um_mol-ini_Xray-rev_enhanced-water_nH-rev_1.0e6yr.eps}
\includegraphics[scale=0.63]{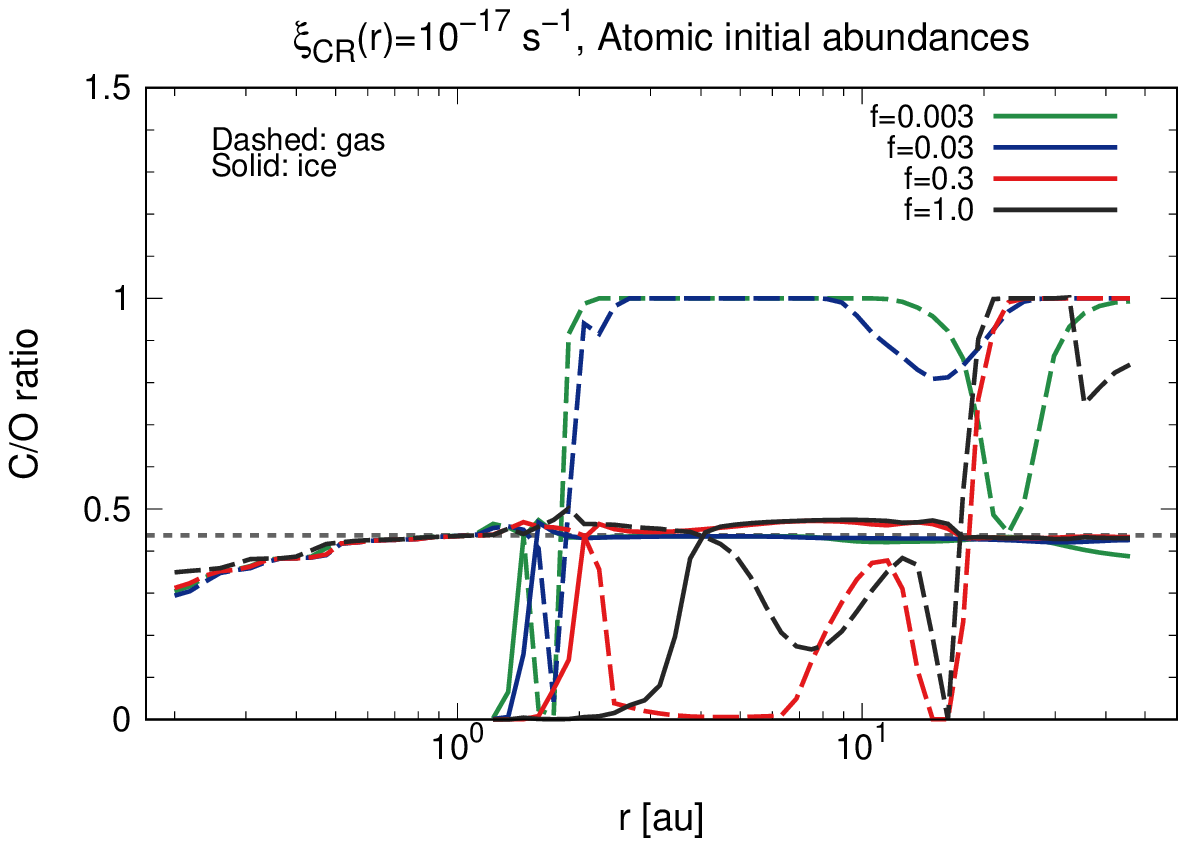}
\includegraphics[scale=0.63]{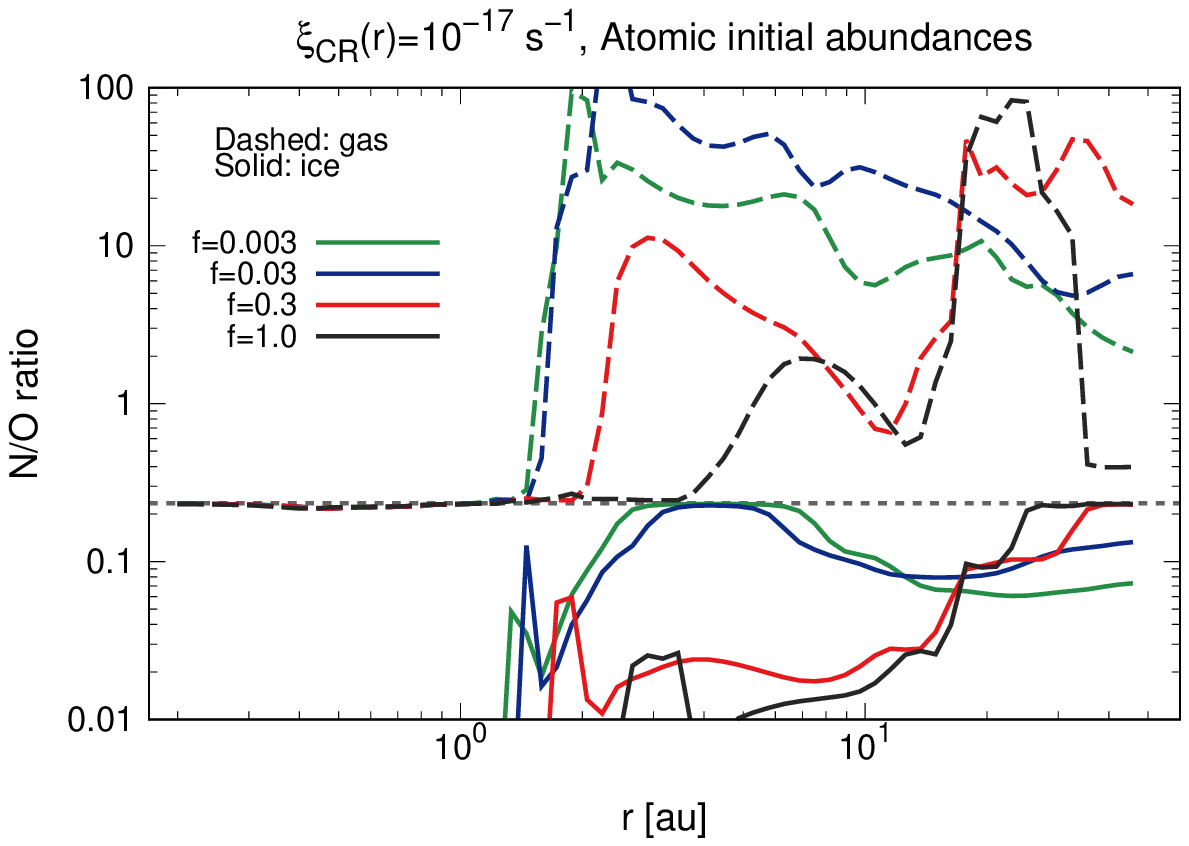}
\includegraphics[scale=0.63]{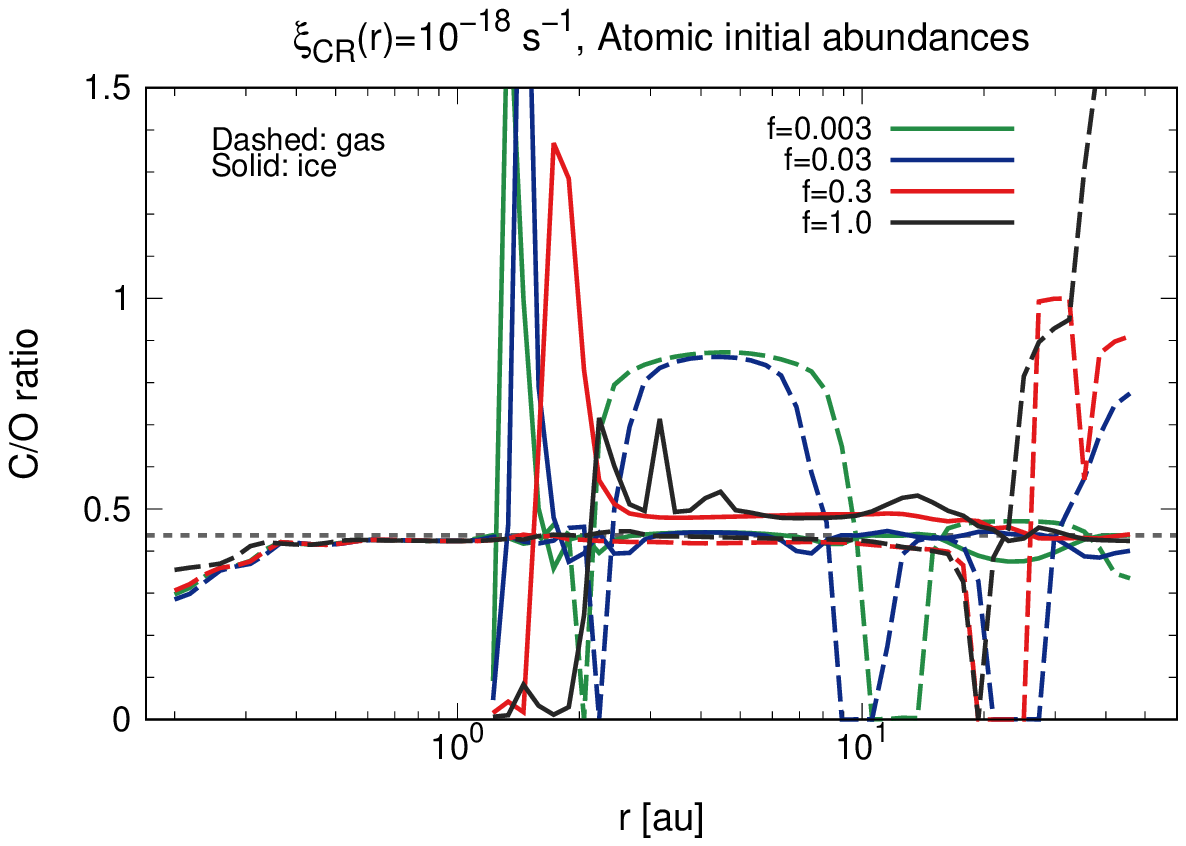}
\includegraphics[scale=0.63]{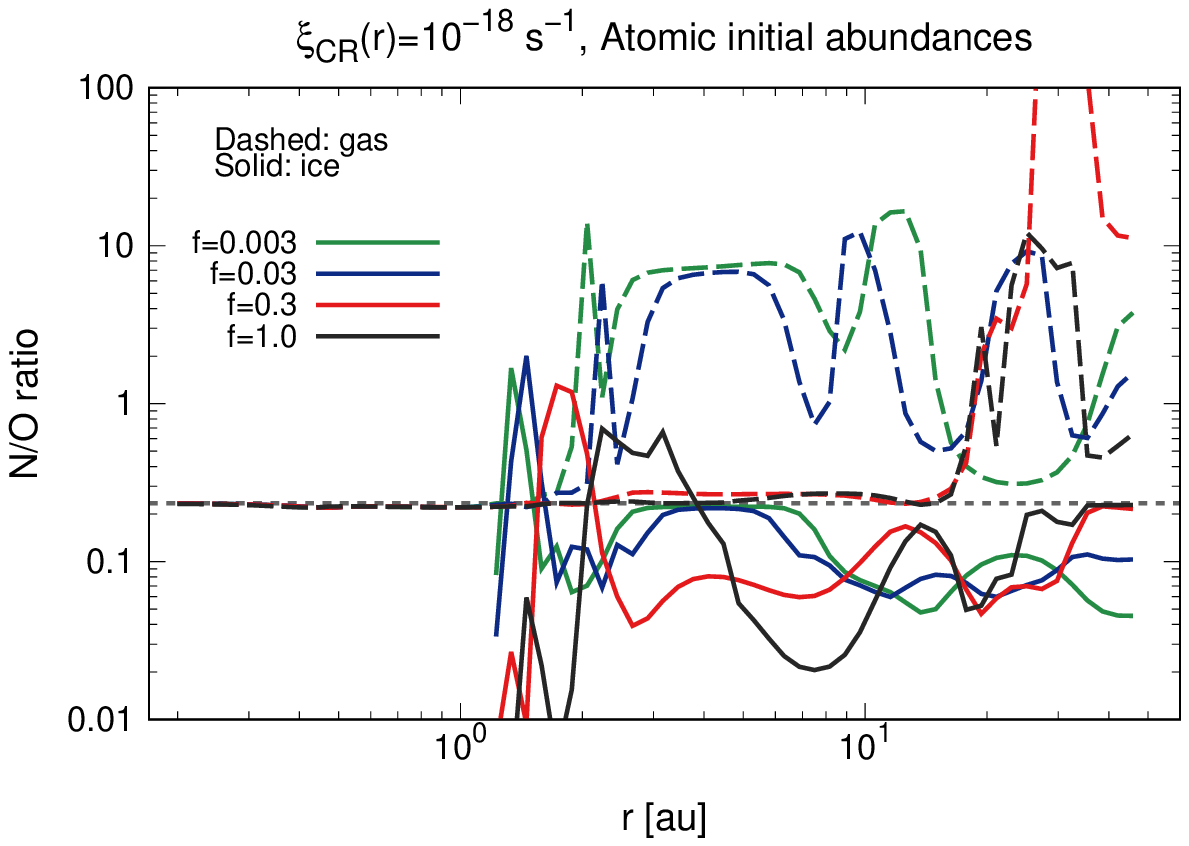}
%%%
%%%
%%%
\end{center}
\vspace{-0.2cm}
\caption{
Same as Figure \ref{Figure10rev_CtoO-NtoO_inheritance}, but for atomic initial abundances (``reset'' scenario).
\\ \\
\vspace{0.2cm}
}\label{Figure11rev_CtoO-NtoO_reset}
\end{figure*}
%%%
The carbon-to-oxygen (C/O) ratios of exoplanet atmospheres 
have been proposed to be a possible tool to link gas-giant exoplanets to their formation sites in the natal protoplanetary disk (e.g., \citealt{Oberg2011, Oberg2016, Madhusudhan2014, Pontoppidan2014, Booth2019, Cridland2020, Notsu2020, Ohno2021, Schneider2021, Turrini2021, Dash2022, Molliere2022}).
This is because the radial-dependent positions of snowlines of abundant oxygen- and carbon-bearing molecules result in systematic radial variations in the C/O ratios in the gas and ice. 
%%. 
However, disk chemistry can affect the C/O ratios in the gas and ice, thus potentially erasing the chemical fingerprint of snowlines in atmospheres \citep{Eistrup2016, Eistrup2018, Notsu2020}.
\citet{Notsu2020} discussed that hot Jupiters with C/O$>1$ can only form between the CO$_{2}$ and CH$_{4}$ snowlines in the non-shadowed disk which has fully inherited interstellar abundances, and where negligible chemistry has occurred because of a low ionisation rate. They also discussed that carbon rich planets are likely rare unless efficient transport of hydrocarbon-rich ices via pebble drift to within the CH$_{4}$ snowline \citep{Booth2017, Booth2019} is a common phenomenon.
We note that disk chemistry significantly affect the C/O ratios, through the change of abundances of hydrocarbons and O$_{2}$ gas abundances (e.g., \citealt{Helling2014, Eistrup2016, Eistrup2018, Notsu2020, Ohno2021}).
\\ \\
The left panels of Figures \ref{Figure10rev_CtoO-NtoO_inheritance} and \ref{Figure11rev_CtoO-NtoO_reset} shows the radial profiles of C/O ratios at t=$10^{6}$ years.
Different color lines show the profiles for different values of the parameter $f$ (=1.0, 0.3, 0.03, and 0.003), respectively.
As shown in our previous studies \citep{Eistrup2016, Eistrup2018, Notsu2020}, the inclusion of chemistry has a significant impact on the disk elemental abundance ratios of both gas and ice.
The ices remain, on the whole, dominated by oxygen (i.e., C/O$<0.5$).
In the non-shadowed disk ($f=1.0$), between the H$_{2}$O and CH$_{4}$ snowlines the gas is carbon rich relative to the initial elemental value ($=0.44$) for molecular initial abundances, whereas gas-phase C/O ratios are $\lesssim0.44$ for atomic initial abundances.
For molecular initial abundances, the gas-phase C/O ratios in these regions are $\sim1.0-1.1$ for the low ionisation rate and 
$\sim0.7-1.0$ for the high ionisation rate.
%%%%. 
In the shadowed disk ($f\leq0.03$), however, the gas-phase C/O ratios in the shadowed region ($r\sim2-10$ au) are almost unity for the high ionisation rate and/or molecular initial abundances (as suggested by \citealt{Ohno2021}) and 0.8 for the low ionisation rate and atomic initial abundances.
This is because CO carries most of the gas-phase C and O there.
In addition, the icy-phase C/O ratios reach the initial elemental value ($=0.44$ in our model) since almost all C and O reservoirs freeze out onto dust grains in the shadowed region ($r\sim3-8$ au).
%%%%%
Thus, if planets acquire their atmospheres from the gas in the disk shadowed region, the atmospheres will have sub-stellar metallicities and 
larger C/O ratios than the initial elemental value ($=0.44$ in our model) regardless of disk initial abundances and ionisation rates.
In addition, the C/O ratios are unity unless the disk has fully atomized initial abundances and negligible chemistry has occurred because of the low ionisation rate.
We note that the atmospheres can be polluted by dissolution of accreting oxygen-rich icy planetesimals/pebbles (e.g., \citealt{Hori2011, Mordasini2016}) and erosion of cores (e.g., \citealt{Moll2017}), potentially lowering C/O ratios in planetary atmospheres.
%%%%%
\\ \\
In the bottom left panel of Figure \ref{Figure11rev_CtoO-NtoO_reset} (the case for low ionisation and atomic initial abundances), the icy-phase C/O ratios just outside the water snowline are larger than the initial elemental value, and they exceed 1.0 for $f\leq0.3$.
The peak position shifts inside with decreasing $f$. 
In this region, H$_{2}$O ice abundances are much smaller ($\sim10^{-6}$) than those in other models ($>10^{-5}$).
In addition, this region is between HCN and CO$_{2}$ snowline, and the HCN ice abundance is larger than those of other molecules (such as H$_{2}$O and H$_{2}$CO) in this region (see Appendix \ref{Asec:A}).
We note that the presence or otherwise of this large peak depends on the relative binding energies of HCN and CO$_{2}$ assumed in the model (see also \citealt{Eistrup2016}).
\\ \\
%%%
Since previous studies suggested that atmospheric N/O ratio is also a useful tracer of planet formation locations \citep{Piso2016, Cridland2020, Turrini2021, Ohno2021}, we also study the radial distributions of N/O ratios in our disk models.
%%%
The right panels of Figures \ref{Figure10rev_CtoO-NtoO_inheritance} and \ref{Figure11rev_CtoO-NtoO_reset} shows the radial profiles of N/O ratios at t=$10^{6}$ years.
The ices remain, on the whole, dominated by oxygen (i.e., N/O$<0.23$).
In the non-shadowed disk ($f=1.0$) and for molecular initial abundances, the gas-phase N/O ratios increase ($>0.4$) outside the water snowline, they gradually increase ($\sim0.4-0.5$ and $\sim1.0$ outside the H$_{2}$O and CO$_{2}$ snowlines, respectively) for the high ionization rate, whereas they are $\sim0.5$ for the low ionization rate. This is because of the significant enhancement of CO$_{2}$ abundances in the disk with a high ionization rate (see Appendix \ref{Asec:A-1}).
In the non-shadowed disk ($f=1.0$) and for atomic initial abundances, the gas-phase N/O ratios do not increase outside the water snowline, since gas-phase abundances of NH$_{3}$, HCN, and NH$_{2}$CHO are at least around an order of magnitude smaller than those for molecular initial abundances (see Appendix \ref{Asec:A-3}).
For the high ionisation rate, they reach around unity outside the CO$_{2}$ snowline.
This is because CO/CO$_{2}$ abundance ratios are smaller in the high ionisation rate case than those in the low ionisation rate case (see Appendix \ref{Asec:A-1}).
%%%
%%%
In addition, the gas-phase N/O ratios exceed unity and reach around 10 just outside the CO snowline in all models, because the binding energy of N$_{2}$ is lower than that of CO (see Table \ref{Table:1}), leading to the higher N$_{2}$ gas abundances around these radii (see also e.g., \citealt{Turrini2021}).
%%% 
\\ \\
In the shadowed disk, the radial profiles of the gas-phase N/O ratios show spatial variations which are much larger than those of the C/O ratios, and thus the N/O ratio would be a useful tracer of the shadowed region in the disks.
For molecular initial abundances, the gas-phase N/O ratios are much larger than unity in the shadowed region ($r\sim3-8$ au), and they are $\sim6-9$ for the low ionization rate and $\sim10-20$ for the high ionization rate.
This is because of the difference in the binding energies between N$_{2}$ and CO, similar to the situation outside the snowline of the non-shadowed disk ($f=1.0$).
In addition, the icy-phase N/O ratios reach the initial elemental value ($=0.23$ in our model) since almost all N and O reservoirs freeze out onto dust grains in the shadowed region ($r\sim3-8$ au).
For atomic initial abundances, the overall pictures are similar, and around the shadowed region the gas-phase N/O ratios are $\gtrsim10$ and icy-phase N/O ratios reach 0.23.
We note that in our results, the icy-phase elemental carbon, oxygen, and nitrogen abundances at $r=5.3$ au are $\sim1.4\times10^{-5}$, $\sim3.2\times10^{-4}$, and $\sim(6.5-7.1)\times10^{-5}$, respectively, which are similar to the initial elemental abundances.
\\ \\
The results for molecular initial abundances and the low ionisation rate are consistent with those in \citet{Ohno2021}, 
although the gas-phase N/O ratios in the shadowed region are much larger ($>100$) than that in our disk model ($\sim10$).
This is likely because they used larger literature values of binding energies of CO and N$_{2}$ ($E_{\mathrm{des}}$(CO)=1180 K and $E_{\mathrm{des}}$(N$_{2}$)=1051 K) than ours, as the gas-phase N/O ratio is sensitive to those values when both CO and N$_{2}$ are frozen.
In addition, \citet{Ohno2021} included several dominant molecules only (i.e., N$_{2}$ and NH$_{3}$ only) and did not include the effects of disk chemical evolution. Moreover, they assumed slightly larger N$_{2}$ abundances ($=3.50\times10^{-5}$) and smaller NH$_{3}$ abundances ($=7.78\times10^{-6}$) than those in our calculations (see Figure \ref{Figure16rev3_appendix}).
Our calculations also investigate the dependance on initial disk conditions.
\\ \\
Therefore, if planets acquire their atmospheres from the gas in the shadowed region ($r\sim3-8$ au in our disk model), they are expected to have the super-stellar N/O ratios of $\gg1$ and sub-stellar metallicities. On the other hand, they are expected to have the stellar N/O ratios and super-solar metallicities if the planetary atmospheres are efficiently polluted by solid components (including the case of Jupiter, see \citealt{Ohno2021}).
\\ \\
As discussed in \citet{Ohno2021}, Saturn, in contrast to Jupiter, may not have the uniform enrichment in the elemental abundances of their atmospheres, if formed in their vicinity of the current orbits.
This is because the current Saturn orbit is outside the shadowed region ($r\sim3-8$ au), and N$_{2}$ has sublimated into the gas-phase.
In the shadowed disk, the icy-phase N/O ratios at around the current orbit of Saturn ($r\sim10$ au) are $\lesssim0.1$ for all models, 
whereas they reach the initial elemental value ($=0.23$ in our model) at around the current orbit of Jupiter.
In addition, regardless of the initial abundances and ionisation rates, the icy-phase N/O ratios at $r\sim10-40$ au are $\lesssim0.1$ in the shadowed disk, whereas in the non-shadowed disk they reach the initial elemental value at $r>24$ au (outside the N$_{2}$ snowline).
Future observations and entry probe missions on Saturn and also outer icy planets (Uranus and Neptune) would help to distinguish the shadow formation scenario proposed in \citet{Ohno2021} from other scenarios, since the elemental abundances such as N in such planets are still uncertain (see e.g., \citealt{Atreya2018, Mandt2020}).
\\ \\
Nitrogen abundances within the atmospheres of exoplanets (such as hot Jupiters) can be constrained by the observations of HCN and NH$_{3}$ (e.g., \citealt{MacDonald2017, Hawker2018, Giacobbe2021}), which are the major nitrogen-bearing species in hot gas-giant atmospheres along with N$_{2}$ (e.g., \citealt{Moses2011, Moses2013}).
Through upcoming observations with the James Webb Space Telescope (JWST) and observations with next-generation facilities (such as ARIEL and ground-based telescopes), it is anticipated that the elemental composition of carbon, oxygen, and nitrogen, and the ratios between them will be determined with much higher precision than currently possible for the atmospheres of many exoplanets such as hot Jupiters (see e.g., \citealt{Tinetti2018, Madhusudhan2019, Changeat2020, Turrini2021Ariel}).
%%%
%
%%
\\ \\
We note that in addition to C/O ratios, the C/H ratios (carbon elemental abundances) of exoplanet atmospheres 
have been also proposed to be a possible tool to link gas-giant exoplanets to their formation sites in the protoplanetary disk (e.g., \citealt{Oberg2011, Madhusudhan2014, Pontoppidan2014, Line2021, Pelletier2021}).
Recently, super-solar and Jupiter-like C/H ratios ($\gtrsim10^{-3}$) have been confirmed for some hot Jupiters (e.g., \citealt{Brogi2019, Gandhi2019, Pelletier2021}).
\citet{Pelletier2021} analyzed thermal emission spectra of a non-transiting hot Jupiter $\tau$ Boo b with high spectral resolutions ($R=\lambda/\Delta\lambda=70000$) and reported that the planet's atmosphere has super-solar C/H ratio and possibly super-solar C/O ratio. 
We suggest that the super-solar C/H and C/O ratios might be explained if the atmosphere was formed near the inner edge of the shadowed region and polluted by solid components, since the solid C/O ratio exceeds unity there when disk chemistry starts from the atomise initial abundances and the cosmic-ray ionization rate is low (see Figure \ref{Figure11rev_CtoO-NtoO_reset}, see also \citealt{Eistrup2016}). 
%%%
\\ \\
%.
Figures \ref{Figure20rev_CtoH_appendix}, \ref{Figure21rev_OtoH_appendix}, and \ref{Figure22rev_NtoH_appendix} in Appendix \ref{Csec:C} show respectively the radial profiles of C/H, O/H, and N/H ratios at t=$10^{6}$ years.
In the shadowed disk ($f\leq0.03$), at $r\sim3-8$ au (around the current orbit of Jupiter), the gas-phase N/H ratios ($\sim10^{-7}-10^{-5}$) are larger than the gas-phase C/H and O/H ratios ($\sim10^{-8}-10^{-6}$) in each model, which produce super-stellar N/O ratios of $\gg1$ (see Figures \ref{Figure10rev_CtoO-NtoO_inheritance} and \ref{Figure11rev_CtoO-NtoO_reset}).
%
%%
%%
%%%
%%\\ \\
\subsection{Implications for the small bodies in the solar system}\label{sec:4-2}
%%%
%%%
\begin{figure*}[hbtp]
\begin{center}
\vspace{1cm}
%\vspace{-0.4cm}
%\plotone{cost.pdf}
%\plotone[scale=0.63]{{20210624_surface-density.eps}
%\plotone[scale=0.63]{{20210624_ngas-T.eps}
\includegraphics[scale=1.0]{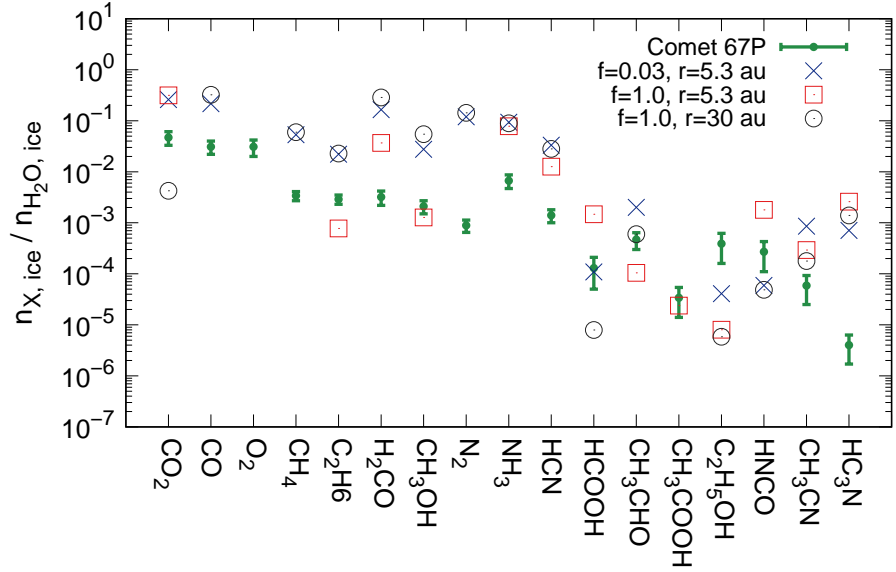}
\end{center}
\vspace{-0.2cm}
\caption{
The fractional molecular ice abundances with respect to water ice $n_{\mathrm{X, ice}}$/$n_{\mathrm{H}_{2}\mathrm{O}, \mathrm{ice}}$ for the coma of comet 67P/Churyumov-Gerasimenko and our standard disk model calculations at t=$10^{6}$ years. 
The comet data (green filled circles and lines) are taken from \citet{Rubin2019a, Rubin2020} and originally derived from in-situ measurements by ROSINA (Rosetta Orbiter Spectrometer for Ion and Neutral Analysis) for the coma of comet 67P in May/June 2015 before perihelion.
%%, 
The blue crosses show the results of our standard model calculations at $r=5.3$ au and for $f=0.03$ (in the shadowed region), the red open squares show those at $r=5.3$ au and for $f=1.0$ (between CO$_{2}$ and CH$_{4}$ snowlines in the non-shadowed disk), and
the black open circles show those at $r=30$ au and for $f=1.0$ (outside CO and N$_{2}$ snowlines in the non-shadowed disk).
We note that the points which are not shown in this Figure indicate low ice abundances ($<10^{-7}$ with respect to water ice) of those species.
In our standard model, we assume molecular initial abundances and $\xi_{\mathrm{CR}}(r)=$$1.0\times10^{-17}$ [s$^{-1}$] (see Sections \ref{sec:3-1rev} and \ref{sec:3-2rev}).\\
\\ \\
%%
%%.
}\label{Figure12rev_Comet-Comparison}
\end{figure*}
%%%
%.
On the basis of our calculations, in the shadowed region the snowline positions of molecules with smaller $E_{\mathrm{des}}$ than that of H$_{2}$O move inward, and even the most volatile species CO and N$_{2}$ freeze-out onto dust grains at around the current orbit of Jupiter ($r\sim3-8$ au), as found in \citet{Ohno2021}.
In the shadowed region, the dust grains at $r\sim3-8$ au are expected to have significant (more than $\sim5-10$ times) amounts of saturated hydrocarbon ices such as CH$_{4}$ and C$_{2}$H$_{6}$, ices of organic molecules (which are mostly saturated) such as H$_{2}$CO, NH$_{2}$CHO, CH$_{3}$OH, and CH$_{3}$NH$_{2}$, in addition to H$_{2}$O, CO, CO$_{2}$, NH$_{3}$, N$_{2}$, HCN, and NH$_{2}$OH ices, compared with those in the non-shadowed disks (mostly ices of H$_{2}$O, CO$_{2}$, NH$_{3}$, and unsaturated hydrocarbon molecules).
In addition, the icy abundances of unsaturated hydrocarbons and other organic molecules in the shadowed disks are much smaller than those in the non-shadowed disks.
%%
\begin{comment}
\end{comment}
%%
%%
%%
\\ \\
Here it is worth discussing whether the presence of the shadowed region influences the chemical composition of small objects, such as primitive comets in the solar system.
If primitive comets formed from the icy dust grains in the shadowed region of the disks (such as $r\sim3-8$ au for $f\leq0.03$), they are expected to include more saturated hydrocarbons and complex organic molecules such as H$_{2}$CO, NH$_{2}$CHO, and CH$_{3}$OH, and less unsaturated organic molecules than those formed in the non-shadowed disks.
\\ \\
Reactions of radicals, which are mainly formed by cosmic-ray induced photodissociation of CH$_{3}$OH ice, are needed to form complex organic molecules in disks (see Sections \ref{sec:3-2-2rev} and \ref{sec:3-2-4rev}), assuming that they are not already formed in the molecular cloud phases.
In addition, \citet{Garrod2008} discussed that (CH$_{2}$OH)$_{2}$ (ethylene glycol) is formed on the dust grains around the water snowline from its precursor radical CH$_{2}$OH, which becomes mobile just as water and other species are beginning to desorb. 
Moreover, icy grains can efficiently coagulate into larger ($\gg$1 mm) dust particles and cm-size pebbles outside the water snowline (see Section \ref{sec:2-1} and e.g., \citealt{Ros2013, Sato2016, Drazkowska2017, Pinilla2017}).
If CH$_{3}$OH rich large dust grains and/or pebbles migrate to inside the shadowed region (such as around the water snowline), and/or if the shadowed region disappears due to the disk evolution, the ice abundances of various complex organic molecules in such dust grains and pebbles are also expected to increase (see also Section \ref{sec:4-4}). 
This is because the temperature of such dust grains and pebbles increases ($>30$ K) 
and radical-radical reactions can proceed efficiently in such warm conditions.
Thus in the shadowed disk, both efficient CH$_{3}$OH ice formation and the formation of ices of further complex organic molecules such as (CH$_{2}$OH)$_{2}$ may be realized due to the dissipation of shadowed structures and/or migration inside the shadowed region, and they can be realized without dust grains and/or pebbles migrating vast distances, compared with the non-shadowed disk.
We suggest that in shadowed disks, complex organic molecules can be formed in situ rather than being fully inherited from molecular clouds.
\\ \\
To date, various complex organic molecules have been detected in comets (such as Hale-Bopp and 67P/Churyumov-Gerasimenko).
Several studies discussed that the molecular abundances in comets are determined by the combination of chemical evolution in the protosolar disk and inheritance from molecular clouds (e.g., \citealt{Mumma2011, Caselli2012, Walsh2014, Eistrup2016, Eistrup2018, Altwegg2017, Altwegg2019, Drozdovskaya2019, Oberg2021}).
\citet{Walsh2014} discussed that grain-surface fractional abundances (relative to water ice) for the outer region of the non-shadowed disk ($T<50$ K) are consistent with abundances derived for comets, suggesting a grain-surface route to the formation of COMs observed in cometary comae.
%%% 
\\ \\
Recently, the abundances of various volatiles and organic molecules (including hydrocarbons) towards comet 67P/Churyumov-Gerasimenko were reported by e.g., \citet{LeRoy2015}, \citet{Rubin2015S, Rubin2015, Rubin2019a, Rubin2020}, \citet{Altwegg2017, Altwegg2019} and \citet{Schuhmann2019}.
They reported that the relative abundances of CO, CO$_{2}$, and C$_{2}$H$_{6}$ with respect to H$_{2}$O are larger than those of other Jupiter-family comets, which might suggest the formation of cometary grains at lower temperature regions (such as below 30 K).
In addition, \citet{Rubin2015S} reported the first cometary detection of N$_{2}$ towards comet 67P, and discussed that the lower N$_{2}$/CO abundance ratio ($\sim3\times10^{-2}$ based on \citealt{Rubin2019a, Rubin2020}) compared with the protostar value ($\sim10^{-1}$) suggested the cometary grains of 67P are formed in cold regions with $\sim24-30$ K.
\citet{Schuhmann2019} also reported the existence of unsaturated hydrocarbons in the coma of comet 67P. 
We note that gas-phase reactions within the CH$_{4}$ snowline are needed to efficiently form unsaturated hydrocarbons (see Sections  \ref{sec:3-2-2rev} and \ref{sec:3-2-4rev}).
%%.
%%%
\\ \\
In Figure \ref{Figure12rev_Comet-Comparison}, we compare the molecular ice abundances (with respect to water ice) of the coma of comet 67P \citep{Rubin2019a, Rubin2020} and our standard disk model calculations (see Sections \ref{sec:3-1rev} and \ref{sec:3-2rev}).
On the basis of Figure \ref{Figure12rev_Comet-Comparison}, the ice abundances of CO, CH$_{4}$, and N$_{2}$ in the coma of comet 67P cannot be explained by the icy grains within the CH$_{4}$ snowline in non-shadowed disks. 
In addition, as for results in the cold regions where both CO and N$_{2}$ are frozen onto dust grains (both in the shadowed and non-shadowed disks), the CO ice abundances are closer to the value of comet 67P than the N$_{2}$ ice abundances.
These results also suggest that the icy grains of comet 67P were formed between the CO and N$_{2}$ snowlines (see above and \citealt{Rubin2015S, Rubin2019a, Rubin2020}).
Such cold regions, where the icy grains of comet 67P are formed, can be located at $r\sim$ a few au for the shadowed disks, whereas they are located at r$>20$ au for the non-shadowed disk.
\\ \\
In the results of our standard disk model calculations, the CO$_{2}$ ice abundances at $r\sim5.3$ au in the shadowed disk are around two orders of magnitude larger than those at $r\sim30$ au in the non-shadowed disk, although the differences in other molecular ice abundances between the two cases are within around one order of magnitude.
This is because CO$_{2}$ ice formation on the dust grains is not efficient in the coldest regions with $T(r)\lesssim20$ K (see Section \ref{sec:3-2-1rev}).
Thus, the CO$_{2}$ ice abundances can be used to distinguish these two regions (the inner cold region ($r\sim3-8$ au, $T(r)\sim20-30$ K) in the shadowed disk and outermost coldest region ($r\sim30$ au, $T(r)\lesssim20$ K) in the non-shadowed disk) as formation sites of icy dust grains.
The CO$_{2}$ ice abundance 
at $r\sim30$ au in the non-shadowed disk is around one orders of magnitude smaller than that in comet 67P, whereas that at $r\sim5.3$ au in the shadowed disk is 4-7 times larger than that in comet 67P.
Thus, for explaining the CO$_{2}$ ice abundance in comet 67P, the formation of CO$_{2}$ ice at $r\sim5.3$ au in the shadowed disk may be more suitable than that at $r\sim30$ au in the non-shadowed disk.
%%%
In addition, if we also include the above discussion about the CO and N$_{2}$ abundances, we suggest that the icy grains of comet 67P may have been formed at the region with $T(r)\sim25$ K, such as the shadowed region.%%%
%%%
\\ \\
%%We note that 
For the model results in the shadowed region (see Figure \ref{Figure12rev_Comet-Comparison}), the abundances of ices such as H$_{2}$CO and CH$_{3}$OH are around 1-2 orders of magnitude larger than those of comet 67P, whereas those of C$_{2}$H$_{5}$OH and CH$_{3}$COOH are more than 1 and 3 orders of magnitude smaller, respectively.
We suggest that if the cometary grains of 67P are originally from the shadowed region, a temperature rise (due to the inward migration and/or dissipation of the shadowed region) after CH$_{3}$OH ice formation in the shadowed region may be needed to trigger radical-radical reactions (see also above) to produce more complex molecules such as C$_{2}$H$_{5}$OH and CH$_{3}$COOH.
Future detailed chemical modeling including such physical evolutions may be needed to construct the formation scenario which explains the ice abundances of all molecules in comet 67P simultaneously.
\\ \\
\citet{Bieler2015} and \citet{Rubin2015} reported that O$_{2}$ ice is abundant in comet 67P and the ratio of O$_{2}$/H$_{2}$O is a few percent (see also \citealt{Luspay-Kuti2022}).
On the basis of Figure \ref{Figure12rev_Comet-Comparison}, our standard disk model is unable to explain
the cometary O$_{2}$ ice abundances 
(see Sections \ref{sec:3-1rev} and \ref{sec:3-2rev}).
According to previous studies (e.g., \citealt{Eistrup2016, Eistrup2018, Taquet2016}) and our calculations, such higher O$_{2}$ ice abundances are possible in disks only if chemical starting conditions were purely atomised and the disk ionisation level was low (see Appendix \ref{Asec:A-1}). In addition, O$_{2}$ is very volatile and freezes-out onto dust grains at $r\sim2-10$ au only in the shadowed disk ($f\leq0.03$, see Appendix \ref{Asec:A-1}).
We note that \citet{Taquet2016} suggest that O$_{2}$ trapping in H$_{2}$O ice at earlier evolutionary stages (such as in the molecular clouds) may also be an explanation.
%%%
%%
%%
\\ \\
Other small objects such as asteroids will have similar abundances (such as CO$_{2}$ and CH$_{3}$OH rich), if they are formed from the dust grains in the shadowed region.
%%.
\citet{Yada2021} reported the results of preliminary analyses of the Hayabusa2 samples returned from C-type asteroid Ryugu, and showed the infrared spectral profile with weak absorptions at 2.7 and 3.4 $\mu$m that imply a carbonaceous composition with indigenous aqueous alteration.
\citet{Kurokawa2022} compared the infrared spectra (including 3.1$\mu$m absorption features of ammoniated phyllosilicate) of main belt asteroids collected by the AKARI space telescope and their models of water-rock reactions, and suggested that multiple large main belt asteroids formed beyond the NH$_{3}$ and CO$_{2}$ snowlines and have been transported to their current locations.
\citet{Fujiya2019} inferred from the measurements of CO$_{2}$ and H$_{2}$O abundances and carbon isotope ratios of carbonate minerals in the Tagish Lake meteorite (an carbonaceous chondrite) that at least some D-type asteroids were formed beyond the CO$_{2}$ snowline.
%.
We note that D-type asteroids are discovered mainly at the outer edge of the main asteroid belt and in the Jupiter Trojan regions \citep{Demeo2014}.
%%%
\citet{Tsuchiyama2021} reported the discovery of primitive CO$_{2}$-bearing fluids in an aqueously altered carbonaceous chondrite (one of the primitive meteorites), and discussed that its parent body was formed outside the CO$_{2}$ snowline and later transported to the inner solar system.
Since the CO$_{2}$ snowline position moves inward in the shadowed disk, the supply of dust grains and small objects with significant amounts of CO$_{2}$ and complex organic molecules to the inner region may be relatively easier than that in the non-shadowed disk.
%%.
We note that
if the shadowed region is maintained for a relatively long time ($t\sim10^{6}$ years), chemical evolution 
%%%% 
may produce dust grains and solid objects with large amounts of CO$_{2}$ ice and ices of complex organic molecules such as H$_{2}$CO and CH$_{3}$OH.
%%.
\\ \\
The solid bodies formed at the Kuiper belt may be CO$_{2}$ rich if we consider the disk shadowing effect. 
This is because at $r>20$ au, the CO$_{2}$ ice abundance in the shadowed disk is around one orders of magnitude larger than that in the non-shadowed disk (see Section \ref{sec:3-2-1rev} and Figure \ref{Figure3_rev_radial}).
\\ \\
We note that the CO$_{2}$ snowline position will be important both for chemistry and dust grain growth in the disk.
Recent laboratory experiments showed that CO$_{2}$ ice is less sticky compared to H$_{2}$O ice \citep{Musiolik2016a, Musiolik2016b, Fritscher2021}. 
Thus, efficient grain growth is only expected between the H$_{2}$O and CO$_{2}$ snowlines \citep{Okuzumi2019, Arakawa2021}.
If the disk has a shadowed region beyond the water snowline, the region where efficient grain growth can take place might become narrower.
\\ \\
\citet{Dartois2013, Dartois2018} found that UltraCarbonaceous Antarctic MicroMeteorites (UCAMMs) have higher C/Si abundance ratios ($\gtrsim10^{2}$) and N/C abundance ratios ($\sim0.05-0.12$) in organic compounds than those in other primitive meteorites and the local interplanetary dust particle (IDPs).
They discussed that UCAMMs might be formed in a cold nitrogen rich environment, such as outside the N$_{2}$ snowline.
We suggest that if the disk has a shadowed region beyond the water snowline, the formation sites of UCAMMs may be located in the inner region of the disk, such as $r\sim3-8$ au.
%%%%
%
%%
\\
\subsection{Implication for the observations of protoplanetary disks}\label{sec:4-3}
\begin{figure*}[hbtp]
\begin{center}
\vspace{1cm}
%\vspace{-0.4cm}
%\plotone{cost.pdf}
%\plotone[scale=0.63]{{20210624_surface-density.eps}
%\plotone[scale=0.63]{{20210624_ngas-T.eps}
\includegraphics[scale=0.68]{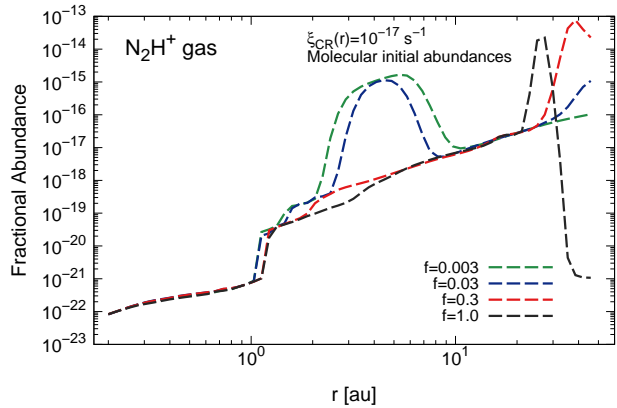}
\end{center}
\vspace{-0.2cm}
\caption{
The radial profiles of gas-phase fractional abundances with respect to total hydrogen nuclei densities at t=$10^{6}$ years for N$_{2}$H$^{+}$ ($n_{\mathrm{N}_{2}\mathrm{H}^{+}}$/$n_{\mathrm{H}}$).
These profiles show the results for $\xi_{\mathrm{CR}}(r)=$$1.0\times10^{-17}$ [s$^{-1}$] and molecular initial abundances (the ``inheritance'' scenario).
%%The dashed and solid lines show the profiles for gaseous and icy molecules, respectively.
The black, red, blue, and green lines show the profiles for different values of the parameter $f$ (=1.0, 0.3, 0.03, and 0.003), respectively.
%.
\\ \\
}\label{Figure13rev_N2H+}
\end{figure*}
%%%
%%%
\begin{comment}
\end{comment}
%%.
%%
Many of the complex (organic) molecules described in Sections \ref{sec:3-2-2rev}-\ref{sec:3-2-4rev} have been observed by previous observations with e.g., ALMA towards hot cores/corinos in high-/low-mass star forming regions, respectively (see e.g., \citealt{Herbst2009, Sakai2013, Jorgensen2020, Yang2021}), and have been detected in the comae of multiple comets (see Section \ref{sec:4-2} and e.g., \citealt{Mumma2011, Walsh2014, Altwegg2019, Drozdovskaya2019, Rubin2020}).
\citet{Lee2019} reported the detections of CH$_{3}$CHO and CH$_{3}$CN, in addition to CH$_{3}$OH, towards Class I disk around FU Ori type young star V883 Ori with ALMA.
In addition, some of these molecules have been observed towards Class II disks by previous infrared observations (C$_{2}$H$_{2}$, e.g., \citealt{Pontoppidan2010}) and ALMA observations, such as H$_{2}$CO and CH$_{3}$OH (see e.g., \citealt{Loomis2015, Walsh2016, Walsh2018, Booth2021a, vanderMarel2021, Guzman2021}), and C$_{3}$H$_{2}$, CH$_{3}$CN, HC$_{3}$N, and HCOOH (see e.g., \citealt{Qi2013, Oberg2015, Bergner2018, Favre2018, Loomis2018, Loomis2020, Ilee2021}).
\\ \\
%%.
\citet{Booth2021a} reported the first detection of CH$_{3}$OH in the disk around a Herbig Ae star, HD 100546.
They reported that the CH$_{3}$OH to H$_{2}$CO abundance ratio is higher ($\gtrsim10$) at the inner edge of the dust ring in the disk ($\sim10-30$ au)
than those ($\sim1-2$) in the outer disk ($>100$ au) and in the TW Hya disk \citep{Walsh2016}.
They discussed that at the inner edge CH$_{3}$OH is proposed to originate from thermal desorption.
%%.
%
\citet{vanderMarel2021} and \citet{Brunken2022} reported the detections of CH$_{3}$OH, H$_{2}$CO, and CH$_{3}$OCH$_{3}$ line emission in the vicinity of the asymmetric dust trap in the disk around a Herbig Ae star Oph IRS 48.
%In addition, \citet{Brunken2022} reported the dections of 
%%
They discussed that these molecules are thermally desorbed from icy dust grains and that such dust traps provide huge icy grain reservoirs in the disk midplane.
%%%
\\ \\
Recent dust continuum survey observations of protoplanetary disks (with e.g., ALMA) have shown that disk substructures, most prominently gas-depleted gaps and dust-rich rings, are common (e.g., \citealt{Andrews2016, Andrews2018, Andrews2020, Dullemond2018, Huang2018, Isella2016, Isella2018, Tsukagoshi2016}).
%.
%%%
In addition, the recent MAPS (Molecules with ALMA at Planet-forming Scales)\footnote{\url{http://alma-maps.info}} survey obtained spatial distributions (with $\Delta$$r\sim10-50$ au) of line emission for CO \citep{Zhang2021} and some other organic molecules (H$_{2}$CO, HCN, CH$_{3}$CN, HC$_{3}$N, and C$_{3}$H$_{2}$, see \citealt{Guzman2021, Oberg2021MAPS, Ilee2021}) towards three T Tauri disks (IM Lup, GM Aur, AS 209) and two Herbig Ae disks (HD 163296, MWC 480).
They found that disk substructures (such as rings and gaps) are also common in molecular gas emission. 
%%%
\\ \\
Since a shadowed region can be formed in the midplanes beyond such dust rings/traps of these disks by blocking the radiation from the central star, we propose that the effects of shadows need to be taken
into account 
when we discuss chemical evolution of complex organic molecules in disks with such dust rings and asymmetric traps.
\citet{Alarcon2020} calculated the temperature and chemical structures of disks with gas-depleted gaps.
They showed that the disk midplane temperature in the gap increases, producing local sublimation of key volatiles, while it decreases in the ring, causing a higher volatile deposition onto the dust grain surfaces.
\citet{Isella2018} and \citet{Okuzumi2022} described that the outer wall of the gap produced by a giant planet receives extra starlight heating and puffs up, throwing a shadow across the disk beyond.
In addition, \citet{Ohashi2022} suggested from their ALMA observations that the dust clumps in the disk around a Class 0/I protostar create the shadowed region outside, resulting in the sudden drop in temperature.
We predict that, as well as the shadowed region beyond the water snowline (discussed in this paper), the freeze-out of molecules onto dust grains and efficient formation of organic molecules are also expected to occur in the shadowed region formed by the presence of a giant planet and/or dust clumps.
%.
\\ \\
In this study, we adopt a 1D physical model of the protosolar disk (T Tauri disk) in order to investigate the effects of shadow structures on the chemical evolution of the disk midplane and compare the results with the molecular composition of small objects in the solar system (see also Section \ref{sec:4-2}). The influence of disk shadowing is expected to be larger in the disk midplane than in the disk surface, since it is easier for direct stellar light to reach the surface than the midplane \citep{Ueda2019, Okuzumi2022}.
We note that the lines from more rare isotopologues trace deeper regions in the disks, but the disk midplane is obscured if the dust emission is optically thick.
In addition to future observations of molecular lines with much higher spatial resolutions towards the disks using e.g., ALMA and ngVLA (next generation Very Large Array), further calculations of disk chemical evolution with 2D physical models of both T Tauri disks and Herbig Ae disks are also important to investigate whether or not the effects of shadows are needed to explain the observed distributions of molecular emission.
Since the dust opacities at the frequencies of ngVLA are lower than those of ALMA, observations with ngVLA will be useful to trace the inner molecular gas abundances, such as within 10 au.
In addition, ngVLA expects to resolve the dust emission at cm wavelengths with $\Delta r\lesssim1$ au, and thus will be able to confirm the variation in dust density profile at the water snowline and presence of the shadowed region beyond the water snowline even in disks around T Tauri stars \footnote{see Okuzumi et al. (2021) in ngVLA-J memo series, \url{https://ngvla.nao.ac.jp/researcher/memo/}}.
\\ \\
We note that future calculations of disk chemical evolution with 2D physical models would be also useful to investigate the effects of shadowing on composition of planetary atmospheres. This is because some recent theoretical studies (e.g., \citealt{Tanigawa2012, Morbidelli2014}) and observational studies (e.g., \citealt{Teague2019}) suggested that the flow of gas into the gap produced by a growing planet is dominated by gas falling vertically from a height of at least one scale height (see e.g., \citealt{Cridland2020v}).
\\ \\
N$_{2}$H$^{+}$ is considered to be a useful probe of the CO and N$_{2}$ snowlines in disks (e.g., \citealt{Qi2013, Qi2019, Aikawa2015, vantHoff2017, Murillo2022}), because it is destroyed by proton transfer to CO,
\begin{equation}\label{Rec18}
\mathrm{N}_{2}\mathrm{H}^{+} + \mathrm{CO} \rightarrow \mathrm{HCO}^{+} + \mathrm{N}_{2}. \\
\end{equation}
\citet{Aikawa2015} suggested from their disk modeling that the N$_{2}$H$^{+}$ abundance can have a peak at the temperature slightly below that of CO sublimation, even if the binding energies of CO and N$_{2}$ are nearly the same.
Figure \ref{Figure13rev_N2H+}
 shows the radial profiles of the N$_{2}$H$^{+}$ abundances for various values of $f$.
The CO freezes-out onto dust grains in the shadowed region, and thus N$_{2}$H$^{+}$ abundances at $r\sim3-8$ au become larger in the shadowed disk ($\sim10^{-15}$ for $f\leq0.03$) than those in the non-shadowed disk ($<10^{-17}$ for $f=1.0$) by several orders of magnitudes.
Thus, N$_{2}$H$^{+}$ line emission may potentially trace the shadowed region of the disk beyond the water snowline, although previous observations of N$_{2}$H$^{+}$ lines do not resolve the spatial scales around the water snowline ($r\sim1-10$ au).
We note that the N$_{2}$H$^{+}$ abundances in the disk midplane are lower than those within the disk surface \citep{Aikawa2015, vantHoff2017}, because of the high densities and low ionization rates.
%%.
Further chemical and radiative-transfer modeling in the 2D disk structures with shadow structures are needed to confirm whether N$_{2}$H$^{+}$ line emission can really be used as observational tracer of such shadowed regions, since the line emission also traces the vertical distributions of CO and N$_{2}$ \citep{Qi2019}, and observations of the disk midplane directly are not easy because of the high optical depths at ALMA wavelengths.
%
%%%
\subsection{Other model caveats}\label{sec:4-4}
In our disk chemical modeling, we implicitly assumed that ices on grains are formed by homogeneous layers regardless of their composition or crystallinity, as most astrochemical models of disks assume.
However, recently \citet{Kouchi2021} suggested from their transmission electron microscopy studies that the macroscopic morphology of icy dust grains are as follows: amorphous H$_{2}$O covered the refractory grain uniformly, CO$_{2}$ nano-crystals were embedded in the amorphous H$_{2}$O, and a polyhedral CO crystal is attached to the amorphous H$_{2}$O.
Such morphology of dust grains would affect the chemical evolution in the disks, since the binding energies and non-thermal desorption rates depend on the chemical composition of the ice mantles on the dust grains (e.g., \citealt{Bertin2016, Cuppen2017, Penteado2017}).
\\ \\
\citet{Heinzeller2011} investigated the effects of physical mass transport phenomena in the radial direction by viscous accretion and in the vertical direction by diffusive turbulent mixing and disk winds. They showed that the gas-phase molecular abundances of such as H$_{2}$O and CH$_{3}$OH are enhanced in the warm surface layer due to the effects of vertical mixing.
\citet{vanderMarel2021} discussed that this vertical transport may be important to explain the observed abundance of CH$_{3}$OH in the disk around IRS 48, in addition to icy dust concentrations at the dust trap.
In addition, we assume that the disk physical structure is steady and that the shadow structure is maintained for 10$^{6}$ years.
We note that the inward migration of solids and/or destruction of the shadow structure after the efficient formation of CH$_{3}$OH ice in the shadowed region may increase the abundances of various complex organic molecules in the inner disks around the water snowline (see also Section \ref{sec:4-2}).
Future observations of organic molecular lines with e.g., ngVLA and infrared telescopes (such as JWST and GREX-PLUS) are expected to constrain the abundances in such inner warm region.\\
%%.
%%
%\\ \\
%%%
%%
%%%%
\section{Conclusions}\label{sec:5c}
In this study, we investigated the radial abundance distributions of dominant carbon-, oxygen-, and nitrogen-bearing molecules and the radial distributions of elemental abundance ratios (C/O and N/O ratios) in the gas and ice of disks with shadow structures.
We used a detailed gas-grain chemical reaction network and calculated chemical structures in the shadowed disk midplane around a T Tauri star (a protosolar-like star).
Gas-phase reactions, thermal and non-thermal gas-grain interactions, and grain-surface reactions were included in our adopted reaction network.
%% .
We investigated the dependance of the disk chemical structures on ionisation rates and initial abundances.
%%.
We discussed the effects of disk shadowing on chemical evolution of complex organic molecules and forming planetary atmospheres. We also compared the results of our calculations with the molecular composition of small bodies in the solar system (such as comets and asteroids) and recent observational results of protoplanetary disks. Our findings can be summarized as follows: 
%! \\ \\ 
\begin{itemize}
\item
In the shadowed disks ($f\leq0.03$), the snowline positions of molecules with smaller $E_{\mathrm{des}}$ than that of H$_{2}$O move inward, and even the most volatile species, CO and N$_{2}$, freeze-out onto dust grains at around the current orbit of Jupiter ($r\sim3-8$ au).
%%. 
Our detailed calculations confirm the results of \citet{Ohno2021} who showed the freeze-out of CO and N$_{2}$ onto dust grains in the shadowed region with using more simplified calculations. 
%%%
Meanwhile, we newly find that the dust grains within the shadowed regions have
significant (more than $5-10$ times) amounts of saturated hydrocarbon ices such as CH$_{4}$ and C$_{2}$H$_{6}$, ices of organic molecules such as H$_{2}$CO, NH$_{2}$CHO, and CH$_{3}$OH, in addition to H$_{2}$O, CO, CO$_{2}$, NH$_{3}$, N$_{2}$, HCN, and NH$_{2}$OH ices, compared with those in the non-shadowed disks (mostly ices of H$_{2}$O, CO$_{2}$, NH$_{3}$, and unsaturated hydrocarbon molecules).
We note that these abundant saturated hydrocarbons and various organic molecules such as H$_{2}$CO, NH$_{2}$CHO, and CH$_{3}$OH were not reported by \citet{Ohno2021}, since they approximated various organic molecules by considering C$_{2}$H$_{6}$ alone.
\\
\item
The icy abundances of C$_{2}$H$_{6}$, HCN, NH$_{2}$OH, and complex organic molecules (which are mostly saturated) such as H$_{2}$CO, CH$_{3}$OH, NH$_{2}$OH, NH$_{2}$CHO, and CH$_{3}$NH$_{2}$ at $r\sim3-8$ au in the shadowed disks are enhanced compared with the values in the non-shadowed disk, although the snowline positions of these molecules are $r\lesssim3$ au regardless of the values of $f$.
We concluded that the sequential hydrogenation reactions of (especially) CO on the cold dust grains 
play a vital role in efficiently forming
these molecules, in addition to reaction pathways starting from the cosmic-ray induced photodissociation of CH$_{3}$OH.
%.
We found that
if the shadowed region is maintained for a relatively long time ($t\sim10^{6}$ years), chemical evolution 
may produce dust grains and solid objects with large amounts of CO$_{2}$ and organic molecular ices (see also Appendix \ref{Bsec:B}).
\\ 
\item
The icy abundances of unsaturated hydrocarbons such as C$_{2}$H$_{2}$, C$_{2}$H$_{4}$, C$_{3}$H$_{2}$, C$_{3}$H$_{4}$, C$_{3}$H$_{6}$ and HC$_{3}$N at $r\sim3-8$ au (just outside their snowlines) are much smaller than those of a non-shadowed disk, since gas-phase chemical reactions especially within the CH$_{4}$ snowline ($\sim15$ au for $f=1.0$ and $\lesssim2$ au for $f\leq0.03$) mainly drive the formation of these molecules.
Moreover, the icy abundances of unsaturated complex organic molecules such as HCOOCH$_{3}$, HCOOH, CH$_{3}$COOH outside the snowlines are much smaller than those of a non-shadowed disk, since grain-surface association of large radical-radical reactions in the warm regions ($T(r)\sim50$ K) are needed for the formation of these molecules.
In addition, the icy abundances of other molecules such as CH$_{3}$CN, CH$_{3}$CHO, CH$_{3}$OCH$_{3}$, and C$_{2}$H$_{5}$OH are larger for $f=0.03$ than those for $f=$1.0, 0.3, and 0.003, since they are determined by the combinations of radical-radical reactions and gas-phase reactions / radical formation reactions from CO and/or CH$_{3}$OH on the dust grain surfaces.
%%.
\\
\item
We also studied the impacts of different ionization rates and initial chemical abundances on disk chemical structures in Appendix \ref{Asec:A}.
We found that CO/CO$_{2}$ abundances become smaller/larger with increasing ionization rates, respectively.
In addition, CH$_{4}$ and C$_{2}$H$_{6}$ gas abundances within their snowline become smaller as the ionization rates become larger.
Abundances of H$_{2}$O and organic molecules are larger for molecular initial abundances than those for atomic initial abundances. 
Moreover, O$_{2}$ ice abundances are $\sim10^{-5}-10^{-4}$ (consistent with the measured cometary abundances) only for atomic initial abundances and the low ionization rates.
O$_{2}$ is very volatile and freezes-out onto dust grains at $r\sim2-10$ au only in the shadowed disk.
\\
\item
In the shadowed region, the gas-phase C/O ratios are almost unity for the high ionisation rate and/or molecular initial abundances and 0.8 for the low ionisation rate and atomic initial abundances.
%%% 
In addition, the radial profiles of the gas-phase N/O ratios show spatial variations which are much larger than those of the C/O ratios, and thus the N/O ratio would be a useful tracer of the shadowed regions of disks.
For molecular initial abundances, the gas-phase N/O ratios are much larger than unity in the shadowed region, and they are $\sim6-9$ for the low ionization rate and $\sim10-20$ for the high ionization rate.
The icy-phase C/O and N/O ratios reach the initial elemental values in the shadowed region.
Therefore, if the planets acquire their atmospheres from the gas in the shadowed region ($r\sim3-8$ au in our disk model), they are expected to have super-stellar N/O ratios of $\gg1$, super-stellar C/O ratios of around unity in most cases, and sub-stellar metallicities.
In contrast, they are expected to have stellar N/O and C/O ratios and super-solar metallicities if the planetary atmospheres are efficiently polluted by solid components (including the case of Jupiter, see \citealt{Ohno2021}). Upcoming and future observations by JWST and ARIEL will constrain such elemental ratios precisely for atmospheres of many exoplanets such as hot Jupiters.
\\
\item
We discussed whether the presence of a shadowed region influences the chemical composition of small objects, such as primitive comets and asteroids in the solar system.
Recently the abundances of various volatiles and organic molecules towards comet 67P/Churyumov-Gerasimenko were reported, and some of the results (including CO, CO$_{2}$, C$_{2}$H$_{6}$, N$_{2}$, and O$_{2}$ abundances) would suggest the formation of cometary grains in lower temperature regions with $T\sim25$ K, which can be found in the shadowed region.
\\
\item
We propose that CO$_{2}$ ice abundances can be used to distinguish the inner cold region ($r\sim3-8$ au, $T(r)\sim20-30$ K) in the shadowed disk and the outermost coldest region ($r\sim30$ au, $T(r)\lesssim20$ K) in the non-shadowed disk as formation sites of icy dust grains.
%%
%%
%%%
Moreover, if CH$_{3}$OH rich large dust grains and/or pebbles migrate to inside the shadowed region (such as around the water snowline), and/or if the shadowed region disappears due to the disk evolution, the ice abundances of various complex organic molecules are also expected to increase. This is because radical-radical reactions may be able to proceed efficiently due to the rapid heating.
%In disk.
\\
\item
N$_{2}$H$^{+}$ line emission could potentially trace the shadowed region of the protoplanetary disks beyond the water snowline, although further modeling and observations with much higher spatial resolution than currently conducted are needed.
%%%
\end{itemize}
On the basis of our calculations, 
%%.
we conclude that a shadowed region allows the recondensation of key volatiles onto dust grains, 
%,
and may explain to some degree the trapping of icy molecules such as CO$_{2}$, CO, N$_{2}$ and O$_{2}$ in the dust grains that formed comet 67P/Churyumov-Gerasimenko.
In addition, in a shadowed disk Jupiter need not to have migrated vast distances to explain its atmospheric composition.
%%. 
In the shadowed disk, both efficient CH$_{3}$OH ice formation and formation of ices of further complex organic molecules such as (CH$_{2}$OH)$_{2}$ driven by rapid heating may be realized without dust grains and/or pebbles migrating vast distances, compared with the non-shadowed disk.
%%.
Thus, we propose that in the shadowed disks various complex organic molecules can be formed in situ rather than being fully inherited from molecular clouds.
Further chemical modeling (such as using 2D physical models of both T Tauri disks and Herbig Ae disks) and comparison with observations of disks (using e.g., ALMA and ngVLA), planetary atmospheres, and small objects (comets and asteroids) in the solar system will further constrain the effects of disk shadowing on chemical evolution.
\\ \\
%\begin{comment}
% Sulfar
%\end{comment}
%%%
\begin{acknowledgments}
This study was started originally based on the discussion during and after the online workshop 2021 on planetary system formation for young scientists in Japan,
%%\footnote{\url{https://kshunta3403.wixsite.com/planet-formation}}
in which S.N.~and T.U.~are two of organizers of the workshop.
We thank the referee for important suggestions and comments. 
We are grateful to Nami Sakai, Yuri Aikawa, and Satoshi Okuzumi for their useful comments.
%%. 
Our numerical studies were carried out on PC cluster at Center for Computational Astrophysics (CfCA), National Astronomical Observatory of Japan (NAOJ). 
S.N.~is grateful for support from RIKEN Special Postdoctoral Researcher Program (Fellowships), and MEXT/JSPS (Japan Society for the Promotion of Science) Grants-in-Aid for Scientific Research (KAKENHI) Grant Numbers JP20K22376, JP20H05845, and JP20H05847.
K.O.~is supported by JSPS Overseas Research Fellowships and JSPS KAKENHI Grant Number JP19K03926.
T.U.~is supported by Grants-in-Aid for JSPS Fellows Grant Number JP19J01929, and acknowledges the support of the DFG-Grant ``Inside: inner regions of protoplanetary disks: simulations and observations'' (FL 909/5-1).
C.W.~acknowledges financial support from the University of Leeds and from the Science and Technology Facilities Council (grant numbers 
%%%%ST/R000549/1 and ST/T000287/1).
ST/T000287/1 and MR/T040726/1).
H.N.~is supported by MEXT/JSPS KAKENHI Grant Numbers JP18H05441, JP19K03910 and JP20H00182, NAOJ ALMA Scientific Research grant No. 2018-10B, and FY2019 Leadership Program at NAOJ.
%%%
\end{acknowledgments}

\software{RADMC-3D \citep{Dullemond2012}}

\appendix 

%%%
\section{The dependance of the disk chemical structures on ionization rates and initial abundances}\label{Asec:A}
\begin{figure*}[hbtp]
\begin{center}
\vspace{0.5cm}
%\vspace{-1cm}
\includegraphics[scale=0.57]{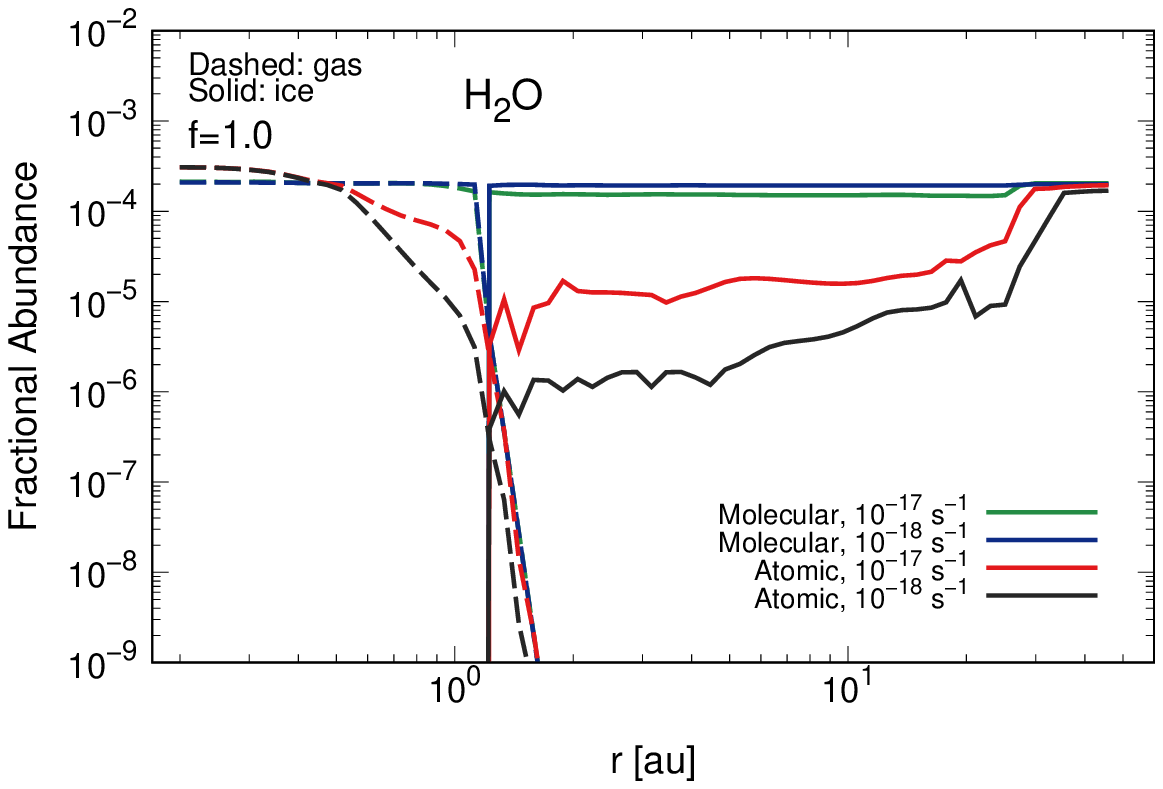}
\includegraphics[scale=0.57]{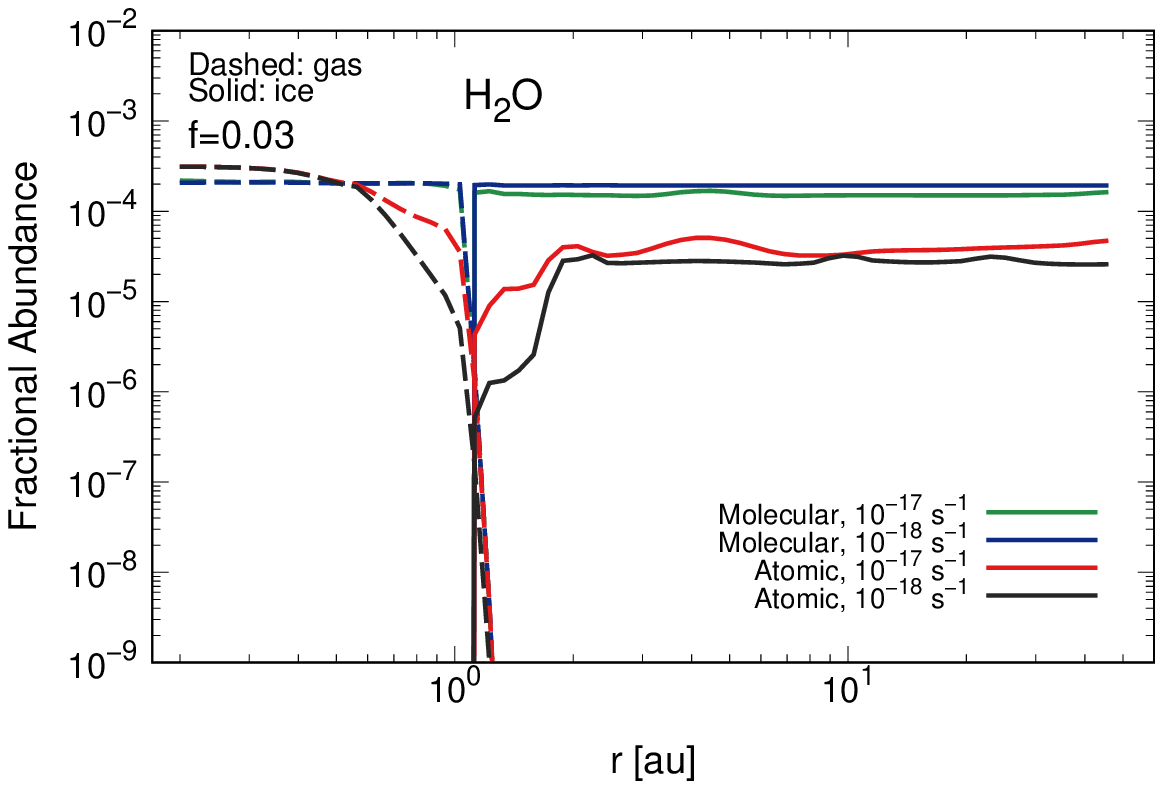}
\includegraphics[scale=0.57]{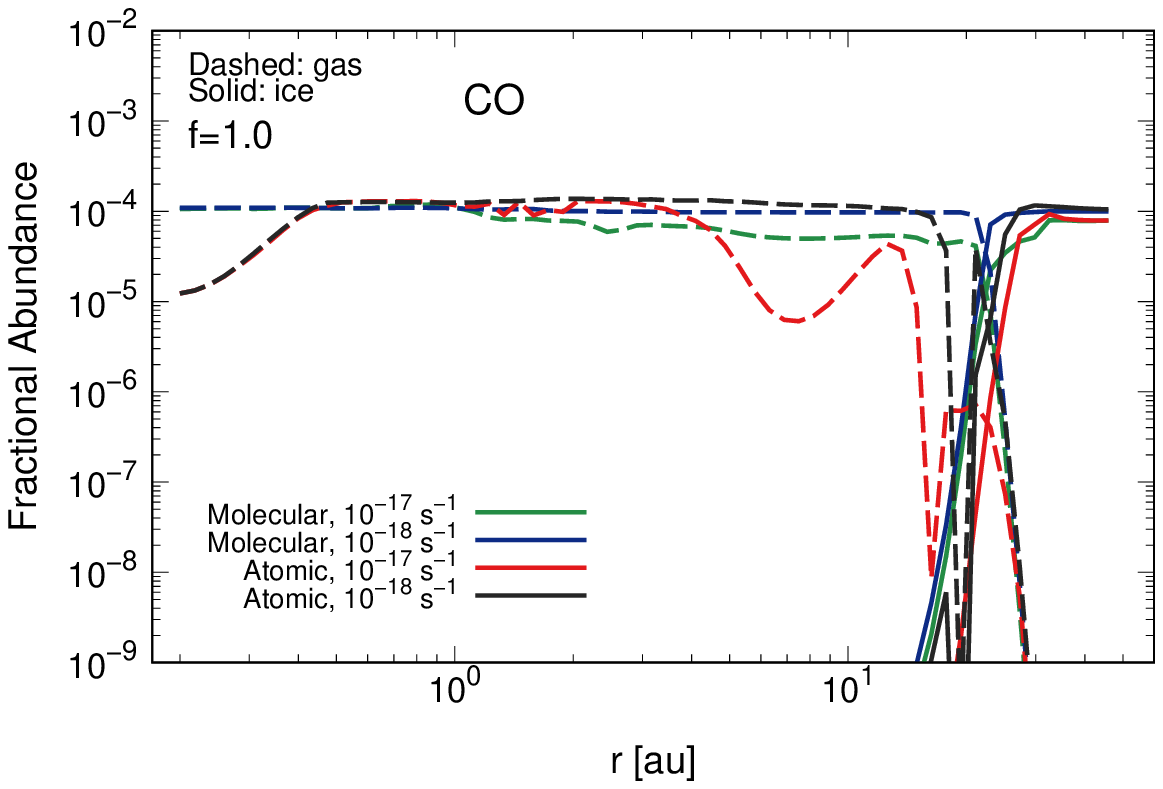}
\includegraphics[scale=0.57]{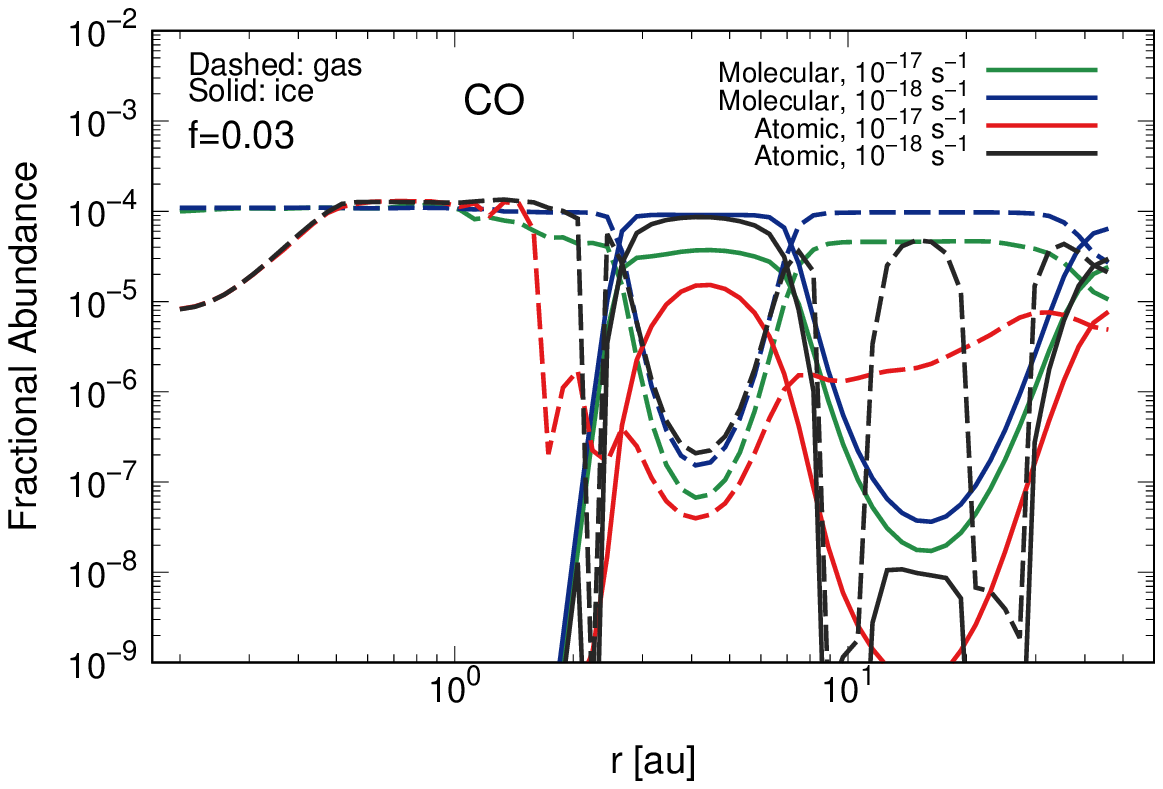}
\includegraphics[scale=0.57]{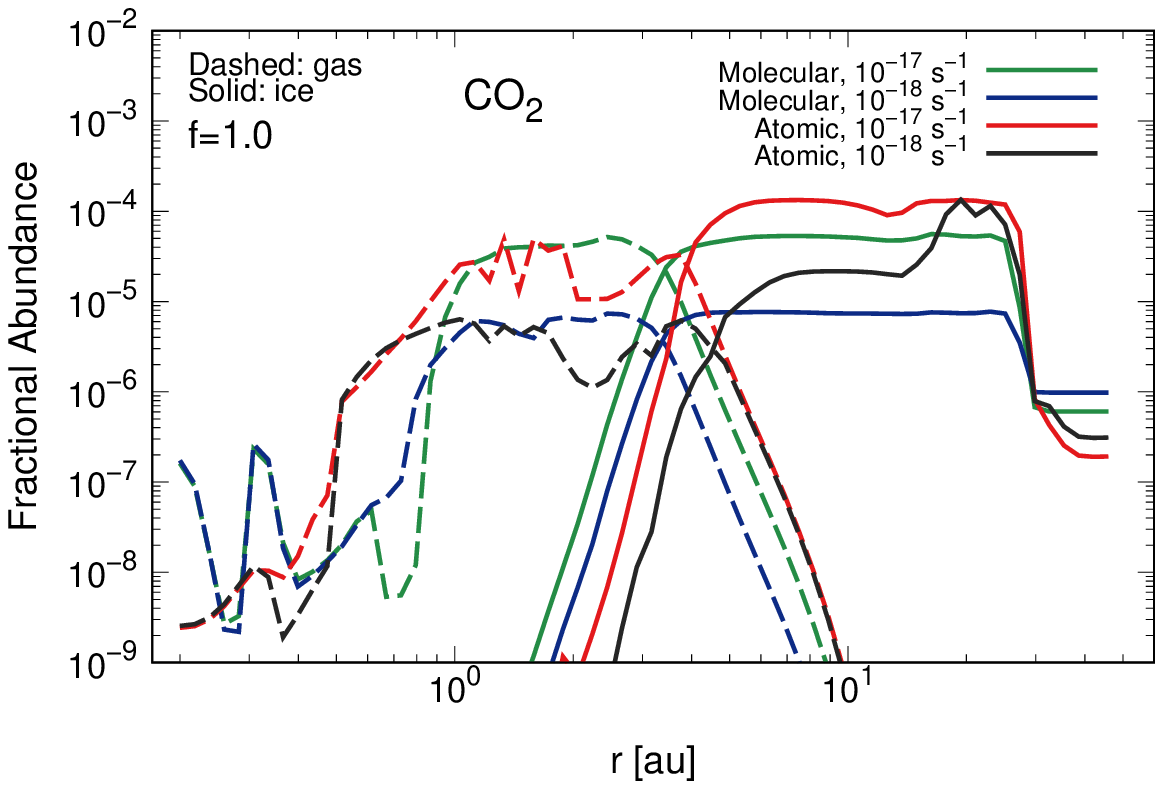}
\includegraphics[scale=0.57]{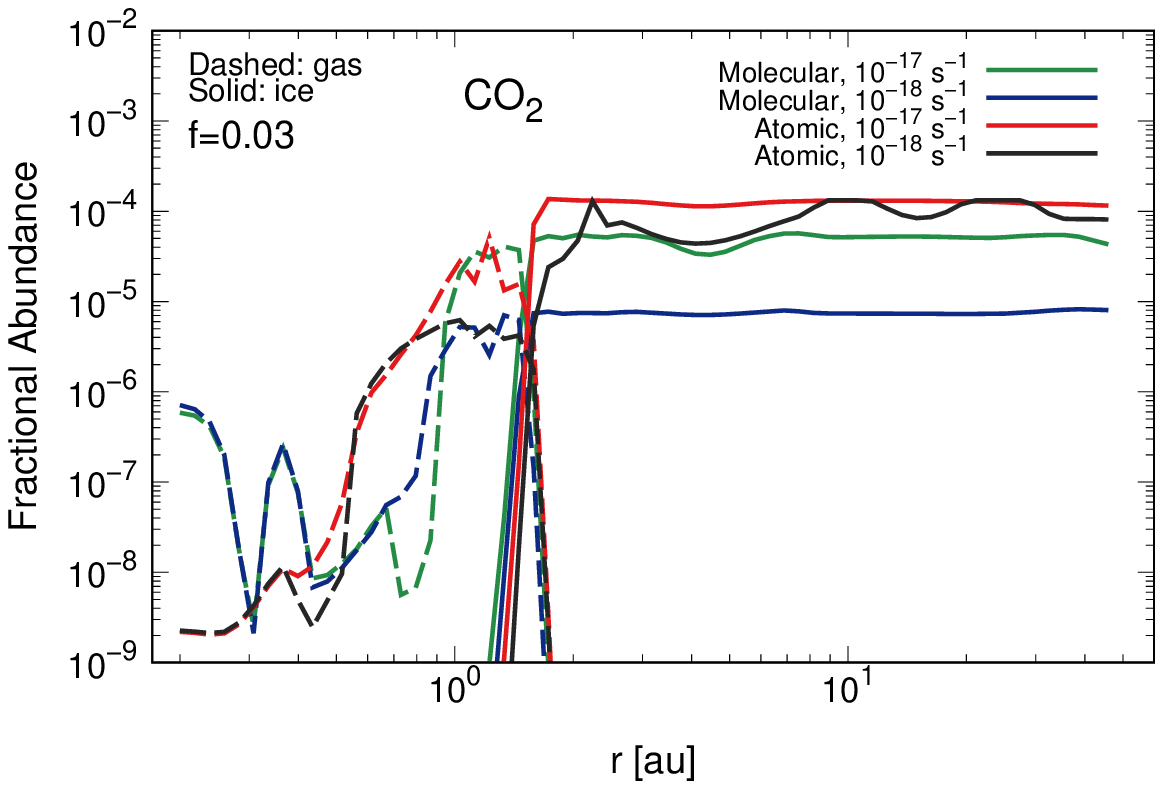}
\includegraphics[scale=0.57]{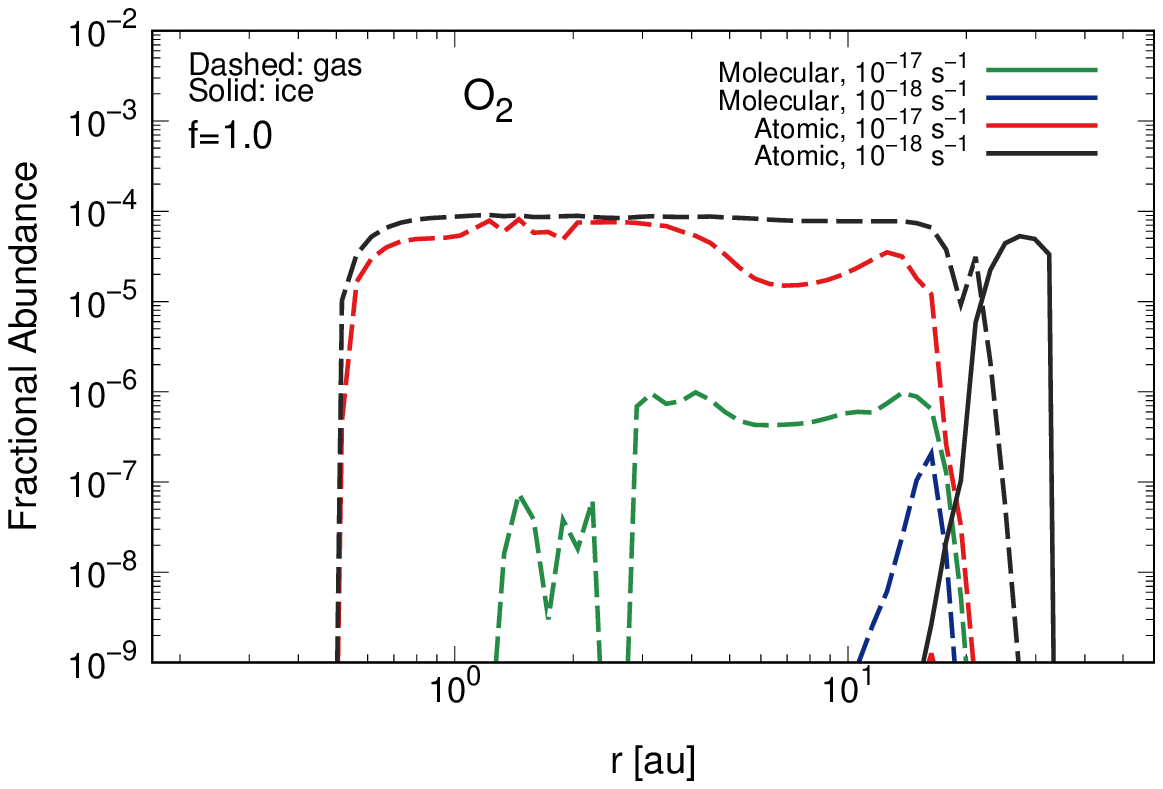}
\includegraphics[scale=0.57]{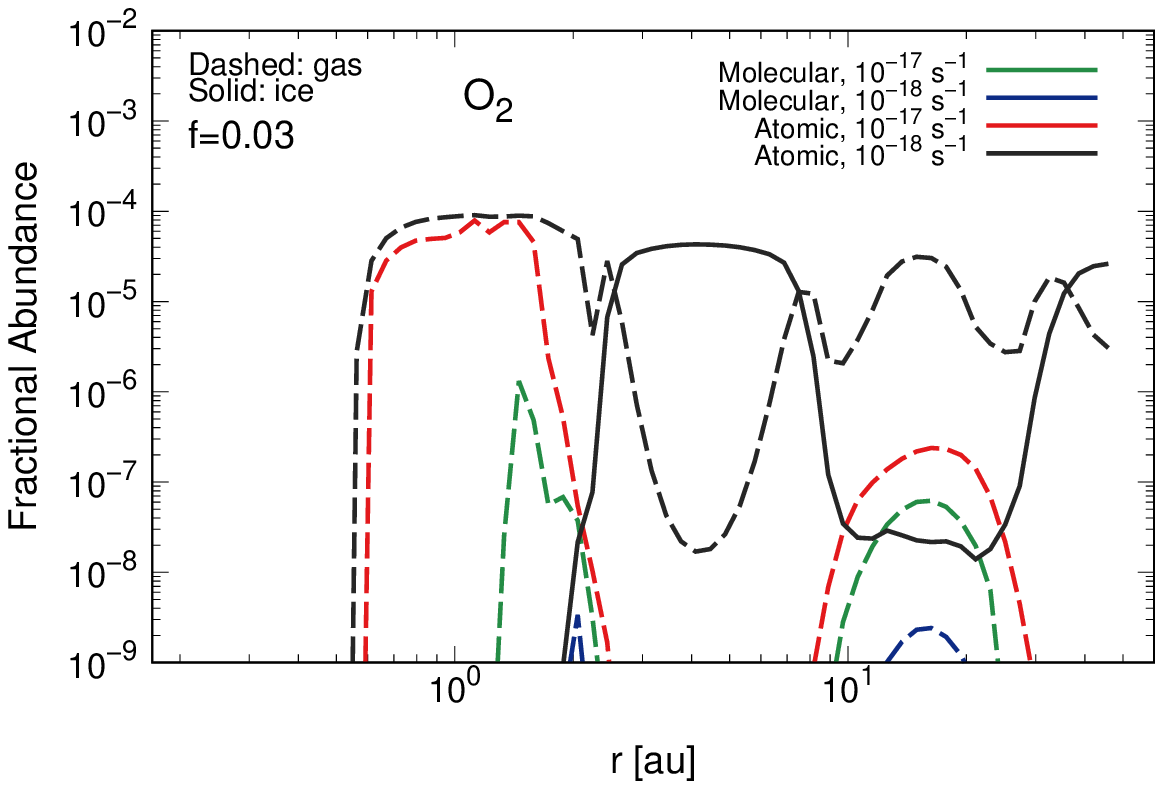}
%\plotone{cost.pdf}
%
\end{center}
\vspace{-0.2cm}
\caption{
The radial profiles of fractional abundances with respect to total hydrogen nuclei densities $n_{\mathrm{X}}$/$n_{\mathrm{H}}$ at t=$10^{6}$ years for H$_{2}$O (top panels), CO (second row panels), CO$_{2}$ (third row panels), and O$_{2}$ (bottom panels).
%and dominant nitrogen-bearing molecules and organic molecules (N$_{2}$, NH$_{3}$, HCN, H$_{2}$CO, and CH$_{3}$OH, right panels).
Left panels show the results for the disk midplane with the monotonically decreasing density and temperature profile ($f$=1.0), and right panels show the results for the shadowed disk midplane ($f$=0.03).
In each panel, the dashed and solid lines show the profiles for gaseous and icy molecules, respectively.
Green and blue lines show the results for molecular initial abundances (the ``inheritance'' scenario), and red and black lines show the results for atomic initial abundances (the ``reset'' scenario).
Green and red lines show the results for the higher cosmic-ray ionization rate $\xi_{\mathrm{CR}}(r)=$$1.0\times10^{-17}$ [s$^{-1}$], 
whereas blue and black lines show the results for the lower cosmic-ray ionization rate $\xi_{\mathrm{CR}}(r)=$$1.0\times10^{-18}$ [s$^{-1}$].
\\ \\
%%.
}\label{Figure14rev3_appendix}
\end{figure*}

\begin{figure*}[hbtp]
\begin{center}
\vspace{1cm}
\includegraphics[scale=0.57]{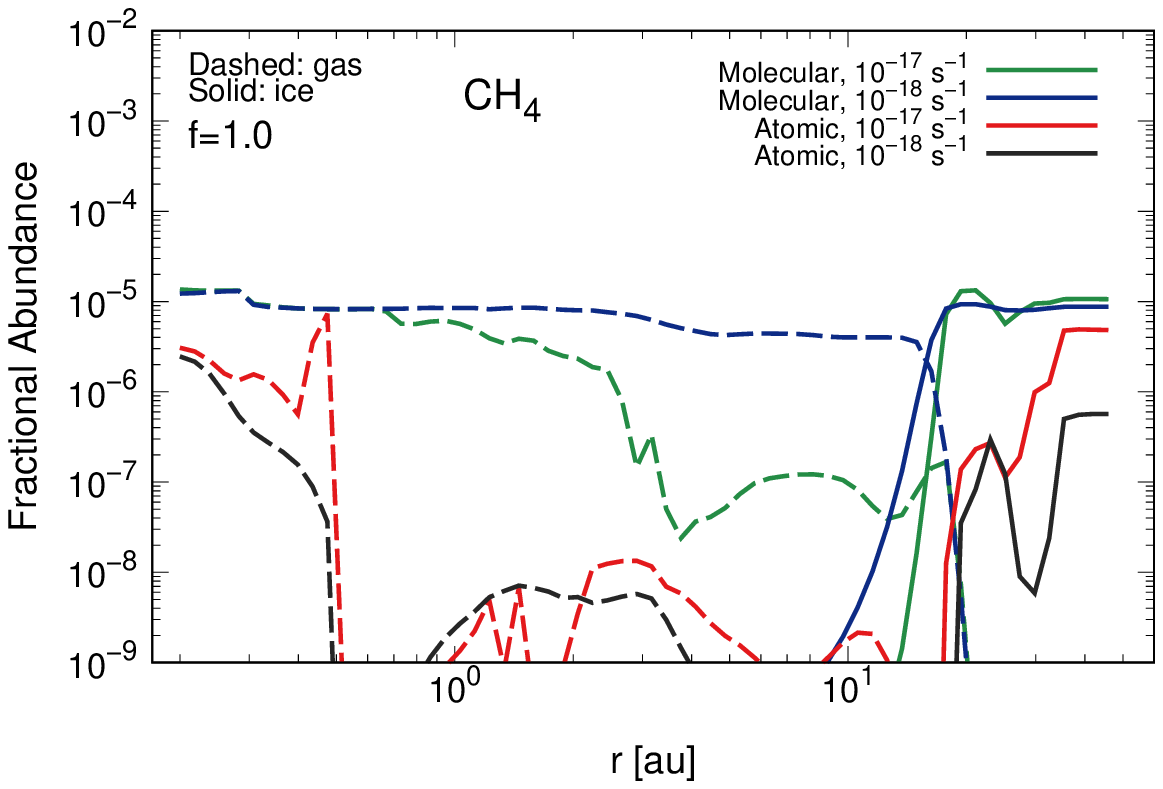}
\includegraphics[scale=0.57]{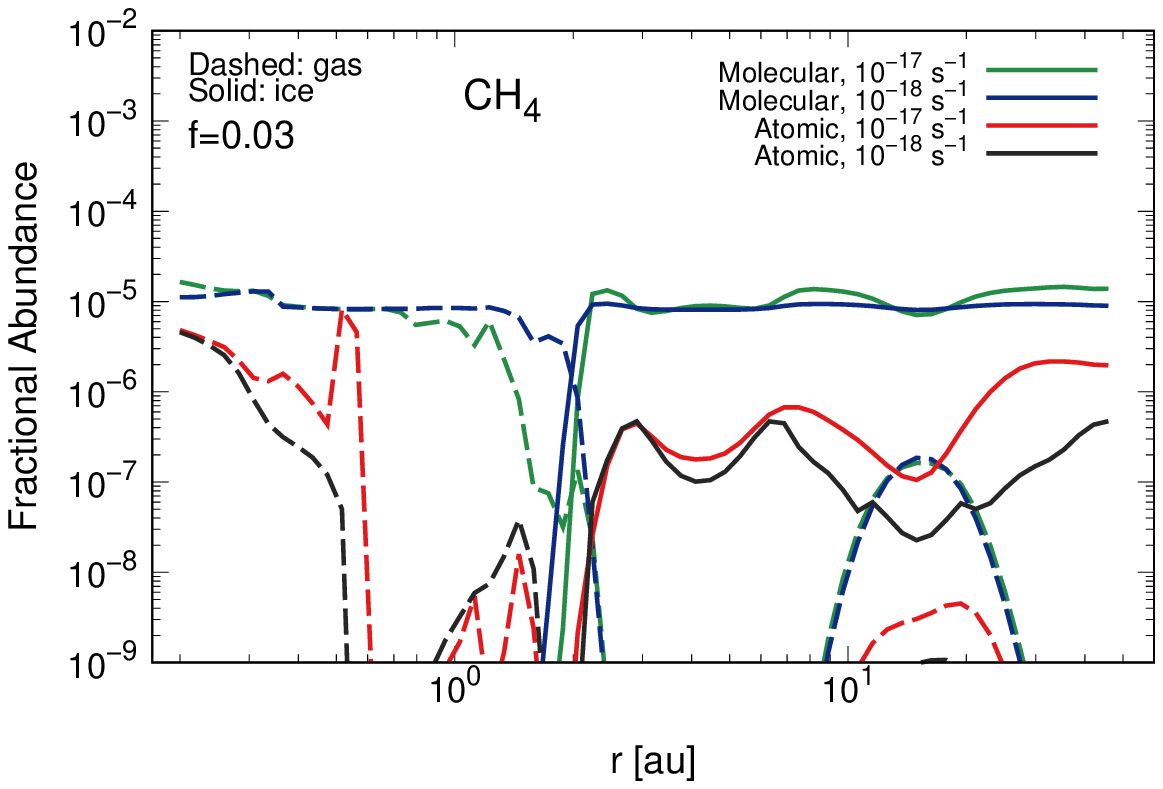}
\includegraphics[scale=0.57]{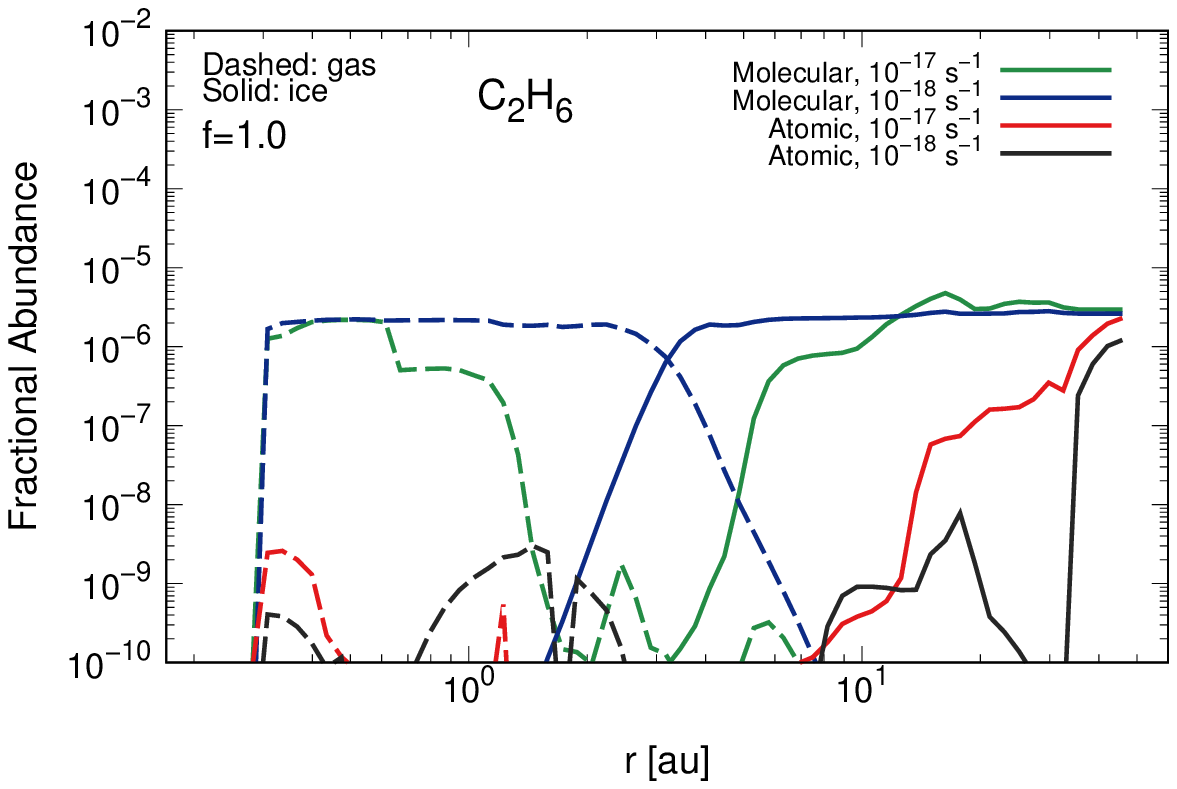}
\includegraphics[scale=0.57]{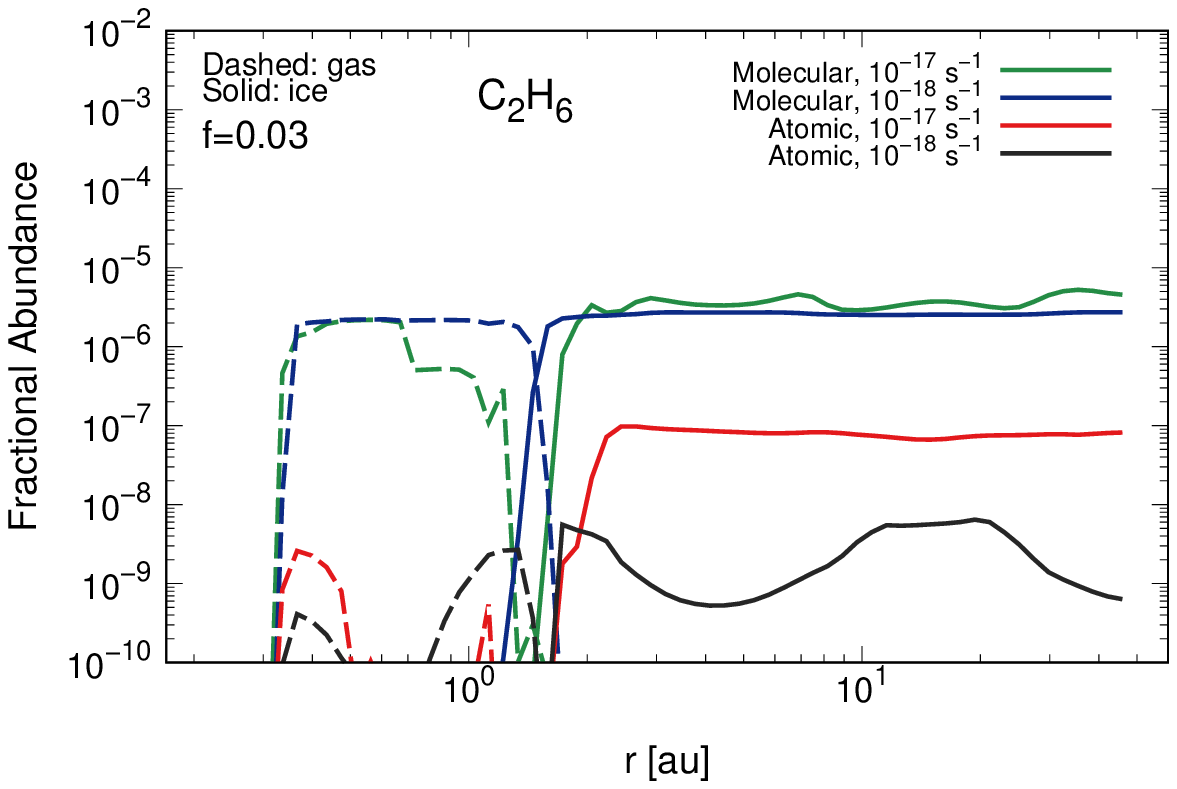}
\includegraphics[scale=0.57]{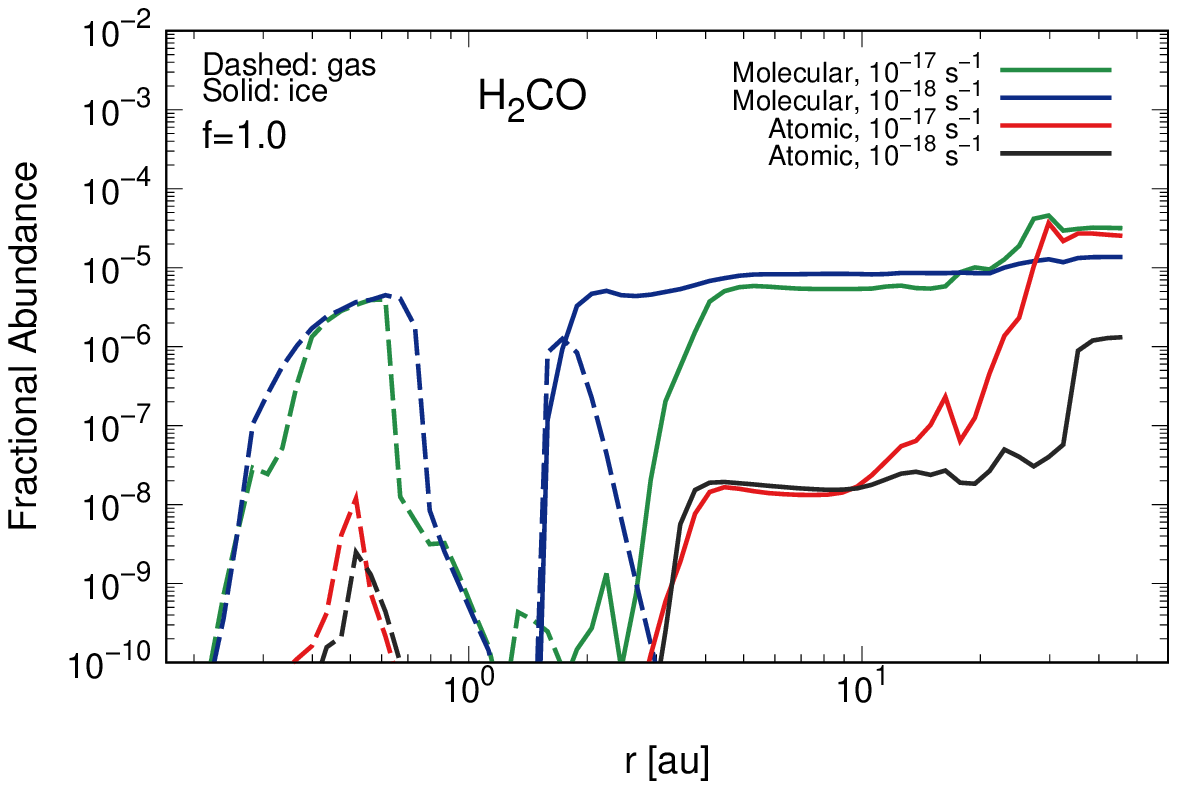}
\includegraphics[scale=0.57]{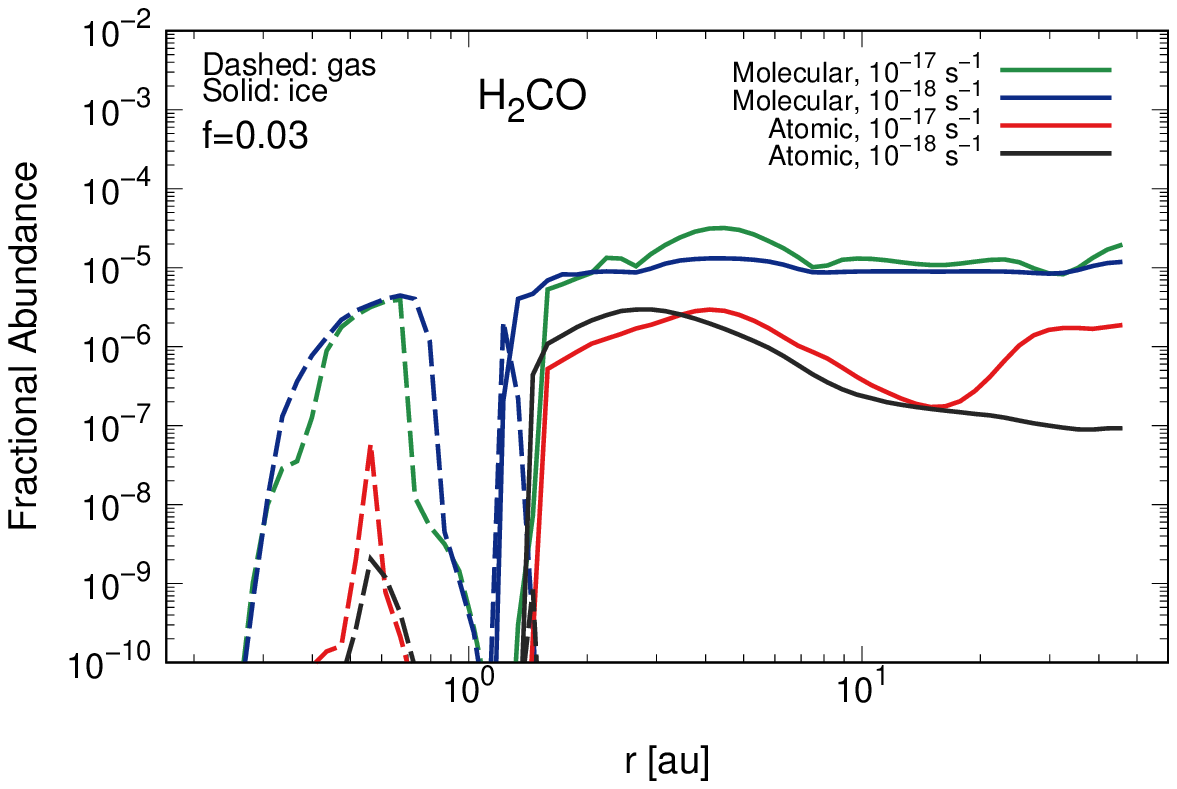}
\includegraphics[scale=0.57]{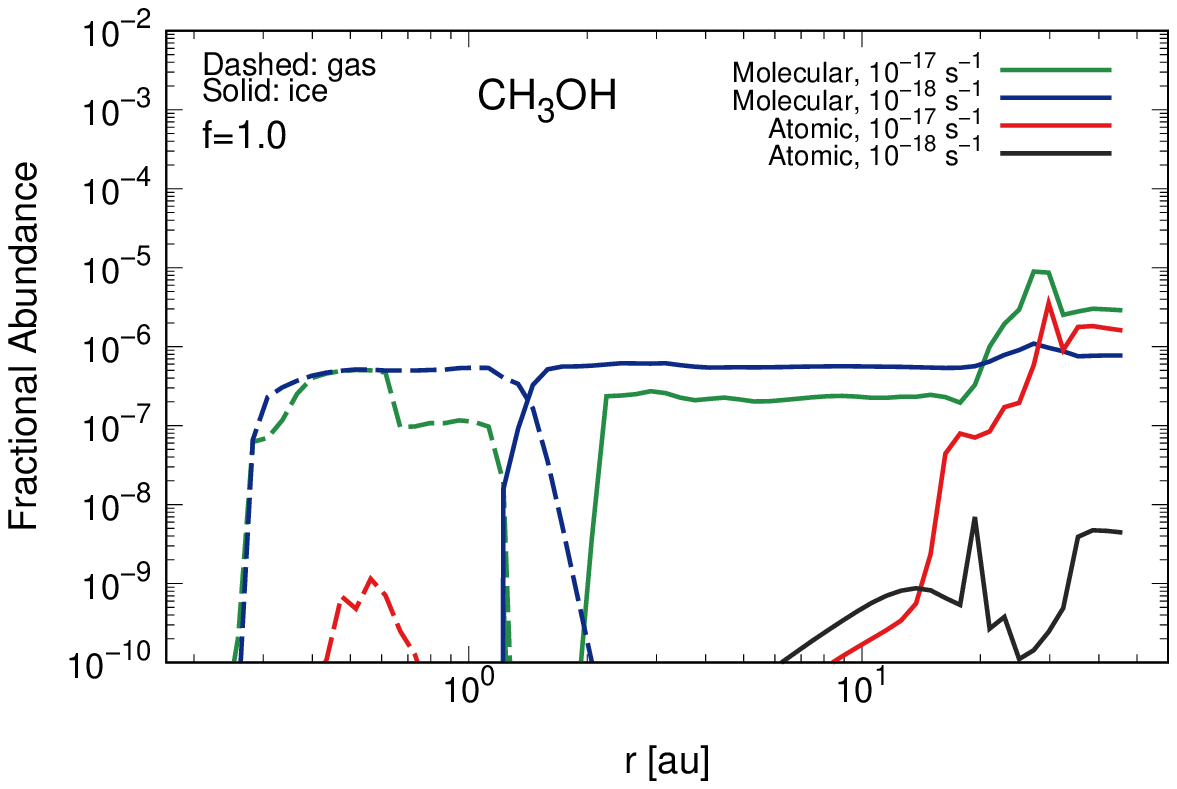}
\includegraphics[scale=0.57]{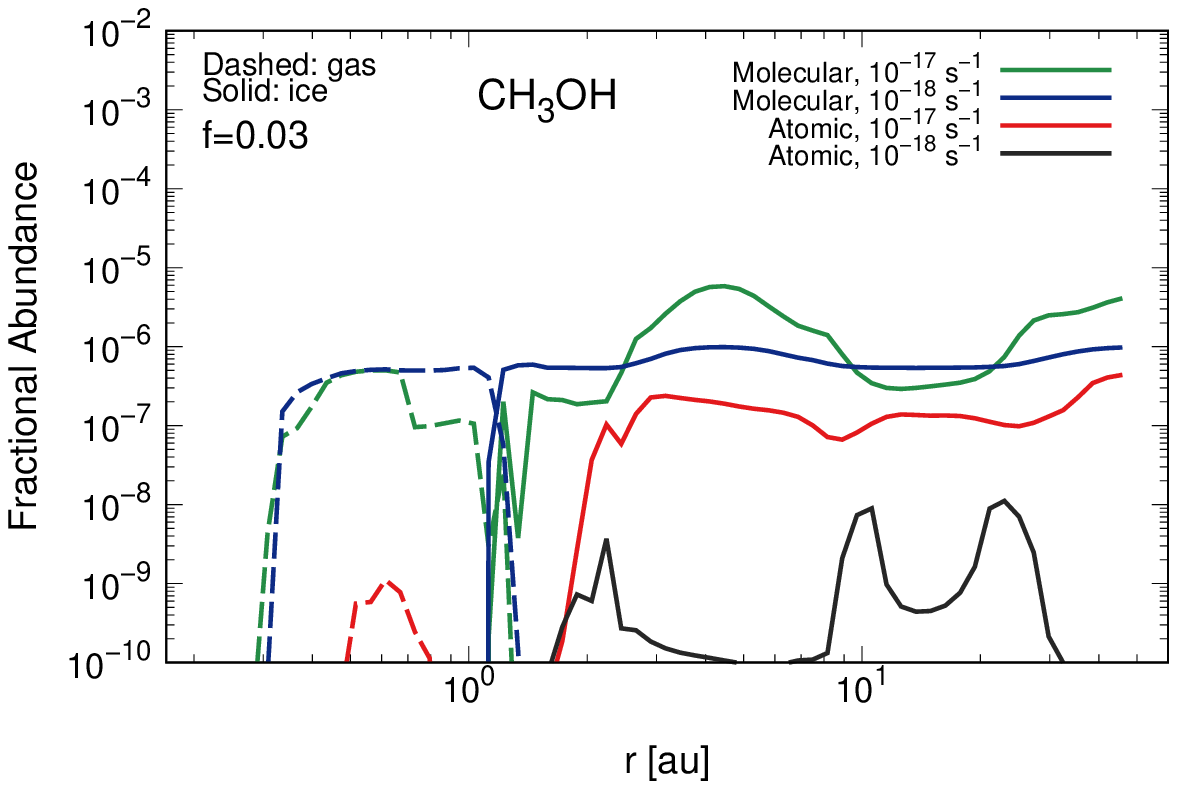}
\end{center}
\vspace{-0.2cm}
\caption{
Same as Figure \ref{Figure14rev3_appendix}, but for CH$_{4}$ (top panels), C$_{2}$H$_{6}$ (second row panels), H$_{2}$CO (third row panels), and CH$_{3}$OH (bottom panels).
\\ \\
%.
}\label{Figure15rev3_appendix}
\end{figure*}

\begin{figure*}[hbtp]
\begin{center}
\vspace{1cm}
\includegraphics[scale=0.57]{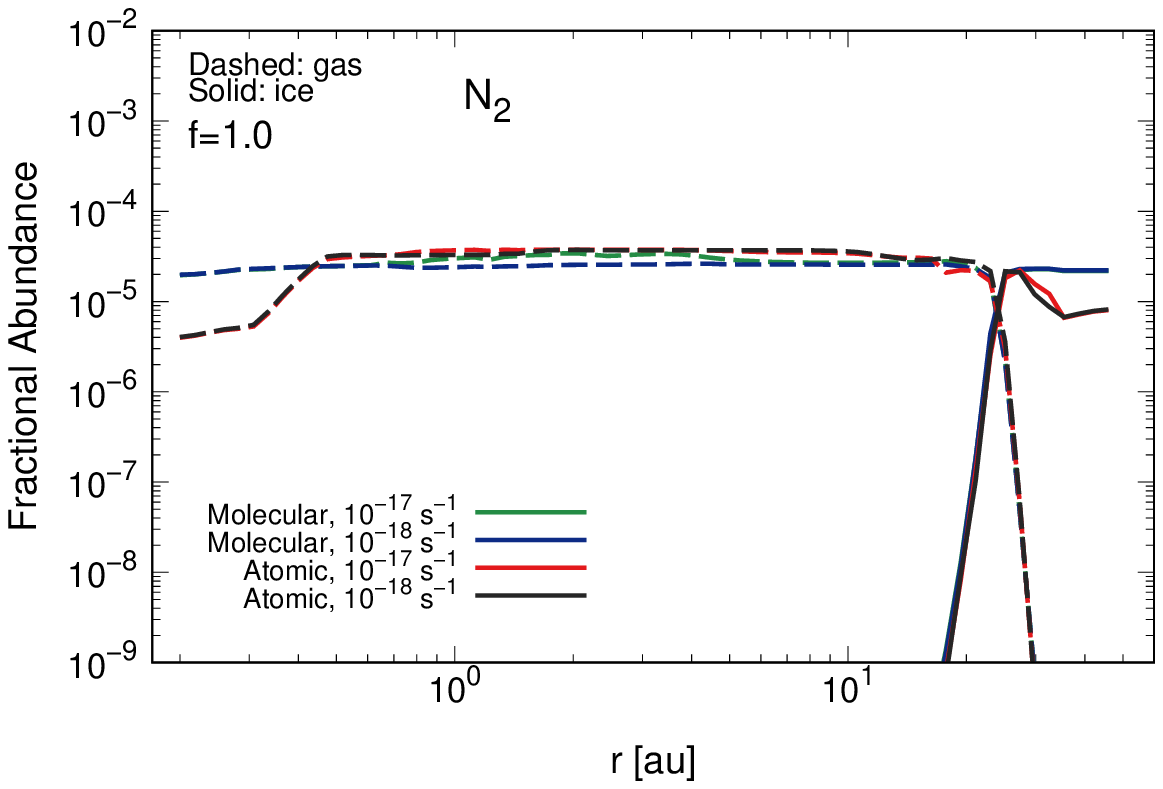}
\includegraphics[scale=0.57]{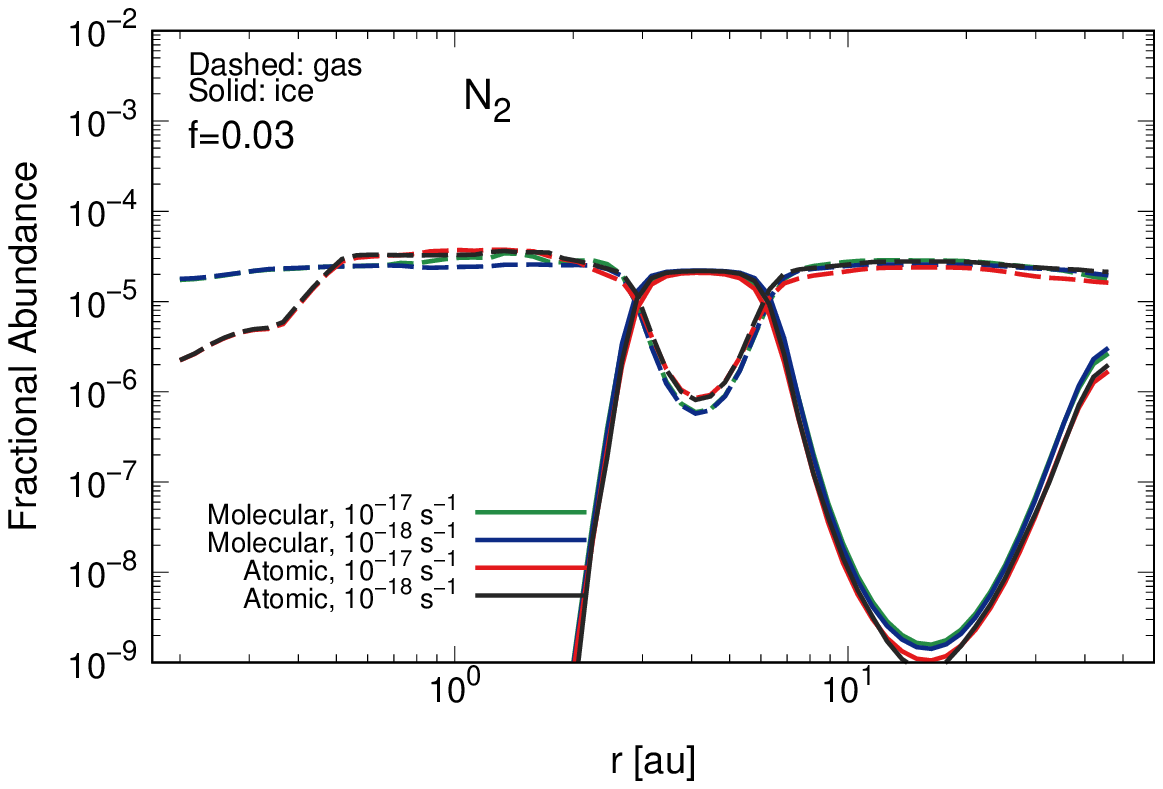}
\includegraphics[scale=0.57]{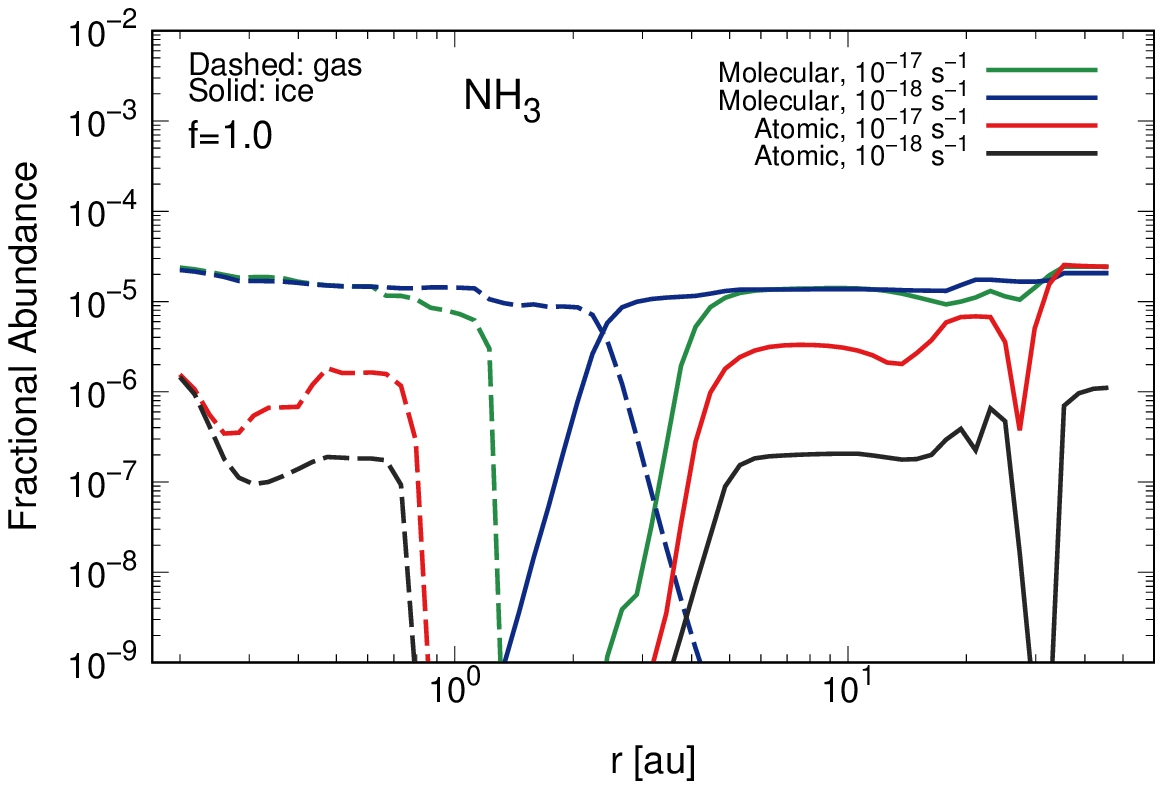}
\includegraphics[scale=0.57]{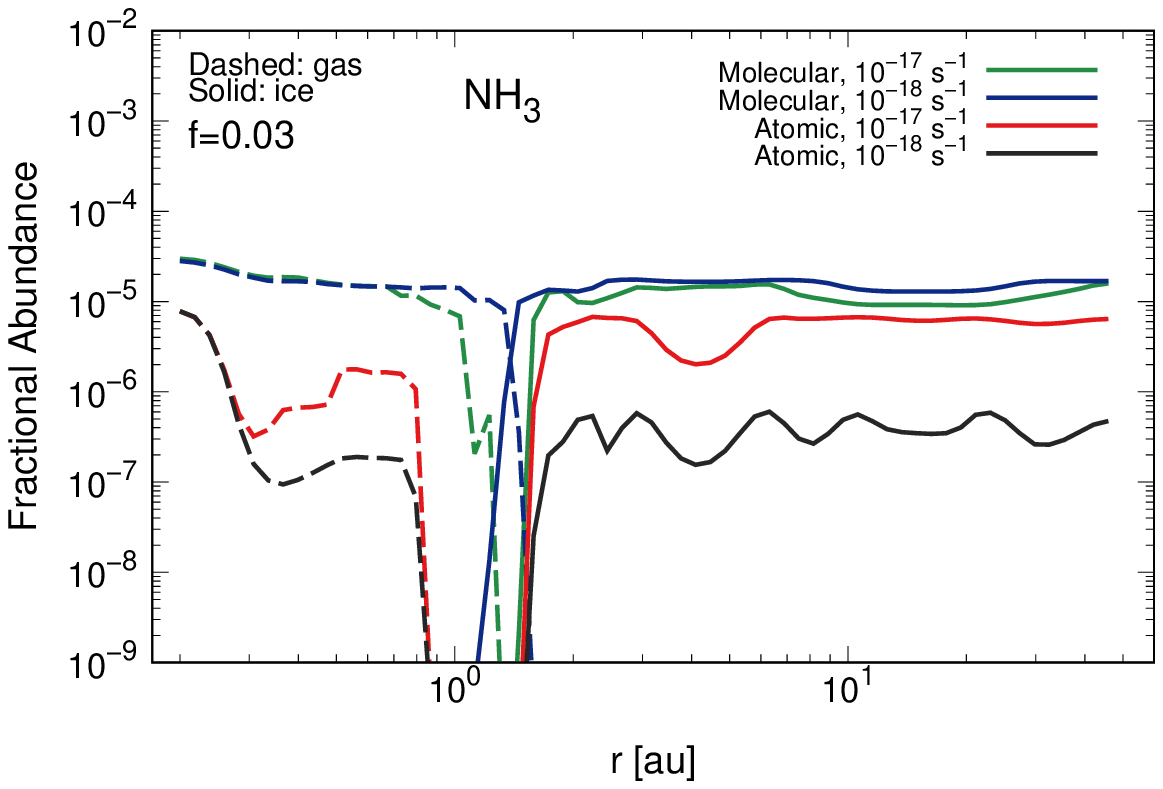}
\includegraphics[scale=0.57]{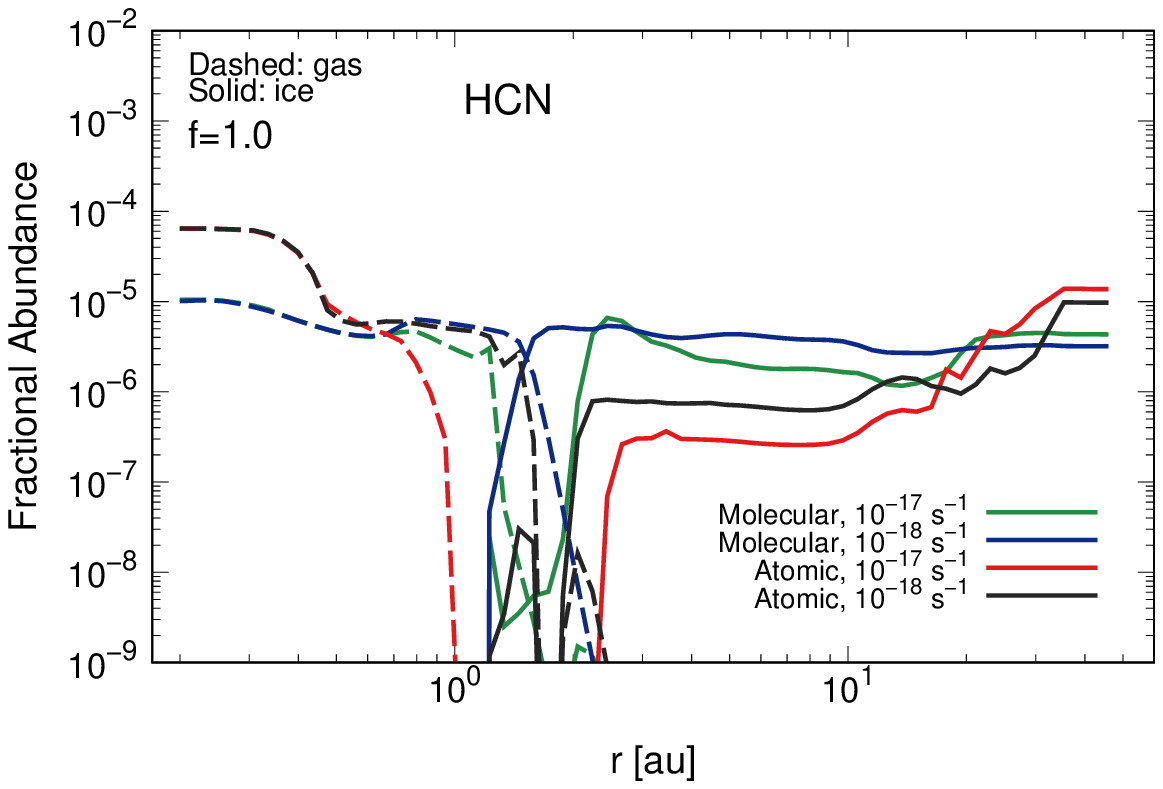}
\includegraphics[scale=0.57]{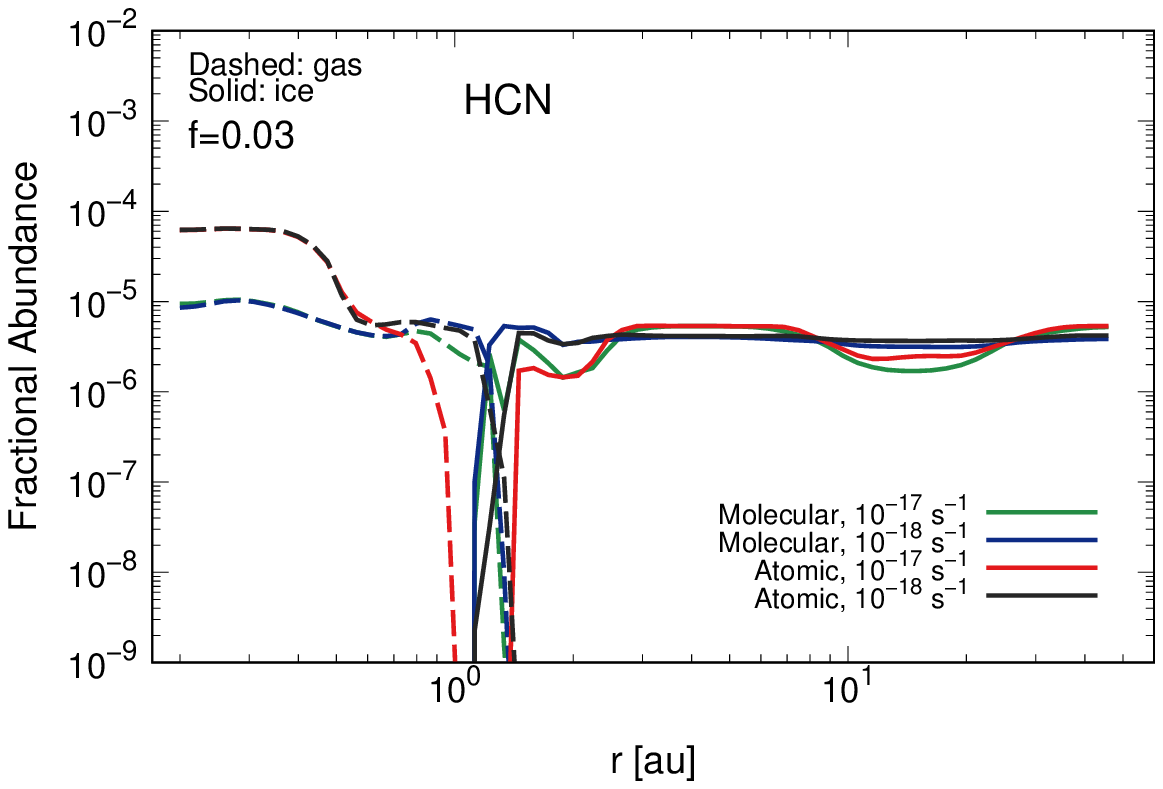}
\includegraphics[scale=0.57]{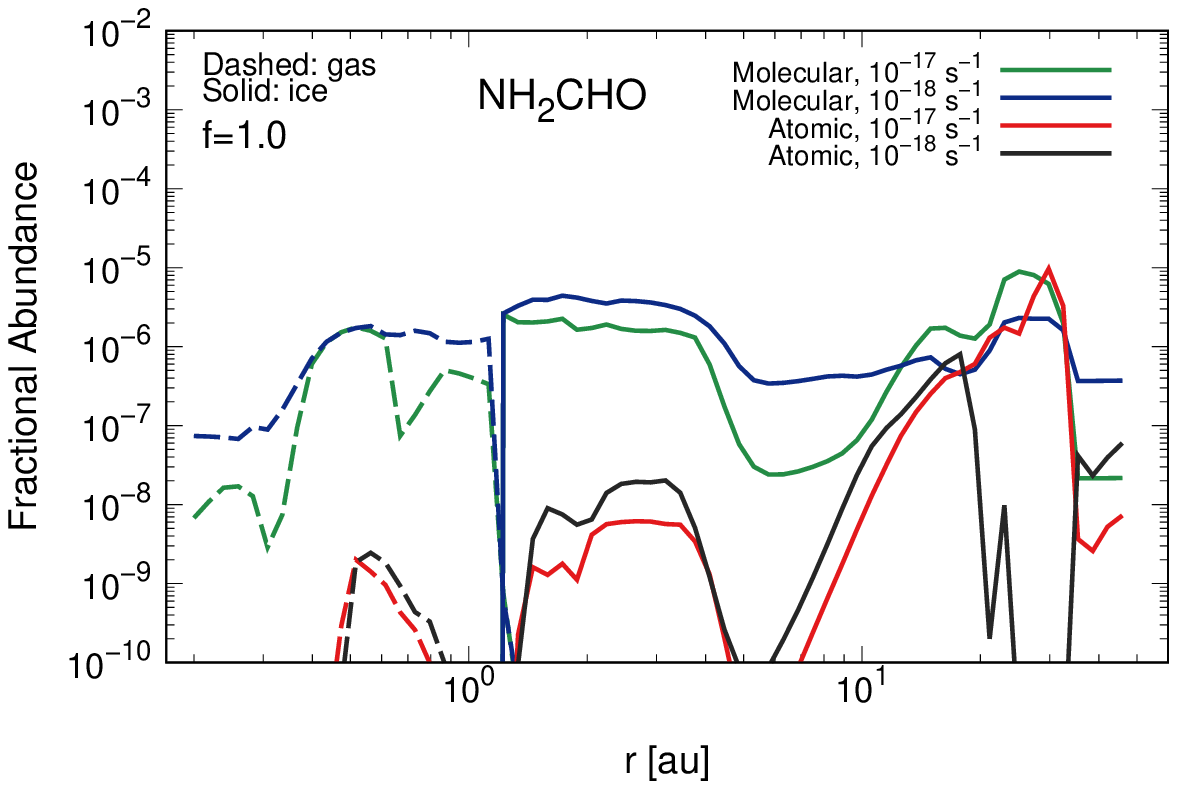}
\includegraphics[scale=0.57]{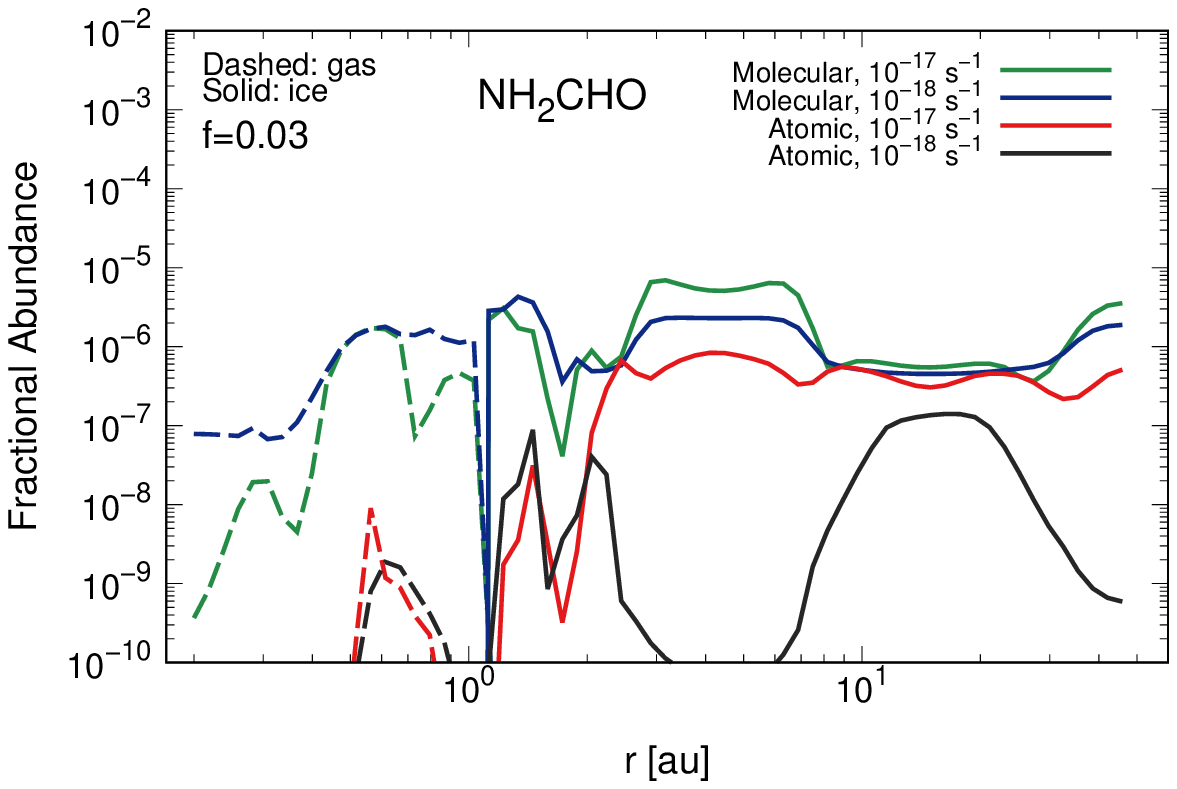}
\end{center}
\vspace{-0.2cm}
\caption{
Same as Figure \ref{Figure14rev3_appendix}, but for N$_{2}$ (top panels), NH$_{3}$ (second row panels), HCN (third row panels), and NH$_{2}$CHO (bottom panels).
%%.
\\ \\
}\label{Figure16rev3_appendix}
\end{figure*}

%%%
%%%
In this Appendix \ref{Asec:A}, we explain the dependance of the disk chemical evolution on disk ionization rates and initial abundances.
Figures \ref{Figure14rev3_appendix}-\ref{Figure16rev3_appendix} shows the radial profiles of fractional abundances at t=$10^{6}$ years
for dominant oxygen-, carbon-, nitrogen-bearing molecules (H$_{2}$O, CO, CO$_{2}$, O$_{2}$, CH$_{4}$, C$_{2}$H$_{6}$, H$_{2}$CO, CH$_{3}$OH, N$_{2}$, NH$_{3}$, HCN, and NH$_{2}$CHO)
in the shadowed and non-shadowed disk midplane ($f=1.0$ and $f=0.03$, respecitively).
%%%%.
In these Figures, we assume either molecular or atomic initial abundances and either low or high ionization rates ($\xi_{\mathrm{CR}}(r)=$$10^{-18}$, $10^{-17}$ [s$^{-1}$]).
We note that Figures \ref{Figure3_rev_radial}-\ref{Figure5_rev_radial} in Section \ref{sec:3-2rev} show the radial abundance profiles for the same molecules with molecular initial abundances and $\xi_{\mathrm{CR}}(r)=10^{-17}$ [s$^{-1}$].
%\\ \\
%
%%%
%%%
%%%
%%%
\subsection{H$_{2}$O, CO, CO$_{2}$, and O$_{2}$}\label{Asec:A-1}
%%%
%%%
Figure \ref{Figure14rev3_appendix} shows the radial profiles of fractional abundances for dominant oxygen-bearing molecules; H$_{2}$O, CO, CO$_{2}$, and O$_{2}$.
For molecular initial abundances, the H$_{2}$O ice abundances outside the water snowline are $\sim(1-2)\times10^{-4}$ (see also Section \ref{sec:3-2-1rev}), which are similar to the assumed value of the initial H$_{2}$O ice abundance ($=1.984\times10^{-4}$).
For atomic initial abundances, they decrease outside the water snowline, and at $r\sim2-10$ au they are $\sim10^{-6}-10^{-5}$ for $f=1.0$ and $\sim(2-5)\times10^{-5}$ for $f\leq0.03$.
In addition, just outside the H$_{2}$O snowline the H$_{2}$O ice abundances are about an order of magnitude larger in the high ionisation case ($\sim10^{-5}$) than those in 
the low ionisation case ($\sim10^{-6}$).
\citet{Eistrup2016} described that in the case of atomic initial abundances, the H$_{2}$O ice is not efficiently produced in the outer disk and the ion-molecule reactions in the gas-phase \citep{Hollenbach2009} are contributing to the formation of water.
Previous studies (e.g., \citealt{Schmalzl2014, Notsu2021, vanDishoeck2021}) discussed that it takes more than around 1 Myr of the pre-stellar phase to produce water ice with an abundance of $\gtrsim10^{-4}$.
\\ \\
For molecular initial abundances, the total CO (gas+ice) abundances outside the water snowline are $\sim10^{-4}$ for the low ionisation rate and $\sim(3-6)\times10^{-5}$ for the high ionisation rate.
This decrease of CO coincides with the overall enhancement of CO$_{2}$ ice
%%% 
outside the CO$_{2}$ snowline.
Under the high ionisation rate, CO destruction pathways of Reactions \ref{Rec6} (icy phase) and \ref{Rec7} (gas phase) are efficient (see Section \ref{sec:3-2-1rev}).
%%%.
In the shadowed disk ($f=0.03$), CO freezes-out onto dust grains at around the current orbit of Jupiter ($r\sim3-8$ au), 
and for atomic initial abundances and the high ionisation rate, a larger decrease of total CO abundance ($\lesssim10^{-5}$) beyond the CO$_{2}$ snowline is shown.
%%, 
%%%
%%%
%%%.
\\ \\
CO$_{2}$ gas abundances between the H$_{2}$O and CO$_{2}$ snowlines and CO$_{2}$ ice abundances outside CO$_{2}$ snowline are larger for the high ionisation rate than those for the low ionisation rate.
In addition, they are also larger for atomic initial abundances than those for molecular initial abundances.
For atomic initial abundances and the high ionisation rate, CO$_{2}$ ice abundances are larger ($\gtrsim10^{-4}$) outside the CO$_{2}$ snowline, with the decreases of CH$_{4}$ gas and total (gas+ice) CO abundances.
We note that in the non-shadowed disk ($f=1.0$), CO$_{2}$ ice abundances at $r>30$ au are significantly smaller ($<10^{-6}$).
This is because this region has the coldest conditions ($T(r)\lesssim20$ K), and the formation of H$_{2}$O ice (Reaction \ref{Rec5}) is faster than that of CO$_{2}$ ice (Reaction \ref{Rec6}) \citep{Eistrup2016}. 
%%%
%%%
%%
%
%%%%
\\ \\
The gas-phase O$_{2}$ abundances are approximately two orders of magnitude larger for atomic initial abundances than those for molecular initial abundances.
%.
The O$_{2}$ gas abundances between the water and O$_{2}$ snowlines are $\sim10^{-5}-10^{-4}$ for atomic initial abundances and $<10^{-6}$ for molecular initial abundances.
This is because in the gas-phase O$_{2}$ is formed from atomic oxygen via Reaction \ref{Rec8} \citep{Walsh2015, Taquet2016, Eistrup2019}, and the initial atomic oxygen abundance is $3.2\times10^{-4}$ for atomic initial abundances whereas that is zero for molecular initial abundances.
In the shadowed disk ($f\leq0.03$), O$_{2}$ freezes-out onto dust grains at $r\sim2-10$ au, and returns to the gas phase at $r>10$ au.
O$_{2}$ ice is mainly formed from atomic oxygen via Reaction \ref{Rec9} \citep{Taquet2016, Eistrup2019}.
For the high ionisation rate, O$_{2}$ ice photodissociation becomes efficient and the released oxygen is contained in other major oxygen-bearing molecules \citep{Eistrup2016, Taquet2016}).
Thus, O$_{2}$ ice abundances are low ($<<10^{-9}$) for molecular initial abundances and/or the high ionisation rates, and they are $\sim10^{-5}-10^{-4}$ only for atomic initial abundances and the low ionisation rates, which is consistent with the results of previous studies (e.g., \citealt{Eistrup2016, Eistrup2019, Taquet2016}).
Such larger abundances are consistent with the measured cometary abundances (\citealt{Bieler2015, Rubin2015}, see also Section \ref{sec:4-2} of this paper).
\\ \\
According to right-hand panels in Figure \ref{Figure14rev3_appendix}, both O$_{2}$ and H$_{2}$O ice abundances are similar ($\sim(3-5)\times10^{-5}$) in the shadowed region ($r\sim2-10$ au) in the disk with atomic initial abundances and low ionisation rates. 
As we explain in Section \ref{sec:4-2}, the ratio of O$_{2}$/H$_{2}$O in comet 67P is a few percent \citep{Bieler2015, Rubin2015}. 
Thus in reality, the models between our assumed parameters (such as partially inheritance initial abundances and/or the ionization rates of $\sim5\times10^{-18}$) might well reproduced the observed O$_{2}$ and H$_{2}$O ice abundance ratios.
\subsection{Other dominant carbon-bearing molecules}\label{Asec:A-2}
%%%
%%%
%%%%
For molecular initial abundances, the CH$_{4}$ and C$_{2}$H$_{6}$ gas abundances within their snowline become smaller as the ionisation rate becomes larger.
This is because CH$_{4}$ and C$_{2}$H$_{6}$ gas are not efficiently formed in these regions, and 
they are destroyed by cosmic-ray-induced photodissociation and ion-molecule reactions (such as CH$_{4}$$+$C$^{+}$).
Thus, carbon is
%%and the carbon is thus 
converted from CH$_{4}$ gas to CO$_{2}$, H$_{2}$CO, and hydrocarbons such as e.g., C$_{2}$H$_{2}$, C$_{2}$H$_{4}$, C$_{3}$H$_{2}$, and C$_{3}$H$_{4}$ (\citealt{Aikawa1999, Eistrup2016, Eistrup2018, Yu2016}, see also Sections \ref{sec:3-2-2rev} and \ref{sec:3-2-4rev}).
%%In addition, 
%%%
In addition, for atomic initial abundances, the CH$_{4}$ and C$_{2}$H$_{6}$ gas abundances within the CH$_{4}$ snowline are much smaller ($<<10^{-8}$) than those for molecular initial abundances.
%%%,.
%%%
For atomic initial abundances, the CH$_{4}$ and C$_{2}$H$_{6}$ ice abundances outside their snowlines ($<10^{-6}$ for CH$_{4}$ ice and $<10^{-6}$ for C$_{2}$H$_{6}$ ice) are much smaller than those for molecular initial abundances ($\sim10^{-5}$ for CH$_{4}$ ice and $\gtrsim10^{-6}$ for C$_{2}$H$_{6}$ ice).
%%%.
%
\\ \\
For molecular initial abundances, H$_{2}$CO and CH$_{3}$OH ice abundances between their snowlines and the CO snowline are larger for the low ionisation rate
($\sim2\times10^{-7}$ for CH$_{3}$OH ice)
than those for the high ionisation rate ($\sim5\times10^{-7}$ for CH$_{3}$OH ice).
We note that cosmic-ray photodissociation of these molecules are efficient in these regions and produce many radicals, which creates more complex organic molecules (see Section \ref{sec:3-2-2rev}).
In addition, H$_{2}$CO and CH$_{3}$OH ice abundances outside the CO snowline (including the shadowed region) are smaller for the low ionisation rate
($\gtrsim10^{-6}$ for CH$_{3}$OH ice)
than those for the high ionisation rate ($\gtrsim3\times10^{-6}$ for CH$_{3}$OH ice).
This is because H$_{2}$CO and CH$_{3}$OH ice are mainly formed by the sequential hydrogenation of CO on the grains surfaces (see Reactions \ref{Rec9} and \ref{Rec10}), and 
the atomic hydrogen needed to hydrogenate CO is an end product of H$_{2}$ ionization (e.g., \citealt{Schwarz2018}).
%%%.
\\ \\
For atomic initial abundances, both H$_{2}$CO and CH$_{3}$OH ice abundances at $r\sim2-10$ au significantly decrease.
The H$_{2}$CO ice abundances at such radii are $<<10^{-7}$ in the non-shadowed disk ($f=1.0$) and $\sim(1-6)\times10^{-6}$ in the shadowed disk ($f\leq0.03$).
The CH$_{3}$OH ice abundances are $<<10^{-8}$ in the non-shadowed disk ($f=1.0$) and $\sim(1-3)\times10^{-7}$ in the shadowed disk ($f\leq0.03$).
\citet{Schwarz2018} discussed that the timescales needed to build a reservoir of CH$_{3}$OH ice are longer than those needed for CO$_{2}$, because 
CO$_{2}$ can form directly from CO ice (see Reactions \ref{Rec6}) while CH$_{3}$OH ice is formed via a series of hydrogenation reactions. 
%%
%%%
%%%\subsection{N$_{2}$, NH$_{3}$, HCN, and NH$_{2}$CHO}\label{Asec:A-3}
\subsection{Dominant nitrogen-bearing molecules}\label{Asec:A-3}
%%%
%%%
Unlike the case for CO, N$_{2}$ gas and ice abundance profiles barely change for different initial abundances and ionisation rates.
%%.
\\ \\
For molecular initial abundances, NH$_{3}$ gas and ice abundances around the NH$_{3}$ snowline decrease with increasing the ionisation rate.
This is because NH$_{3}$ is converted into N$_{2}$ both through ion-molecule reactions as well as through cosmic-ray-induced photoreactions \citep{Eistrup2016}.
In addition, for molecular initial abundances, NH$_{3}$ ice abundances at $r>5$ au are $\sim1\times10^{-5}$.
%%%%.
In contrast, for atomic initial abundances NH$_{3}$ ice abundances at $r>5$ au decrease ($\sim10^{-6}-10^{-5}$ for the high ionisation rate and $\sim10^{-7}-10^{-6}$ for the low ionisation rate).
\citet{Schwarz2014} and \citet{Eistrup2016} noted that under atomic initial abundances, atomic N quickly forms N$_{2}$ gas.
\\ \\
Between the HCN and CO snowlines, HCN ice abundances are larger for molecular initial abundances ($\sim10^{-6}-10^{-5}$) than those for atomic initial abundances ($\sim10^{-7}-10^{-6}$).
In contrast, HCN ice abundances just outside the CO snowline (including the shadowed region) are similar ($\sim(3-5)\times10^{-6}$) both for the atomic and molecular initial abundances.
This is because HCN is formed through the gas-phase reaction of HCO with N atom, with subsequent freeze-out onto dust grains, where HCO is formed by hydrogenation of CO on the dust grain surface (see Section \ref{sec:3-2-3rev} and e.g., \citealt{Aikawa1999, Eistrup2016}).
\\ \\
Between the NH$_{2}$CHO and CO snowlines, NH$_{2}$CHO ice abundances are larger for molecular initial abundances ($\sim10^{-6}-10^{-5}$ at $r\sim1-5$ au) than those for atomic initial abundances ($\sim10^{-9}-10^{-8}$ at $r\sim1-5$ au).
Just outside the CO snowline (including the shadowed region), NH$_{2}$CHO ice abundances for molecular initial abundances are around $10^{-6}-10^{-5}$.
In addition, NH$_{2}$CHO ice abundances for atomic initial abundances significantly increase with increasing the ionisation rate.
We find that NH$_{2}$CHO ice can form via the subsequent hydrogenation of OCN in the cold region (see Reaction \ref{Rec12} in Section \ref{sec:3-2-3rev} and \citealt{Garrod2008, Walsh2014, Lopez-Sepulcre2015}), and the atomic hydrogen is an end product of H$_{2}$ ionozation.
%%%.
%\\ \\ \\
%%%
%%%
%%%
%%
\section{Time evolution of molecular abundances}\label{Bsec:B}
\begin{figure*}[hbtp]
\begin{center}
\vspace{0.5cm}
%\vspace{-1cm}
%\plotone{cost.pdf}
%\plotone[scale=0.63]{{20210624_surface-density.eps}
%\plotone[scale=0.63]{{20210624_ngas-T.eps}
\includegraphics[scale=0.57]{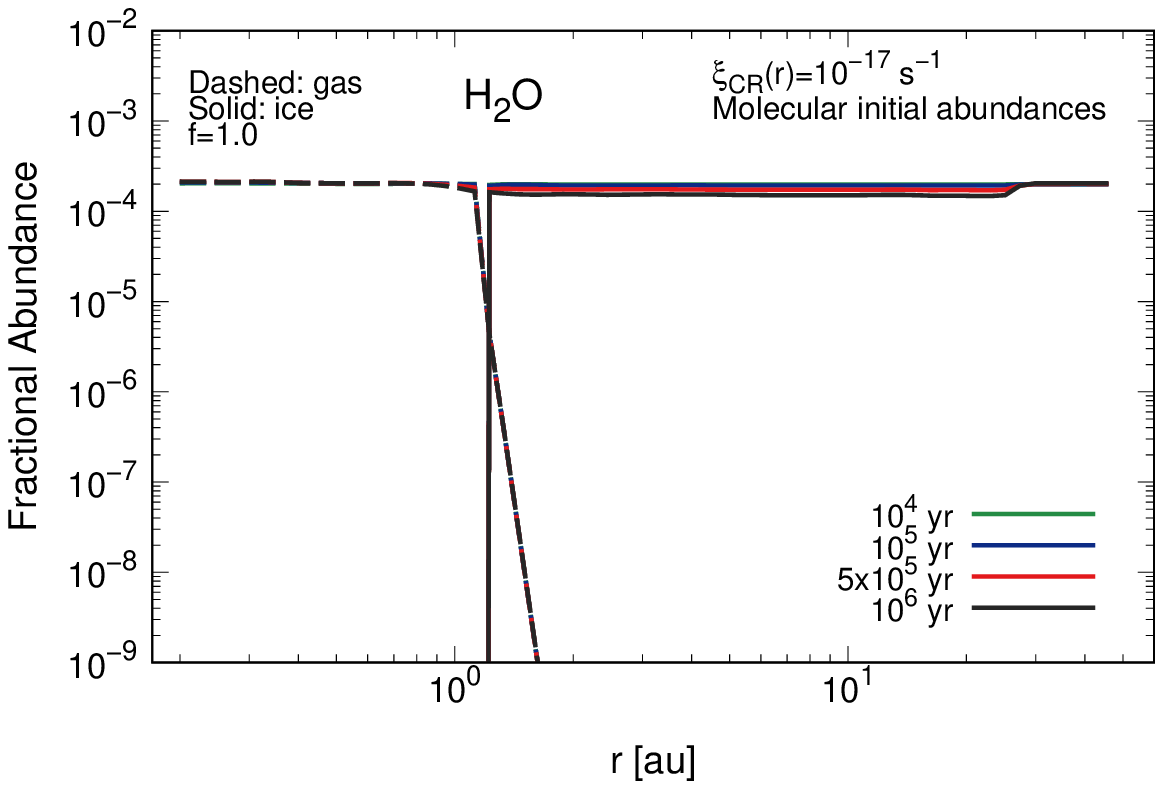}
\includegraphics[scale=0.57]{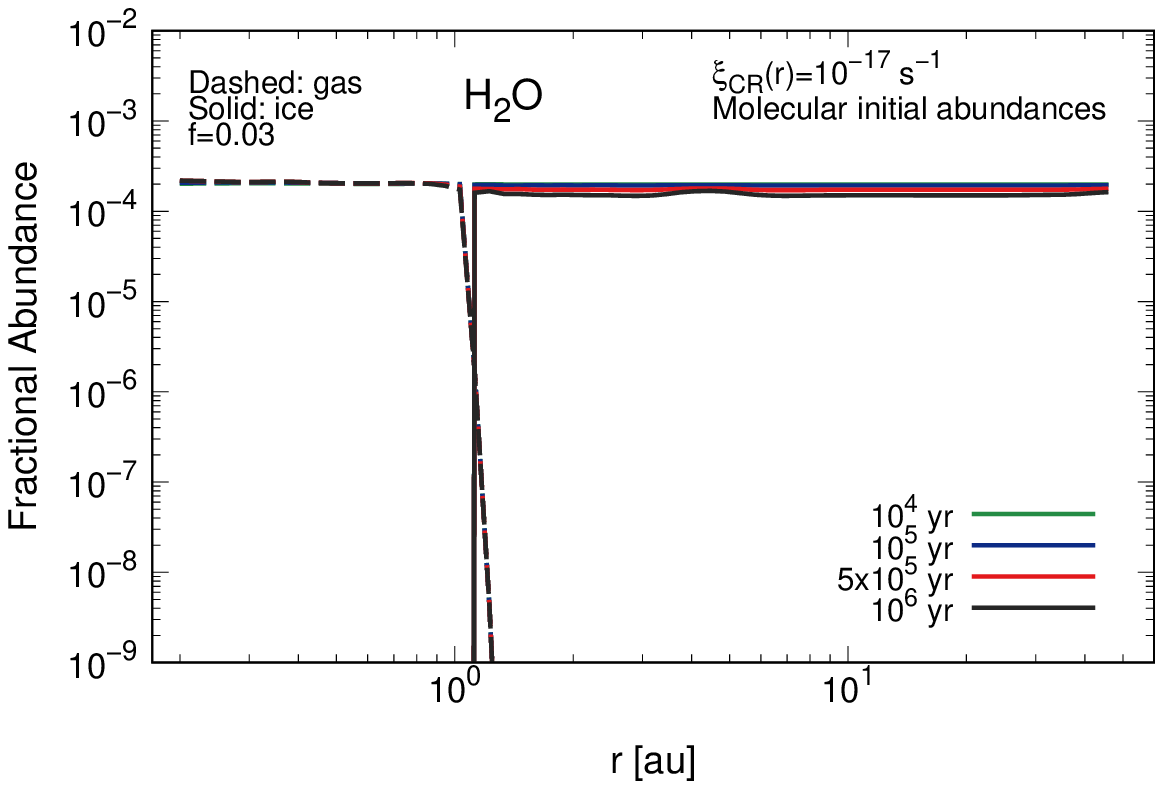}
\includegraphics[scale=0.57]{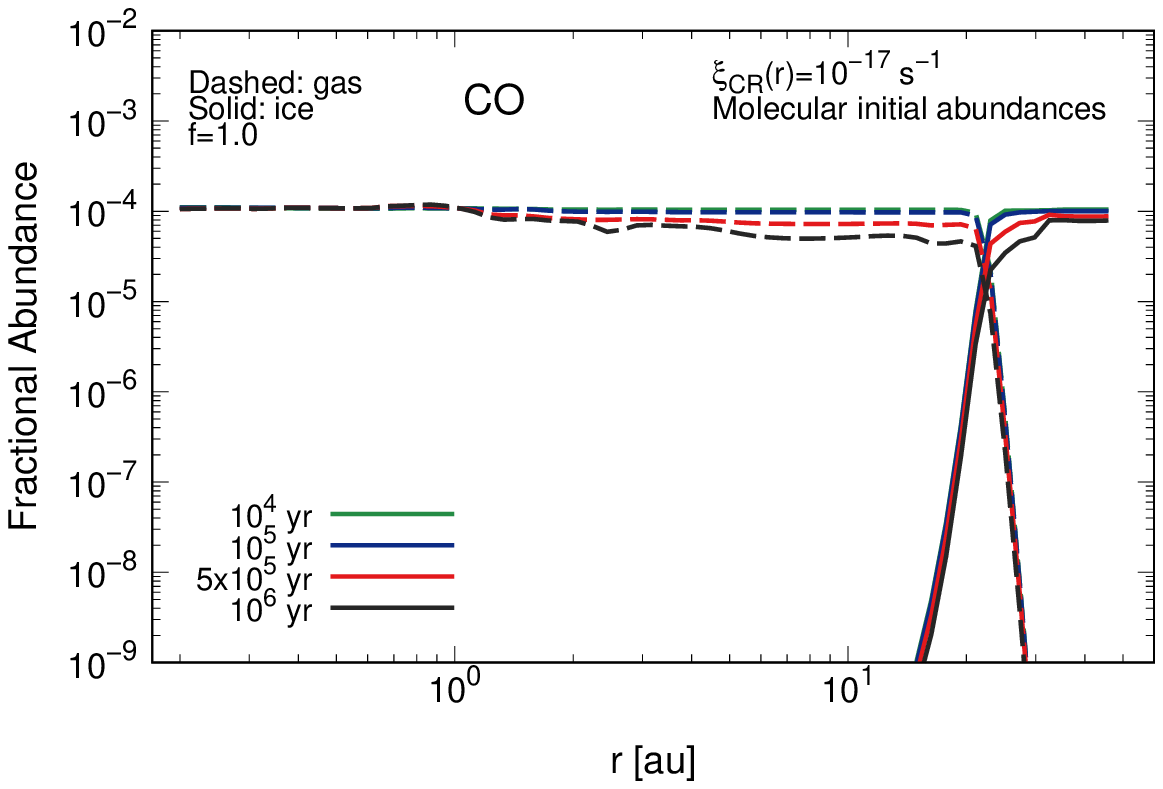}
\includegraphics[scale=0.57]{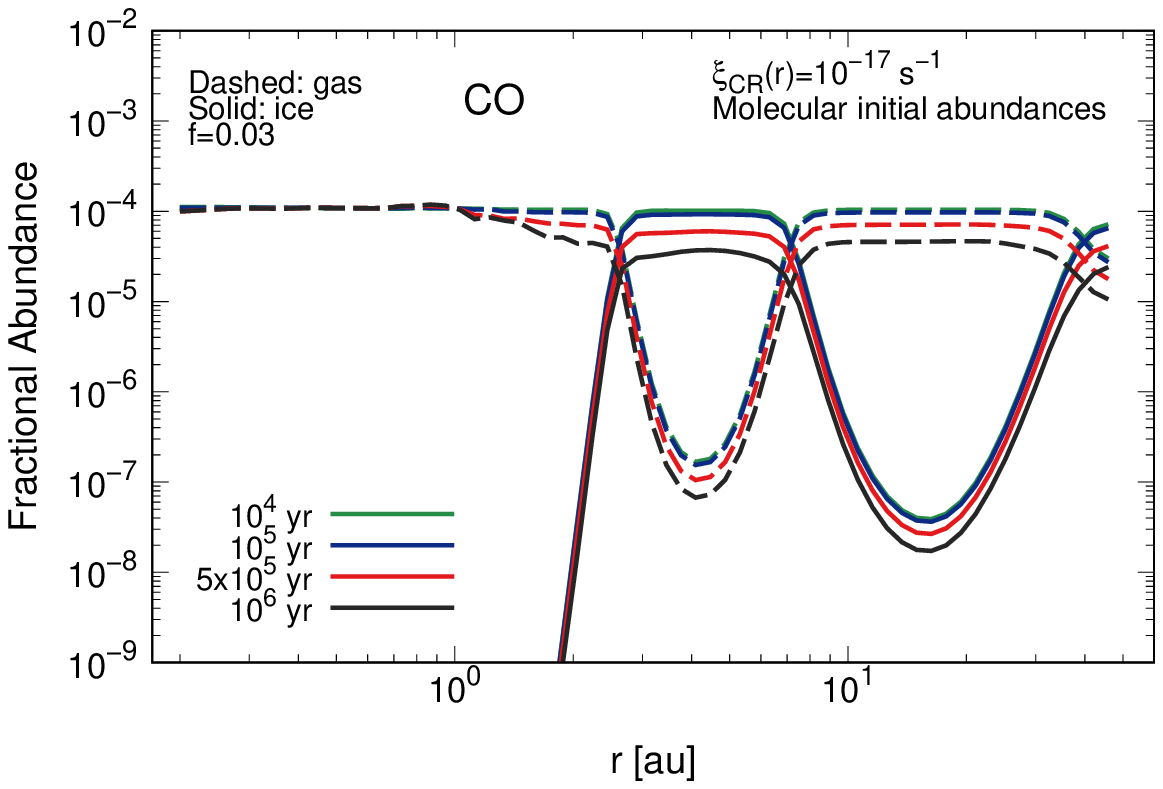}
\includegraphics[scale=0.57]{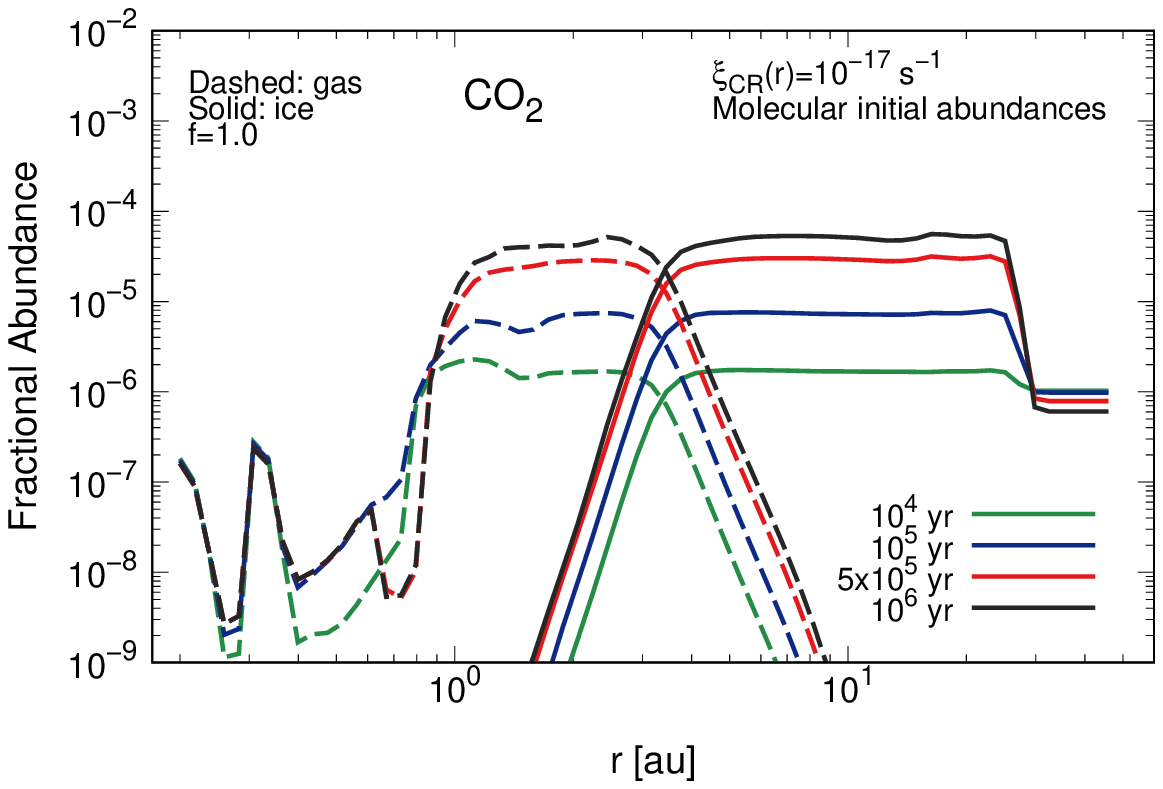}
\includegraphics[scale=0.57]{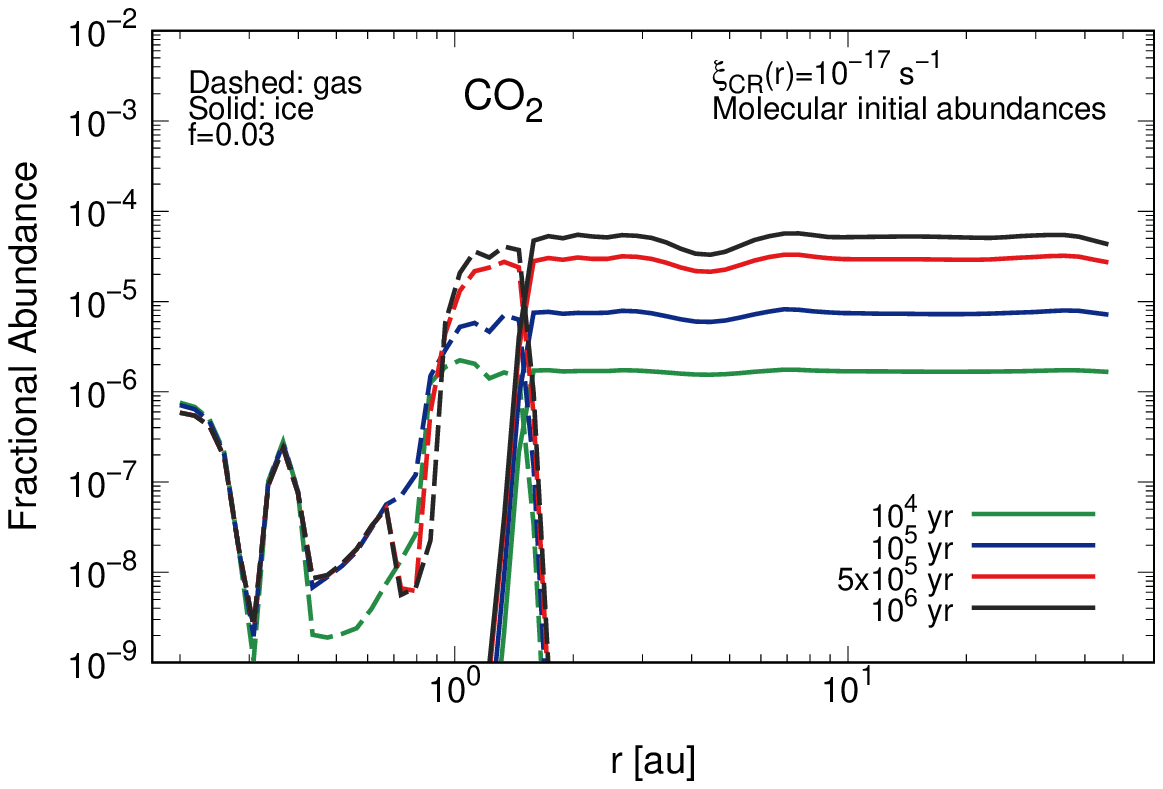}
\includegraphics[scale=0.57]{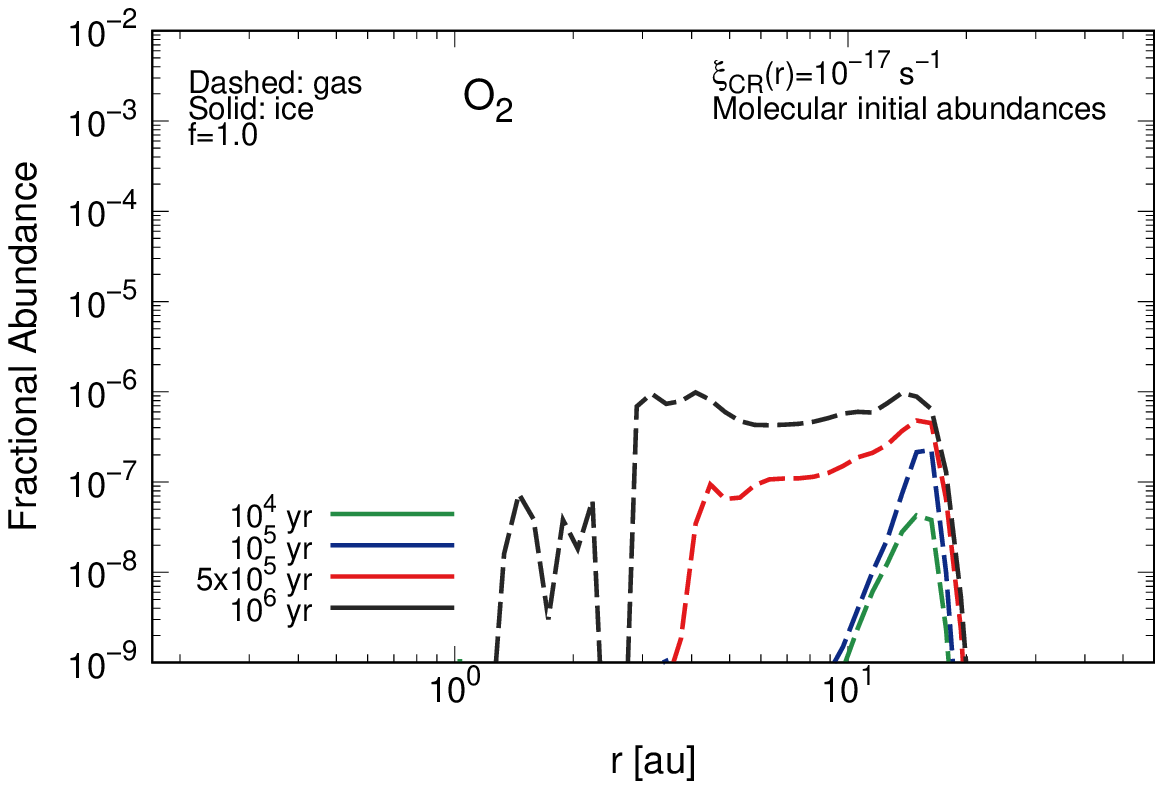}
\includegraphics[scale=0.57]{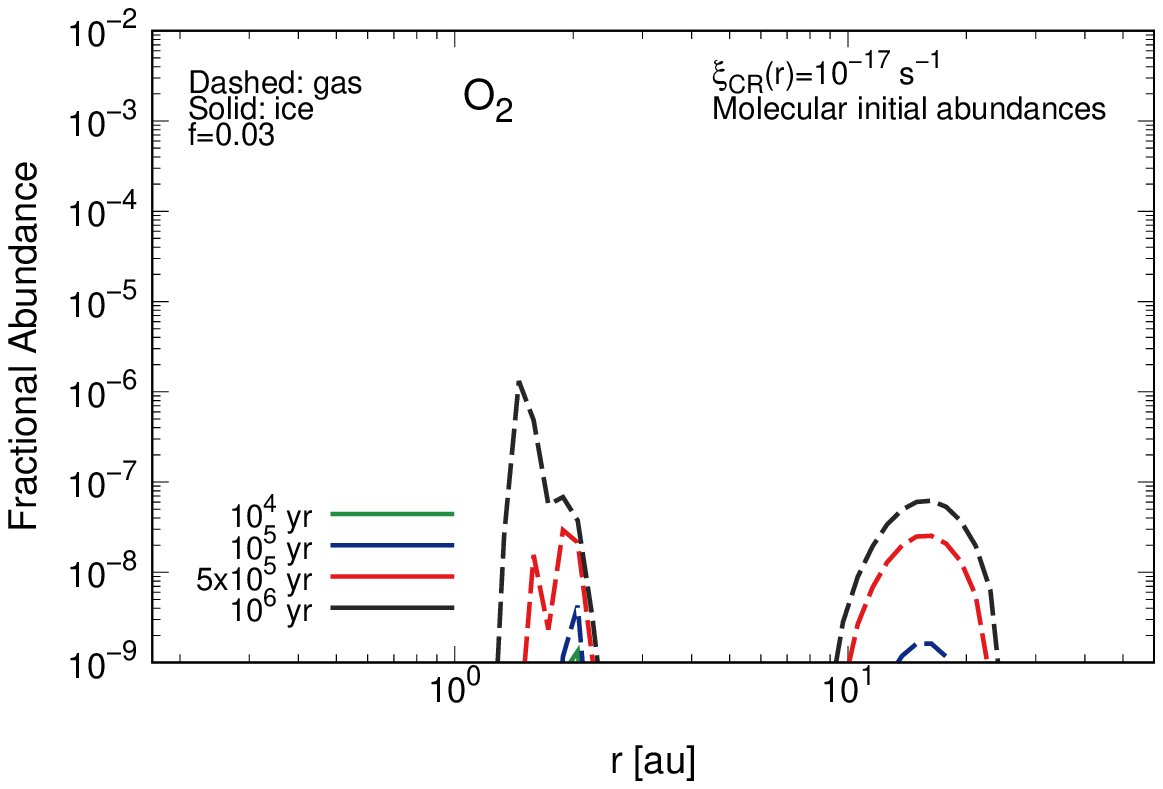}
\end{center}
\vspace{-0.2cm}
\caption{
The time evolution of the radial profiles of fractional abundances with respect to total hydrogen nuclei densities for H$_{2}$O (top panels), CO (second row panels), CO$_{2}$ (third row panels), and O$_{2}$ (bottom panels).
The dashed and solid lines show the profiles for gaseous and icy molecules, respectively.
The green, blue, red, and black lines show the profiles for different evolutional stage (t=10$^{4}$, 10$^{5}$, 5$\times10^{5}$, and 10$^{6}$ years), respectively.
These panels show the results when assuming molecular initial abundances (the ``inheritance'' scenario) and for $\xi_{\mathrm{CR}}(r)=$$1.0\times10^{-17}$ [s$^{-1}$].
Left panels show those for the disk midplane with the monotonically decreasing temperature and density structure ($f=1.0$), and right panels show the results for the shadowed disk midplane ($f=0.03$). 
%%.
\\ \\
}\label{Figure17rev3_time_appendix}
\end{figure*}
%%%
%%%
\begin{figure*}[hbtp]
\begin{center}
\vspace{1cm}
%\vspace{-1cm}
%\plotone{cost.pdf}
%\plotone[scale=0.63]{{20210624_surface-density.eps}
%\plotone[scale=0.63]{{20210624_ngas-T.eps}
\includegraphics[scale=0.57]{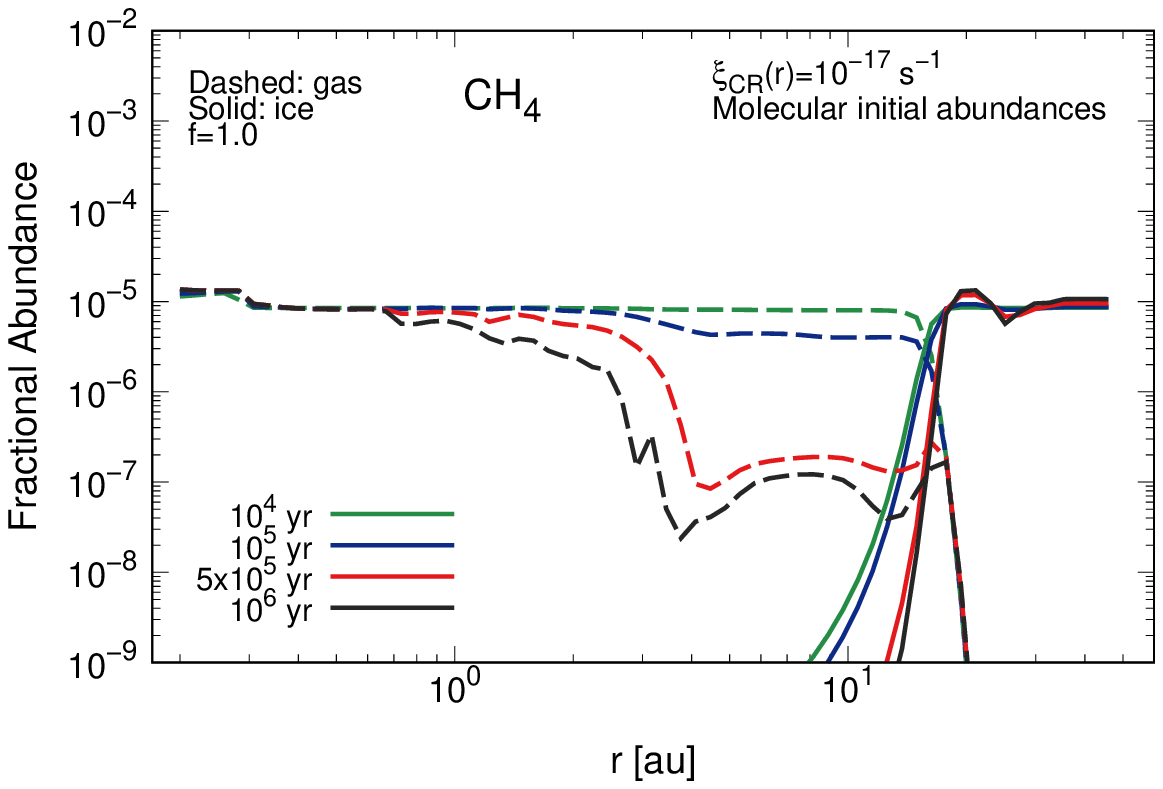}
\includegraphics[scale=0.57]{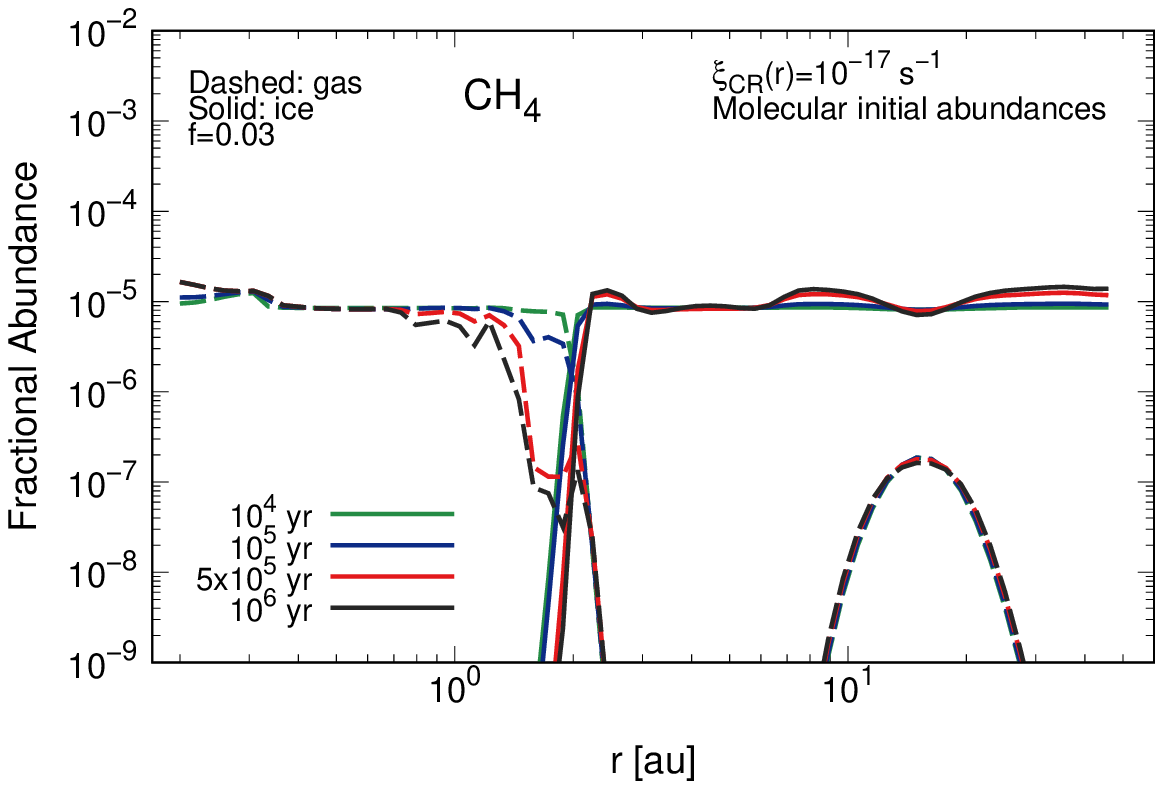}
\includegraphics[scale=0.57]{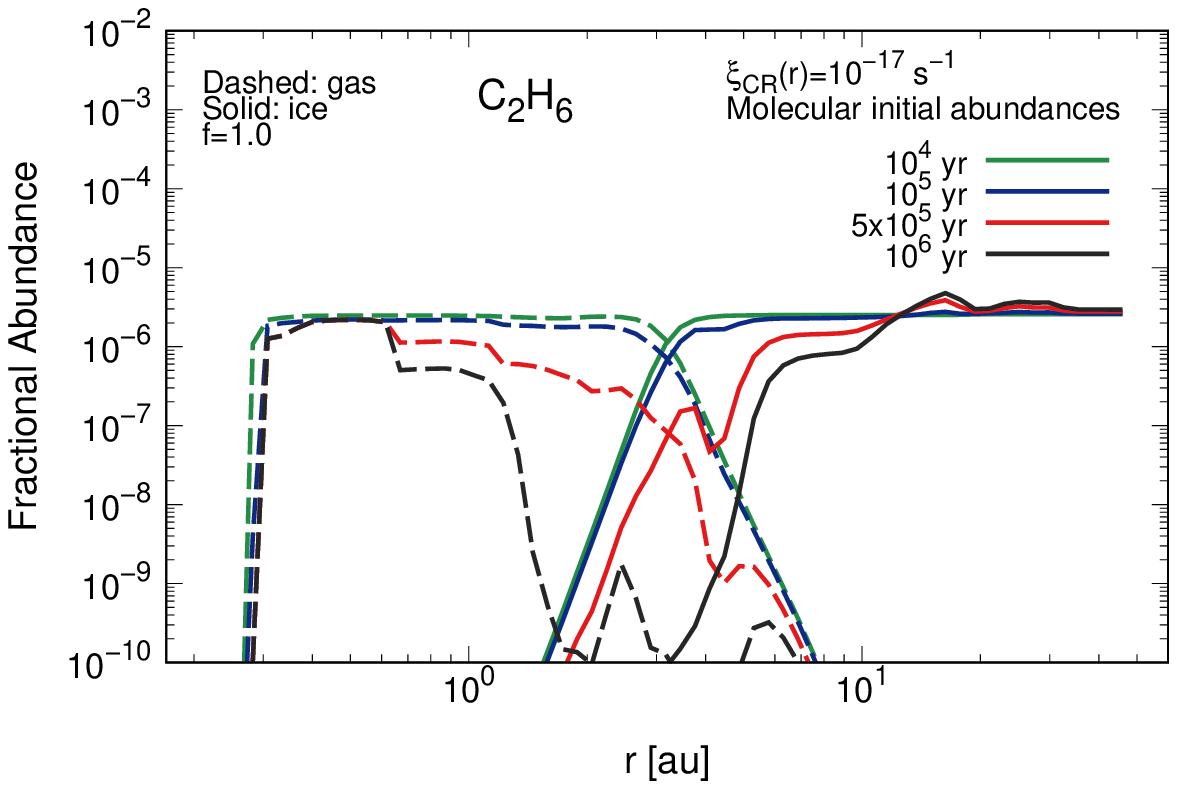}
\includegraphics[scale=0.57]{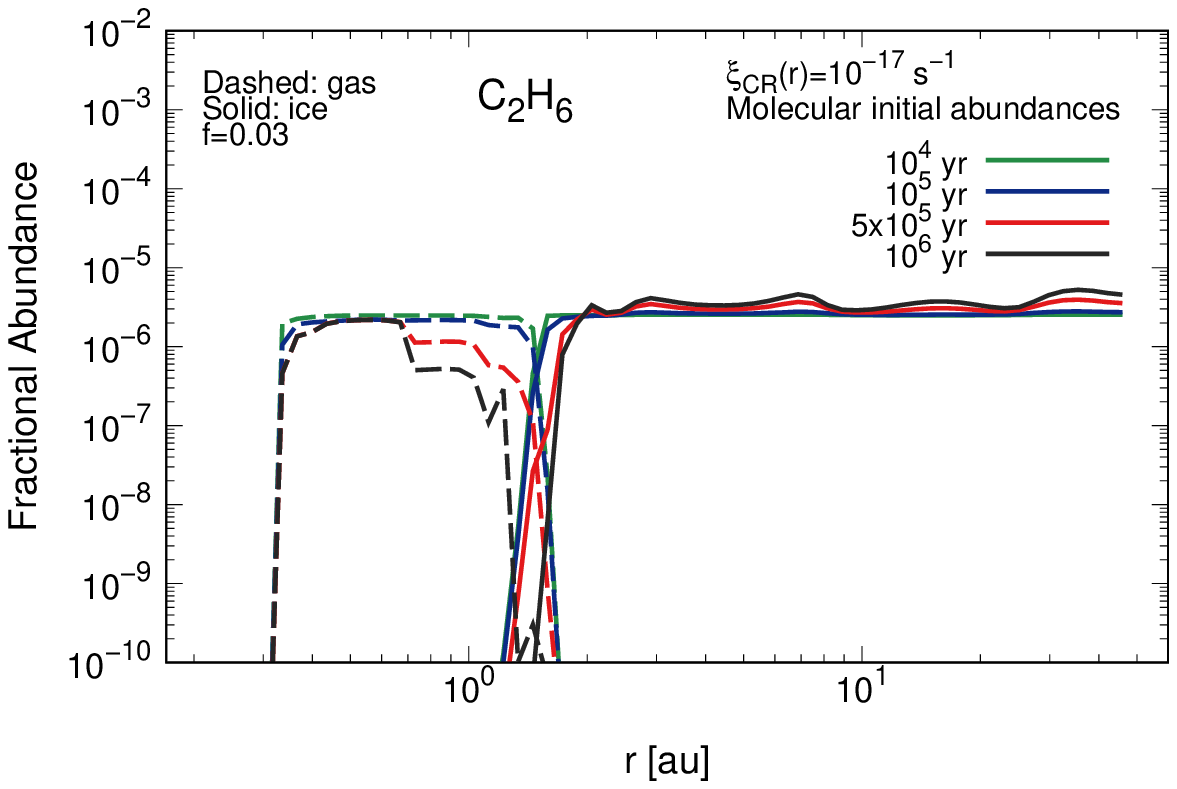}
\includegraphics[scale=0.57]{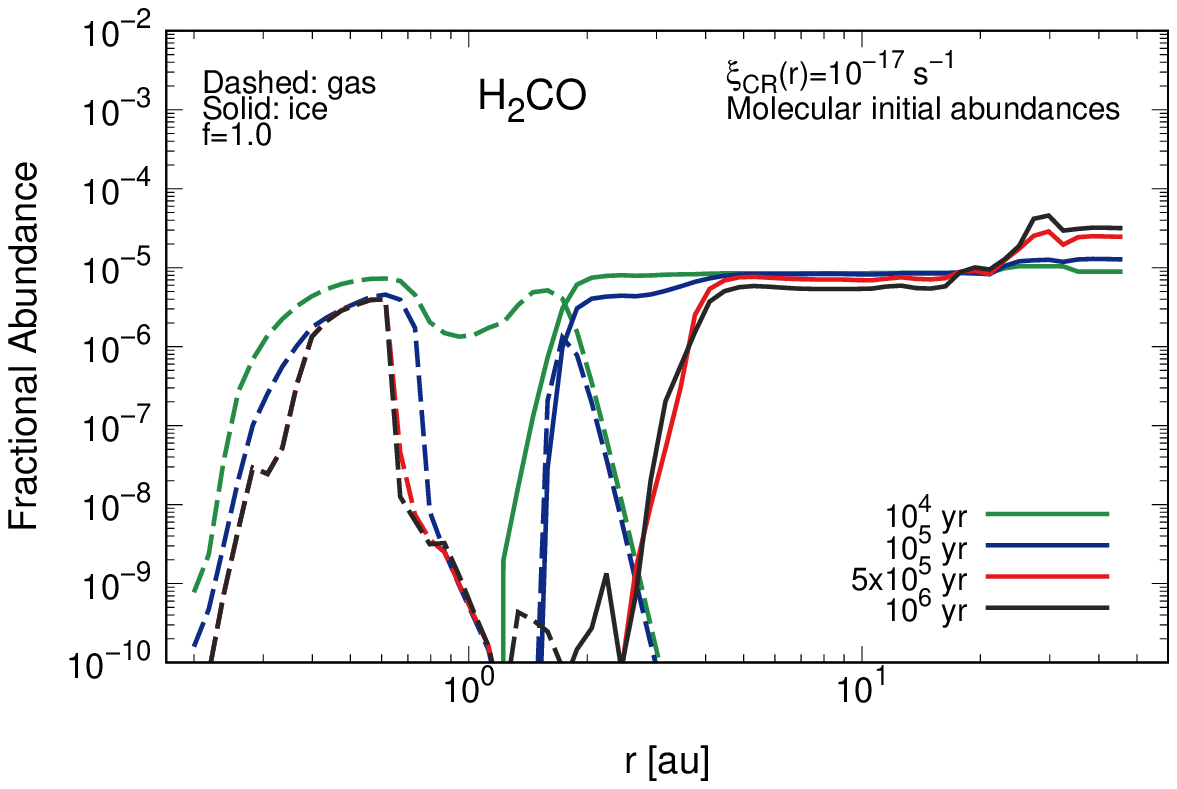}
\includegraphics[scale=0.57]{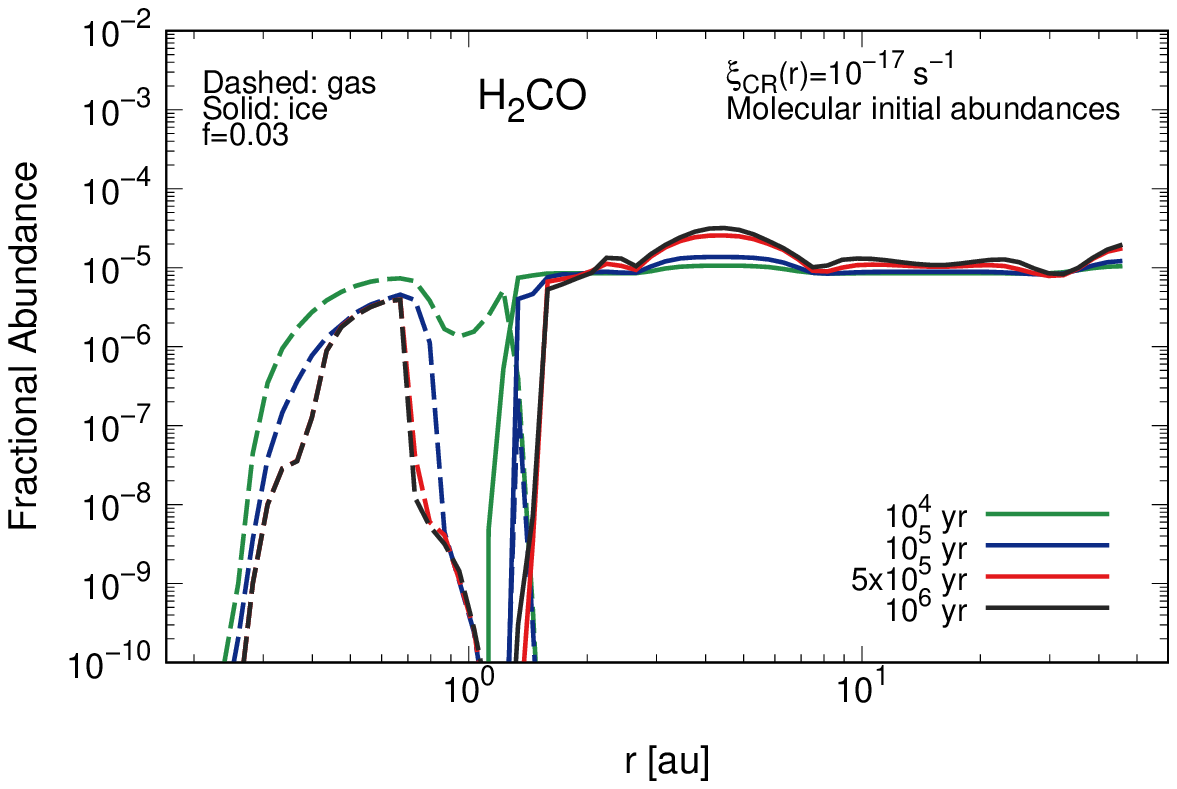}
\includegraphics[scale=0.57]{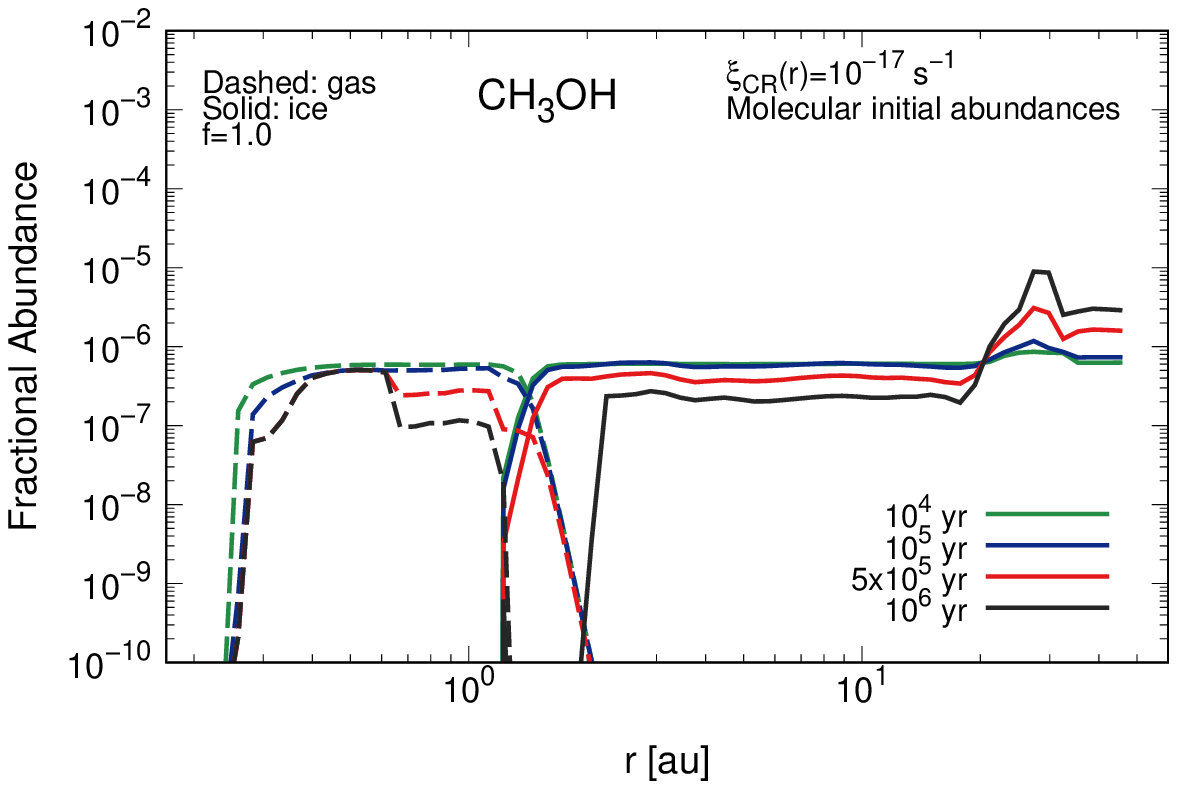}
\includegraphics[scale=0.57]{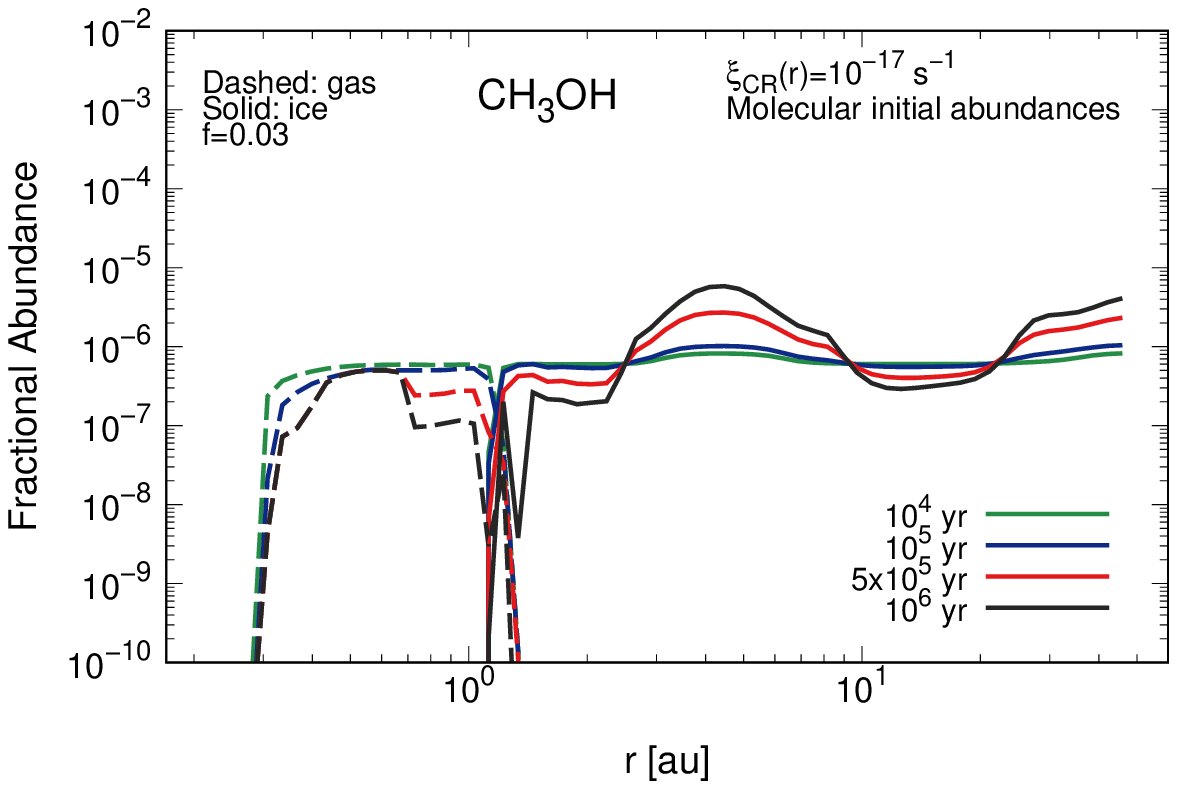}
\end{center}
\vspace{-0.2cm}
\caption{
Same as Figure \ref{Figure17rev3_time_appendix}, but for CH$_{4}$ (top panels), C$_{2}$H$_{6}$ (second row panels), H$_{2}$CO (third row panels), and CH$_{3}$OH (bottom panels).
%.
\\ \\
}\label{Figure18rev3_time_appendix}
\end{figure*}
%%%%
\begin{figure*}[hbtp]
\begin{center}
\vspace{1cm}
%\vspace{-1cm}
%
\includegraphics[scale=0.57]{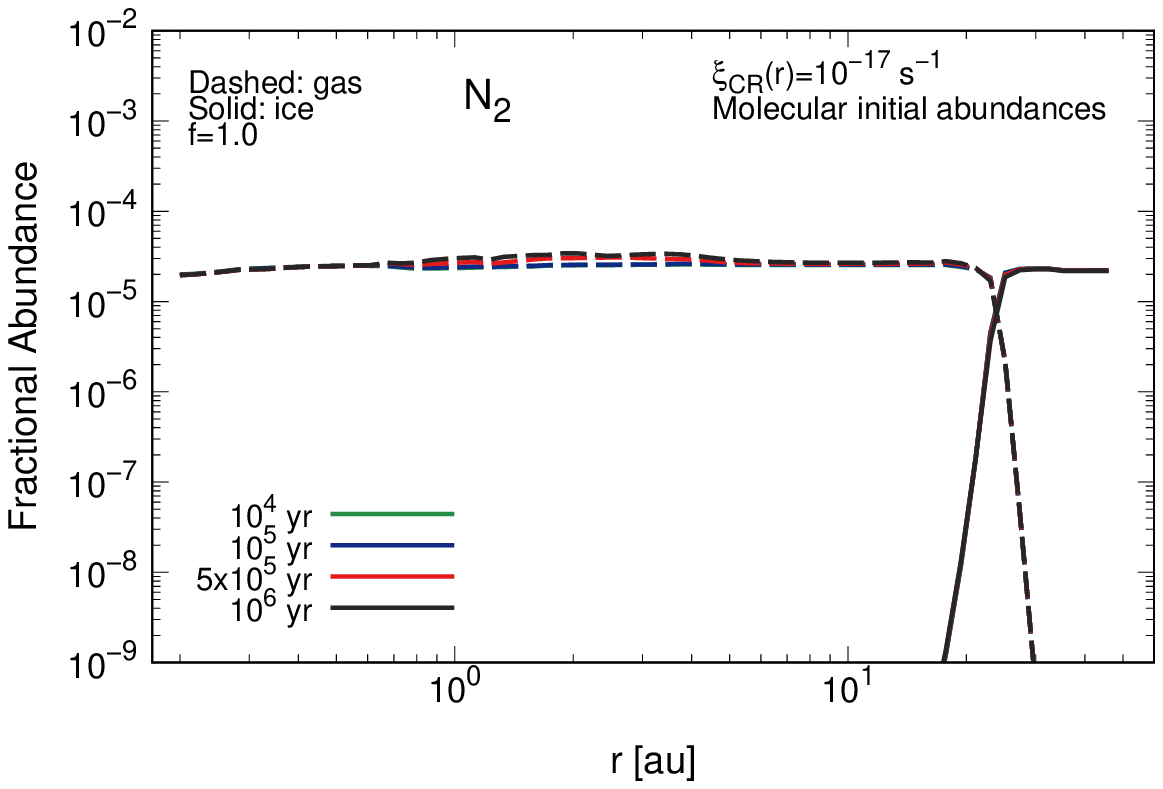}
\includegraphics[scale=0.57]{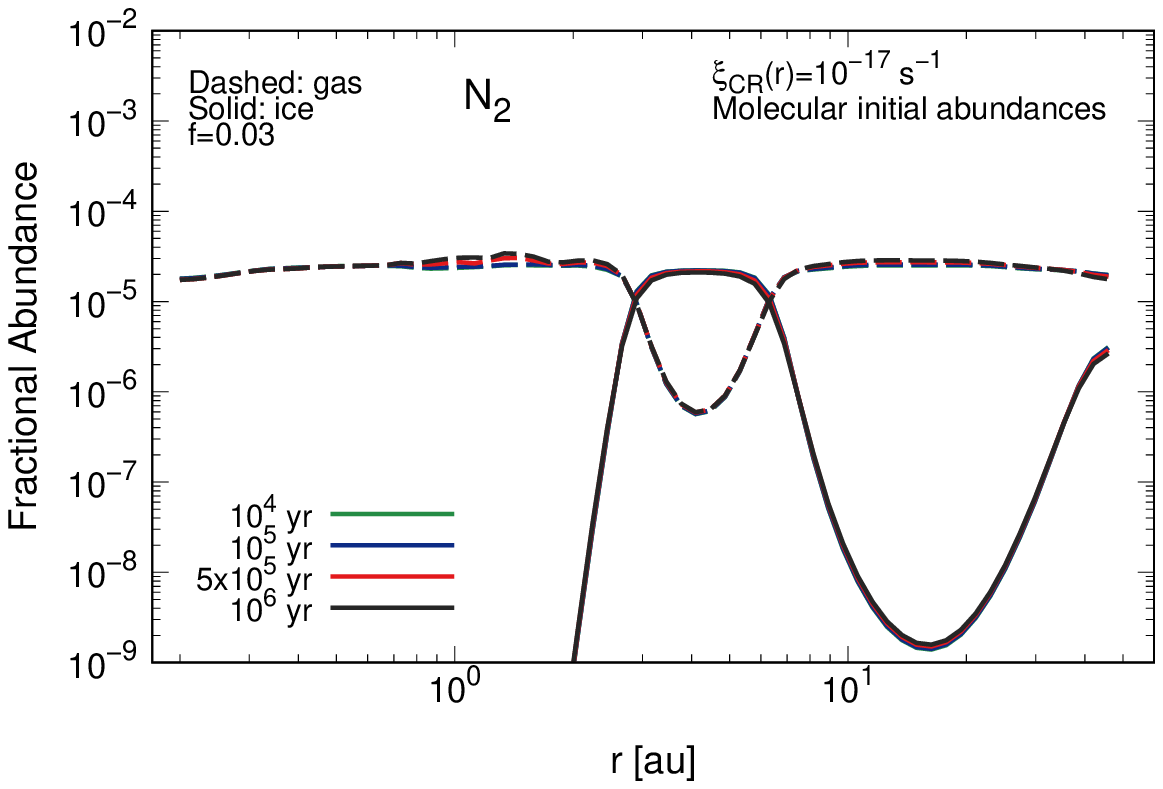}
\includegraphics[scale=0.57]{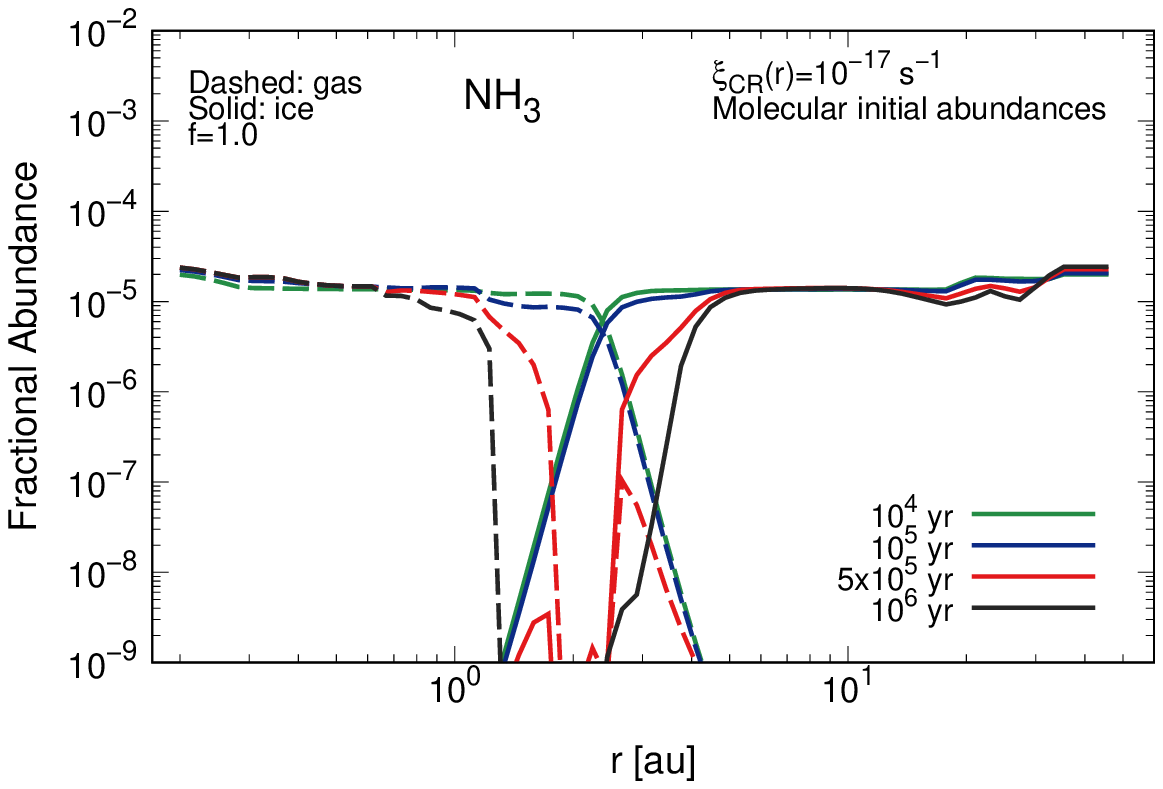}
\includegraphics[scale=0.57]{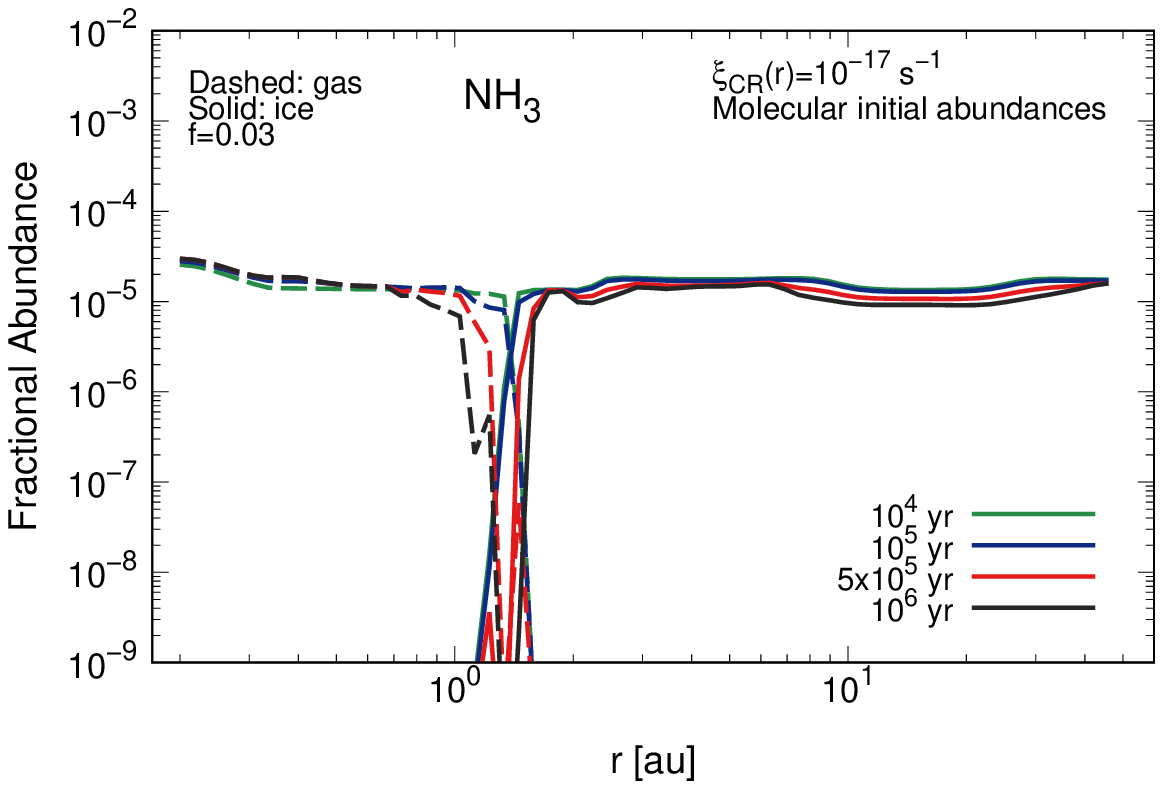}
\includegraphics[scale=0.57]{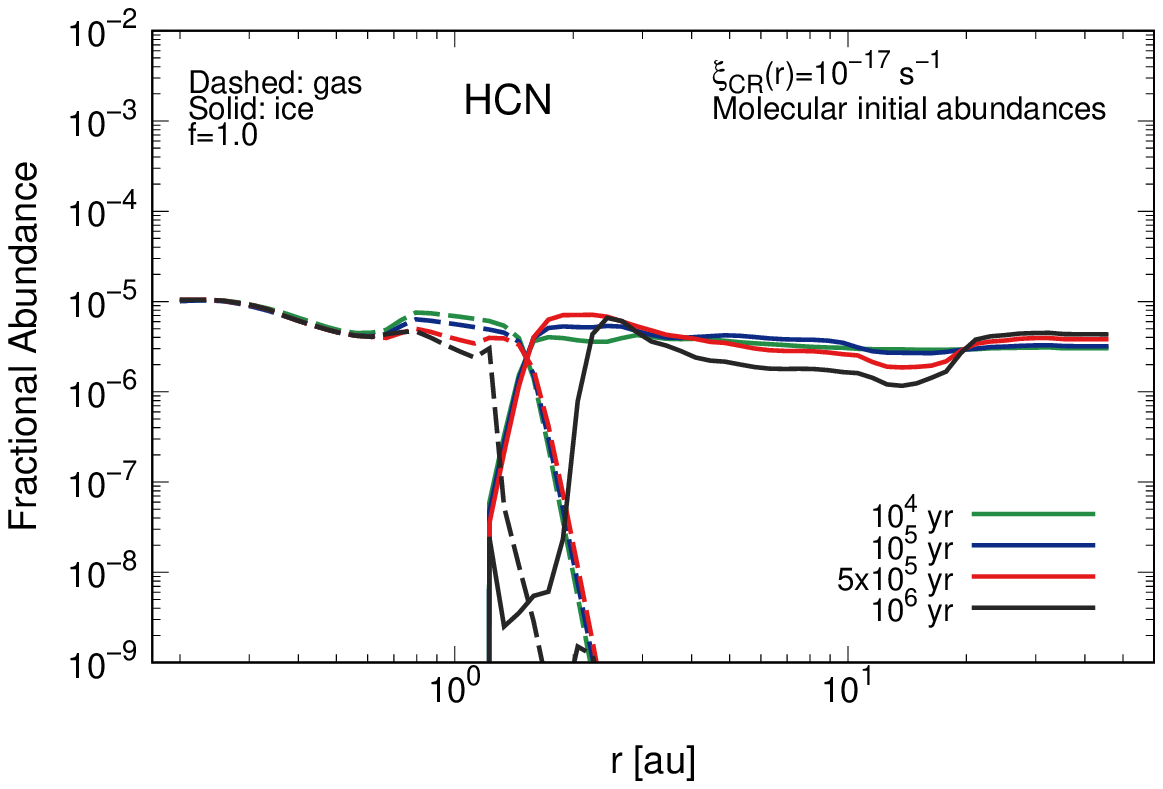}
\includegraphics[scale=0.57]{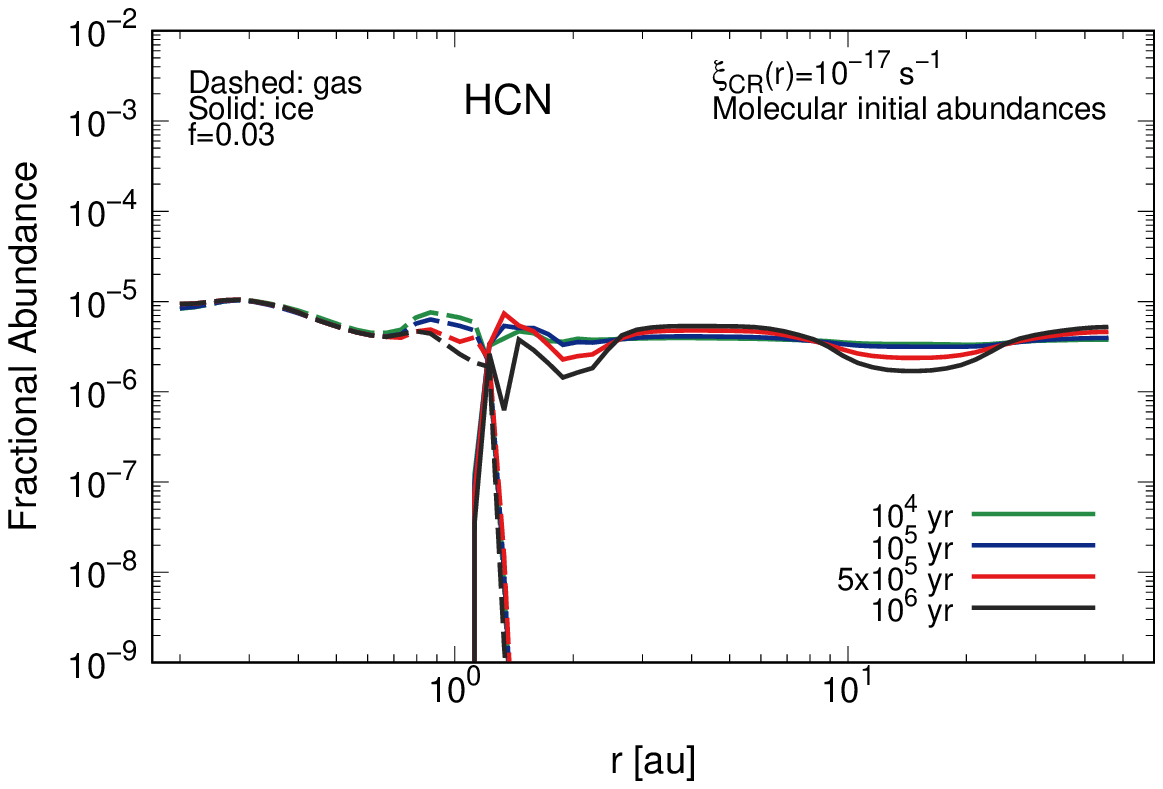}
\includegraphics[scale=0.57]{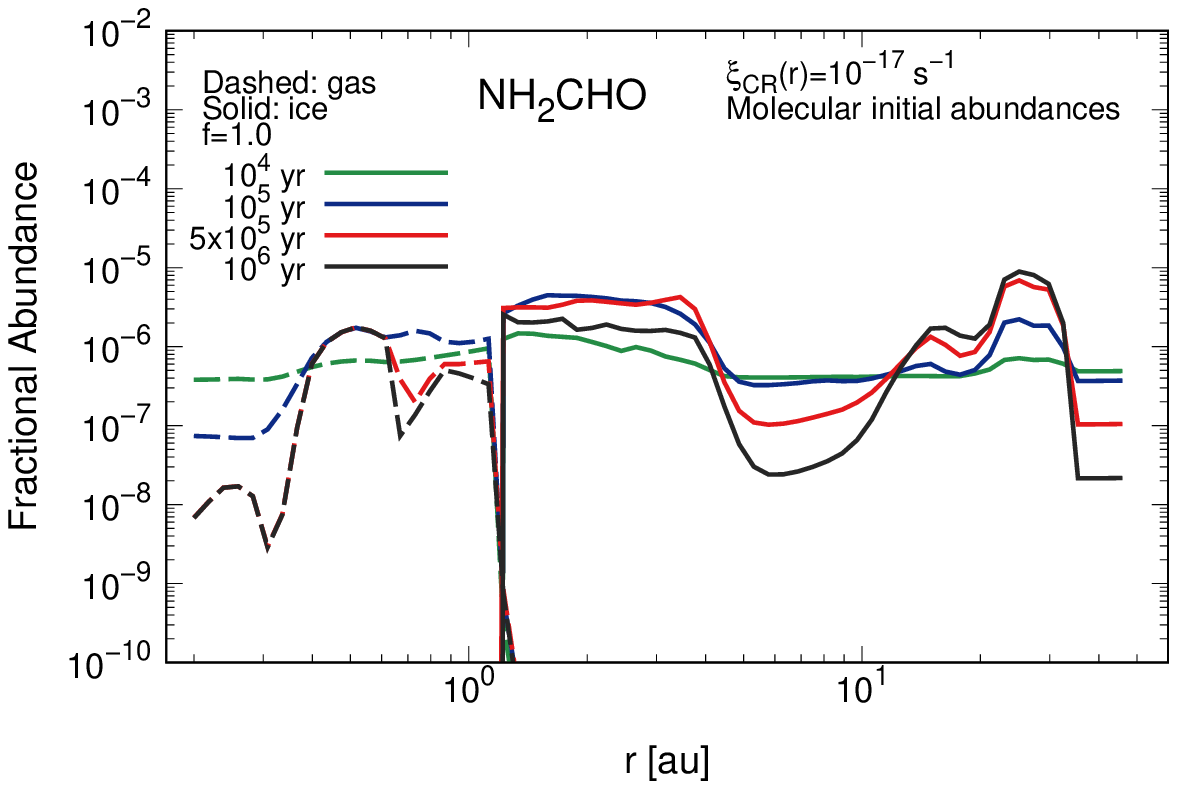}
\includegraphics[scale=0.57]{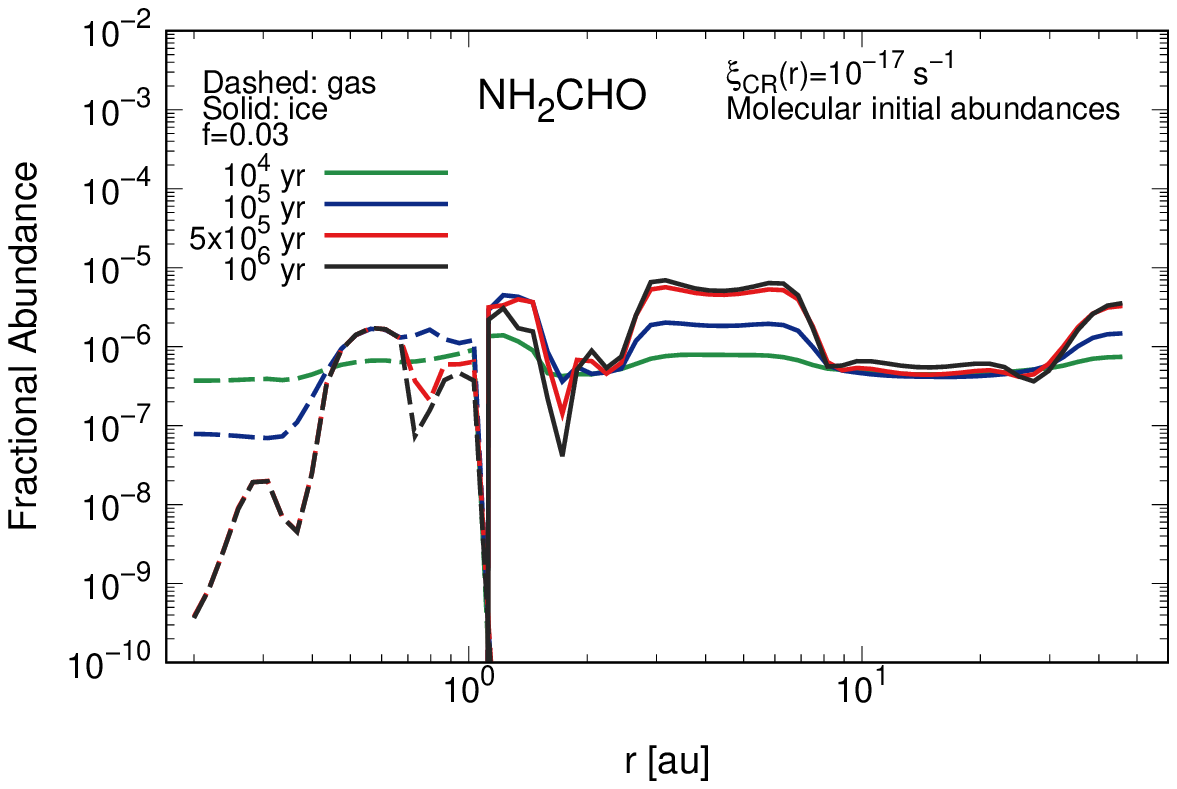}
\end{center}
\vspace{-0.2cm}
\caption{
%The time evolution of the radial profiles of fractional abundances with respect to total hydrogen nuclei densities for 
Same as Figure \ref{Figure17rev3_time_appendix}, but for 
N$_{2}$ (top panels), NH$_{3}$ (second row panels), HCN (third row panels), and NH$_{2}$CHO (bottom panels).
%%
%%%.
\\ \\
}\label{Figure19rev3_time_appendix}
\end{figure*}
Figures \ref{Figure17rev3_time_appendix}-\ref{Figure19rev3_time_appendix} show the time evolution of the radial profiles of fractional abundances for dominant oxygen-, carbon-, nitrogen-bearing molecules (H$_{2}$O, CO, CO$_{2}$, O$_{2}$, CH$_{4}$, C$_{2}$H$_{6}$, H$_{2}$CO, CH$_{3}$OH, N$_{2}$, NH$_{3}$, HCN, and NH$_{2}$CHO)
in the shadowed and non-shadowed disk midplane ($f=1.0$ and $f=0.03$, respectively), for molecular initial abundances and a high ionization rate ($\xi_{\mathrm{CR}}(r)=$$10^{-17}$ [s$^{-1}$]).
These initial conditions are same as those in Sections \ref{sec:3-1rev} and \ref{sec:3-2rev}.
\\ \\
According to these Figures, H$_{2}$O and N$_{2}$ abundances do not change with time (see Table \ref{Table:2}).
O$_{2}$ gas abundances inside its snowline increase with time (from $<<10^{-9}$ at $t\lesssim10^{5}$ years to $\sim10^{-6}$ at $t\sim10^{6}$ years).
\\ \\
Both CO gas and ice abundances (inside and outside the CO snowline, respectively) decrease with time (from $\sim10^{-4}$ at $t\lesssim10^{5}$ years to $\sim(3-5)\times10^{-5}$ at $t\sim10^{6}$ years).
%%%
In contrast, both CO$_{2}$ gas and ice abundances (inside and outside the CO$_{2}$ snowline, respectively) increase with time (from $<<10^{-5}$ at $t<10^{5}$ years to $\sim5\times10^{-5}$ at $t\sim10^{6}$ years), 
%%%%.
\citet{Eistrup2018} discussed that the decreasing abundance of CO gas and increasing abundance of CO$_{2}$ ice with time between CO$_{2}$ and CO snowlines can be explained by CO gas collisions with the grains, followed by fast reactions with OH (faster than CO can desorb) that produce CO$_{2}$ on dust grain surfaces.
\citet{Bosman2018} discussed that for $\xi_{\mathrm{CR}}(r)=$$1.0\times10^{-17}$ [s$^{-1}$], the CO abundance can be reduced by two orders of magnitude in $t\sim(2-3)\times10^{6}$ years.
%%%
\\ \\
As previous studies have shown (see e.g., \citealt{Eistrup2018}), CH$_{4}$ gas abundances within its snowline decrease with time (from $\gtrsim5\times10^{-6}$ at $t\lesssim10^{5}$ years to $\lesssim10^{-7}$ at $t\gtrsim5\times10^{5}$ years) and convert to CO$_{2}$ (see Section \ref{sec:3-2-2rev}).
In addition, we show that abundances of other molecules such as C$_{2}$H$_{6}$, H$_{2}$CO, CH$_{3}$OH, and NH$_{3}$ decrease with time around their own snowlines.
We predict that in these regions gas-phase destruction processes (ion-molecule reactions and cosmic-ray-induced photodissociation) and cosmic-ray-induced photodesorption/photodissociation overcome the production processes in gas and on the dust grain surfaces, respectively.
\\ \\
In the region where CO is frozen-out onto dust grains (including the shadowed region for $f\leq0.03$), the abundances of e.g., H$_{2}$CO, CH$_{3}$OH, and NH$_{2}$CHO increase with time because of the grain-surface reactions
(from $\lesssim10^{-6}$ at $t\lesssim10^{5}$ years to $\gtrsim5\times10^{-6}$ at $t\sim10^{6}$ years for CH$_{3}$OH).
On the basis of these results, the timescales of gas-phase reactions (such as ion-molecule reactions and cosmic-ray-induced photodissociation) and cosmic-ray-induced photodesorption/photodissociation are generally shorter than those of the grain-surface formation reactions, such as sequential hydrogenation reactions.
\\ \\
The HCN ice abundances are slightly enhanced (from $\sim3\times10^{-6}$ at $10^{4}$ years to $\sim5\times10^{-6}$ at $10^{6}$ years) in the region where CO is frozen-out onto dust grains (including the shadowed region for $f\leq0.03$).
As we describe in Section \ref{sec:3-2-3rev}, HCN ice is efficiently formed within the cold regions where CO freezes out onto dust grains. \citet{Schwarz2014} and \citet{Eistrup2018} described that HCN ice is efficiently produced by a few $\times10^{6}$ years.
%%%.
\\ \\
On the basis of these results, in the shadowed region ($r\sim3-8$ au) the icy abundances of CO$_{2}$ and organic molecules such as H$_{2}$CO, CH$_{3}$OH, and NH$_{2}$CHO become larger with time. 
%%.
Thus, we find that
if the shadowed region is maintained for a relatively long time ($t\sim10^{6}$ years), chemical evolution 
may produce dust grains and solid objects with large amounts of CO$_{2}$ and organic molecular ices (see also Section \ref{sec:4-2}).
In addition, the ice abundances of these molecules can be a clue in constraining the formation age of solid bodies in the shadowed region.
\\ \\
We note that in our modeling, we assume a constant central stellar luminosity and a constant viscous accretion heating (see Section \ref{sec:2-1}). 
\citet{Kunitomo2021} calculated that in the case of in the case of a T Tauri star with with mass $M_{\mathrm{*}}$=1.0$M_{\bigodot}$, the central star luminosity $L_{\mathrm{*}}$
gradually decreases from $\sim5$$L_{\bigodot}$ at  t$\sim10^{5}$ years to $\lesssim$$L_{\bigodot}$ at  t$\gtrsim3\times10^{6}$ years. 
In addition, the viscous accretion heating determines the temperature in the inner disk (see Section \ref{sec:2-1}). 
Thus, the disk temperature in the inner disk (around the water snowline) does not change very much within the timescale of $10^{6}$ years by evolving stellar luminosity.
\\ \\
In addition, in reality the efficiency of viscous accretion heating also changes with time. 
\citet{Oka2011} described the relation between the midplane temperature and mass accretion rate: $T\propto\dot{M}^{1/4}$. Thus, based on \citet{Oka2011}, the water snowline position expects to move inward (from $\sim5$ au to $\sim1.3$ au) as the mass accretion rate becomes around ten times smaller with time (from $t\sim10^{5}$ years (Class 0-I disk) to $t\sim10^{6}$ years (Class II disk)). 
%%.
%%
%\clearpage
%%
\section{The C/H, O/H, and N/H ratios}\label{Csec:C}
\begin{figure*}[hbtp]
\begin{center}
\vspace{-4cm}
%\hspace{1cm}
%\plotone{cost.pdf}
%
%\includegraphics[scale=0.57]{2022.6.17r2_CtoH_t1_ds_0.1um_mol-ini_Xray-rev_enhanced-water_nH-rev_1.0e6yr.eps}
%\includegraphics[scale=0.57]{2022.6.17r2_CtoH_t2_ds_0.1um_mol-ini_Xray-rev_enhanced-water_nH-rev_1.0e6yr.eps}
%\includegraphics[scale=0.57]{2022.6.17r2_CtoH_t3_ds_0.1um_atomic-ini_Xray-rev_enhanced-water_nH-rev_1.0e6yr.eps}
%\includegraphics[scale=0.57]{2022.6.17r2_CtoH_t4_ds_0.1um_atomic-ini_Xray-rev_enhanced-water_nH-rev_1.0e6yr.eps}
%
\includegraphics[scale=0.8]{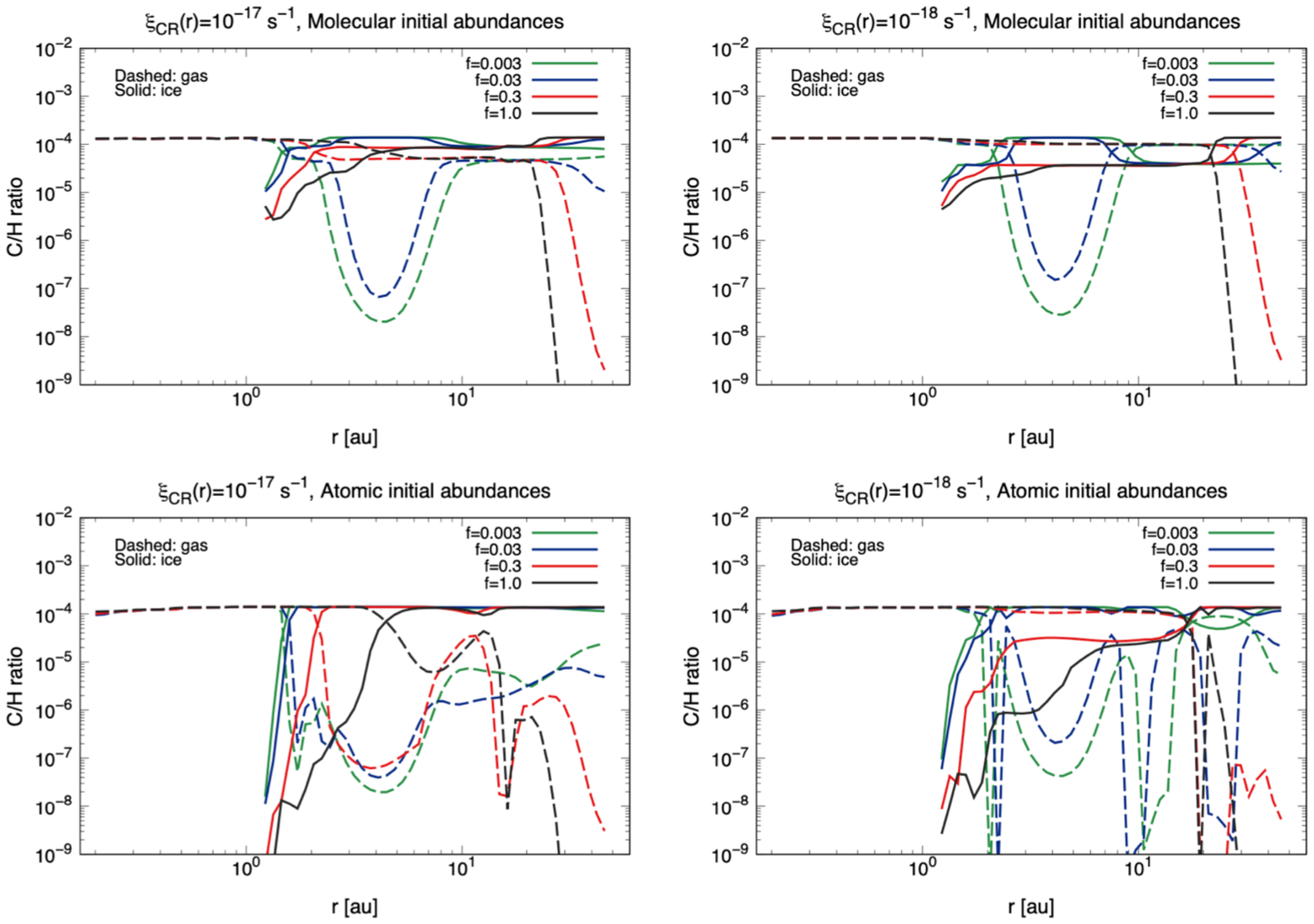}
%%%
%%%
%%%
\end{center}
\vspace{-3cm}
\caption{
The radial profiles of C/H ratios at t=$10^{6}$ years.
The dashed and solid lines show the profiles for gaseous and icy molecules, respectively.
The black, red, blue, and green lines show the profiles for different values of the parameter $f$ (=1.0, 0.3, 0.03, and 0.003), respectively.
Top panels show the results when assuming molecular initial abundances (the ``inheritance'' scenario), whereas bottom panels show those when assuming atomic initial abundances (``reset'' scenario).
Left panels show the results for $\xi_{\mathrm{CR}}(r)=$$1.0\times10^{-17}$ [s$^{-1}$], whereas right panels show the results for $\xi_{\mathrm{CR}}(r)=$$1.0\times10^{-18}$ [s$^{-1}$]. The initial elemental C/H ratio is $1.40\times10^{-4}$.
%%.
%%%
\\ \\
%\vspace{0.2cm}
}\label{Figure20rev_CtoH_appendix}
\end{figure*}
\begin{figure*}[hbtp]
\begin{center}
\vspace{-4cm}
%\hspace{1cm}
%\plotone{cost.pdf}
%
%\includegraphics[scale=0.57]{2022.7.19r2_OtoH_t1_ds_0.1um_mol-ini_Xray-rev_enhanced-water_nH-rev_1.0e6yr.eps}
%\includegraphics[scale=0.57]{2022.7.19r2_OtoH_t2_ds_0.1um_mol-ini_Xray-rev_enhanced-water_nH-rev_1.0e6yr.eps}
%\includegraphics[scale=0.57]{2022.7.19r2_OtoH_t3_ds_0.1um_atomic-ini_Xray-rev_enhanced-water_nH-rev_1.0e6yr.eps}
%\includegraphics[scale=0.57]{2022.7.19r2_OtoH_t4_ds_0.1um_atomic-ini_Xray-rev_enhanced-water_nH-rev_1.0e6yr.eps}
%%%
\includegraphics[scale=0.8]{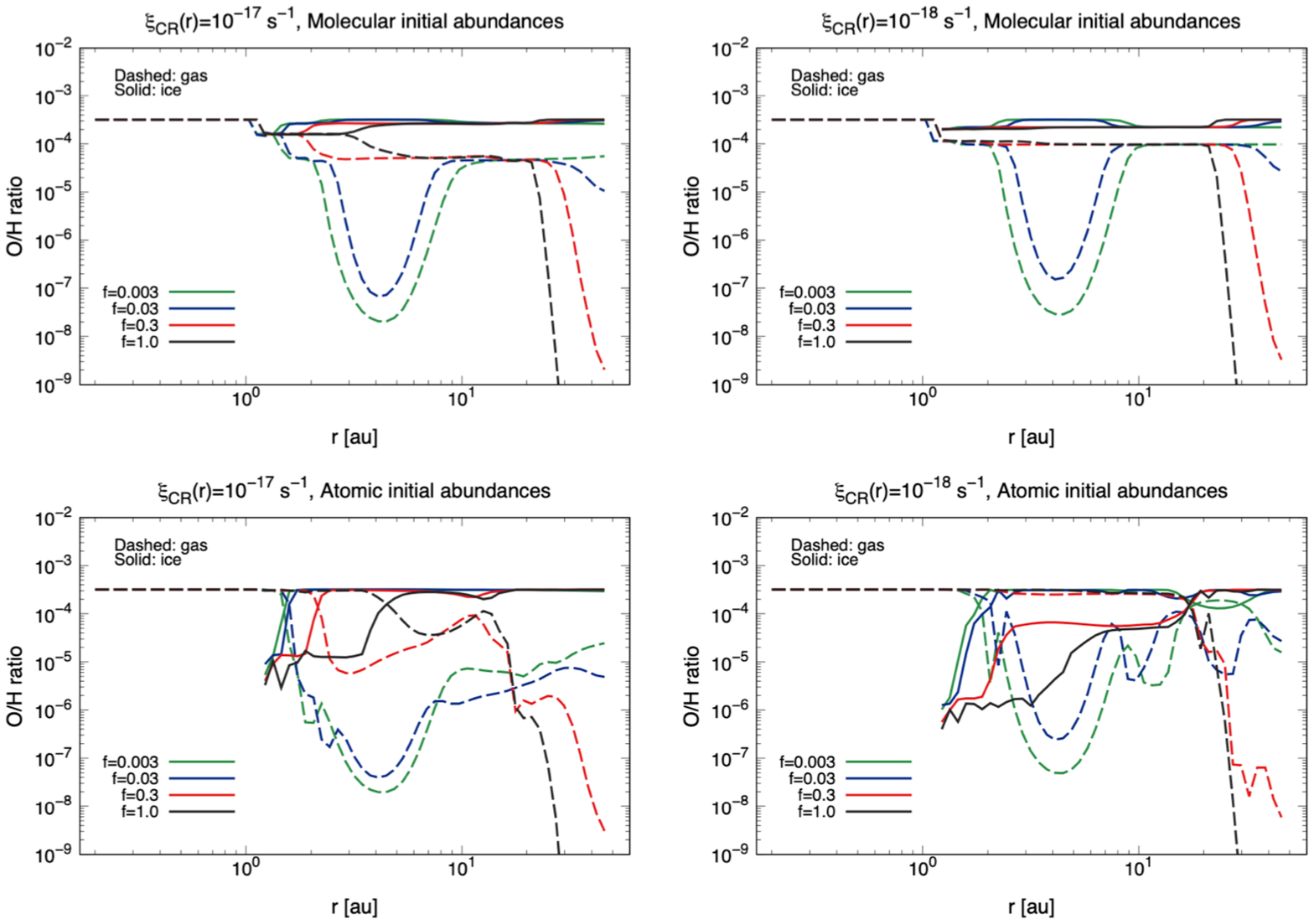}
%%%
%%%
%%%
\end{center}
\vspace{-3cm}
\caption{
Same as Figure \ref{Figure20rev_CtoH_appendix}, but for the radial profiles of O/H ratios at t=$10^{6}$ years.
The initial elemental O/H ratio is $3.20\times10^{-4}$.
%\vspace{0.2cm}
\\ \\
}\label{Figure21rev_OtoH_appendix}
\end{figure*}
%%%
\begin{figure*}[hbtp]
\begin{center}
\vspace{-4cm}
%\hspace{1cm}
%\plotone{cost.pdf}
%
%\includegraphics[scale=0.57]{2022.7.19r2_NtoH_t1_ds_0.1um_mol-ini_Xray-rev_enhanced-water_nH-rev_1.0e6yr.eps}
%\includegraphics[scale=0.57]{2022.7.19r2_NtoH_t2_ds_0.1um_mol-ini_Xray-rev_enhanced-water_nH-rev_1.0e6yr.eps}
%\includegraphics[scale=0.57]{2022.7.19r2_NtoH_t3_ds_0.1um_atomic-ini_Xray-rev_enhanced-water_nH-rev_1.0e6yr.eps}
%\includegraphics[scale=0.57]{2022.7.19r2_NtoH_t4_ds_0.1um_atomic-ini_Xray-rev_enhanced-water_nH-rev_1.0e6yr.eps}
%%%
\includegraphics[scale=0.8]{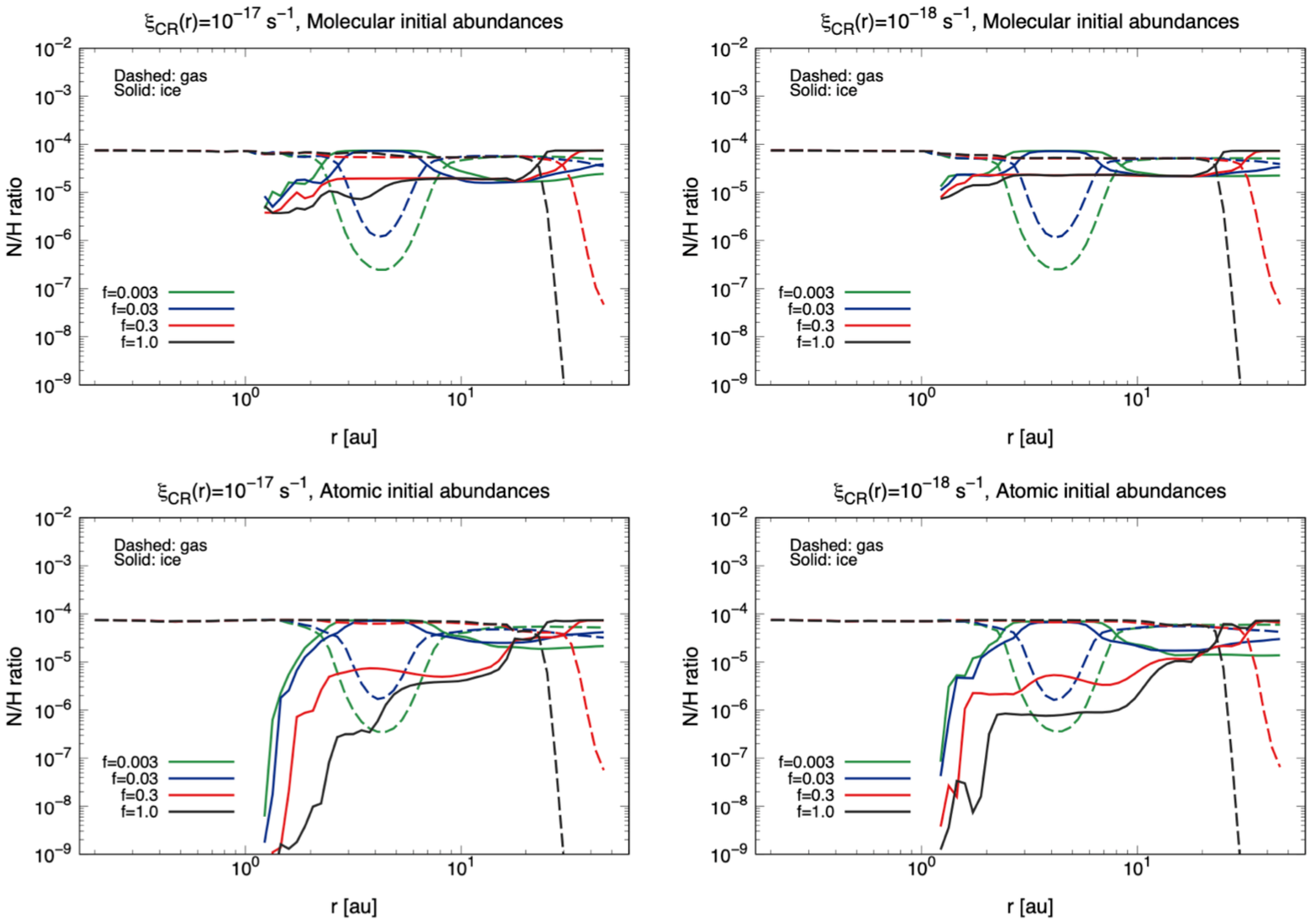}
%%%
%%%
%%%
\end{center}
\vspace{-3cm}
\caption{
Same as Figure \ref{Figure20rev_CtoH_appendix}, but for the radial profiles of N/H ratios at t=$10^{6}$ years.
The initial elemental N/H ratio is $7.50\times10^{-5}$.
%\vspace{0.2cm}
\\ \\
}\label{Figure22rev_NtoH_appendix}
\end{figure*}
%%%
Figures \ref{Figure20rev_CtoH_appendix}, \ref{Figure21rev_OtoH_appendix}, and \ref{Figure22rev_NtoH_appendix} show respectively the radial profiles of C/H, O/H, and N/H ratios at t=$10^{6}$ years.
Different color lines show the profiles for different values of the parameter $f$ (=1.0, 0.3, 0.03, and 0.003), respectively.
As described in our previous studies for the case of non-shadowed disk (see Section 2.3 and Figure 2 of \citealt{Notsu2020}), the inclusion of chemistry has a significant impact on the disk elemental abundances of both gas and ice.
\\ \\
In the non-shadowed disk ($f=1.0$), for $\xi_{\mathrm{CR}}(r)=$$1.0\times10^{-18}$ [s$^{-1}$], the gas-phase C/H and O/H ratios are respectively $\sim10^{-4}$ and $\sim(1-3)\times10^{-4}$ at $r\lesssim22$ au (within the CO snowline).
In addition, for $\xi_{\mathrm{CR}}(r)=$$1.0\times10^{-17}$ [s$^{-1}$], the gas-phase C/H and O/H ratios are respectively $\sim10^{-4}$ and $\gtrsim10^{-4}$ within the CO$_{2}$ snowline 
and respectively $\sim10^{-6} - 5\times10^{-5}$ and $\sim10^{-5} - 10^{-4}$ between the CO$_{2}$ and CO snowlines.
This is because under the high ionisation rate ($\xi_{\mathrm{CR}}(r)=$$1.0\times10^{-17}$ [s$^{-1}$]), CO and CH$_{4}$ gas are converted to less volatile molecules, such as CO$_{2}$, H$_{2}$CO, and hydrocarbons (see Sections \ref{sec:3-2-1rev} and \ref{sec:3-2-2rev}).
Outside the CO snowline ($r>22$ au), both gas-phase C/H and O/H abundances are $\ll10^{-7}$.
The icy-phase C/H and O/H ratios increase with increasing $r$, and reach the similar value as the initial elemental abundance outside the CO snowline.
\\ \\
In the non-shadowed disk ($f=1.0$), the gas-phase N/H ratios are $\sim(5-7)\times10^{-5}$ at $r\lesssim24$ au (within the N$_{2}$ snowline).
Outside the N$_{2}$ snowline ($r>24$ au), the gas-phase N/H abundances are $\ll10^{-7}$.
The icy-phase N/H ratios increase with increasing $r$, and reach the similar value as the initial elemental abundance outside the N$_{2}$ snowline.
\\ \\
In the shadowed disk ($f\leq0.03$), at $r\sim3-8$ au (around the current orbit of Jupiter), the gas-phase C/H and O/H ratios are both $\sim10^{-8}-10^{-6}$.
In addition, at $r\sim3-8$ au (around the current orbit of Jupiter), the gas-phase N/H ratios ($\sim10^{-7}-10^{-5}$) are larger than the gas-phase C/H and O/H ratios in each model, which produce super-stellar N/O ratios of $\gg1$ (see Figures \ref{Figure10rev_CtoO-NtoO_inheritance} and \ref{Figure11rev_CtoO-NtoO_reset}).
The icy-phase C/H, O/H, and N/H ratios reach the similar values as the initial elemental abundances.
%
%%
%%

\begin{comment}
\bibliography{sample631}{}
\bibliographystyle{aasjournal}


%%\listofchanges

\end{comment}

\end{document}